\documentclass{jpp}
\usepackage{graphicx}
\usepackage{epstopdf, epsfig}
\usepackage{graphicx}
\usepackage{color}
\usepackage{amsmath}

\newcommand{\refe}[1]{(\ref{#1})}
\newcommand{\bE}{\boldsymbol{E}}
\newcommand{\bB}{\boldsymbol{B}}
\newcommand{\bv}{\boldsymbol{v}}
\newcommand{\bx}{\boldsymbol{x}}

\newcommand{\bun}{\hat{\boldsymbol{b}}}
\newcommand{\eun}{\hat{\boldsymbol{e}}}
\newcommand{\vbar}{\overline{v}}
\newcommand{\lambdabar}{\overline{\lambda}}
\newcommand{\xibar}{\overline{\xi}}
\newcommand{\rmd}{\mathrm{d}}
\newcommand{\bkappa}{\boldsymbol{\kappa}}

\shorttitle{Finite orbit width effects in large aspect ratio stellarators}
\shortauthor{V. d'Herbemont, F.I. Parra, I. Calvo and J.L. Velasco}

\title{Finite orbit width effects in large aspect ratio stellarators}

\author{Vincent d'Herbemont\aff{1, 2}
  Felix I. Parra\aff{3}
  \corresp{\email{fparradi@pppl.gov}},
  Iv\'an Calvo\aff{4},
  \and Jos\'e Luis Velasco\aff{4}}

\affiliation{\aff{1}Rudolf Peierls Centre for Theoretical Physics, University of Oxford, Oxford OX1 3PU, United Kingdom
\aff{2}Mines ParisTech, 75 272 Paris, France
\aff{3}Princeton Plasma Physics Laboratory, Princeton, NJ 08540, USA
\aff{4}Laboratorio Nacional de Fusi\'on, CIEMAT, 28040 Madrid, Spain}

\begin{document}

\maketitle

\begin{abstract}
New orbit averaged equations for low collisionality neoclassical fluxes in large aspect ratio stellarators with mirror ratios close to unity are derived. The equations retain finite orbit width effects by employing the second adiabatic invariant $J$ as a velocity space coordinate and they have been implemented in the orbit-averaged neoclassical code KNOSOS \citep{velasco20, velasco21}. The equations are used to study the $1/\nu$ regime and the lower collisionality regimes. For generic large aspect ratio stellarators with mirror ratios close to unity, as the collision frequency decreases, the $1/\nu$ regime transitions directly into the $\nu$ regime, without passing through a $\sqrt{\nu}$ regime. An explicit formula for the neoclassical fluxes in the $\nu$ regime is obtained. The formula includes the effect of particles that transition between different types of wells. While these transitions produce stochastic scattering independent of the value of the collision frequency in velocity space, the diffusion in real space remains proportional to the collision frequency. The $\sqrt{\nu}$ regime is only recovered in large aspect ratio stellarators close to omnigeneity: large aspect ratio stellarators with large mirror ratios and optimized large aspect ratio stellarators with mirror ratios close to unity. Neoclassical transport in large aspect ratio stellarators with large mirror ratios can be calculated with the orbit-averaged equations derived by \cite{calvo17}. In these stellarators, the $\sqrt{\nu}$ regime exists in the collisionality interval $\epsilon |\ln \epsilon| \ll \nu_{ii} R  a/\rho_i v_{ti} \ll 1/\epsilon$. In optimized large aspect ratio stellarators with mirror ratios close to unity, the $\sqrt{\nu}$ regime occurs in an interval of collisionality that depends on the deviation from omnigeneity $\delta$: $\delta^2 |\ln \delta | \ll \nu_{ii} R  a/\rho_i v_{ti} \ll 1$. Here, $\nu_{ii}$ is the ion-ion collision frequency, $\rho_i$ and $v_{ti}$ are the ion gyroradius and thermal speed, and $a$ and $R$ are the minor and major radius.
\end{abstract}

\section{Introduction}  \label{sec:intro}
In stellarators, trapped particles can move a significant distance away from their initial flux surface even in the absence of collisions or turbulent fluctuations. Due to these large orbits, stellarator collisional transport at low collision frequencies \citep{kovrizhnykh84} is of the order of or larger than the turbulent transport, dominating energy transport in the core \citep{dinklage13, dinklage18}. 

The width of the trapped particle orbits is of the order of the size of the stellarator unless the stellarator is (i) optimized \citep{calvo17}, i.e. close to omnigeneous \citep{cary97a, cary97b, parra15a}, or (ii) the stellarator has a small inverse aspect ratio $\epsilon := a/R \ll 1$ \citep{ho87}, where $R$ and $a$ are the characteristic values of the major and minor radii of the stellarator, respectively. 

In the case of large aspect ratio stellarators, the ion orbit width is determined by the balance between the component of the $\bE \times \bB$ drift that is tangential to the flux surface, and the component of the ion curvature and $\nabla B$ drifts that is perpendicular to the flux surface -- the large $\bE \times \bB$ drift is mostly parallel to flux surfaces and its small radial component is comparable or smaller than the average radial component of the curvature and $\nabla B$ drifts \citep{calvo17}. 

To estimate the width of the ion orbits, we assume that the electric field is the gradient of an electric potential, and that the potential has a variation of order $T_i/e$ across the minor radius $a$, where $T_i$ is the ion temperature and $e$ is the proton charge. The lowest order value of the radial electric field is set by the need to maintain ambipolarity. The $\bE \times \bB$ drift is of order 
\begin{equation}
\bv_E \sim \rho_{i*} v_{ti}, 
\end{equation}
whereas the curvature and $\nabla B$ drifts are smaller by a factor of $\epsilon$ because they are proportional to the gradient of the magnetic field $\bB$, $|\nabla \bB| \sim B/R$,
\begin{equation}
\bv_{Mi} \sim \epsilon \rho_{i*} v_{ti}.
\end{equation}
Here 
\begin{equation}
\rho_{i*}:= \frac{\rho_i}{a} \ll 1 
\end{equation}
is the normalized ion gyroradius, $\rho_i := v_{ti}/\Omega_i$ is the ion gyroradius, $v_{ti} := \sqrt{2T_i/m_i}$ is the ion thermal speed, $\Omega_i := Z_ieB / m_i c$ is the ion gyrofrequency, $B := |\bB|$ is the magnitude of the magnetic field, $Z_i e$ and $m_i$ are the ion charge and mass, respectively, and $c$ is the speed of light. The radial motion due to the drifts is not secular because it averages out once the $\bE \times \bB$ drift has moved the particle several times around the stellarator. The typical length of the orbit parallel to the flux surfaces is of order $a$, giving a characteristic orbital time of the order of
\begin{equation}
\frac{a}{|\bv_E|} \sim \frac{1}{\rho_{i*}} \frac{a}{v_{ti}}.
\end{equation}
During this time interval, the radial component of the curvature and $\nabla B$ drifts (and, in some cases, the $\bE \times \bB$ drift) leads to an orbit width of order
\begin{equation} \label{eq:wso}
w \sim \frac{a}{|\bv_E|} |\bv_{Mi}| \sim \epsilon a,
\end{equation}
that is, the width of these orbits is smaller than the characteristic size of the stellarator, although it is much larger than the typical width of orbits in tokamaks, of order $\rho_i$.

For small collision frequencies (see equation~\refe{eq:nustarorder} for a precise ordering for the collision frequency), the large width of the trapped particle orbits has called into question the validity of local models for neoclassical transport. Here, `local' refers to models that calculate neoclassical fluxes through a surface of interest using only the electric field, the magnetic field and certain radial gradients of the magnetic field at that flux surface, whereas `global' codes need the electric and magnetic field of the flux surfaces neighboring the flux surface of interest.  The most naive way to obtain a local model is to zero out the radial component of the drifts in certain terms of the drift kinetic equation \citep{sugama16, paul17}, but it has been noted that there are different ways in which this could be done, none of them necessarily consistent \citep{paul17}. Moreover, global neoclassical codes \citep{satake06} have shown that neoclassical fluxes depend on parameters that do not appear in simplified drift kinetic models without radial drifts, such as the magnetic shear \citep{matsuoka15, huang17}. \cite{calvo17} used closeness to omnigeneity to derive local orbit-averaged equations without having to artificially zero out the radial magnetic drifts. In this article, we use another expansion parameter, the inverse aspect ratio, to derive a different set of self-consistent local orbit-averaged equations that is valid for a wide class of stellarators: large aspect ratio stellarators with a mirror ratio close to unity. In our derivation, we do not start by assuming that the distribution function is Maxwellian or that the problem can be solved using a local equation, but we derive these properties from the expansion. The equations presented in this article coincide with the low collisionality limit of the equations in DKES \citep{hirshman86} to lowest order in the small inverse aspect ratio expansion, but differ to higher order. The radial energy flux derived from the new equations in this paper, calculated using a modified version of the code KNOSOS \citep{velasco20, velasco21}, has been shown to be close to the energy flux calculated by DKES in several experimentally relevant configurations \citep{velasco21}. 

There is a subtlety in the derivation of the orbit-averaged equations for large aspect ratio stellarators with mirror ratios close to unity. Given the smallness of the orbit width in $\epsilon$, it is tempting to neglect the radial drifts when calculating the lowest order particle motion. However, in large aspect ratio stellarators with mirror ratios close to unity, the radial displacement of the particles is sufficiently large to affect the trapped-particle motion to lowest order. Indeed, trapped particles in this type of large aspect ratio stellarators have very small parallel velocities of order $\sqrt{\epsilon} v_{ti}$, and small changes in energy of order $\epsilon T_i$ affect their trajectories, causing trapped particles to become passing and vice versa. Radial displacements of order $\epsilon a$ are small compared to the size of the stellarator, but they lead to changes in energy of order $\epsilon T_i$ due to the work done by the radial electric field. Our new equations keep the necessary finite orbit width effects by using the second adiabatic invariant as a velocity space coordinate. 

Our derivation of a local model is valid for collisionalities as small as 
\begin{equation} \label{eq:nustarorder}
\nu_{i*} \sim \rho_{i*} \ll 1, 
\end{equation}
where 
\begin{equation}
\nu_{i*} := \frac{R\nu_{ii}}{v_{ti}} 
\end{equation}
is the collisionality, 
\begin{equation}
\nu_{ii} := \frac{4\sqrt{\pi}}{3} \frac{Z_i^4 e^4 n_i \ln \Lambda}{m_i^{1/2} T_i^{3/2}}
\end{equation}
is the ion-ion collision frequency \citep{braginskii58}, $n_i$ is the ion density, and $\ln \Lambda$ is the Coulomb logarithm. We analyze the behavior of our new equations for large aspect ratio stellarators with mirror ratio close to unity in the limit $\nu_{i*} \gg \rho_{i*}$, in which we recover the $1/\nu$ regime \citep{ho87}, and in the limit $\nu_{i*} \ll \rho_{i*}$. Surprisingly, for $\nu_{i*} \ll \rho_{i*}$, a rigorous expansion of our equations does not lead to the $\sqrt{\nu}$ regime \citep{galeev69} for generic large aspect ratio stellarators with mirror ratio close to unity. Instead, the limit $\nu_{i*} \ll \rho_{i*}$ gives the $\nu$ regime \citep{mynick83}. In this regime, particles follow their collisionless orbits for long times, moving away from their initial flux surface a distance of order $\epsilon a$, as explained above. Particles can only move to a flux surface further away than $\epsilon a$ by having several collisions interrupt their orbits, thus leading to a radial flux that is proportional to the collision frequency.  Importantly, trapped particles remain a distance of order $\epsilon a$ away from their initial flux surface even when they undergo transitions between different types of wells and these transitions stochastize their motion \citep{beidler87}. To treat these transitions between different types of wells, we do not need to introduce in the equations the transition probabilities calculated by \cite{cary86}.

There is a class of stellarators for which the $\sqrt{\nu}$ regime exists for $\nu_{i*} \ll \rho_{i*}$: stellarators close to omnigeneity \citep{calvo17}. In stellarators far from ominigeneity, the transitions between different types of wells of certain trapped particles smear out the $\sqrt{\nu}$ velocity space boundary layer \citep{mynick83}. We show that large aspect ratio stellarators with large mirror ratios are close to omnigeneity and hence neoclassical transport in them can be calculated using the equations derived by \cite{calvo17}. We also consider large aspect ratio stellarators with mirror ratios close to unity that are close to omnigeneity, finding results that are consistent with our previous work on this area \citep{calvo18a}.

Throughout the paper we focus on ion transport. In section~\ref{sec:dkequation} we remind the reader of the kinetic equations for a general stellarator in the limit $\nu_{i*} \sim \rho_{i*} \ll 1$. In section~\ref{sec:MHDepsilon} we discuss the MHD equilibrium equations for $\epsilon \ll 1$, making a distinction between large aspect ratio stellarators with mirror ratios close to unity and large aspect ratio stellarators with large mirror ratios. Most of the rest of the paper is dedicated to large aspect ratio stellarators with mirror ratios close to unity, with the only exception being section~\ref{sub:largemirror}. In section~\ref{sec:dkequationJ} we propose a new set of velocity space coordinates that are necessary to simplify the expansion in $\epsilon \ll 1$ for large aspect ratio stellarators with mirror ratios close to unity, and in section~\ref{sec:fiepsilon} we finally perform the expansion in $\epsilon$ for the ion distribution function and the electric potential in this type of large aspect ratio stellarators. In sections~\ref{sec:1nuregime} and \ref{sec:nuregime} we study the cases $\nu_{i*} \gg \rho_{i*}$ and $\nu_{i*} \ll \rho_{i*}$ for large aspect ratio stellarators with mirror ratios close to unity. In section~\ref{sec:omnigeneity} we consider large aspect ratio stellarators close to omnigeneity. We divide our discussion of large aspect ratio stellarators close to omnigeneity into two parts: in one we show that large aspect ratio stellarator with large mirror ratios are close to omnigeneity, and in the other, we study optimized large aspect ratio stellarators with mirror ratios close to unity. We conclude in section~\ref{sec:conclusion}.

\section{Drift kinetic equation for ions in a generic stellarator} \label{sec:dkequation}

We assume $a \sim R$ in this section, but we keep the distinction between $a$ and $R$ in our estimates in preparation for the expansion in small inverse aspect ratio in sections~\ref{sec:MHDepsilon}, \ref{sec:dkequationJ}, \ref{sec:fiepsilon} and \ref{sec:omnigeneity}.  

We assume that the magnetic field $\bB$ is constant in time and hence the electric field $\bE$ satisfies $\bE(\bx, t) = - \nabla \phi (\bx, t)$. The electric potential $\phi$ is of order $T_i/e$, has a characteristic lengthscale of the order of the minor radius $a$, and is determined by the quasineutrality equation
\begin{equation}
\label{quasineutrality}
Z_i \int f_i (\bx, \bv, t) \, \mathrm{d}^3 v = \int f_{e} (\bx, \bv, t) \, \mathrm{d}^3 v,
\end{equation}
where $f_i (\bx, \bv, t)$ and $f_e(\bx, \bv, t)$ are the distribution functions of ions and electrons, $\bx$ and $\bv$ are the particle's Cartesian position and velocity, and $t$ is time. Throughout the paper, we assume that the electrons can be modeled with the modified Maxwell-Boltzmann response
\begin{equation} \label{eq:maxwboltz}
\int f_e (\bx, \bv, t) \, \rmd^3 v = \hat{n}_e(r(\bx), t)\exp\Bigg(\frac{e\phi(\bx, t)}{T_e (r(\bx), t)}\Bigg),
\end{equation}
where $T_e (r, t)$ is the temperature of the electrons and $\hat{n}_e(r, t)$ is the density of electrons in the absence of $\phi$. Note that $\hat{n}_e$ and $T_e$ are flux functions that only depend on the flux surface label $r(\bx)$. We define $r$ more carefully below.

We use the drift kinetic equation \citep{hazeltine73} to obtain the ion distribution function. To describe velocity space, we choose the coordinates $\{ \mathcal{E}, \mu, \sigma, \varphi \}$, where $\mathcal{E} :=v^2/2 +Z_i e \phi /m_i$ is the total energy per unit mass, $\mu:=v_\perp^2 /2B$ is the magnetic moment, $\sigma$ is the sign of the parallel velocity and $\varphi$ is the gyrophase, defined such that
\begin{equation} \label{eq:eperp}
\frac{\bv_\perp}{v_\perp} = \eun_\perp (\bx, \varphi) := \cos \varphi\, \eun_1 (\bx) + \sin \varphi \, \eun_2 (\bx).
\end{equation}
Here $\bv_\perp$ is the component of $\bv$ perpendicular to the magnetic field, $v_\perp := |\bv_\perp|$, and $\eun_1(\bx)$ and $\eun_2(\bx)$ are two unit vectors that form an orthonormal basis with the unit vector parallel to the magnetic field $\bun(\bx) = \bB/B$ and satisfy $\eun_1 \times \eun_2 = \bun$. In the coordinates $\{ \mathcal{E}, \mu, \sigma, \varphi \}$, the velocity space volume element is
\begin{equation}
\rmd^3 v = \frac{\rmd \mathcal{E}\, \rmd \mu\, \rmd \varphi}{|(\nabla_v \mathcal{E} \times \nabla_v \mu) \cdot \nabla_v \varphi|} = \frac{B}{|v_\||} \, \rmd \mathcal{E}\, \rmd \mu\, \rmd \varphi,
\end{equation}
where 
\begin{equation} \label{eq:vpardef}
v_\| (\bx, \mathcal{E}, \mu, \sigma) := \sigma \sqrt{2 \left ( \mathcal{E} - \mu B(\bx) - \frac{Z_i e \phi (\bx)}{m_i} \right )}
\end{equation}
is the parallel velocity.

The distribution function can be split into its gyroaverage, $\overline{f}_i := (2\pi)^{-1} \int_0^{2\pi} f_i\, \rmd \varphi$, and the gyrophase-dependent piece $\tilde{f}_i := f_i - \overline{f}_i$. The gyrophase-dependent piece is of order $\rho_{i*} \overline{f}_i$ and can be neglected in the limit $\rho_{i*} \sim \nu_{i*}$ because the fluxes of particles and energy depend to lowest order on the much larger gyroaveraged piece $\overline{f}_i$, unlike the fluxes in the neoclassical regimes of moderate collisionality ($\nu_{i*} \sim 1$) in which the piece of the gyoraveraged distribution function that matters for transport is small in $\rho_{i*}$ (see the discussion at the end of this section for more details). Since the ion-electron collisions can be neglected within an expansion in the electron-to-ion mass ratio, the equation for $\overline{f}_i$ is
\begin{equation}
\label{dk} 
 \partial_t \overline{f}_i + \dot{\bx} \cdot \nabla \overline{f}_i + \dot{\mathcal{E}}\, \partial_\mathcal{E} \overline{f}_i + \dot{\mu}\, \partial_\mu \overline{f}_i= C_{ii}[\overline{f}_i, \overline{f}_i] (1 + O(\rho_{i*}^2)) +  \overline{S}_i (1 + O(\rho_{i*} )) .
\end{equation}
Here $S_i$ is a source term representing fueling and heating, and $\overline{S}_i$ is the gyroaverage of $S_i$. The particle motion can be split into parallel motion and perpendicular drifts,
\begin{equation} \label{eq:xdot}
\dot{\bx} := \left ( v_\parallel + \frac{m_i c \mu}{Z_i e} \bun \cdot \nabla \times \bun \right ) \bun + \mathbf{v}_{Mi} + \mathbf{v}_E + O(\rho_{i*}^2 v_{ti}),
\end{equation}
where
\begin{equation}
\mathbf{v}_{Mi} := \frac{1}{\Omega_i} \bun \times (v_\parallel^2 \, \boldsymbol{\kappa} + \mu \nabla B)
\end{equation}
are the curvature and $\nabla B$ drifts, collectively known as magnetic drift, $\boldsymbol{\kappa} := \bun \cdot \nabla \bun$ is the curvature of the magnetic field lines, and
\begin{equation}
\mathbf{v}_{E}:=\frac{c}{B} \bun \times \nabla \phi
\end{equation}
is the $\bE\times \bB$ drift. The time derivative of the total energy is
\begin{equation} \label{eq:Edot}
\dot{\mathcal{E}} := \frac{Z_i e}{m_i} \partial_t \phi + O \left ( \rho_{i*}^2 \frac{v_{ti}^3}{a} \right ).
\end{equation}
The time derivative of the magnetic moment is
\begin{equation} \label{eq:mudot}
\dot{\mu} := v_\| \bun \cdot \nabla \left ( \frac{v_\| \mu}{\Omega_i} \bun \cdot \nabla \times \bun \right ) + O \left ( \rho_{i\ast}^2 \frac{v_{ti}^3}{a B} \right ).
\end{equation}
The ion-ion collision operator $C_{ii}$ is a Fokker-Planck collision operator, 
\begin{equation} \label{eq:CiiRosen}
C_{ii} [ f_a, f_b ] := \gamma_{ii} \nabla_v \cdot \left ( \nabla_v \nabla_v H [ f_b] \cdot \nabla_v f_a - f_a \nabla_v L [f_b] \right ),
\end{equation}
where $\gamma_{ii} := 2\pi Z_i^4 e^4 \ln \Lambda/m_i^2$, and the Rosenbluth potentials $H$ and $L$ \citep{rosenbluth57} are the functionals
\begin{equation}
H[f](\bv) := \int f(\bv^\prime) |\bv - \bv^\prime|\, \rmd^3 v^\prime
\end{equation}
and
\begin{equation}
L[f] (\bv) := 2 \int \frac{f(\bv^\prime)}{|\bv - \bv^\prime|}\, \rmd^3 v^\prime.
\end{equation}
Note that, in our notation, the first argument of $C_{ii}$ refers to the distribution function that is evaluated at the velocity $\bv$ of interest, whereas the second argument refers to the distribution function that is integrated to obtain the Rosenbluth potentials. In the coordinates $\{ \mathcal{E}, \mu, \sigma, \varphi \}$, the Fokker-Planck collision operator is given by
\begin{align} \label{eq:CiiEmu}
 C_{ii}[f_a , f_b] = \gamma_{ii} |v_\| | \, \partial_\mathcal{E} \left [ \frac{1}{|v_\| |} \left ( H_{\mathcal{E} \mathcal{E}} [ f_b ] \,  \partial_\mathcal{E} f_a + H_{\mathcal{E} \mu} [ f_b ] \, \partial_\mu f_a - L_\mathcal{E} [ f_b ] \, f_a \right ) \right ] \nonumber\\ + \gamma_{ii} |v_\| | \, \partial_\mu \left [ \frac{1}{|v_\| |} \left ( H_{\mu \mathcal{E}} [ f_b ] \,  \partial_\mathcal{E} f_a + H_{\mu \mu} [ f_b ] \, \partial_\mu f_a - L_\mu [ f_b ] \, f_a \right ) \right ],
\end{align}
where we have used the fact that $f_a$ does not depend on the gyrophase $\varphi$, and we have defined $H_{pq} [f] := \nabla_v p \cdot \nabla_v \nabla_v H [f] \cdot \nabla_v q$ and $L_p [ f] := \nabla_v p \cdot \nabla_v L [f]$, with $p = \mathcal{E}, \mu$ and $q = \mathcal{E}, \mu$. Note that, in equations~\refe{eq:xdot}, \refe{eq:Edot}, \refe{eq:mudot} and in the right side of equation~\refe{dk}, we have indicated the size of terms associated to the gyrophase-dependent piece of the distribution function $\tilde{f}_i \sim \rho_{i\ast} \overline{f}_i$ that we have neglected -- the errors in the collision operator $C_{ii}$ are smaller than expected due to its gyrotropy.

Instead of the Cartesian coordinates $\bx$, it is convenient to use spatial coordinates that conform to the shape of the magnetic field. From here on, we use $\{ r, \alpha, l \}$, where $r$ is a flux surface label with units of length and of the order of the minor radius $a$, $\alpha$ is a poloidal angle that labels magnetic field lines on a given flux surface, and $l$ is the arc-length of the magnetic field line and is used to determine the position along the field line. In these coordinates, the magnetic field can be written as 
\begin{equation} \label{eq:Brhoalpha}
\mathbf{B} =\Psi_t'(r) \nabla r \times \nabla \alpha,
\end{equation}
where $\Psi_t(r)$ is the toroidal magnetic flux within the flux surface $r$ divided by $2\pi$, and $\Psi_t^\prime := \rmd \Psi_t/\rmd r$. In these coordinates, the unit vector $\bun$ is given by
\begin{equation}
\bun = \partial_l \bx,
\end{equation}
and the element of volume is
\begin{equation} \label{eq:d3x}
\rmd^3 x = \frac{\rmd r\, \rmd \alpha\, \rmd l}{|(\nabla r \times \nabla \alpha) \cdot \nabla l |} = \frac{\Psi_t^\prime}{B}\, \rmd r\, \rmd \alpha\, \rmd l.
\end{equation} 

\begin{figure}
\begin{center}
\includegraphics[width=8cm]{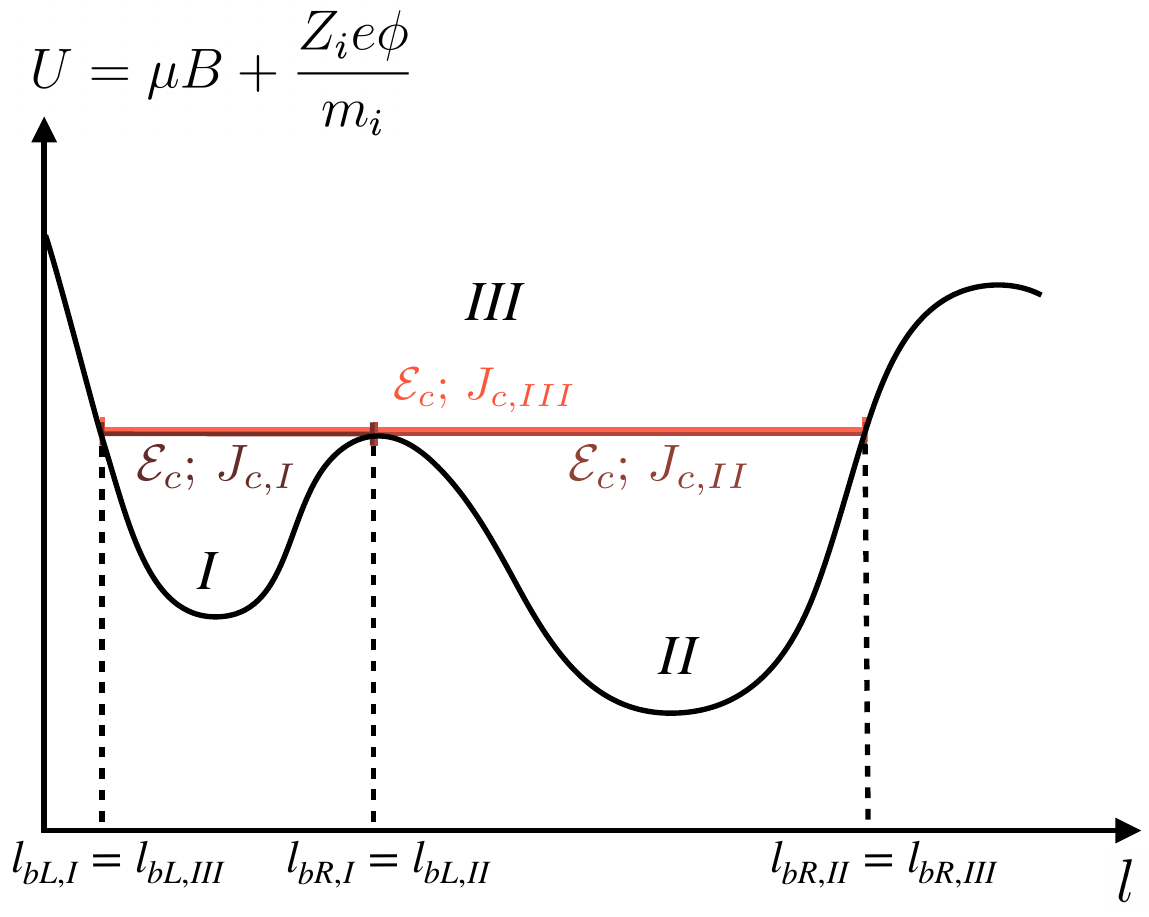}
\end{center}
\caption{\label{fig:exampleU} Sketch of the effective potential $U := \mu B + Z_i e \phi/m_i$ as a function of $l$. }
\end{figure}

The equations for general stellarators with $\nu_{i*} \sim \rho_{i*}$ were derived in section 3.1 of \cite{calvo17}. Here, we generalize the work done in \cite{calvo17}, and change the presentation in places to make the derivation of the large aspect ratio stellarator equations easier. We remind the reader that in this section we assume $\epsilon \sim 1$, but that we will perform a subsidiary expansion in $\epsilon \ll 1$ in the rest of the paper. We expand the drift kinetic system of equations in $\rho_{i*}\ll 1$ assuming $\nu_{i*} \sim \rho_{i*}$ and 
\begin{equation} \label{eq:dtSieps1}
\partial_t \sim \frac{S_i}{f_i} \sim \nu_{ii} \sim \rho_{i\ast} \frac{v_{ti}}{R}. 
\end{equation}
Our assumption for the size of the time derivative and the source $S_i$ might be surprising, but it is justified by the fact that the final equation is consistent. Physically, these estimates are the result of the particle orbits being comparable to the size of the device. Thus, the time derivative and the source must compensate for both direct particle losses ($\partial_t \sim S_i/f_i \sim \rho_{i\ast} v_{ti}/R$) and, for particles in confined orbits, losses due to collisions ($\partial_t \sim S_i/f_i \sim \nu_{ii}$). With these assumptions, we can write $\overline{f}_i$ as
\begin{equation}
\label{dk1}
\overline{f}_{i} = \overline{f}_{i}^{(0)} + \overline{f}_{i}^{(1)} + \dots,
\end{equation}
with $\overline{f}_{i}^{(n)}\sim \rho_{i*}^n \overline{f}_{i}^{(0)}$. To lowest order in $\rho_{i*}$, equation~\refe{dk} gives
\begin{equation}
\label{rho1}
v_{\parallel}\, \partial_{l} \overline{f}_{i}^{(0)} =0.
\end{equation}
To solve this equation, we need to distinguish between passing and trapped particles. The function 
\begin{equation}
U(r, \alpha, l, \mu, t) := \mu B (r, \alpha, l) + \frac{Z_i e \phi  (r, \alpha, l, t)}{m_i} 
\end{equation}
is an effective potential for the motion parallel to the magnetic field line. If $\mathcal{E}$ is larger than the maximum of $U$ on a flux surface, $U_M (r, \mu, t)$, the parallel velocity in \refe{eq:vpardef} never vanishes and the particle is a passing particle. If $\mathcal{E}$ is smaller than $U_M (r, \mu, t)$, the parallel velocity vanishes at least at two bounce points, $l_{bL,W} (r, \alpha, \mathcal{E}, \mu, t)$ and $l_{bR,W}(r, \alpha, \mathcal{E}, \mu, t)$, defined by $\mathcal{E} - U(r, \alpha, l_{bL,W}, \mu, t) = 0 = \mathcal{E} - U(r, \alpha, l_{bR,W}, \mu, t)$ (the subscripts $L$ and $R$ refer to `left' and `right', respectively; see figure~\ref{fig:exampleU}). Note that for given values of $\mathcal{E}$ and $\mu$, trapped particle can be located inside several different $U$ wells. We will use the discrete index $W$ to distinguish between these wells, where $W$ takes Roman numeral values (see figure~\ref{fig:exampleU}).

On an ergodic flux surface where a single field line connects any two points, equation~\refe{rho1} implies that $\overline{f}_{i}^{(0)}$ must be independent of $\alpha$ for passing particles. For trapped particles, due to continuity at the bounce points $l_{bL,W}$ and $l_{bR,W}$ and equation~\refe{rho1}, $\overline{f}_i^{(0)}$ cannot depend on $\sigma$. Using these conditions, we write $\overline{f}_i^{(0)}$ as
\begin{equation}
\label{trapped-passing}
\overline{f}_{i}^{(0)} = \left \{ 
\begin{array}{l l}
g_{i, W}(r, \alpha, \mathcal{E}, \mu, t) &\mathrm{for}\,\, \mathcal{E} \leq U_M(r, \mu, t),  \\
h_{i} (r, \mathcal{E}, \mu, \sigma, t) & \mathrm{for}\,\, \mathcal{E} > U_M(r, \mu, t),
\end{array}
\right. 
\end{equation}
where $g_{i,W}$ is defined only in the trapped particle region, $\mathcal{E} \leq U_M(r, \mu, t)$, and $h_i$ is defined only in the passing particle region, $\mathcal{E} > U_M(r, \mu, t)$. In most of this article, we will consider stellarators in which the effective potential $U$ only reaches the maximum value $U_M(r, \mu, t)$ at a finite number of points on the flux surface $r$ (the exception is section~\ref{sec:omnigeneity}, where we discuss a case with contours $U = U_M$ that are lines that wrap around the flux surface: the omnigeneous stellarator). In an ergodic flux surface where the value $U_M$ is only reached at a finite number of points, there is one barely-trapped particle orbit with $\mathcal{E} = U_M$ that covers the entire flux surface between its two bounce points (except for possibly a subset of points that has no area, i.e. a segment of magnetic field line of finite length might connect two of the maxima of $U$, but not all of them). If a surface-covering barely-trapped particle orbit did not exist, we would be able to join all the points with $U = U_M$ with a single magnetic field line that closes on itself, contradicting the initial assumption that the surface is ergodic. We denote the well index of this surface-covering barely-trapped particle as $W_\mathrm{bt}$. For $W = W_\mathrm{bt}$ and $\mathcal{E} = U_M$, $g_{i,W}$ does not depend on $\alpha$, and we can impose the boundary conditions
\begin{equation} \label{eq:condJinftyv1}
g_{i,W_\mathrm{bt}} (r, \alpha, U_M(r, \mu, t), \mu, t) = h_i (r, U_M(r, \mu, t), \mu, \sigma, t)
\end{equation}
and
\begin{equation} \label{eq:diffcondJinftyv1}
\partial_\mathcal{E} g_{i,W_\mathrm{bt}} (r, \alpha, U_M(r, \mu, t), \mu, t) = \partial_\mathcal{E} h_i (r, U_M(r, \mu, t), \mu, \sigma, t).
\end{equation}
These conditions imply that $h_i$ cannot depend on $\sigma$ at $\mathcal{E} = U_M$.

To next order in $\rho_{i*}$, equation~\refe{dk} gives
\begin{align}
\label{rho2}
v_\parallel\, & \partial_l \left ( \overline{f}_{i}^{(1)} + \frac{v_\| \mu}{\Omega_i} \bun \cdot \nabla \times \bun \, \partial_\mu \overline{f}_i^{(0)} \right ) + \partial_t \overline{f}_i^{(0)} + \frac{Z_i e}{m_i} \partial_t \phi \, \partial_\mathcal{E} \overline{f}_i^{(0)} \nonumber\\  &+ (\mathbf{v}_{E}+\mathbf{v}_{Mi}) \cdot (\nabla \alpha\, \partial_{\alpha} \overline{f}_{i}^{(0)} + \nabla r\, \partial_{r} \overline{f}_{i}^{(0)})=C_{ii}[ \overline{f}_{i}^{(0)}, \overline{f}_{i}^{(0)}] + \overline{S}_i.
\end{align}
We proceed to eliminate $\overline{f}_i^{(1)}$ from the equation. For trapped particles, we divide equation~\refe{rho2} by $|v_\parallel |$, sum over the two possible values of $\sigma$ and integrate over $l$ between bounce points to obtain
\begin{align}
\label{rho2orbit}
\partial_t g_{i,W} & + \frac{Z_i e}{m_i} \left \langle \partial_t \phi \right \rangle_{\tau, W} \, \partial_\mathcal{E} g_{i,W} + \langle (\mathbf{v}_{E}+\mathbf{v}_{Mi}) \cdot \nabla \alpha \rangle_{\tau, W}\, \partial_{\alpha} g_{i,W} \nonumber\\ & + \langle (\mathbf{v}_{E}+\mathbf{v}_{Mi}) \cdot \nabla r  \rangle_{\tau, W}\, \partial_{r} g_{i,W} = \langle C_{ii}[g_{i,W} ,\overline{f}_{i}^{(0)}] \rangle_{\tau, W} + \langle \overline{S}_i \rangle_{\tau, W},
\end{align}
where we have used the transit average
\begin{equation} \label{eq:transitave}
\langle \ldots \rangle_{\tau, W}:= \frac{1}{\tau_W} \sum_{\sigma} \int_{l_{bL,W}}^{l_{bR,W}}\frac{(\ldots)}{|v_{\parallel}|}\,\mathrm{d}l,
\end{equation}
and 
\begin{equation}
\label{tau}
\tau_W :=2\int^{l_{bR,W}}_{l_{bL,W}}\frac{\mathrm{d}l}{|v_{\parallel}|},
\end{equation}
is the period of a trapped particle orbit. For passing particles, we divide equation~\refe{rho2} by $|v_\||$ and we integrate over $l$ and $\alpha$ to find
\begin{equation}
\label{passingfi0}
 \left \langle \frac{B}{|v_{\parallel}|} \right \rangle_\mathrm{fs} \partial_t h_i + \frac{Z_i e}{m_i} \left \langle \frac{B}{|v_{\parallel}|} \partial_t \phi \right \rangle_\mathrm{fs} \, \partial_\mathcal{E} h_i + \left \langle \frac{B}{|v_{\parallel}|} C_{ii}[h_i, \overline{f}_{i}^{(0)}] \right \rangle_\mathrm{fs} = \left \langle \frac{B}{|v_{\parallel}|} \overline{S}_i \right \rangle_\mathrm{fs},
\end{equation}
where we have defined the flux surface average
\begin{equation} \label{eq:fsavedef}
\langle \ldots \rangle_\mathrm{fs} := \frac{1}{V^\prime} \int_0^{2\pi} \rmd \alpha \int_0^{L(r, \alpha)} \rmd l \, \frac{\Psi_t^\prime}{B} (\ldots).
\end{equation}
Here, $L(r,\alpha)$ is the length along the magnetic field line between the two points where the magnetic field line crosses the curve defined by $l=0$, and 
\begin{equation}
V^\prime(r) := \int_0^{2\pi} \rmd \alpha \int_0^{L(r, \alpha)} \rmd l \, \frac{\Psi_t^\prime}{B} 
\end{equation}
is the derivative with respect to $r$ of the volume $V(r)$ contained within the flux surface $r$. To obtain equation~\refe{passingfi0}, we have written the radial component of the drifts as $(\bv_E + \bv_{Mi}) \cdot \nabla r = (v_\|/\Omega_i) \nabla \cdot ( v_\| \bun \times \nabla r)$ to find
\begin{equation} \label{eq:avepassingdrift}
\left \langle \frac{B}{|v_{\parallel}|} (\bv_E + \bv_{Mi}) \cdot \nabla r \right \rangle_\mathrm{fs} = 0.
\end{equation}

Equation~\refe{rho2orbit} cannot be used at values of $\mathcal{E}$ that are junctures of three or more types of wells. We show an example of such a value of $\mathcal{E}$ in figure~\ref{fig:exampleU}.  At these junctures, particles can and, in most cases, will transition from one type of well to another. In general, the value of $\mathcal{E}$ at which several types of well coincide, $\mathcal{E} = \mathcal{E}_c (r, \alpha, \mu, t)$,  depends on $r$, $\alpha$, $\mu$ and $t$. Around these values of $\mathcal{E}$, there are boundary layers, thin in $\mathcal{E}$, where the dependence of $\overline{f}_i$ on $l$ cannot be neglected (see, for example, \cite{nemov99, calvo14}). These boundary layers impose continuity in $g_{i,W}$ across these junctures. The derivatives of $g_{i,W}$ with respect $\mathcal{E}$ and $\mu$ are not necessarily continuous. The derivatives of $g_{i, W}$ with respect to $\mathcal{E}$ and $\mu$ on the different wells of a juncture are related to each other by two conditions: the combination $\partial_\mu \mathcal{E}_c \, \partial_\mathcal{E} g_{i,W} + \partial_\mu g_{i,W}$ is continuous across a juncture, and the collisional flux in velocity space across a juncture must be conserved. The combination $\partial_\mu \mathcal{E}_c \, \partial_\mathcal{E} g_{i,W} + \partial_\mu g_{i,W}$ is continuous at $\mathcal{E} = \mathcal{E}_c (r, \alpha, \mu, t)$ because $g_{i,W}$ is continuous at $\mathcal{E} = \mathcal{E}_c (r, \alpha, \mu, t)$. In the example of figure~\ref{fig:exampleU}, continuity of $g_{i, W}$ at $\mathcal{E} = \mathcal{E}_c (r, \alpha, \mu, t)$ imposes
\begin{equation} \label{eq:contjunct}
g_{i, I} (r, \alpha, \mathcal{E}_c (r, \alpha, \mu, t), \mu, t) = g_{i, II} (r, \alpha, \mathcal{E}_c (r, \alpha, \mu, t), \mu, t) = g_{i, III} (r, \alpha, \mathcal{E}_c (r, \alpha, \mu, t), \mu, t).
\end{equation}
for all $\mu$. Differentiating this expression with respect to $\mu$, we find 
\begin{align}
\partial_\mu \mathcal{E}_c & (r, \alpha, \mu, t) \, \partial_\mathcal{E} g_{i, I} (r, \alpha, \mathcal{E}_c (r, \alpha, \mu, t), \mu, t) + \partial_\mu g_{i, I} (r, \alpha, \mathcal{E}_c (r, \alpha, \mu, t), \mu, t) \nonumber \\ & = \partial_\mu \mathcal{E}_c (r, \alpha, \mu, t)\, \partial_\mathcal{E} g_{i, II} (r, \alpha, \mathcal{E}_c (r, \alpha, \mu, t), \mu, t) + \partial_\mu g_{i, II} (r, \alpha, \mathcal{E}_c (r, \alpha, \mu, t), \mu, t) \nonumber \\ & = \partial_\mu \mathcal{E}_c (r, \alpha, \mu, t)\, \partial_\mathcal{E} g_{i, III} (r, \alpha, \mathcal{E}_c (r, \alpha, \mu, t), \mu, t) + \partial_\mu g_{i, III} (r, \alpha, \mathcal{E}_c (r, \alpha, \mu, t), \mu, t),
\end{align}
that is, the combination $\partial_\mu \mathcal{E}_c\, \partial_\mathcal{E} g_{i, W} + \partial_\mu g_{i, W}$ is continuous. The other condition for the derivatives of $g_{i, W}$ at $\mathcal{E} = \mathcal{E}_c (r, \alpha, \mu, t)$ is conservation of particle number in phase space. For example, for the case represented in figure~\ref{fig:exampleU}, one needs to calculate the particles that are leaving wells $I$ and $II$ due to collisions, and then enforce that they enter well $III$. This velocity space flux continuity condition is manipulated in Appendix~\ref{app:junctures} to give the following relation between the derivatives of $g_{i, W}$ with respect of $\mathcal{E}$ on different sides of the juncture:
\begin{align}
\label{rho2transition}
 \tau_I & \left [ \langle H_{\mathcal{E} \mathcal{E}}[ \overline{f}_i^{(0)}]\rangle_{\tau, I} - 2  \langle H_{\mathcal{E} \mu}[ \overline{f}_i^{(0)}]\rangle_{\tau, I} \, \partial_\mu \mathcal{E}_c +  \langle H_{\mu \mu}[ \overline{f}_i^{(0)}]\rangle_{\tau, I} \, (\partial_\mu \mathcal{E}_c)^2 \right ] \partial_\mathcal{E} g_{i, I} \nonumber\\ & + \tau_{II} \left [  \langle H_{\mathcal{E} \mathcal{E}}[ \overline{f}_i^{(0)}]\rangle_{\tau, II} - 2  \langle H_{\mathcal{E} \mu}[ \overline{f}_i^{(0)}]\rangle_{\tau, II} \, \partial_\mu \mathcal{E}_c+  \langle H_{\mu \mu}[ \overline{f}_i^{(0)}] \rangle_{\tau, II} \, (\partial_\mu \mathcal{E}_c)^2 \right ] \partial_\mathcal{E} g_{i, II} \nonumber\\ = & \tau_{III} \left [  \langle H_{\mathcal{E} \mathcal{E}}[ \overline{f}_i^{(0)}]\rangle_{\tau, III} - 2 \langle H_{\mathcal{E} \mu}[ \overline{f}_i^{(0)}]\rangle_{\tau, III} \, \partial_\mu \mathcal{E}_c + \langle H_{\mu \mu} [\overline{f}_i^{(0)}] \rangle_{\tau, III} \, (\partial_\mu \mathcal{E}_c)^2 \right ] \partial_\mathcal{E} g_{i, III}.
\end{align}
The relation between the derivatives $\partial_\mu g_{i,W}$ on each side of the juncture can be obtained from equation~\refe{rho2transition} by using the fact that the combination $\partial_\mu \mathcal{E}_c \, \partial_\mathcal{E} g_{i,W} + \partial_\mu g_{i,W}$ is continuous.

Equations~\refe{quasineutrality}, \refe{eq:maxwboltz}, \refe{rho2orbit}, \refe{passingfi0} and \refe{rho2transition} are the same as equations (31), (33) and (37) of \cite{calvo17} but for the inclusion of sources and time derivatives, and a different treatment of the split of $\overline{f}_i^{(0)}$ between trapped and passing particles. These equations are radially non-local and lead to very large transport and to a non-Maxwellian distribution function. In \cite{calvo17}, closeness to omnigeneity was employed to derive radially local equations for a near-Maxwellian distribution function, but here we will use an expansion in the small inverse aspect ratio $\epsilon$.

Equations~\refe{quasineutrality}, \refe{eq:maxwboltz}, \refe{rho2orbit}, \refe{passingfi0} and \refe{rho2transition} are noticeably different from the usual neoclassical equations \citep{hinton76}, derived assuming $\nu_{i*} \sim 1$. In the limit $\nu_{i\ast} \sim 1$, to lowest order in $\rho_{i\ast}$, the gyroaveraged ion distribution is a stationary Maxwellian with density $n_i$ and temperature $T_i$ that only depend on the flux label $r$, $\overline{f}_i^{(0)} = f_{Mi}$, and the electrostatic potential is a flux function to lowest order in $\rho_\ast$, $\phi (r, \alpha, l, t) = \phi^{(0)}(r, t) + \phi^{(1)} (r, \alpha, l, t) + \ldots$ The next order corrections in $\rho_{i*}$, $\overline{f}_i^{(1)}$ and $\phi^{(1)}$, are determined by 
\begin{equation} \label{eq:typicalneo}
v_\| \, \partial_l \overline{f}_i^{(1)} - C_{ii}^{\ell} [ \overline{f}_i^{(1)}] = - \bv_{Mi} \cdot \nabla r \, \partial_r f_{Mi}
\end{equation}
and the quasineutrality equation, respectively. Here, $C_{ii}^{\ell}$ is the linearized collision operator, discussed further in section~\ref{sub:gi1}. The density and temperature in the Maxwellian $f_{Mi}$ are calculated using particle and energy conservation equations. The neoclassical fluxes in these conservation equations are integrals of $\overline{f}_i^{(1)}\bv_{Mi} \cdot \nabla r$, and hence scale as $\rho_{i\ast}^2$. These neoclassical fluxes give a typical time scale for changes in density and temperature of $\partial_t \sim \rho_{i\ast}^2 \nu_{ii}$. For comparison, in a generic stellarator with $\nu_{i\ast} \sim \rho_{i\ast}$, equations~\refe{quasineutrality}, \refe{eq:maxwboltz}, \refe{rho2orbit}, \refe{passingfi0} and \refe{rho2transition} show that the distribution function need not be close to a Maxwellian, and the typical time scale for transport is $\partial_t \sim \nu_{ii} \sim \rho_{i\ast} v_{ti}/R$. The difference between the orderings $\nu_{i\ast} \sim 1$ and $\nu_{i\ast} \sim \rho_{i\ast}$ is due to the typical radial separation between the position of a particle and the flux surface $r$ that the particle started at. For $\nu_{i\ast} \sim 1$, particles collide often, and hence the radial drift does not have time to act and move the particle away from its initial radial position more than a distance of the order of the ion gyroradius $\rho_i$ between collisions. For smaller collision frequencies ($\rho_{i*} \ll \nu_{i*} \ll 1$), particles in a stellarator drift out distances of order $\rho_i/\nu_{i\ast}$ \citep{ho87}, giving higher and higher transport as the collision frequency decreases until eventually, for $\nu_{i\ast} \sim \rho_{i\ast}$, the separation between the initial radial position of the particle and its typical position becomes of the order of the minor radius of the device, $a$. In this regime, orbits are as large as the device, and transport occurs by either direct losses, giving the typical timescale $\partial_t \sim \rho_{i\ast} v_{ti}/R$, or, for particles in confined orbits, by collisions, giving $\partial_t \sim \nu_{ii}$. By expanding in closeness to omnigeneity \citep{calvo17} or in the small inverse aspect ratio (this article), one can recover that the distribution function is close to a Maxwellian, and that the radial flux of particles and energy is determined by a higher order correction to that Maxwellian. The equations for these higher corrections in general look similar to equation~\refe{eq:typicalneo}, but the parallel streaming term is replaced by the drift in the $\alpha$-direction. The correction to the Maxwellian in this case does not scale with $\rho_{i\ast}$ as the expansion parameter is not $\rho_{i\ast}$ but closeness to omnigeneity or the inverse aspect ratio. One can devise equations for the correction to the Maxwellian that recover both orderings $\nu_{i\ast} \sim 1$ and $\nu_{i\ast} \sim \rho_{i\ast}$ by including both parallel streaming and drifts in the $\alpha$-direction -- we show in Appendix~\ref{app:DKES} that the equations in DKES are an example of this, recovering both the $\nu_{i\ast} \sim 1$ and $\nu_{i\ast} \sim \rho_{i\ast}$ limits for large aspect ratio stellarators.

\section{MHD equilibria in large aspect ratio stellarators} \label{sec:MHDepsilon}

In the coordinates $\{ r, \alpha, l \}$, a large aspect ratio stellarator shape is
\begin{equation} \label{eq:expansionr}
\bx (r, \alpha, l) = \bx_0(l) + \bx_1(r, \alpha, l) + \bx_2(r, \alpha, l) + \ldots,
\end{equation}
where $\bx_0(l) \sim R$ is the magnetic axis, and $\bx_n (r, \alpha, l) \sim \epsilon^n R$. We assume that 
\begin{equation}
\partial_r \sim \frac{1}{a}, \quad \partial_\alpha \sim 1, \quad \partial_l \sim \frac{1}{R}.
\end{equation}
Note that the expansion in \refe{eq:expansionr} is not the \cite{garren91} polynomial expansion because we are not assuming that $\bx_n(r, \alpha, l)$ is proportional to $r^n$. The \cite{garren91}  expansion is a particular case of the expansion used here.

The values that $\bx(r, \alpha, l)$ can take are constrained by the definition of arc length,
\begin{equation} \label{eq:arc}
\left | \partial_l \bx \right |^2 = 1,
\end{equation}
and by the MHD force balance equation,
\begin{equation} \label{eq:MHDeq}
\nabla_\perp \left ( P + \frac{B^2}{8\pi} \right ) = \frac{B^2}{4\pi} \bun \cdot \nabla \bun,
\end{equation}
where $\nabla_\perp := \nabla - \bun \bun \cdot \nabla$ is the projection of the gradient in the plane perpendicular to the magnetic field, and $P(r)$ is the total plasma pressure, which is a flux function. We project equation~\refe{eq:MHDeq} on $\partial_r \bx$ and $\partial_\alpha \bx$ to obtain
\begin{equation} \label{eq:MHDeqpsi}
P^\prime + \partial_r \left ( \frac{B^2}{8\pi} \right ) - \partial_l \left ( \frac{B^2}{8\pi} \right ) \partial_l \bx \cdot \partial_r \bx  = \frac{B^2}{4\pi} \partial^2_{ll} \bx \cdot \partial_r \bx
\end{equation}
and
\begin{equation} \label{eq:MHDeqalpha}
\partial_\alpha \left ( \frac{B^2}{8\pi} \right ) - \partial_l \left ( \frac{B^2}{8\pi} \right ) \partial_l \bx \cdot \partial_\alpha \bx = \frac{B^2}{4\pi} \partial^2_{ll} \bx \cdot \partial_\alpha \bx,
\end{equation}
where $P^\prime := \rmd P/\rmd r$. To solve these equations, we need to obtain the magnitude of the magnetic field $B$ from $\bx(r, \alpha, l)$. Using equation~\refe{eq:d3x}, we find that the magnitude of the magnetic field is given by
\begin{equation} \label{eq:BJacob}
B = \frac{\Psi_t^\prime}{( \partial_r \bx \times \partial_\alpha \bx ) \cdot \partial_l \bx}.
\end{equation}

We expand the MHD equilibrium equations in $\epsilon \ll 1$ by assuming that the plasma pressure is sufficiently small to satisfy
\begin{equation}
\beta := \frac{8\pi P}{B^2} \lesssim \epsilon \ll 1.
\end{equation}
We are particularly interested in the magnitude of the magnetic field, given by
\begin{equation} \label{expansionB}
B(r, \alpha, l) = B_0(r, \alpha, l) + B_1(r, \alpha, l) + \ldots,
\end{equation}
where $B_n \sim \epsilon^n B_0 \ll 1$. To lowest order in $\epsilon$, the MHD equations~\refe{eq:MHDeqpsi} and \refe{eq:MHDeqalpha} become $\partial_r ( B_0^2/8\pi ) = 0$ and $\partial_\alpha ( B_0^2/8\pi ) = 0$. The solution to these equations is that $B_0(l)$ can only be a function of $l$. As a result, the lowest order version of equation~\refe{eq:BJacob}, 
\begin{equation} \label{eq:BJacob0}
\left ( \partial_r \bx_1 \times \partial_\alpha \bx_1 \right ) \cdot \bun_0 = \frac{\Psi_t^\prime(r)}{B_0(l)}
\end{equation}
cannot depend on $\alpha$. Here, $\bun_0 (l) := \rmd \bx_0/\rmd l$ is the unit vector parallel to the magnetic axis. Note that condition~\refe{eq:BJacob0} implies that
\begin{equation}
\Psi_t^\prime \sim B_0 a.
\end{equation}
Condition~\refe{eq:BJacob0} limits the choice of $\bx_1(r, \alpha, l)$. The function $\bx_1 (r, \alpha, l)$ must satisfy two constraints in addition to satisfying equation~\refe{eq:BJacob0}: the first order correction to equation~\refe{eq:arc} and the conservation of electric current. These two extra constraints will not be needed for rest of the article, but we give them in Appendix~\ref{app:MHDeq} for completeness. For the rest of this paper, we only need to know that the three scalar constraints discussed above can be satisfied by choosing the three components of $\bx_1 (r, \alpha, l)$ wisely, i.e. the large aspect ratio expansion is self-consistent.

The first order correction $B_1$ can be calculated using MHD force balance. Keeping only the first order terms in $\epsilon$ in equations~\refe{eq:MHDeqpsi} and \refe{eq:MHDeqalpha}, we find
\begin{equation} \label{eq:MHDeqpsi0}
P^\prime + \frac{B_0}{4\pi} \partial_r B_1 - \frac{B_0}{4\pi} \frac{\rmd B_0}{\rmd l} \bun_0 \cdot \partial_r \bx_1 = \frac{B_0^2}{4\pi} \frac{\rmd^2 \bx_0}{\rmd l^2} \cdot \partial_r \bx_1
\end{equation}
and
\begin{equation} \label{eq:MHDeqalpha0}
\frac{B_0}{4\pi} \partial_\alpha B_1 - \frac{B_0}{4\pi} \frac{\rmd B_0}{\rmd l} \bun_0 \cdot \partial_\alpha \bx_1 = \frac{B_0^2}{4\pi} \frac{\rmd^2 \bx_0}{\rmd l^2} \cdot \partial_\alpha \bx_1.
\end{equation}
Equations \refe{eq:MHDeqpsi0} and \refe{eq:MHDeqalpha0} can be integrated to find
\begin{equation} \label{eq:B1}
B_1 (r, \alpha, l) = B_0(l) \bkappa_0 (l) \cdot \bx_1 (r, \alpha, l) + \frac{\rmd B_0(l)}{\rmd l} \bun_0(l) \cdot \bx_1(r, \alpha, l) - \frac{4\pi P(r)}{B_0(l)},
\end{equation}
where $\bkappa_0 (l) := \rmd^2 \bx_0/\rmd l^2$ is the curvature of the magnetic axis.

For a given $\bx_1$, we can calculate the components of the drifts that we need to solve equation~\refe{rho2orbit}. In a general stellarator, the radial and $\alpha$ components of the magnetic drift are
\begin{equation} \label{eq:vMr}
\bv_{Mi} \cdot \nabla r = - \frac{m_i c}{Z_i e \Psi_t^\prime} \left ( v_\|^2 \partial_\alpha \bx \cdot \partial_{ll}^2 \bx + \mu\, \partial_\alpha B - \mu\, \partial_l B \, \partial_\alpha \bx \cdot \partial_l \bx \right )
\end{equation}
and
\begin{equation} \label{eq:vMalpha}
\bv_{Mi} \cdot \nabla \alpha = \frac{m_i c}{Z_i e \Psi_t^\prime} \left ( v_\|^2 \partial_r \bx \cdot \partial_{ll}^2 \bx + \mu\, \partial_r B - \mu\, \partial_l B \, \partial_r \bx \cdot \partial_l \bx \right ),
\end{equation}
and the same components of the $\bE \times \bB$ drift are
\begin{equation} \label{eq:vEr}
\bv_E \cdot \nabla r = - \frac{c}{\Psi_t^\prime} \left ( \partial_\alpha \phi - \partial_l \phi \, \partial_\alpha \bx \cdot \partial_l \bx \right )
\end{equation}
and
\begin{equation} \label{eq:vEalpha}
\bv_E \cdot \nabla \alpha = \frac{c}{\Psi_t^\prime} \left ( \partial_r \phi - \partial_l \phi \, \partial_r \bx \cdot \partial_l \bx \right ).
\end{equation}
To lowest order in $\epsilon \ll 1$, the expressions for the magnetic drift become
\begin{equation} \label{eq:vMreps1}
\bv_{Mi} \cdot \nabla r = - \frac{m_i c (v_\|^2 + \mu B_0)}{Z_i e B_0 \Psi_t^\prime} \partial_\alpha \left ( B_1 - \frac{\rmd B_0}{\rmd l} \bun_0 \cdot \bx_1 \right ) + O\left (\epsilon^2 \rho_{i*} v_{ti} \right ) \sim \epsilon \rho_{i\ast} v_{ti}
\end{equation}
and
\begin{align} \label{eq:vMalphaeps1}
\bv_{Mi} \cdot \nabla \alpha = & \frac{m_i c (v_\|^2 + \mu B_0)}{Z_i e B_0 \Psi_t^\prime} \partial_r \left ( B_1 - \frac{\rmd B_0}{\rmd l} \bun_0 \cdot \bx_1 \right ) + \frac{4\pi m_i c P^\prime v_\|^2}{Z_i e B_0^2 \Psi_t^\prime} \nonumber\\ &+ O\left (\epsilon^2 \rho_{i*} \frac{v_{ti}}{a} \right ) \sim \epsilon \rho_{i\ast} \frac{v_{ti}}{a},
\end{align}
where we have used equation~\refe{eq:B1} to write the magnetic drift components as derivatives of $B_1$. Similarly, the radial and $\alpha$ components of the $\bE \times \bB$ drift are
\begin{equation} \label{eq:vEreps1}
\bv_E \cdot \nabla r = - \frac{c}{\Psi_t^\prime} \partial_\alpha \phi + O\left (a |\partial_l \ln \phi| \rho_{i*} v_{ti} \right ) \sim \rho_{i\ast} v_{ti}
\end{equation}
and
\begin{equation} \label{eq:vEalphaeps1}
\bv_E \cdot \nabla \alpha = \frac{c}{\Psi_t^\prime} \partial_r \phi  + O\left ( |\partial_l \ln \phi| \rho_{i*} v_{ti} \right ) \sim \rho_{i\ast} \frac{v_{ti}}{a}
\end{equation}
to lowest order in $\epsilon \ll 1$. For the $\bE \times \bB$ drift, we have emphasized that the size of the first order corrections in $\epsilon$ is proportional to the derivative of $\phi$ with respect to $l$. The fact that the next order corrections only depend on $\partial_l \phi$ is important because we show in section~\ref{sec:dkequationJ} that the potential is a flux function to lowest order, making these corrections even smaller than first order in $\epsilon$.

For most of this article (sections \ref{sec:dkequationJ} - \ref{sec:nuregime} and section~\ref{sub:optimized}), we will focus on large aspect ratio stellarators with constant $B_0$, that is, $\rmd B_0/\rmd l = 0$. According to equation~\refe{eq:BJacob0}, stellarators with constant $B_0$ must have flux surfaces such that the area of a cut of a flux surface $r$ through a plane perpendicular to the magnetic axis cannot depend on the position along the magnetic axis. Indeed, this area is given by
\begin{equation}
A(r, l) := \int_0^r \rmd r^\prime \int_0^{2\pi} \rmd \alpha \, \left [ \partial_{r} \bx_1 (r^\prime, \alpha, l) \times \partial_\alpha \bx_1 (r^\prime, \alpha, l) \right ] \cdot \bun_0 (l) = \frac{2\pi \Psi_t(r)}{B_0}.
\end{equation}
From here on, we refer to these large aspect ratio stellarators as stellarators with mirror ratios close to unity because the ratio between the maximum and the minimum of $B$ on a flux surface (mirror ratio) is $1 + O(\epsilon) \simeq 1$. We focus on large aspect ratio stellarators with mirror ratios close to unity for two reasons: (i) their description requires careful analysis and an unintuitive choice of velocity space coordinates, and (ii) these stellarators with mirror ratio close to unity are extremely common -- see, for example, the maps of magnetic field magnitude $B$ in \cite{beidler11} that show mirror ratios in the interval $1.05 - 1.2$. 

To have a mirror ratio significantly different from unity in a large aspect ratio stellarator, $B_0$ must depend on $l$. In this case, the mirror ratio is $B_{0,M}/B_{0,m}$, where $B_{0, M}$ and $B_{0, m}$ are the maximum and minimum of $B_0(l)$, respectively. We are not aware of any large aspect ratio stellarators with mirror ratios significantly different from unity that have been built. Despite this fact, we will study large aspect ratio stellarators with mirror ratios significantly different from unity (from here on, ``with large mirror ratios" for short) in section~\ref{sub:largemirror}, where we will consider $B_0(l)$ to be a general function of $l$. This type of stellarator is always close to omnigeneous and hence one can use the formalism developed by \cite{calvo17} to calculate neoclassical transport in them.

\section{New velocity space coordinates for large aspect ratio stellarators with mirror ratios close to unity} \label{sec:dkequationJ}
We first consider the possibility of the potential $\phi(\bx, t)$ being very different from a flux function, that is, $\partial_\alpha \phi \neq 0$ and $\partial_l \phi \neq 0$. We show that this is not possible in a large aspect ratio stellarator with mirror ratios close to unity, that is, large aspect ratio stellarators with constant $B_0$. If $\phi$ is not a flux function, the variation of $v_\|$ within a flux surface is dominated by the variation of $\phi$,
\begin{equation} \label{eq:vparbadapprox}
v_\| \simeq \sigma \sqrt{ 2 \left ( \mathcal{E} - \mu B_0 - \frac{Z_i e \phi(r, \alpha, l, t)}{m_i} \right )}.
\end{equation}
Thus, for a general $\phi$, trapped particles satisfy $\mathcal{E} \leq U_M \simeq \mu B_0 + Z_i e \phi_M(r, t)/m_i$, where $\phi_M(r, t)$ is the maximum of $\phi$ on the flux surface. The parallel velocity of trapped particles is of order $v_{ti}$, and the fraction of trapped particles is of order unity. In this case, the quasineutrality equation \refe{quasineutrality} becomes
\begin{align} \label{eq:QNgeneral}
4\pi Z_i \int_0^\infty \rmd \mu &\int_{\mu B_0 + Z_i e \phi/m_i}^{\mu B_0 + Z_i e \phi_M/m_i} \rmd \mathcal{E}\, \frac{B_0}{\sqrt{ 2 ( \mathcal{E} - \mu B_0 - Z_i e \phi/m_i )}} \sum_{W \in \mathcal{W}} g_{i,W} (r, \alpha, \mathcal{E}, \mu, t) \nonumber\\ & + 2\pi Z_i \sum_\sigma \int_0^\infty \rmd \mu \int_{\mu B_0 + Z_i e \phi_M/m_i}^\infty \rmd \mathcal{E}\, \frac{B_0 h_i (r, \mathcal{E}, \mu, \sigma, t)}{\sqrt{ 2 ( \mathcal{E} - \mu B_0 - Z_i e \phi/m_i )}} \nonumber\\ = & \hat{n}_e (r, t) \exp \left ( \frac{e\phi}{T_e(r, t)} \right )
\end{align}
to lowest order in $\epsilon$. The function $g_{i, W}$ is defined for $\mathcal{E} \leq U_M$, but it is zero for the values of $\mathcal{E}$ outside of well $W$ (in the example of figure \ref{fig:exampleU}, $g_{i, I}$ and $g_{i, II}$ are zero for $\mathcal{E} > \mathcal{E}_c$, and $g_{i, III}$ is zero for $\mathcal{E} < \mathcal{E}_c$). Then, the sum of $g_{i,W}$ over the well index $W$ gives a continuous function of $\mathcal{E}$. Note that the sum over $W$ is performed over a subset $\mathcal{W}$ of all possible wells. Set $\mathcal{W}$ depends on the location where the quasineutrality is being evaluated because any well $W$ has a limited range of values of $l$. In the example in figure~\ref{fig:exampleU}, $l$ in well $I$ is between $l_{bL, I}$ and $l_{bR, I}$, and $l$ in well $II$ is between $l_{bL, II}$ and $l_{bR, II}$. When the ion density is evaluated for $l \in [l_{bL, I}, l_{bR, I}]$, set $\mathcal{W}$ should include wells $I$ and $III$, but exclude well $II$. Note that set $\mathcal{W}$ is independent of $l$ in a finite region of $l$ around most points in the stellarator (e.g. in the case of figure~\ref{fig:exampleU}, set $\mathcal{W}$ includes wells $I$ and $III$ for $l \in [l_{bL, I}, l_{bR, I}]$). Thus, the quasineutrality equation~\refe{eq:QNgeneral} only depends on $\phi$, $r$ and $\alpha$ around most spatial points, giving a solution $\phi$ that can only depend on $r$ and $\alpha$, that is, $\partial_l \phi = 0$. As we are considering only ergodic flux surfaces, $\partial_l \phi = 0$ implies that $\partial_\alpha \phi = 0$, and hence $\phi$ is a flux function, $\phi = \phi(r)$. Note that this is an arbitrary flux function because we can choose it at will using the free function $\hat{n}_e (r, t)$.

Since the electric potential is a flux function to lowest order in $\epsilon$, we write it as
\begin{equation} \label{expansionphi}
\phi(r, \alpha, l, t)  =\phi_0(r, t) + \phi_{3/2}(r, \alpha, l, t) + \dots,
\end{equation}
where $\phi_0(r, t) \sim T_i/e$ is a flux function, and $\phi_{3/2}(r, \alpha, l, t) \sim \epsilon^{3/2} T_i/e$.  We show that the correction to the lowest order flux function is small in $\epsilon^{3/2}$ in section~\ref{sub:gi1}. Due to expansion \refe{expansionphi}, equation~\refe{eq:vparbadapprox} is a bad approximation for trapped particles. Instead, we need to use
\begin{equation}
v_\| = \sigma \sqrt{ 2 \left (\mathcal{E}_1  - \mu B_1(r, \alpha, l) - \frac{Z_i e \phi_{3/2}(r, \alpha, l, t)}{m_i} \right )} [1 + O(\epsilon)].
\end{equation}
where the quantity $\mathcal{E}_1 := \mathcal{E} - \mu B_0 - Z_i e \phi_0(r, t)/m_i$ must be of order $\epsilon v_{ti}^2$ for trapped particles -- otherwise, $v_\|$ would not vanish. Thus, the characteristic size of the parallel velocity of trapped particles is $v_\| \sim \sqrt{\epsilon} v_{ti}$, and the fraction of trapped particles is of order $\sqrt{\epsilon}$. 

Before expanding the ion distribution function in $\epsilon$, we need new velocity space coordinates for trapped and passing particles. We discuss the velocity space coordinates for trapped and passing particles in subsections~\ref{sub:trappedcoord} and \ref{sub:passingcoord}, respectively.

\subsection{Velocity space coordinates for trapped particles} \label{sub:trappedcoord}
Due to the smallness of $v_\|$ and to the expansion in equation~\refe{expansionphi}, the magnetic and $\bE \times \bB$ drifts in equations~\refe{eq:vMreps1}, \refe{eq:vMalphaeps1}, \refe{eq:vEreps1} and \refe{eq:vEalphaeps1} simplify to
\begin{equation} \label{eq:vdreps}
(\bv_E + \bv_{Mi}) \cdot \nabla r = - \frac{m_i c}{Z_i e \Psi_t^\prime} \left ( \mu \partial_\alpha B_1 + \frac{Z_i e}{m_i} \partial_\alpha \phi_{3/2} \right ) + O \left ( \epsilon^2 \rho_{i\ast} v_{ti} \right ) \sim \epsilon \rho_{i\ast} v_{ti}
\end{equation}
and
\begin{equation} \label{eq:vdalphaeps}
(\bv_E + \bv_{Mi}) \cdot \nabla \alpha = \frac{c \phi_0^\prime}{\Psi_t^\prime} [1 + O ( \epsilon  ) ] \sim\rho_{i\ast} \frac{v_{ti}}{a}
\end{equation}
for trapped particles. Here $\phi_0^\prime := \partial_r \phi_0$. 

\begin{figure}
\centering
\includegraphics[width=9cm]{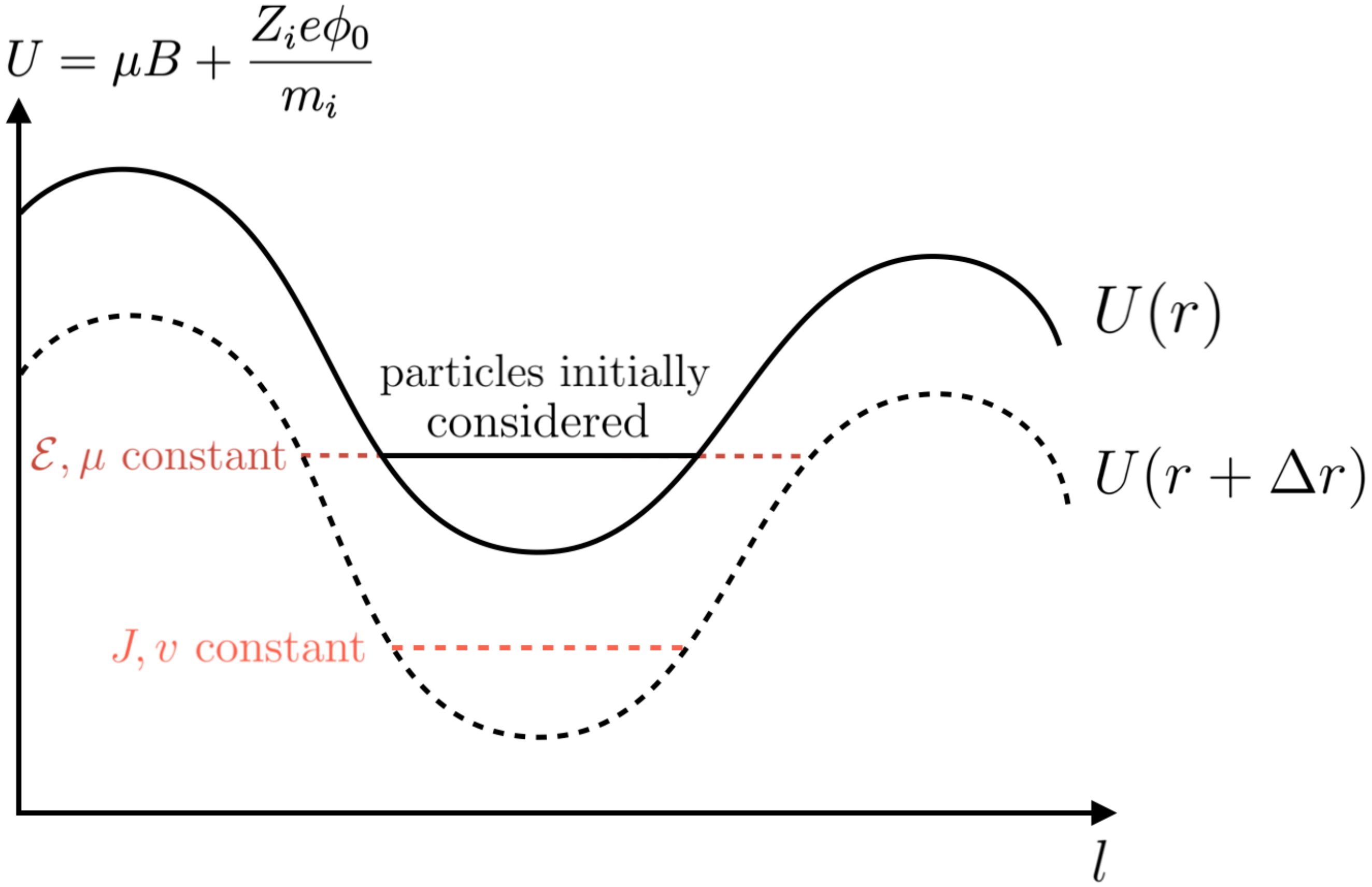}
\caption{Trajectories of trapped particles in a given $U$ well in the $(\mathcal{E}, l)$ plane when one moves from a given flux surface to a neighboring one keeping either $J$ or $\mathcal{E}$ constant. The shape of the $U$ well does not change much whereas the whole well moves up and down due to the change in potential $\phi$.}
\label{fig:JorEconstant}
\end{figure}

Since the radial component of the drifts is small in $\epsilon$, it is tempting to neglect the term proportional to the radial drift in equation~\refe{rho2orbit}. Unfortunately, this term cannot be neglected because the radial derivative of $g_{i,W}$ is very large,
\begin{equation} \label{assessment1}
\partial_r \ln g_{i,W} (r, \alpha, \mathcal{E}, \mu, t) \sim \frac{1}{\epsilon a}.
\end{equation}
In order to understand estimate \refe{assessment1}, one has to keep in mind that the potential changes significantly with radius. Indeed, the change in potential due to radial displacement $\Delta r \sim \epsilon a$ is $\Delta r\, \phi_0^\prime \sim \epsilon T_i/e$, and this change in potential energy means that $v_\parallel$ has to change by $\sqrt{Z_i e \Delta r\, \phi_0^\prime /m_i} \sim \sqrt{\epsilon} v_{ti}$ if we keep $\mathcal{E}$ constant when varying $r$. Hence, surprisingly, trapped particles with the same value of $\mathcal{E}$ and separated only by $\Delta r \sim \epsilon a$ occupy very different heights with respect to the minimum of the $U$ well. This situation is represented in figure~\ref{fig:JorEconstant}. The difference in the height of the particle with respect to the minimum of the $U$ well leads to the large radial derivative in equation~\refe{assessment1}. Note that the characteristic length of $g_{i,W}$, $\Delta r \sim \epsilon a$, is of the order of the width $w$ of the particle orbits in \refe{eq:wso}, indicating that we need to keep the radial drifts in equation~\refe{rho2orbit}.

Importantly, the radial derivative of $g_{i,W}$ is very large because we are holding $\mathcal{E}$ fixed and, as a result, the radial derivative of $g_{i,W}$ is related to the radial derivative of $\phi_0$. In contrast with the electric potential, the magnitude of the magnetic field does not change much if $r$ is changed by $\Delta r \sim \epsilon a$ because its characteristic length of variation is $R$. Indeed, the change in $B$ is $\Delta r\, \partial_r B \simeq \Delta r\, \partial_r B_1 \sim \epsilon^2 B_0$. The lack of rapid variation in $B$ suggest using velocity space coordinates that, when held constant, do not change the height of the trapped particle with respect to the minimum of the $B$ well. The coordinates $v$ and 
\begin{equation} \label{eq:lambdadef}
\lambda :=\frac{v_{\perp}^2}{v^2B} 
\end{equation}
are often used because the variation of $v$ along a orbit is small in $\epsilon$ and $\lambda$ is equal to $1/B$ at the bounce points. To see that $v$ does not change much during an orbit, note that in a large aspect ratio stellarator, the parallel velocity of a trapped particle is very small, giving $v \simeq v_\perp \simeq \sqrt{2\mu B_0}$, where $B_0$ is constant. Conversely, $\lambda$ is not constant along trajectories. Writing $\lambda$ as
\begin{equation} \label{eq:lambdaEmu}
\lambda = \frac{\mu}{\mathcal{E}-Z_i e \phi/m_i} = \frac{\mu}{\mathcal{E}-Z_i e \phi_0/m_i} [ 1 + O(\epsilon^{3/2})]
\end{equation}
shows that $\lambda$ changes by an amount of order $\epsilon/B_0$ along the orbit because of the variation of $\phi_0$. Changes of order $\epsilon/B_0$ in $\lambda$ are important for trapped particles because the interval of $\lambda$ values where we find trapped particles, $\lambda \in [B_M^{-1}, B_m^{-1}]$, has a length of order $\epsilon/B_0$. Here $B_m(r)$ and $B_M(r)$ are the minimum and maximum of $B$ on flux surface $r$, respectively. 

Instead of the usual coordinates $v$ and $\lambda$, we propose two other coordinates. In section~\ref{sec:fiepsilon}, we will discover that $v$ is not constant to a sufficiently high order in $\epsilon$ for one of the calculations that we perform: $v = \sqrt{2 ( \mathcal{E} - Z_i e \phi/m_i )}$ causes problems because it introduces dependence on $l$ through $\phi_{3/2}$. Due to this limitation, we choose 
\begin{equation} \label{eq:vbardef}
\vbar (r, \mathcal{E}, t) := \sqrt{2 \left ( \mathcal{E} - \frac{Z_i e \phi_0(r, t)}{m_i} \right )}
\end{equation}
to be one of our velocity space coordinates. Regarding $\lambda$, using it as a velocity space coordinate is inconvenient because it is not constant in time. Fortunately, there is another quantity that gives the same information as $\lambda$ and is constant in time: the second adiabatic invariant 
\begin{align} \label{eq:Jdef}
J_W (r, \alpha, \mathcal{E}, \mu, t) &:= 2\int_{l_{bL,W}}^{l_{bR,W}}|v_{\parallel}| \, \mathrm{d}l \nonumber\\ & = 2\int_{l_{bL,W}}^{l_{bR,W}} \sqrt{2\left ( \mathcal{E} - \mu B (r, \alpha, l) - \frac{Z_i e \phi(r, \alpha, l, t)}{m_i} \right ) }  \, \mathrm{d}l.
\end{align}
We use the second adiabatic invariant as a velocity space coordinates for trapped particles. Employing $\partial_\mathcal{E} J_W = \tau_W \sim \epsilon^{-1/2} R/ v_{ti}$ and $\partial_\mu J_W = - \tau_W \langle B \rangle_{\tau, W}  \sim \epsilon^{-1/2} B_0 R/v_{ti}$ to find
\begin{equation} \label{eq:gradvJbar}
\nabla_v J_W = \partial_\mathcal{E} J_W\, \nabla_v \mathcal{E} + \partial_\mu J_W\, \nabla_v \mu = \tau_W v_\| \bun + \left (1 - \frac{\langle B \rangle_{\tau, W}}{B} \right ) \tau_W \bv_\perp,
\end{equation}
we can calculate the velocity space volume element,
\begin{equation} \label{eq:d3vvJvarphi}
\rmd^3 v = \frac{\rmd \vbar\, \rmd J\, \rmd \varphi}{|(\nabla_v \vbar \times \nabla_v J_W) \cdot \nabla_v \varphi|} = \frac{\vbar B}{\tau_W |v_\|| \langle B \rangle_{\tau, W}} \, \rmd \vbar\, \rmd J\, \rmd \varphi = \frac{\vbar}{\tau_W |v_\||} \, \rmd \vbar\, \rmd J\, \rmd \varphi\, [ 1 + O(\epsilon)],
\end{equation}
Note that we use the symbol $J_W$ when the second adiabatic invariant is considered a function of $r$, $\alpha$, $\mathcal{E}$, $\mu$, $t$ and the type of well $W$, whereas we employ the symbol $J$ without the subscript $W$ when the second adiabatic invariant is a coordinate.

Previous work \citep{hazeltine81, calvo17} has used $J$ as a coordinate, but as a replacement for the radial coordinate $r$ instead of as a replacement for the pitch angle variable $\lambda$. Note that the three quantities $\mathcal{E}$, $\mu$ and $J$ are all desirable coordinates because they are constant along particle trajectories. One of the three variables works as a proxy for the radial position of the particle, whereas the other two represent velocity space. In 3D magnetic fields close to omnigeneity \citep{hazeltine81, calvo17}, $J$ is a flux function to lowest order and hence it can be used to replace $r$. For trapped particles in large aspect ratio stellarators with mirror ratios close to unity, the kinetic energy associated to the parallel velocity is negligible compared to the perpendicular kinetic energy $\mu B \simeq \mu B_0$, giving $\mathcal{E} \simeq \mu B_0 + Z_i e\phi_0 (r, t)/m_i$. As a result, the total energy determines the radial position $r$ for a given $\mu$ through the potential energy $Z_i e \phi_0(r, t)/m_i$, and $\mu$ and $J$ are the velocity space coordinates -- note that $\vbar \simeq \sqrt{2\mu B_0}$. We make the relation between the total energy and the radial position obvious in subsection~\ref{sub:deeplytrapped}, where we discuss the motion of the deeply trapped particles to demonstrate the advantages of our formulation. 

We remind the reader that we are changing from the coordinates $\mathcal{E}$ and $\mu$ to the coordinates $\vbar$ and $J$ to reduce the size of the derivative $\partial_r \ln g_{i,W}$ from $(\epsilon a)^{-1}$ to $a^{-1}$. We will not be able to show that the derivative of $g_{i,W}$ with respect to $r$ holding $\vbar$ and $J$ constant is small in $\epsilon$ until section~\ref{sub:gi1}, but we can now show that if
\begin{equation}
\partial_r \ln g_{i,W} (r, \alpha, \vbar, J, t) \sim \frac{1}{a}, 
\end{equation}
the derivative of $g_{i,W}$ with respect to $r$ holding $\mathcal{E}$ and $\mu$ fixed is as large as estimated in equation~\refe{assessment1}. Indeed, by using the chain rule, we find
\begin{align}
\partial_r \ln g_{i,W} (r, \alpha, \mathcal{E}, \mu, t) = \partial_r \ln g_{i,W} (r, \alpha, \vbar, J, t) + \partial_r \vbar(r, \mathcal{E}, t)\, \partial_{\vbar} \ln g_{i,W} (r, \alpha, \vbar, J, t) \nonumber\\ + \partial_r J_W (r, \alpha, \mathcal{E}, \mu, t)\, \partial_J \ln g_{i,W} (r, \alpha, \vbar, J, t). 
\end{align}
Noting that 
\begin{equation}
\partial_r \vbar = - \frac{Z_i e \phi_0^\prime}{m_i \vbar} \sim \frac{v_{ti}}{a},
\end{equation}
\begin{equation}
\partial_r J_W = - 2 \int_{l_{bL,W}}^{l_{bR,W}} \frac{\mu \partial_r B + Z_i e \partial_r \phi/m_i}{|v_\||}\, \rmd l \simeq - \frac{Z_i e \phi_0^\prime \tau_W}{m_i} \sim \frac{v_{ti}}{\epsilon^{3/2}},
\end{equation}
\begin{equation} \label{eq:dvbargiestimate}
\partial_{\vbar} \ln g_{i,W} (r, \alpha, \vbar, J, t) \sim \frac{1}{v_{ti}}
\end{equation}
and
\begin{equation} \label{eq:dJgiestimate}
\partial_J \ln g_{i,W} (r, \alpha, \vbar, J, t) \sim \frac{1}{\sqrt{\epsilon} v_{ti} R},
\end{equation}
we obtain estimate~\refe{assessment1}. 

\subsection{Velocity space coordinates for passing particles} \label{sub:passingcoord}
For passing particles, the variable $J$ cannot be defined. For most passing particles, $v \simeq \vbar := \sqrt{2(\mathcal{E} - Z_i e \phi_0/m_i)}$ and $v_\| \simeq \sigma \sqrt{2(\mathcal{E} - \mu B_0 - Z_i e \phi_0/m_i)}$ are approximate constants of the motion. For this reason, we use the coordinates $\vbar$ and $\xi := v_\|/v$ for passing particles. The velocity space volume element in these variables is
\begin{equation}
\rmd^3 v = \frac{\rmd \vbar\, \rmd \xi\, \rmd \varphi}{| (\nabla_v \vbar \times \nabla_v \xi) \cdot \nabla_v \varphi |} = \vbar v\, \rmd \vbar\, \rmd \xi\, \rmd \varphi = \vbar^2 \, \rmd \vbar\, \rmd \xi\, \rmd \varphi\, [1 + O(\epsilon^{3/2})].
\end{equation}

For the few passing particles with $|\xi| \sim \sqrt{\epsilon} \ll 1$, $\xi$ is not approximately constant, but we show in Appendix~\ref{app:barelypassing} that this region of phase space can be treated as a boundary condition for the passing particle distribution function $h_i$ at $\xi = 0$. Appendix~\ref{app:barelypassing} should be read after having gone through sections~\ref{sub:Maxwellian} and \ref{sub:gi1} and Appendix~\ref{app:Maxwellian}.

\section{Ion distribution function and potential in low collisionality large aspect ratio stellarators with mirror ratios close to unity} \label{sec:fiepsilon}

We proceed to expand equations~\refe{rho2orbit}, \refe{passingfi0} and \refe{rho2transition} in the inverse aspect ratio $\epsilon \ll 1$ assuming that $B_0$ is independent of $l$ and that $\rho_{i*} \sim \nu_{i*}$. Based on estimates~\refe{eq:vdreps} and \refe{eq:vdalphaeps} for the components of the $\bE \times \bB$ and magnetic drifts, we will show in section~\ref{sub:transport} that we need to order the time derivative and the source as
\begin{equation} \label{eq:Sieps}
\partial_t \sim \frac{S_i}{f_i} \sim \epsilon^{3/2} \nu_{ii}. 
\end{equation}
This estimate is the result of a subsidiary expansion in $\epsilon \ll 1$ of the radially global equations for generic stellarators in section~\ref{sec:dkequation}. Note that it is consistent with assumption~\refe{eq:dtSieps1} in section~\ref{sec:dkequation} when $\epsilon \sim 1$.

We expand $g_{i,W}$ and $h_i$ in $\epsilon \ll 1$ assuming that $\rho_{i*} \sim \nu_{i*}$ and using the estimates in equation~\refe{eq:Sieps}. For $g_{i,W}$, we find
\begin{equation} \label{eq:giexpansion}
g_{i,W} = g_{i,0, W} + g_{i, 1, W} + g_{i,3/2, W} + g_{i,2, W} \ldots,
\end{equation}
where $g_{i, n, W} \sim \epsilon^n g_{i, 0, W}$. For $h_i$, the expansion gives
\begin{equation} \label{eq:hiexpansion}
h_i = h_{i,0} + h_{i,3/2} + \ldots,
\end{equation}
where $h_{i,n} \sim \epsilon^n h_{i,0}$. The corrections $g_{i, 3/2, W}$, $g_{i, 2, W}$ and $h_{i.3/2}$ are important because their size determines the boundary conditions for $g_{i, 1, W}$, but in the end we do not need to calculate them. 

We need to consider half integer powers of $\epsilon$ in our expansion for two reasons: (i) the size of the trapped particle region in velocity space is small by a factor of order $\sqrt{\epsilon}$, and (ii) boundary condition~\refe{eq:diffcondJinftyv1} introduces these half-integer powers naturally, as we will see shortly. 

We organize the calculation as follows. In section~\ref{sub:Maxwellian} we argue that the lowest order solution is a Maxwellian. In section~\ref{sub:gi1} we obtain the equation and boundary conditions for $g_{i, 1, W}$. Finally, in section~\ref{sub:transport} we obtain the transport equations for the ion density and temperature. In section~\ref{sub:deeplytrapped}, we summarize the equations, we compared them to those implemented in DKES \citep{hirshman86} and we illustrate how they work by discussing the behavior of deeply trapped particles.

\subsection{Lowest order ion distribution function} \label{sub:Maxwellian}

The proof that the distribution is a stationary Maxwellian to lowest order in $\epsilon$ does not follow the derivation used in standard neoclassical theory -- for examples of the usual procedure, see section V.4 of \cite{hinton76} or section 3.1 of \cite{calvo17}. In the typical derivation, the radial drifts are small and can be neglected in the lowest order kinetic equation. By calculating certain moment of this simplified lowest order kinetic equation, one obtains an entropy equation for an infinitesimal volume between two flux surfaces close to each other. In this equation, the entropy production due to collisions within the infinitesimal volume cannot be compensated by any outward flow of entropy because the radial drifts are negligible. Consequently, in steady state, the average entropy production must vanish to lowest order, leading to a distribution function close to Maxwellian. In large aspect ratio stellarators with mirror ratios close to unity, we cannot follow this procedure because, for $\nu_{i*} \sim \rho_{i*}$, the typical frequency associated to the radial drift, $\langle (\bv_E + \bv_{Mi}) \cdot \nabla r \rangle_{\tau, W}/a \sim \rho_{i*} v_{ti}/R$, is comparable to the collision frequency $\nu_{ii}$. As a result, trapped particles move a significant radial distance in the time that the distribution function evolves towards a Maxwellian. This motion can, in principle, disrupt the natural evolution of the distribution function towards a Maxwellian.

We proceed to argue that the distribution function is close to a Maxwellian in large aspect ratio stellarator with mirror ratios close to unity even though the trapped particle radial drifts are large. Collisions can drive the distribution function close to a Maxwellian because only trapped particles drift off flux surfaces. The number of trapped particles is small by $\sqrt{\epsilon}$ and, for this reason, despite the considerable radial displacements of trapped particles, the radial flux of entropy due to trapped particles is not sufficient to compensate for the large entropy production associated to a distribution function far from Maxwellian. Moreover, trapped particles themselves are close to Maxwellian despite their large displacements because they exchange momentum and energy with passing particles much faster than the ion-ion collision frequency $\nu_{ii}$ would suggest. Indeed, due to the small parallel velocity of trapped particles, $v_\| \sim \sqrt{\epsilon} v_{ti}$, grazing collisions can detrap them, leading to an effective collision frequency $\nu_{ii}/\epsilon \gg \rho_{i*} v_{ti}/R$. 

Appendix~\ref{app:Maxwellian} contains the calculations that show that the lowest order distribution function is Maxwellian. In this Appendix, we first argue that collisions force the lowest order trapped particle distribution function $g_{i,0,W} (r, \alpha, \vbar, J, t)$ to be independent of $\alpha$ and $J$. Continuity at the trapped-passing particle boundary then imposes that the trapped particle distribution function be the passing particle distribution function $h_{i, 0} (r, \vbar, \xi, t)$ at small pitch angles $\xi \sim \sqrt{\epsilon} \ll 1$, 
\begin{equation} \label{eq:gi0sol}
g_{i, 0, W} (r, \alpha, \vbar, J, t) = h_{i,0} (r, \vbar, 0, t).
\end{equation}
When we examine the barely passing region of space, we find that, to the order of interest, the collisional flux from the trapped particle region to the passing particle region is negligible. Thus, passing particles collide with each other without exchanging significant momentum and energy with trapped particles and as a result their distribution function $h_{i, 0}$ becomes Maxwellian. The trapped particles, however, have a role to play. The lack of collisional flux imposes that the derivative of $h_{i, 0} (r, \vbar, \xi, t)$ with respect to $\xi$ at small values of $\xi \sim \sqrt{\epsilon} \ll 1$ must be close to zero. Thus, the Maxwellian must have zero average flow, that is, the friction between trapped and passing particles damps the average flow to subsonic levels. Finally, noting that the passing particle distribution function $h_{i, 0}$ cannot depend on $\alpha$ or $l$, we find that $h_{i,0}$ is a Maxwellian with density $n_i$ and temperature $T_i$ that are flux functions,
\begin{equation} \label{eq:hifM}
h_{i,0} (r, \vbar, \xi, t) = f_{Mi} (r, \vbar, t) := n_i (r, t) \left ( \frac{m_i}{2\pi T_i(r, t)} \right )^{3/2} \exp \left ( - \frac{m_i \vbar^2}{2T_i(r, t)} \right ).
\end{equation}
The lowest order trapped particle distribution function $g_{i, 0, W}$ is also a Maxwellian because of equation~\refe{eq:gi0sol},
\begin{equation} \label{eq:gifM}
g_{i, 0, W} (r, \alpha, \vbar, J, t) = f_{Mi} (r, \vbar, t).
\end{equation}

\subsection{Corrections to the lowest order solutions for the ion distribution function and the electric potential} \label{sub:gi1}
Since the lowest order distribution function is a Maxwellian, from here on we need to use the linearized collision operator $C_{ii}^{\ell}[f] :=C_{ii}[f, f_{Mi}] + C_{ii}[f_{Mi}, f]$. The linearized Fokker-Planck collision operator is composed of two terms: a differential part $C_{ii, D}^{\ell}$ and an integral part $C_{ii, I}^{\ell}$, 
\begin{equation}
C_{ii}^{\ell}[f] = C_{ii, D}^{\ell}[f] + C_{ii, I}^{\ell}[f].
\end{equation}
The differential part of the linearized collision operator is
\begin{equation}
\label{pitch_energy}
C_{ii, D}^{\ell}[f] := \nabla_v \cdot \Bigg[\frac{\nu_{ii, \perp} f_{Mi}}{4}(v^2\mathbf{I}-\mathbf{vv}) \, \cdot \, \nabla_v \Bigg( \frac{f}{ f_{Mi}} \Bigg) + \frac{\nu_{ii, \parallel} f_{Mi}}{2} \mathbf{vv} \, \cdot \, \nabla_v \Bigg( \frac{f}{ f_{Mi}} \Bigg) \Bigg],
\end{equation}
where $\mathbf{I}$ is the unit matrix,
\begin{equation} \label{eq:nupitch}
\nu_{ii,\perp}(r, v, t) := \frac{4 \gamma_{ii}}{v^3} \partial_v H [ f_{Mi} ] = \frac{3\sqrt{2\pi} \nu_{ii}}{2} \frac{v_{ti}^3}{v^3} \Big[ \mathrm{erf}(v/v_{ti}) - \chi(v/v_{ti})\Big ]
\end{equation}
is the pitch angle scattering frequency, and
\begin{equation} \label{eq:nuE}
\nu_{ii,\parallel}(r, v, t):= \frac{2 \gamma_{ii}}{v^2} \partial^2_{vv} H [ f_{Mi} ] = \frac{3\sqrt{2\pi} \nu_{ii}}{2} \frac{v_{ti}^3}{v^3} \chi(v/v_{ti})
\end{equation}
is the energy diffusion frequency. Here, the function $H[f_{Mi}] (r, v, t)$ only depends on the velocity through its magnitude $v$, $\mathrm{erf}(x) :=(2/\sqrt{\pi})\int_0^x \exp(-s^2)\,\mathrm{d}s$ is the error function, and 
\begin{equation}
\chi (x) := \frac{1}{2x^2} \left [ \mathrm{erf}(x)-\frac{2x}{\sqrt{\pi}}\exp(-x^2) \right ].
\end{equation}
The integral part of the collision operator is
\begin{equation}
\label{Cintegral}
C_{ii,I}^{\ell}[f] := \gamma_{ii} \nabla_v \cdot \left (  \nabla_v \nabla_v H[f] \cdot \nabla_v f_{Mi} - f_{Mi} \nabla_v L[f] \right ).
\end{equation}

Using the linearized collision operator, we proceed to obtain an equation and boundary conditions for $g_{i, 1, W}$, to discuss the equation for the passing particle distribution function, and to show that the correction to the lowest order potential $\phi_0(r)$ is of order $\epsilon^{3/2} T_i/e$.

\subsubsection{Equation for $g_{i, 1, W}$}

We first rewrite equation~\refe{rho2orbit} using the coordinates $\vbar$ and $J$, and the fact that $g_{i, 0, W} = f_{Mi}$ and $h_{i, 0} = f_{Mi}$. We neglect the time derivatives and the source $S_i$ because of the estimates in equation~\refe{eq:Sieps}. We also use the fact that the derivatives with respect to $J$, $\partial_J \sim \epsilon^{-1/2}/v_{ti} R$, are larger by $\epsilon^{-1/2}$ than the derivatives with respect to $\vbar$, $\partial_{\vbar} \sim v_{ti}^{-1}$. This difference in size is particularly important for the linearized collision operator that can be approximated by 
\begin{equation}
C_{ii, D}^\ell [g_{i,1,W} ] \simeq \nabla_v J_W\cdot \partial_J \Bigg[ \left ( \frac{\nu_{ii, \perp}}{4}(v^2\mathbf{I}-\mathbf{vv})  + \frac{\nu_{ii, \parallel}}{2} \mathbf{vv} \right ) \cdot \nabla_v J_W \, \partial_J g_{i, 1, W} \Bigg ].
\end{equation}
Finally, we take into account that, according to equations~\refe{eq:vdreps} and \refe{eq:vdalphaeps}, the component of the drifts in the radial direction is much smaller than the component in the $\alpha$ direction. With all these considerations, to lowest order in $\epsilon$, equation~\refe{rho2orbit} becomes 
\begin{align} \label{eq:eqgi1v2}
\langle (\bv_{E}+\bv_{Mi}) & \cdot \nabla \alpha \rangle_{\tau, W}\, \left ( \partial_\alpha g_{i,1,W} + \partial_\alpha J_W \, \partial_J g_{i,1,W} \right ) \nonumber\\ &+ \langle (\bv_{E}+\bv_{Mi}) \cdot \nabla r \rangle_{\tau, W}\, \left ( \partial_r f_{Mi} + \partial_r \vbar \, \partial_{\vbar} f_{Mi} + \partial_r J_W \, \partial_J g_{i,1,W} \right ) \nonumber \\ = & \left \langle \nabla_v J_W\cdot \partial_J \Bigg[ \left ( \frac{\nu_{ii, \perp}}{4}(v^2\mathbf{I}-\mathbf{vv})  + \frac{\nu_{ii, \parallel}}{2} \mathbf{vv} \right ) \cdot \nabla_v J_W \, \partial_J g_{i, 1, W} \Bigg ] \right \rangle_{\tau, W} ,
\end{align}
where the derivatives of $g_{i,W}$ with respect to $r$ and $\alpha$ are performed holding $\vbar$ and $J$ fixed, whereas the derivatives of $\vbar$ and $J_W$ with respect to the same variables are performed holding $\mathcal{E}$ and $\mu$ constant. To simplify equation~\refe{eq:eqgi1v2}, we need to use the fact that $J$ is an adiabatic invariant. Employing the exact expressions \refe{eq:vMr} and \refe{eq:vEr}, we find 
\begin{equation} \label{eq:dalphaJ}
\langle (\bv_E + \bv_{Mi}) \cdot \nabla r \rangle_{\tau, W} = \frac{m_i c}{Z_i e \Psi_t^\prime \tau_W}\, \partial_\alpha J_W; 
\end{equation}
similarly, using \refe{eq:vMalpha} and \refe{eq:vEalpha}, we obtain 
\begin{equation} \label{eq:drJ}
\langle (\bv_E + \bv_{Mi}) \cdot \nabla \alpha \rangle_{\tau, W} = - \frac{m_i c}{Z_i e \Psi_t^\prime \tau_W}\, \partial_r J_W. 
\end{equation}
Thus, the second adiabatic invariant satisfies
\begin{equation} \label{eq:dJdt0}
\langle (\mathbf{v}_{E}+\mathbf{v}_{Mi}) \cdot \nabla \alpha \rangle_{\tau, W}\, \partial_{\alpha} J_W + \langle (\mathbf{v}_{E}+\mathbf{v}_{Mi}) \cdot \nabla r  \rangle_{\tau, W}\, \partial_{r} J_W  = 0.
\end{equation}
Employing the lowest order expression~\refe{eq:vdreps}, we find 
\begin{equation} \label{eq:dvbardt}
\langle (\mathbf{v}_{E}+\mathbf{v}_{Mi}) \cdot \nabla r  \rangle_{\tau, W}\, \partial_{r} \vbar = \frac{c \phi_0^\prime}{\vbar \Psi_t^\prime} \left \langle \frac{\vbar^2}{2B_0} \partial_\alpha B_1 + \frac{Z_i e}{m_i} \partial_\alpha \phi_{3/2} \right \rangle_{\tau, W} [ 1 + O(\epsilon) ] \sim \epsilon \rho_{i*} \frac{v_{ti}^2}{a}.
\end{equation}
With this result, equation~\refe{eq:dJdt0}, and employing the fact that $\nabla_v J_W \simeq \tau_W v_\| \bun$ for trapped particles (see equation~\refe{eq:gradvJbar}), equation~\refe{eq:eqgi1v2} can be rewritten as
\begin{equation} \label{eq:eqgi1v3}
\frac{c \phi_0^\prime}{\Psi_t^\prime} \partial_\alpha g_{i, 1, W} - \frac{\vbar^2 \nu_{ii,\perp}(r, \vbar, t)}{4}\, \partial_{J} \left ( \tau_W J \, \partial_{J} g_{i, 1, W} \right ) = \frac{m_i c \phi_0^\prime \vbar^2}{2 Z_i e B_0 \Psi_t^\prime} \left \langle \partial_\alpha B_1 \right \rangle_{\tau, W} \Upsilon_i f_{Mi},
\end{equation}
where 
\begin{equation} \label{eq:Upsilon}
\Upsilon_i (r, \vbar, t) := \frac{n_i^\prime}{n_i} + \frac{Ze\phi_0^\prime}{T_i} + \Bigg(\frac{m_i \vbar^2}{2 T_i}-\frac{3}{2} \Bigg) \frac{T_i^\prime}{T_i},
\end{equation}
and the quantities $n_i^\prime$ and $T_i^\prime$ are $\partial_r n_i$ and $\partial_r T_i$, respectively. Importantly, note that $r$ only appears as a parameter in equation~\refe{eq:eqgi1v2}, and hence, the derivative of $g_{i, 1, W}$ with respect to $r$ is determined by the variation of the coefficients in equation~\refe{eq:eqgi1v2}, giving $\partial_r \ln g_{i, 1, W} \sim a^{-1}$, as announced in section~\ref{sub:trappedcoord}.

We can rewrite equation~\refe{eq:eqgi1v3} in a more convenient form. Using the variable
\begin{equation} \label{eq:lambdabardef}
\lambdabar (r, \mathcal{E}, \mu, t) := \frac{\mu}{\mathcal{E} - Z_i e \phi_0(r, t)/m_i} = \lambda + O \left ( \frac{\epsilon^{3/2}}{B_0} \right ),
\end{equation}
the coordinate $J$ can be expressed as
\begin{equation} \label{eq:Jbarlambda}
J_W (r, \alpha, \vbar, \lambdabar) \simeq 2 \vbar \int_{l_{bL,0,W}}^{l_{bR,0,W}} \sqrt{1 - \lambdabar B_0 - \frac{B_1(r, \alpha, l)}{B_0} }\, \rmd l
\end{equation}
to lowest order in $\epsilon$. Here, $l_{bL,0,W} (r, \alpha, \lambdabar)$ and $l_{bR,0,W} (r, \alpha, \lambdabar)$ are the approximate bounce points, determined by the equations $B_1(r, \alpha, l_{bL,0,W} ) /B_0 = 1 - \lambdabar B_0 = B_1(r, \alpha, l_{bR,0,W} ) /B_0$. We can invert equation~\refe{eq:Jbarlambda} to obtain $\lambdabar_W (r, \alpha, \vbar, J, t)$ as a function of $r$, $\alpha$, $\vbar$, $J$ and the well index $W$. We can also calculate $\partial_\alpha \lambdabar_W$ by differentiating equation~\refe{eq:Jbarlambda} with respect to $\alpha$ holding $r$, $\vbar$ and $J$ constant,
\begin{equation}
0 \simeq - \int_{l_{bL,0,W}}^{l_{bR,0,W}} \frac{B_0^{-1} \partial_\alpha B_1}{\sqrt{1 - \lambdabar_W B_0 - B_1/B_0}}\, \rmd l - B_0\, \partial_\alpha \lambdabar_W \int_{l_{bL,0,W}}^{l_{bR,0,W}} \frac{1}{\sqrt{ 1 - \lambdabar_W B_0 - B_1/B_0}}\, \rmd l.
\end{equation}
This expression gives
\begin{equation} \label{eq:avedalphaB1}
\left \langle \partial_\alpha B_1 \right \rangle_{\tau, W} \simeq - B_0^2 \, \partial_\alpha \lambdabar_W.
\end{equation}
We can use this result to rewrite equation~\refe{eq:eqgi1v3} as
\begin{equation} \label{eq:eqgi1}
\frac{c \phi_0^\prime}{\Psi_t^\prime} \partial_\alpha g_{i, 1, W} - \frac{\vbar^2 \nu_{ii,\perp}(r, \vbar, t)}{4}\, \partial_{J} \left ( \tau_W J \, \partial_{J} g_{i, 1, W} \right ) = - \frac{c \phi_0^\prime}{\Psi_t^\prime} \partial_\alpha r_{i,1, W} \Upsilon_i f_{Mi},
\end{equation}
where
\begin{equation} \label{eq:r1def}
r_{i, 1, W} (r, \alpha, \vbar, J, t) := \frac{m_i B_0 \vbar^2}{2Z_i e \phi_0^\prime(r, t)} \left [ \lambdabar_W (r, \alpha, \vbar, J, t) - \lim_{J \rightarrow \infty} \lambdabar_{W_\mathrm{bt}} (r, \alpha, \vbar, J, t) \right ] \sim \epsilon a
\end{equation}
is the lowest order radial displacement of the particle. We have defined $r_{i, 1, W}$ such that it goes to zero at $J \rightarrow \infty$. This limit corresponds to the surface-filling barely trapped particle with $\mathcal{E} = U_M(r, \mu, t)$ and $W = W_\mathrm{bt}$. Note that $\lambdabar_W (r, \alpha, \vbar, J, t)$ does not depend on $\alpha$ for $J \rightarrow \infty$.

In addition to equation~\refe{eq:eqgi1}, we need the conditions to be imposed in junctures of several types of wells. With the new variables $\vbar$ and $J$, these junctures of different types of wells happen at particular values of $J$, $J = J_{c,W} (r, \alpha, \vbar, t)$, which depend on $r$, $\alpha$, $\vbar$, $t$ and the type of well $W$. Note, for example, the juncture in figure~\ref{fig:exampleU}: the value of $J$ for the juncture is different for each type of well. These values of $J$ are related by 
\begin{equation} \label{eq:Jjuncturecond}
J_{c, I} (r, \alpha, \vbar, t) + J_{c, II} (r, \alpha, \vbar, t) = J_{c, III} (r, \alpha, \vbar, t).
\end{equation}
As we noted in section~\ref{sec:dkequation}, the function $g_{i, W}$ is continuous across the juncture, but the derivatives $\partial_\mathcal{E} g_{i, W}$ and $\partial_\mu g_{i, W}$ are discontinuous. There are two conditions that we use to relate the discontinuous derivatives on different sides of the juncture. On the one hand, the combination $\partial_\mu \mathcal{E}_c \, \partial_\mathcal{E} g_{i, W} + \partial_\mu g_{i, W}$, which is the derivative along the boundary $\mathcal{E} = \mathcal{E}_c (r, \alpha, \mu, t)$, is continuous. On the other hand, the discontinuous derivatives $\partial_\mathcal{E} g_{i, W}$ around the juncture are related to each other by equation~\refe{rho2transition}. In the new variables $\vbar$ and $J$, the derivative along the boundary $J = J_{c, W} (r, \alpha, \vbar, t)$ is $\partial_{\vbar} g_{i, 1, W} + \partial_{\vbar} J_{c, W}\, \partial_J g_{i, 1, W}$, and hence this combination of $\partial_{\vbar} g_{i, 1, W}$ and $\partial_J g_{i, 1, W}$ is continuous across the juncture. To finish our discussion of the junctures, we need to rewrite equation~\refe{rho2transition} in the new coordinates $\vbar$ and $J$. We first note that expression~\refe{rho2transition} vanishes to lowest order in $\epsilon$ because $g_{i,0,W}$ and $h_{i, 0}$ are Maxwellians. Thus, equation~\refe{rho2transition} becomes
\begin{align} \label{eq:eqgi1transitionv1}
 \tau_I & \left [ \langle H_{\mathcal{E} \mathcal{E}}[ f_{Mi} ]\rangle_{\tau, I} - 2  \langle H_{\mathcal{E} \mu}[ f_{Mi} ]\rangle_{\tau, I} \, \partial_\mu \mathcal{E}_c +  \langle H_{\mu \mu}[ f_{Mi} ]\rangle_{\tau, I} \, (\partial_\mu \mathcal{E}_c)^2 \right ] \partial_\mathcal{E} g_{i, 1, I} \nonumber\\ & + \tau_{II} \left [  \langle H_{\mathcal{E} \mathcal{E}}[ f_{Mi} ]\rangle_{\tau, II} - 2  \langle H_{\mathcal{E} \mu}[ f_{Mi} ]\rangle_{\tau, II} \, \partial_\mu \mathcal{E}_c+  \langle H_{\mu \mu}[ f_{Mi} ] \rangle_{\tau, II} \, (\partial_\mu \mathcal{E}_c)^2 \right ] \partial_\mathcal{E} g_{i, 1, II} \nonumber\\ = & \tau_{III} \left [  \langle H_{\mathcal{E} \mathcal{E}}[ f_{Mi} ]\rangle_{\tau, III} - 2 \langle H_{\mathcal{E} \mu}[ f_{Mi}  ]\rangle_{\tau, III} \, \partial_\mu \mathcal{E}_c + \langle H_{\mu \mu} [ f_{Mi} ] \rangle_{\tau, III} \, (\partial_\mu \mathcal{E}_c)^2 \right ] \partial_\mathcal{E} g_{i, 1, III}.
\end{align}
Note that the perturbations due to $H_{pq} [ h_{i,3/2} ]$ and $H_{pq} [ g_{i, 1, W} ]$ can be neglected because they are small in $\epsilon$ --  in the case of $H_{pq} [ g_{i, 1, W} ]$ because the integral that gives $H_{pq} [ g_{i,1,W} ]$ is over a region of velocity space small by $\sqrt{\epsilon}$. Using the definitions~\refe{eq:nupitch} and \refe{eq:nuE}, we find $H_{\mathcal{E}\mathcal{E}} = \nabla_v \mathcal{E} \cdot \nabla_v \nabla_v H [ f_{Mi} ] \cdot \nabla_v \mathcal{E} = v^2\, \partial_{vv}^2 H [ f_{Mi}] = v^4 \nu_{ii,\|}/2\gamma_{ii}$, $H_{\mathcal{E}\mu} = \nabla_v \mathcal{E} \cdot \nabla_v \nabla_v H [ f_{Mi} ] \cdot \nabla_v \mu = (v_\perp^2/B)\, \partial_{vv}^2 H [ f_{Mi}] = v^2 v_\perp^2 \nu_{ii,\|}/2\gamma_{ii} B$ and $H_{\mu\mu} = \nabla_v \mu\cdot \nabla_v \nabla_v H [ f_{Mi} ] \cdot \nabla_v \mu = (v_\|^2 v_\perp^2/v^3 B^2) \, \partial_v H [f_{Mi}] + (v_\perp^4/v^2 B^2) \, \partial_{vv}^2 H [ f_{Mi}] = v_\|^2 v_\perp^2 \nu_{ii, \perp}/4\gamma_{ii} B^2 + v_\perp^4 \nu_{ii,\|}/2\gamma_{ii} B^2$. Moreover, $\mathcal{E}_c (r, \alpha, \mu, t)$ for the example in figure~\ref{fig:exampleU} is given by 
\begin{equation} \label{eq:EcLAR}
\mathcal{E}_c (r, \alpha, \mu, t) = \mu B_{lM} (r, \alpha) + \frac{Z_i e\phi_0 (r, t)}{m_i} + O(\epsilon^{3/2} v_{ti}^2)
\end{equation}
to lowest order in $\epsilon$. Here $B_{lM} (r, \alpha)$ is the local maximum of $B$ on magnetic field line $(r, \alpha)$ on which the juncture occurs. Then, $\partial_\mu \mathcal{E}_c = B_{lM} + O(\epsilon^{3/2} B_0)$. As a result of all these considerations, we find
\begin{align} \label{eq:HcoeffLARv1}
H_{\mathcal{E} \mathcal{E}}[ f_{Mi} ] - & 2  H_{\mathcal{E} \mu}[ f_{Mi} ] \, \partial_\mu \mathcal{E}_c + H_{\mu \mu}[ f_{Mi} ] \, (\partial_\mu \mathcal{E}_c)^2 = \frac{v_\|^2 v_\perp^2 B_{lM}^2 \nu_{ii, \perp}}{4\gamma_{ii} B^2} \nonumber\\ & + \frac{\nu_{ii,\|}}{2\gamma_{ii}} \left ( v_\|^2 + v_\perp^2 \left ( 1 - \frac{B_{lM}}{B} \right ) \right )^2.
\end{align}
Note that $v_\| \sim \sqrt{\epsilon} v_{ti}$ and that $1 - B_{lM}/B \sim \epsilon$. Then, expression~\refe{eq:HcoeffLARv1} simplifies to
\begin{equation}
H_{\mathcal{E} \mathcal{E}}[ f_{Mi} ] - 2  H_{\mathcal{E} \mu}[ f_{Mi} ] \, \partial_\mu \mathcal{E}_c + H_{\mu \mu}[ f_{Mi} ] \, (\partial_\mu \mathcal{E}_c)^2 = \frac{v_\|^2 \vbar^2 \nu_{ii, \perp} (r, \vbar, t)}{4\gamma_{ii}} \left [ 1 + O(\epsilon) \right ].
\end{equation}
With this result and the fact that $\tau_W \langle v_\|^2 \rangle_{\tau, W} = J$, we can simplify equation~\refe{eq:eqgi1transitionv1} to 
\begin{equation} \label{eq:eqgi1transitionv2}
 \frac{\vbar^2 \nu_{ii, \perp}}{4\gamma_{ii}} \left ( J_{c, I} \lim_{J \rightarrow J_{c, I}} \partial_\mathcal{E} g_{i, 1, I} + J_{c, II} \lim_{J \rightarrow J_{c, II}} \partial_\mathcal{E} g_{i, 1, II} \right ) = \frac{\vbar^2 \nu_{ii, \perp}}{4\gamma_{ii}} J_{c, III} \lim_{J \rightarrow J_{c, III}} \partial_\mathcal{E} g_{i, 1, III} .
\end{equation}
To obtain a final expression, we need to rewrite $\partial_\mathcal{E} g_{i, 1, W} (r, \alpha, \mathcal{E}, \mu, t)$ as a linear combination of the derivatives of $g_{i, 1, W} (r, \alpha, \vbar, J, t)$ with respect to $\vbar$ and $J$. Using $\partial_\mathcal{E} g_{i, 1, W} = \partial_\mathcal{E} \vbar \, \partial_{\vbar} g_{i, 1, W} + \partial_\mathcal{E} J_W\, \partial_J g_{i, 1, W}$, $\partial_\mathcal{E} \vbar = \vbar^{-1} \sim v_{ti}^{-1}$, $\partial_\mathcal{E} J_W = \tau_W \sim \epsilon^{-1/2} R/v_{ti}$, $\partial_{\vbar} g_{i,1,W} \sim g_{i,1,W}/v_{ti}$ and $\partial_J g_{i,1,W} \sim \epsilon^{-1/2} g_{i, 1, W}/v_{ti}R$, we can approximate $\partial_\mathcal{E} g_{i, 1, W} \simeq \tau_W\, \partial_J g_{i,1,W}$ to lowest order in $\epsilon$. This approximation might seem to be in contradiction with the fact that the two terms in the combination $\partial_{\vbar} g_{i,1,W} + \partial_{\vbar} J_{c, W}\, \partial_J g_{i, 1, W}$ are of the same order, a property that we have used earlier in this paragraph. Note, however, that the function $J_{c, W} (r, \alpha, \vbar, t)$ does not have a large derivative with respect to $\vbar$ -- indeed, when $\phi_{3/2}$ is neglected, $J_{c, W}$ is simply proportional to $\vbar$, giving $\partial_{\vbar} J_{c, W} \sim \epsilon^{1/2} R$, which should be compared with the scaling with $\epsilon$ of $\partial_\mathcal{E} J_W = \tau_W \sim \epsilon^{-1/2} R/v_{ti}$. Using $\partial_\mathcal{E} g_{i, 1, W} \simeq \tau_W\, \partial_J g_{i,1,W}$, equation~\refe{eq:eqgi1transitionv2} finally becomes
\begin{equation} \label{eq:eqgi1transition}
J_{c, I} \lim_{J \rightarrow J_{c, I}}  \tau_I\, \partial_J g_{i,1,I} + J_{c, II} \lim_{J \rightarrow J_{c, II}}  \tau_{II}\, \partial_J g_{i,1,II} = J_{c, III} \lim_{J \rightarrow J_{c, III}}  \tau_{III}\, \partial_J g_{i,1,III}.
\end{equation}
This equation gives the relationship among the derivatives $\partial_J g_{i, 1, W}$ on different sides of the juncture. The orbit periods $\tau_W$ diverge at junctures because particles spend a logarithmically large time at local maxima of $U$, where the velocity $v_\|$ vanishes. For this reason, we have to consider the discontinuities of the combination $\tau_W\, \partial_J g_{i, 1, W}$ instead of the discontinuities of $\partial_J g_{i, 1, W}$.

Condition~\refe{eq:condJinftyv1} determines the boundary condition for $g_{i, 1, W}$ at $J \rightarrow \infty$. According to expansion~\refe{eq:hiexpansion}, there is no correction to $h_{i,0}$ of order $\epsilon f_{Mi}$, and as a result, the boundary condition for $g_{i, 1, W}$ is
\begin{equation} \label{eq:gi1bc}
\lim_{J \rightarrow \infty} g_{i, 1, W_\mathrm{bt}} (r, \alpha, \vbar, J, t) = 0.
\end{equation}
We discuss this boundary condition in more detail in Appendix~\ref{app:barelypassing}.

\subsubsection{Correction to the passing particle distribution function} \label{subsub:passing32}
To obtain boundary condition~\refe{eq:gi1bc}, we had to use the fact that the first correction to $h_{i,0}$ is $h_{i,3/2}$. Indeed, had there been a correction to the passing particle distribution function of order $\epsilon f_{Mi}$, we could not have imposed condition~\refe{eq:gi1bc}.

We proceed to show that expansion~\refe{eq:hiexpansion} is consistent and that, indeed, the largest correction to $h_{i,0}$ is of order $\epsilon^{3/2} h_{i,0}$. The correction to $h_{i,0}$ can be driven by the integral contribution of the trapped particle distribution function $g_{i, 1, W}$ to the linearized collision operator, by the time derivatives of $n_i$ and $T_i$, by the source $S_i$ and by boundary condition~\refe{eq:diffcondJinftyv1} that requires that the derivatives of $g_{i,W}$ and $h_i$ with respect to $\bv$ are continuous at the trapped-passing boundary. The contribution of $g_{i, 1, W}$ to the integral piece of the linearized collision operator is $C_{ii, I}^{\ell} [ g_{i, 1, W} ] \sim \epsilon^{3/2} \nu_{ii} f_{Mi}$ because $g_{i, 1, W}$ is defined in a region of velocity space small by $\sqrt{\epsilon}$. As a result, along with the time derivatives of $n_i$ and $T_i$ and the source $S_i$, $C^\ell_{ii, I} [ g_{i, 1, W} ]$ drives a piece of $h_i$ that is of order $\epsilon^{3/2} f_{Mi}$, as demonstrated by the expansion of equation~\refe{passingfi0} in $\epsilon \ll 1$,
\begin{equation} \label{eq:eqhi32}
C_{ii}^{\ell} [ h_{i,3/2} ] + \langle C_{ii, I}^{\ell} [ g_{i, 1, W} ] \rangle_\mathrm{fs} + \left [ \frac{\partial_t n_i}{n_i} + \frac{\partial_t T_i}{T_i}  \left ( \frac{m_i \vbar^2}{T_i} - \frac{3}{2} \right ) \right ] f_{Mi} = \langle \overline{S}_i \rangle_\mathrm{fs},
\end{equation}
where we have used $B \simeq B_0$ and the fact that $v_\|$ is independent of $\alpha$ and $l$ for most passing particles. Recall that the time derivatives in equation~\refe{passingfi0} are performed holding $\mathcal{E}$ and $\mu$ fixed, and that $\overline{f}_i^{(0)}$ in the term $\langle (B/|v_\| |) C_{ii} [h_i, \overline{f}_i^{(0)}] \rangle_\mathrm{fs}$ in equation~\refe{passingfi0} includes both the passing and trapped particle distribution functions $h_i$ and $g_{i, W}$. Thus, the integral collisional contribution $\langle C_{ii, I}^{\ell} [ g_{i, 1, W} ] \rangle_\mathrm{fs}$ is a result of the term $\langle (B/|v_\| |) C_{ii} [h_i, \overline{f}_i^{(0)}] \rangle_\mathrm{fs}$ in equation~\refe{passingfi0}.

To finish our discussion of the correction to the passing particle distribution function, we need to consider the boundary condition~\refe{eq:diffcondJinftyv1}. This condition establishes that the derivatives with respect to $\bv$ of the trapped and passing particle distribution functions must be continuous across the trapped-passing boundary. It can also be viewed as a flux continuity condition: the collisional flux driven by the trapped particle distribution function across the trapped-passing boundary must be the flux into the passing particle region, driving a correction to the passing particle distribution. The collisional flux across the trapped-passing boundary $J \rightarrow \infty$ driven by $g_{i, 1, W}$ is, according to the collision operator in equation~\refe{eq:eqgi1}, $- (\vbar^2 \nu_{ii, \perp}/2)\, \tau_W J\, \partial_J g_{i, 1, W}$ for $J \rightarrow \infty$. If this flux were different from zero, it would drive a correction to the passing particle distribution function of order $\epsilon^{1/2} f_{Mi}$. We can obtain this result by imposing continuity of derivatives across the trapped-passing boundary. Due to $v_\| \sim \sqrt{\epsilon} v_{ti}$, $\nabla_v g_{i,1,W} \sim \sqrt{\epsilon} f_{Mi}/v_{ti}$. This gradient must be equal to the gradient of the correction to the passing particle distribution function, $\nabla_v (h_i - f_{Mi}) \sim (h_i - f_{Mi})/v_{ti}$, giving the incorrect estimate $h_i - f_{Mi} \sim \sqrt{\epsilon} f_{Mi}$, as announced. This estimate is invalid because there is no net collisional flux across the trapped-passing boundary due to $g_{i, 1, W}$, a property that we prove below. Moreover, we will see that there is no flux due to $g_{i, 3/2, W}$ and hence the collisional flux across the trapped-passing boundary is due to the gradients in velocity space of $g_{i, 2, W}$. Since $\nabla_v g_{i, 2, W} \sim \epsilon^{3/2} f_{Mi}/v_{ti}$, the correction to the passing particle distribution function is indeed $h_{i, 3/2}$ (see Appendix~\ref{app:barelypassing} for more detail on how the flux continuity condition across the barely passing region leads to this result). These arguments show that boundary condition~\refe{eq:gi1bc} for $g_{i,1, W}$ is justified because the largest correction to the passing particle distribution function is indeed of order $\epsilon^{3/2} f_{Mi}$. We proceed to show that $\tau_W\, J\, \partial_J g_{i, 1, W}$ and $\tau_W\, J\, \partial_J g_{i, 3/2, W}$ vanish at the trapped-passing boundary $J \rightarrow \infty$, that is, that neither $g_{i, 1, W}$ nor $g_{i, 3/2, W}$ drive a collisional flux from the trapped particle region into the passing particle region.

\begin{figure}
\begin{center}
\includegraphics[width=13cm]{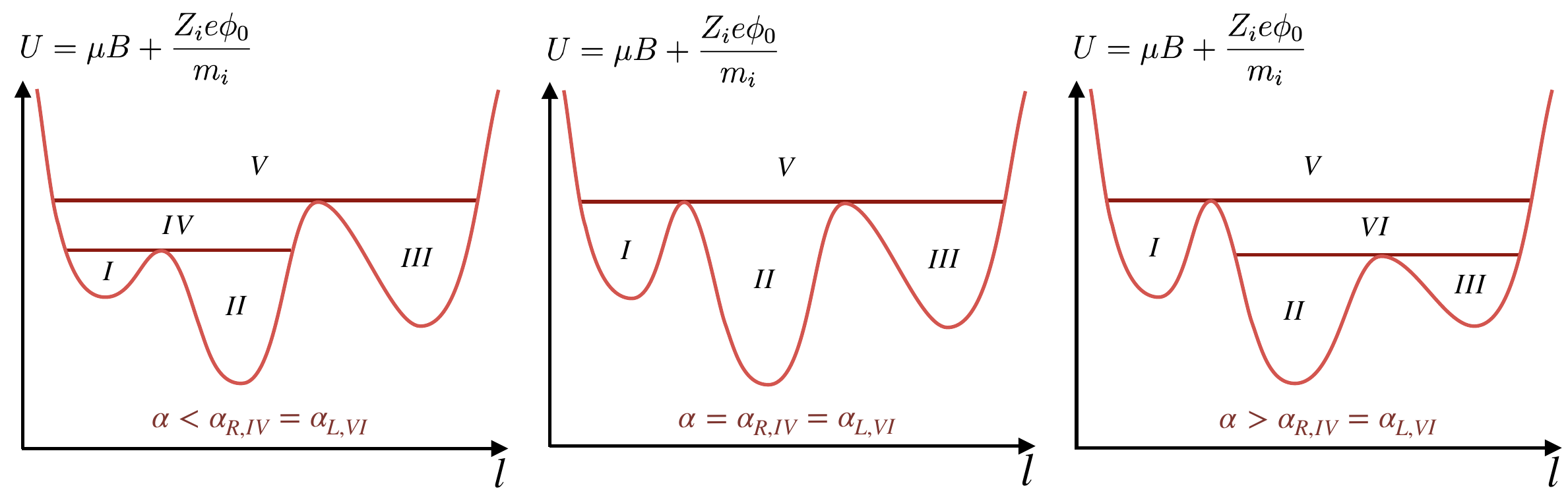}
\end{center}

\caption{\label{fig:alphaLR} Example of limits $\alpha_{L, W}$ and $\alpha_{R, W}$. Well $IV$ disappears at $\alpha = \alpha_{R, IV}$, and at that same value of $\alpha$, a new well $VI$ appears, giving $\alpha_{L, VI} = \alpha_{R, IV}$.}
\end{figure}

We start by showing 
\begin{equation} \label{eq:dJgi1Jinfty}
\lim_{J \rightarrow \infty} \tau_{W_\mathrm{bt}} J \, \partial_{J} g_{i, 1, W_\mathrm{bt}} (r, \alpha, \vbar, J, t) = 0.
\end{equation}
We first integrate equation~\refe{eq:eqgi1} over the region in $J$ where orbits in well $W$ with given values of $\alpha$ and $\vbar$ exist, $J \in [ J_{m, W} (r, \alpha, \vbar, t), J_{M, W} (r, \alpha, \vbar, t)]$,
\begin{align} \label{eq:intdJgi1}
\frac{c \phi_0^\prime}{\Psi_t^\prime}& \partial_\alpha \left [ \int_{J_{m, W}}^{J_{M, W}} \Big ( g_{i, 1, W} + r_{i, 1, W}\, \Upsilon_i f_{Mi} \Big )\, \rmd J \right ] \nonumber\\ & - \frac{c \phi_0^\prime}{\Psi_t^\prime} \partial_\alpha J_{M, W}\, \Big [ g_{i, 1, W}(r, \alpha, \vbar, J_{M, W}, t ) + r_{i, 1, W}(r, \alpha, \vbar, J_{M, W}, t )\, \Upsilon_i f_{Mi} \Big ] \nonumber\\ & + \frac{c \phi_0^\prime}{\Psi_t^\prime} \partial_\alpha J_{m, W}\, \Big [ g_{i, 1, W}(r, \alpha, \vbar, J_{m, W}, t ) +  r_{i, 1, W}(r, \alpha, \vbar, J_{m, W}, t )\, \Upsilon_i f_{Mi} \Big ] \nonumber\\ & - \frac{\vbar^2 \nu_{ii, \perp}}{4} \Big ( J_{M, W} \lim_{J \rightarrow J_{M, W}} \tau_W  \, \partial_{J} g_{i, 1, W} - J_{m, W} \lim_{J \rightarrow J_{m, W}} \tau_W  \, \partial_{J} g_{i, 1, W} \Big ) = 0.
\end{align}
The second and third terms in this equation are included to cancel the derivatives with respect to $\alpha$ of the limits of the integral in the first term. The values of $J_{m, W}$ and $J_{M, W}$ are either $0$, $\infty$ or values at which there is a juncture between different wells. We proceed to integrate equation~\refe{eq:intdJgi1} for all the values of $\alpha$ allowed in well $W$ for a given $\vbar$. The interval $[ \alpha_{L, W}(r, \vbar, t), \alpha_{R, W}(r, \vbar, t) ]$ is the region in $\alpha$ where orbits in well $W$ with a given value of $\vbar$ exist. The limits $\alpha_{L, W}$ and $\alpha_{R, W}$ exist for two different reasons: 
\begin{itemize}
\item either well $W$ closes at $\alpha_{L, W}$ and $\alpha_{R, W}$ because $J_{m, W} = J_{M, W}$, or

\item well $W$ extends to all values of $\alpha$ and hence $\alpha_{L, W} = 0$ and $\alpha_{R, W} = 2\pi$.
\end{itemize}
In figure~\ref{fig:alphaLR}, we give an example in which one of the wells, well $IV$, disappears ($J_{m, IV} = J_{M, IV}$ at $\alpha = \alpha_{R, IV}$) and a new well, well $VI$, appears ($J_{m, VI} \neq J_{M, VI}$ for $\alpha > \alpha_{R, IV}$). Not all cases in which a well disappears are as straightforward as the example in figure~\ref{fig:alphaLR}. We discuss a pathological example in Appendix~\ref{app:problemjuncture}, where we also show that choices can be made such that either $J_{m, W} = J_{M, W}$ or the distribution function is periodic at $\alpha_{L, W}$ and $\alpha_{R, W}$. In either case, integrating equation~\refe{eq:intdJgi1} over $\alpha$, we find
\begin{align} \label{eq:intdalphadJgi1}
- & \frac{c \phi_0^\prime}{\Psi_t^\prime} \int_{\alpha_{L, W}}^{\alpha_{R, W}} \Big [ \partial_\alpha J_{M, W}\, \Big ( g_{i, 1, W}(r, \alpha, \vbar, J_{M, W}, t ) + r_{i, 1, W}(r, \alpha, \vbar, J_{M, W}, t )\, \Upsilon_i f_{Mi} \Big ) \nonumber\\ & \quad - \partial_\alpha J_{m, W}\, \Big ( g_{i, 1, W}(r, \alpha, \vbar, J_{m, W}, t ) + r_{i, 1, W}(r, \alpha, \vbar, J_{m, W}, t ) \, \Upsilon_i f_{Mi} \Big ) \Big ] \, \rmd \alpha \nonumber\\ & \quad- \frac{\vbar^2 \nu_{ii, \perp}}{4} \int_{\alpha_{L, W}}^{\alpha_{R, W}} \Big ( J_{M, W} \lim_{J \rightarrow J_{M, W}} \tau_W  \, \partial_{J} g_{i, 1, W} - J_{m, W} \lim_{J \rightarrow J_{m, W}} \tau_W  \, \partial_{J} g_{i, 1, W} \Big )\, \rmd \alpha \nonumber \\ & = 0.
\end{align}
Summing over all possible well indices $W$, and using the fact that, at the juncture of several wells, $g_{i, 1, W}$ is continuous and $\tau_W\, \partial_J g_{i, 1, W}$ satisfies equation~\refe{eq:eqgi1transition}, several terms cancel and we find the result in equation~\refe{eq:dJgi1Jinfty}. To illustrate the different cancellations that lead to equation~\refe{eq:dJgi1Jinfty}, we consider the juncture of wells $I$, $II$ and $IV$ in figure~\ref{fig:alphaLR}, characterized by $J_{c, I}$, $J_{c, II}$ and $J_{c, IV} = J_{c, I} + J_{c, II}$. For well $I$, we find $J_{m, I} = 0$ and $J_{M, I} = J_{c, I}$. Similarly, for well $II$, we have $J_{m, II} = 0$ and $J_{M, II} = J_{c, II}$. For well $IV$ we only know the minimum value $J_{m, IV} = J_{c, IV}$. In the sum of equation~\refe{eq:intdalphadJgi1} over all wells, the terms proportional to $J_{m, I}$, $J_{m, II}$ and their derivatives with respect to $\alpha$ vanish. The contributions from the juncture of wells $I$, $II$ and $IV$ in figure~\ref{fig:alphaLR} are then
\begin{align} \label{eq:excancel}
&- \frac{c \phi_0^\prime}{\Psi_t^\prime} \Bigg [ \int_{\alpha_{L, I}}^{\alpha_{R, I}} \partial_\alpha J_{c, I}\, \Big ( g_{i, 1, I}(r, \alpha, \vbar, J_{c, I}, t ) +  r_{i, 1, I}(r, \alpha, \vbar, J_{c, I}, t )\, \Upsilon_i f_{Mi} \Big ) \, \rmd \alpha \nonumber \\ & \quad + \int_{\alpha_{L, II}}^{\alpha_{R, II}} \partial_\alpha J_{c, II}\, \Big ( g_{i, 1, II}(r, \alpha, \vbar, J_{c, II}, t ) + r_{i, 1, II}(r, \alpha, \vbar, J_{c, II}, t ) \, \Upsilon_i f_{Mi} \Big ) \, \rmd \alpha \nonumber \\ & \quad - \int_{\alpha_{L, IV}}^{\alpha_{R, IV}} \partial_\alpha J_{c, IV}\, \Big ( g_{i, 1, IV}(r, \alpha, \vbar, J_{c, IV}, t ) + r_{i, 1, IV}(r, \alpha, \vbar, J_{c, IV}, t ) \, \Upsilon_i f_{Mi} \Big ) \, \rmd \alpha \Bigg ] \nonumber\\ &- \frac{\vbar^2 \nu_{ii, \perp}}{4}  \Bigg [ \int_{\alpha_{L, I}}^{\alpha_{R, I}} J_{c, I} \lim_{J \rightarrow J_{c, I}} \tau_I  \, \partial_{J} g_{i, 1, I} \, \rmd \alpha + \int_{\alpha_{L, II}}^{\alpha_{R, II}} J_{c, II} \lim_{J \rightarrow J_{c, II}} \tau_{II}  \, \partial_{J} g_{i, 1, II} \, \rmd \alpha \nonumber \\ & \quad - \int_{\alpha_{L, IV}}^{\alpha_{R, IV}}  J_{c, IV} \lim_{J \rightarrow J_{c, IV}} \tau_{IV}  \, \partial_{J} g_{i, 1, IV} \, \rmd \alpha \Bigg ] + \ldots \nonumber \\ & = 0.
\end{align}
The ellipsis points $\ldots$ here indicates that there are more terms corresponding to other junctures that we have not included in the equation. The first three lines of equation~\refe{eq:excancel} cancel each other because of continuity of $g_{i, 1, W}$ and $r_{i, 1, W}$ across the juncture and the fact that $\partial_\alpha J_{c, I} + \partial_\alpha J_{c, II} = \partial_\alpha J_{c, IV}$. The displacement $r_{i, 1, W}$ is continuous across junctures because each juncture has a single value of $\lambdabar$, $\lambdabar = \lambdabar_c (r, \alpha, \vbar)$. Lines four and five of equation~\refe{eq:excancel} vanish because of condition~\refe{eq:eqgi1transition}.

The correction $g_{i, 3/2, W}$ to the trapped particle distribution function is shown to be independent of $\alpha$ and $J$ in Appendix~\ref{app:Etangential}. Using $\vbar$ instead of $v$ is crucial to obtain this result (see Appendix~\ref{app:Etangential} for more details on the role of $\vbar$ in the derivation). By continuity across the trapped-passing boundary, we also find that
\begin{equation} \label{eq:gi32sol}
g_{i, 3/2, W} (r, \alpha, \vbar, J, t) = h_{i,3/2} (r, \vbar, 0, t).
\end{equation}
Thus, $\tau_W\, J\, \partial_J g_{i, 3/2, W}$ is zero at the trapped-passing boundary. 

\subsubsection{Electric potential $\phi_{3/2}$}
To determine the piece of the electric potential that is not a flux function, we use quasineutrality~\refe{quasineutrality}. To calculate the integral of $\overline{f}_i^{(0)}$ over velocity space, we employ $\vbar \simeq v + Z_i e \phi_{3/2}/m_i v$ to perform a Taylor expansion around $v$, finding
\begin{align} \label{eq:vbarTaylor}
g_{i,W} (r, \alpha, \vbar, J, t) = f_{Mi} (r, v, t) + g_{i, 1, W} (r, \alpha, v, J, t) + O(\epsilon^{3/2} f_{Mi})
\end{align}
and
\begin{equation}
h_i (r, \vbar, \xi, t) = f_{Mi} (r, v, t) + h_{i,3/2} (r, v, \xi, t) - \frac{Z_i e \phi_{3/2}(r, \alpha, l, t)}{T_i(r)} f_{Mi}(r, v, t)  + O(\epsilon^2 f_{Mi}).
\end{equation}
Then, the lowest order quasineutrality equation~\refe{quasineutrality} gives
\begin{equation}
\left ( \frac{e n_e}{T_e} + \frac{Z_i^2 n_i}{T_i} \right ) \phi_{3/2} = Z_i \int \sum_{W \in \mathcal{W}} g_{i, 1, W} \, \rmd^3 v,
\end{equation}
showing that the next order correction to the flux function $\phi_0$ is indeed small by $\epsilon^{3/2}$, as predicted in section~\ref{sec:dkequationJ}. Here, $n_e \simeq \hat{n}_e \exp ( e\phi_0/T_e)$, and $\int g_{i, 1, W} \, \rmd^3 v \sim \epsilon^{3/2} n_i$ because, again, the region of velocity space where $g_{i, 1, W}$ is defined is small by $\sqrt{\epsilon}$. Note that we need not include $\int h_{i,3/2}\, \rmd^3 v$ in the quasineutrality equation because this integral does not depend on $\alpha$ or $l$ to lowest order in $\epsilon$, and hence can be absorbed into the definition of $n_i(r)$.

\subsection{Ion transport equations} \label{sub:transport}
We finish by integrating equations~\refe{rho2orbit} and \refe{passingfi0} to find equations for $n_i$ and $T_i$. Before we integrate, we rewrite the equations in a convenient form. Using equations~\refe{eq:dalphaJ} and \refe{eq:drJ} and employing $Z_i e \langle \partial_t \phi \rangle_{\tau, W}/m_i = - \tau_W^{-1} \partial_t ( \tau_W \langle v_\|^2 \rangle_{\tau, W})$ and $\partial_\mathcal{E} ( \tau_W \langle v_\|^2 \rangle_{\tau, W}) = \tau_W$, equation~\refe{rho2orbit} can be written as
\begin{align} \label{rho2orbitconserv}
\frac{1}{\tau_W} \partial_t  &  \left ( \tau_W g_{i,W} \right )+ \frac{1}{\tau_W} \partial_\mathcal{E} \left ( \frac{Z_i e \tau_W}{m_i} \left \langle \partial_t \phi \right \rangle_{\tau, W} \, g_{i,W} \right ) \nonumber\\ & + \frac{1}{\tau_W} \partial_\alpha \left ( \tau_W \langle (\mathbf{v}_{E}+\mathbf{v}_{Mi}) \cdot \nabla \alpha \rangle_{\tau, W}\, g_{i,W} \right ) \nonumber\\ & + \frac{1}{\tau_W} \partial_r \left ( \tau_W \langle (\mathbf{v}_{E}+\mathbf{v}_{Mi}) \cdot \nabla r  \rangle_{\tau, W}\, g_{i,W} \right ) = \langle C_{ii}[g_{i,W} ,\overline{f}_{i}^{(0)}] \rangle_{\tau, W} + \langle \overline{S}_i \rangle_{\tau, W},
\end{align}
Using $Z_i e \langle (B/|v_\| |)\, \partial_t \phi \rangle_\mathrm{fs}/m_i = - \partial_t \langle B |v_\|| \rangle_\mathrm{fs}$ and $\partial_\mathcal{E} \langle B |v_\|| \rangle_\mathrm{fs} = \langle B/|v_\|| \rangle_\mathrm{fs}$, equation~\refe{passingfi0} can be written as
\begin{equation} \label{passingfi0conserv}
\partial_t \left ( \left \langle \frac{B}{|v_{\parallel}|} \right \rangle_\mathrm{fs} \, h_i \right ) + \partial_\mathcal{E} \left ( \frac{Z_i e}{m_i} \left \langle \frac{B}{|v_{\parallel}|} \partial_t \phi \right \rangle_\mathrm{fs} \, h_i \right ) + \left \langle \frac{B}{|v_{\parallel}|} C_{ii}[h_i, \overline{f}_{i}^{(0)}] \right \rangle_\mathrm{fs} = \left \langle \frac{B}{|v_{\parallel}|} \overline{S}_i \right \rangle_\mathrm{fs}.
\end{equation}

Multiplying equation~\refe{rho2orbitconserv} by $\tau_W$ and equation~\refe{passingfi0conserv} by 1, integrating over $\alpha$, $\mathcal{E}$, $\mu$ and $\varphi$, and summing over both signs of $\sigma$ and over the well index $W$, we find ion particle conservation equation
\begin{equation} \label{eq:contni}
\partial_t n_i + \frac{1}{V^\prime} \partial_r \left ( V^\prime \Gamma_i \right ) = \left \langle \int S_i\, \rmd^3 v \right \rangle_\mathrm{fs},
\end{equation}
where
\begin{equation}
\Gamma_i := 4\pi \left \langle \int_0^\infty \rmd \mu \int_U^{U_M} \rmd \mathcal{E} \, \frac{B}{|v_\| |} \sum_{W \in \mathcal{W}} g_{i,W} \langle (\bv_E + \bv_{Mi} ) \cdot \nabla r \rangle_{\tau, W} \right \rangle_\mathrm{fs}
\end{equation}
is the particle flux. Note that the passing particle piece $h_i$ does not contribute due to property~\refe{eq:avepassingdrift}. Using the expansion in equation~\refe{eq:giexpansion}, the integration variables $v \simeq \vbar$ and $J$ and the lowest order expression for the radial drifts in equation~\refe{eq:vdreps}, and employing equation~\refe{eq:avedalphaB1} to relate the drifts to $r_{i, 1, W}$, defined in equation~\refe{eq:r1def}, the particle flux becomes
\begin{equation} \label{eq:Gammaidef}
\Gamma_i \simeq \frac{2\pi c \phi_0^\prime}{B_0 V^\prime} \int_0^\infty \rmd v \sum_W \int_{\alpha_{L, W}}^{\alpha_{R, W}} \rmd \alpha \int_{J_{m, W}}^{J_{M, W}} \rmd J\, v g_{i, 1, W}\, \partial_\alpha r_{i, 1, W} \sim \epsilon^{5/2} \rho_{i*} n_i v_{ti}.
\end{equation}
To obtain the order of magnitude estimate, we have used $\phi_0^\prime \sim T_i/ea$, $V^\prime \sim Ra$, $J \sim \sqrt{\epsilon} v_{ti} R$ and $r_{i, 1, W} \sim \epsilon a$. The order of magnitude estimate in equation~\refe{eq:Gammaidef} justifies estimates~\refe{eq:Sieps}. 

The estimate for the size of the particle flux in equation~\refe{eq:Gammaidef} can also be obtained from a random walk argument. In the introduction, in equation~\refe{eq:wso}, we argued that trapped particle orbits had a radial width $w \sim \epsilon a$ due to the smallness of the magnetic drift compared to the $\bE \times \bB$ drift, $|\bv_M|/|\bv_E| \sim \epsilon$. These orbits are interrupted by collisions that have an effective collision frequency $\nu_{ii}/\epsilon$, causing a radial random walk with steps of length $w$ every time $\epsilon/\nu_{ii}$. The effective collision frequency $\nu_{ii}/\epsilon$ is larger than $\nu_{ii}$ because small angle collisions can easily detrap particles with $v_\| \sim \sqrt{\epsilon} v_{ti}$ by providing a small amount of parallel momentum. Since we assume $\nu_{i*} \sim \rho_{i*}$, $\nu_{ii}/\epsilon \sim \rho_{i*} v_{ti}/a$. Thus, the random walk diffusion coefficient associated to trapped particle orbits is $D \sim \sqrt{\epsilon} w^2 \rho_{i*} v_{ti}/a$. The factor of $\sqrt{\epsilon}$ in the estimate of the diffusion coefficient is due to the fact that only trapped particles, which are a fraction of order $\sqrt{\epsilon} \ll 1$ of the total number of particles, participate in the diffusion. Using $D \sim \epsilon^{5/2} \rho_{i*} a v_{ti}$, $\Gamma_i \sim D |\nabla n_i| \sim D n_i/a$, we obtain the estimate in equation~\refe{eq:Gammaidef}.

Multiplying equations~\refe{rho2orbitconserv} and \refe{passingfi0conserv} by $m_i \mathcal{E}$ and integrating over velocity space, we find the ion energy conservation equation
\begin{align} \label{eq:energTiv1}
\partial_t \left ( \frac{3}{2} n_i T_i + Z_i e n_i \phi_0 \right ) & + \frac{1}{V^\prime} \partial_r \left [ V^\prime \left ( Q_i + Z_i e \phi_0 \Gamma_i \right ) \right ] = Z_i e n_i \partial_t \phi_0 \nonumber \\ & + \left \langle \int \left ( \frac{1}{2} m_i v^2 + Z_i e \phi_0 \right ) S_i\, \rmd^3 v \right \rangle_\mathrm{fs},
\end{align}
where
\begin{equation} \label{eq:Qidef}
Q_i \simeq \frac{\pi m_i c \phi_0^\prime}{B_0 V^\prime} \int_0^\infty \rmd v  \sum_W \int_{\alpha_{L, W}}^{\alpha_{R, W}} \rmd \alpha \int_{J_{m, W}}^{J_{M, W}} \rmd J\, v^3 g_{i, 1, W}\, \partial_\alpha r_{i, 1, W} \sim \epsilon^{5/2} \rho_{i*} n_i T_i v_{ti}
\end{equation}
is the ion energy flux. Note that, using equation~\refe{eq:contni}, equation~\refe{eq:energTiv1} can also be written as
\begin{equation} \label{eq:energTiv2}
\partial_t \left ( \frac{3}{2} n_i T_i \right ) + \frac{1}{V^\prime} \partial_r \left ( V^\prime Q_i \right ) = - Z_i e \phi_0^\prime \Gamma_i + \left \langle \int \frac{1}{2} m_i v^2\, S_i\, \rmd^3 v \right \rangle_\mathrm{fs}.
\end{equation}

The expressions for the particle and energy fluxes given in equations~\refe{eq:Gammaidef} and \refe{eq:Qidef} would seem to suggest that both fluxes are proportional to the radial electric field $\phi_0^\prime$. This is, however, not the case as $r_{i,1,W}$, defined in equation~\refe{eq:r1def}, is inversely proportional to $\phi_0^\prime$. Thus, $\phi_0^\prime$ only enters in the fluxes through its influence on the correction to the distribution function $g_{i, 1, W}$. 

\subsection{Summary} \label{sub:deeplytrapped}
To summarize, for $\rho_{i*} \sim \nu_{i*}$, the ion distribution function is Maxwellian to lowest order. The density and temperature of this Maxwellian can be calculated using particle and energy conservation equations once the particle and energy fluxes in equations~\refe{eq:Gammaidef} and \refe{eq:Qidef} are obtained. To calculate these fluxes, we must obtain the correction $g_{i, 1, W}$ that is only defined in the trapped particle region. The equation for $g_{i, 1, W}$ is equation~\refe{eq:eqgi1}. This equation must be solved along with boundary condition~\refe{eq:gi1bc} and equation~\refe{eq:eqgi1transition} for the junctures of different types of wells. 

Importantly, due to the presence of the small correction $\phi_{3/2}$ to the potential, we had to use the velocity space coordinates $\vbar$ and $J$, defined in equations~\refe{eq:vbardef} and \refe{eq:Jdef}, and the approximate pitch angle variable $\lambdabar$, defined in equation~\refe{eq:lambdabardef}. These variables were crucial to show that boundary condition~\refe{eq:gi1bc} applies, but once this is done, we can use the approximations $\vbar \simeq v$ and $\lambdabar \simeq \lambda$, where $\lambda$ is defined in equation~\refe{eq:lambdadef}. From here on, we replace $\vbar$ with $v$ and $\lambdabar$ with $\lambda$.

The same set of equations that we have obtained can be derived from the kinetic equation implemented in DKES \citep{hirshman86} in the limit given by $\rho_{i*} \sim \nu_{i*} \ll 1$ and $\epsilon \ll 1$. The procedure to derive equations~\refe{eq:eqgi1}, \refe{eq:eqgi1transition} and \refe{eq:gi1bc} from the DKES kinetic equation is similar to the method described in section~\ref{sec:dkequation} and in this section. We sketch the derivation in Appendix~\ref{app:DKES}. There are three differences with our derivation that are worth mentioning.
\begin{itemize}
\item The DKES equations assume from the start that the lowest order distribution is Maxwellian and that the potential is an exact flux function. 

\item We use the second adiabatic invariant as a variable because it remains constant as the particle moves. However, the DKES kinetic equation does not ensure that the second adiabatic invariant remains constant. Instead, the DKES kinetic equation maintains the quantity
\begin{equation} \label{eq:JDKES}
\hat{J}_W := 2 \int_{l_{bL,W}}^{l_{bR,W}} B |v_\| |\, \rmd l
\end{equation}
constant. In the expansion in $\epsilon \ll 1$, this quantity is approximately proportional to $J$, $\hat{J} \simeq B_0 J$.

\item The perturbation to the passing particle distribution function $h_i$ calculated by DKES is of order $\epsilon^{3/2} f_{Mi}$, but unlike the physical correction $h_{i,3/2}$, determined by equation~\refe{eq:eqhi32}, the DKES correction to the passing particle distribution function is not driven by the time derivatives of density, potential and temperature, the heating and fueling sources, or the integral terms of the collision operator. In DKES, the perturbation to the passing particle distribution function is only driven by the collisional flux from and to the trapped particle region because DKES only evolves the perturbation to the Maxwellian and uses a simple pitch-angle scattering operator without integral contributions. 

\end{itemize}
The three previous differences mean that, even though the DKES kinetic equation leads to the same equations as the full kinetic equation to lowest order in $\rho_{i*} \sim \nu_{i*} \ll 1$ and $\epsilon \ll 1$, the higher order equations given by DKES will be very different from the physical ones. The higher order equations merit further study because they might be important even for small values of $\epsilon$: for example, $g_{i, 1, W}$ is not zero at the trapped-passing boundary, but of order $\sqrt{\epsilon} g_{i, 1, W}$, and this boundary value is determined by the perturbation to the passing particle distribution function, $h_i - f_{Mi}$, that we have neglected. DKES cannot determine $h_i - f_{Mi}$ in the limit given by $\rho_{i*} \sim \nu_{i*} \ll 1$ and $\epsilon \ll 1$.

\begin{figure}
\begin{center}
\includegraphics[width=12cm]{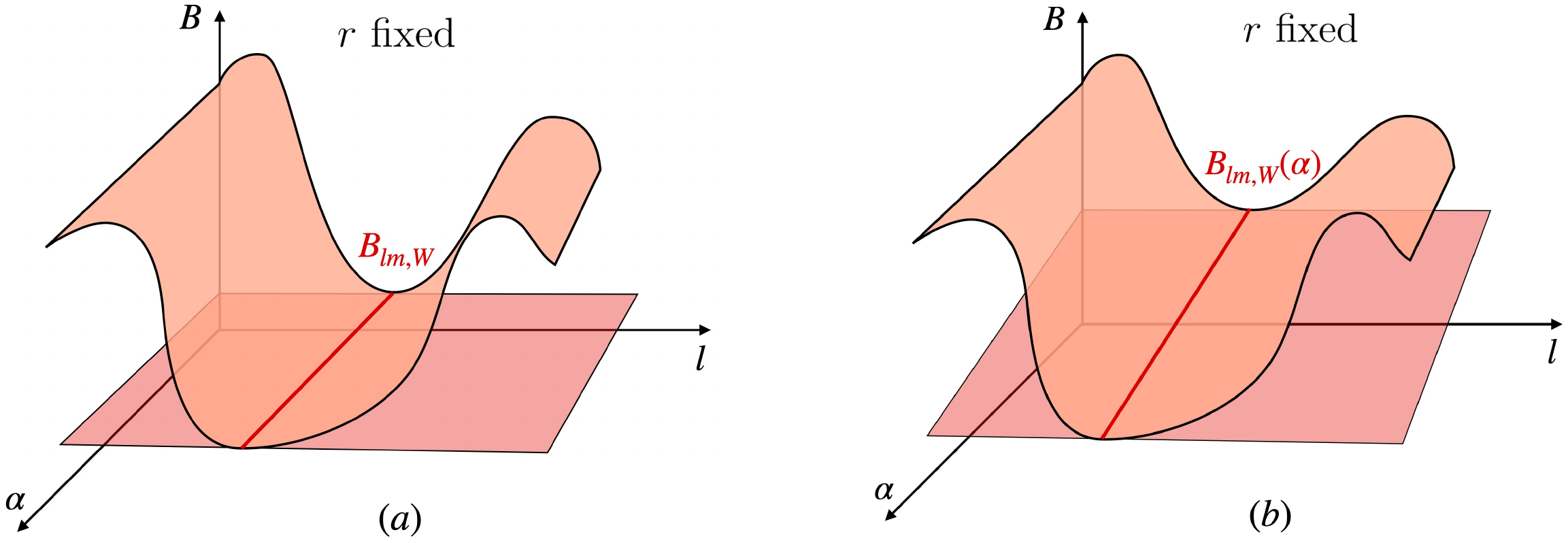}
\end{center}
\caption{(a) A particle deeply trapped in well $W$ can move across field lines along the red line in the particular case in which the local minima of $B$, $B_{lm, W} (r, \alpha)$, are independent of $\alpha$. The local minima of $B$ in omnigeneous magnetic fields behave in this manner. (b) In the more general case in which $B_{lm, W}$ varies with $\alpha$, deeply trapped particles cannot move along the red line to precess around the flux surface unless they are allowed to move radially as well.}
\label{fig:deeplytrapped}
\end{figure}

Finally, to demonstrate the advantages of equations~\refe{eq:eqgi1}, \refe{eq:eqgi1transition} and \refe{eq:gi1bc}, we consider the deeply trapped particles. These particles were singled out as problematic for, for example, the first version of the local orbit-averaged code KNOSOS \citep{velasco20}. The first version of KNOSOS was valid only for stellarators close to omnigeneity \citep{calvo17}, and as a result, the radial displacement of particles is neglected to lowest order in the expansion in closeness to omnigeneity. This approximation is valid for most particles, but fails for deeply and barely trapped particles because these particles cannot move across magnetic field lines within the same flux surface if they do not move simultaneously along $\nabla r$ (across flux surfaces).

We proceed to explain this problem in detail. First, we consider a particle deeply trapped in a given magnetic well $W$ on a given flux surface $r$. The deeply trapped particles have a small $J$ and hence must have a parallel velocity close to zero to keep $J$ constant. Their total energy per unit mass is, to lowest order in $\epsilon$,
\begin{equation} \label{deeplytrapped}
\mathcal{E} = \mu B_{lm, W} (r, \alpha) + \frac{Z_ie}{m_i}\phi_0(r, t),
\end{equation}
where $B_{lm, W}(r, \alpha)$ denotes the local minimum of $B$ in well $W$ along the magnetic field line determined by $r$ and $\alpha$. Since the total energy $\mathcal{E}$ is a constant of the motion, a deeply trapped particle can move across magnetic field lines of the same flux surface $r$ only if the value of $B_{lm, W}$ remains the same while it does so, as in the case represented in figure~\ref{fig:deeplytrapped}(a). In the case of omnigeneity, treated by \cite{cary97a, cary97b} and \cite{parra15a}, $B_{lm, W}$ remains the same for different values of $\alpha$ on a given flux surface. Thus, deeply trapped particles precess around the flux surface without moving radially.

In a generic large aspect ratio stellarator with mirror ratio close to unity, the deeply trapped particles cannot move across magnetic field lines on a given flux surface, as shown on figure~\ref{fig:deeplytrapped}(b). However, since the electric potential $\phi_0$ varies with $r$, it is possible for $v_\parallel$ to be approximately equal to zero after a displacement $\Delta \alpha$ if the particle moves across flux surfaces by a distance $\Delta r$. Using equation \refe{deeplytrapped}, we find that $\Delta \alpha$ and $\Delta r$ must satisfy
\begin{equation} \label{deeplytrapped2}
0=\mu\, \partial_{\alpha}B_{lm, W}(\alpha,r) \Delta \alpha  + \frac{Z_ie \phi_0^\prime(r, t)}{m_i} \Delta r,
\end{equation}
where we have neglected the radial derivative of $B_{lm, W}(r, \alpha)$ because it is small by a factor of $\epsilon$ compared to the radial derivative of $\phi_0$. This is consistent with the velocities $\mathrm{d}\alpha/\mathrm{d}t$ and $\mathrm{d}r / \mathrm{d}t$ obtained from the transit average drifts in equations~\refe{eq:vdreps} and \refe{eq:vdalphaeps},
\begin{equation} \label{dalpha/dt}
\frac{\mathrm{d}\alpha}{\mathrm{d}t} \simeq \frac{c \phi_0^\prime}{\Psi_t^\prime}
\end{equation}
and
\begin{equation} \label{dr/dt}
\frac{\mathrm{d}r}{\mathrm{d}t} \simeq -\frac{m_i c v^2}{2Z_i e B_0  \Psi_t^\prime}\partial_{\alpha}B_{1,lm, W},
\end{equation}
where $B_{1,lm, W}(r, \alpha)$ is the local minimum of $B_1$ in well $W$ along the magnetic field line determined by  $r$ and $\alpha$. Moreover, equation~\refe{deeplytrapped2} is also consistent with the definition of the radial displacement $r_{i, 1, W}$ given in \refe{eq:r1def}. Indeed, to lowest order in $\epsilon$, $\lambda = B_{lm, W}^{-1}$ for deeply trapped particles, and $\lambda \rightarrow B_M^{-1}$ for $J \rightarrow \infty$, giving
\begin{equation} \label{eq:ri1dt}
r_{i, 1, W} \simeq \frac{m_i v^2}{2Z_i e \phi_0^\prime} \frac{B_{1,M} (r) - B_{1, lm, W}(r, \alpha)}{B_0}.
\end{equation}
This equation is a solution to equations~\refe{dalpha/dt} and \refe{dr/dt}, and it shows how the particle moves radially to increase and decrease its electric energy to keep its total energy constant.

\begin{figure}
\centering
\includegraphics[width=7cm]{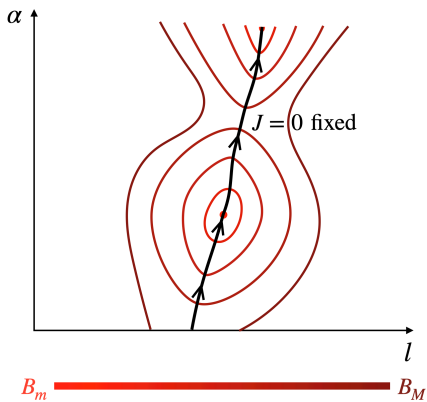}
\caption{Map of the magnetic field magnitude $B$ in an $(l,\alpha)$ plane. In black with arrows, we sketch the path followed by particles with $J=0$ in a stellarator with $\phi_0^\prime > 0$ and $\Psi_t^\prime > 0$. The coordinate $r$ is not constant, and its variation around its average value has the opposite sign to the value of $B_1 = B - B_0$, as shown in equation~\refe{eq:ri1dt}.}
\label{fig:deeplytrappedpath}
\end{figure}

We can draw the path followed by a deeply trapped particle with $J=0$ on an $(l,\alpha)$ plane, sketched in figure~\ref{fig:deeplytrappedpath} for $\phi_0^\prime > 0$ and $\Psi_t^\prime > 0$. The particle moves towards increasing $\alpha$ following the local minima $B_{lm, W}$. Simultaneously, the particle moves back and forth across flux surfaces, increasing $r$ when $\partial_{\alpha}B < 0$ or decreasing $r$ when $\partial_{\alpha}B > 0$. It is interesting to note that
\begin{equation}
\frac{\Delta r}{\Delta \alpha} \sim \epsilon a,
\end{equation}
showing that the oscillating displacement across flux surfaces is much smaller that the precession around it. Equation~\refe{eq:eqgi1} reproduces this motion for deeply trapped particles.

\section{The $1/\nu$ regime in large aspect ratio stellarators with mirror ratios close to unity} \label{sec:1nuregime}

In this section, we briefly consider the limit $\nu_{i*} \gg \rho_{i*}$. In this limit, we can neglect the term $(c \phi_0^\prime/\Psi_t^\prime)\, \partial_\alpha g_{i, 1, W}$ in equation~\refe{eq:eqgi1}, giving to lowest order
\begin{equation} \label{eq:eqgi1nu}
- \frac{v^2 \nu_{ii,\perp}(v)}{4}\, \partial_{J} \left ( \tau_W J \, \partial_{J} g_{i, 1, W} \right ) \simeq - \frac{c \phi_0^\prime}{\Psi_t^\prime} \, \partial_\alpha r_{i, 1, W} \Upsilon_i f_{Mi}.
\end{equation}
This equation can be integrated using boundary condition~\refe{eq:gi1bc} and equation~\refe{eq:eqgi1transition} at the junctures of several types of wells. The final result is a trapped particle distribution function of order $g_{i, 1, W} \sim \epsilon \rho_{i*} f_{Mi}/\nu_{i*}$. Using this result in equations~\refe{eq:Gammaidef} and \refe{eq:Qidef} for the particle and energy fluxes, we obtain
\begin{equation} \label{eq:fluxes1nu}
\Gamma_i \sim \frac{\rho_{i*}}{\nu_{i*}} \epsilon^{5/2} \rho_{i*} n_i v_{ti}, \quad Q_i \sim \frac{\rho_{i*}}{\nu_{i*}} \epsilon^{5/2} \rho_{i*} n_i T_i v_{ti}.
\end{equation}
These are the typical $1/\nu$ regime order-of-magnitude estimates \citep{ho87}. 

Estimates~\refe{eq:fluxes1nu} can be obtained using a random walk argument. For $\nu_{i*} \gg \rho_{i*}$, a typical trapped particle moves radially due to the radial magnetic drift $\bv_{Mi} \cdot \nabla r \sim \epsilon \rho_{i*} v_{ti}$ until a collision detraps it. As a result, particles move a distance $\bv_{Mi} \cdot \nabla r/\nu_\mathrm{eff}$ between collisions, where $\nu_\mathrm{eff} \sim \nu_{ii}/\epsilon$ is the effective collision frequency for trapped particles. After each detrapping collision, a trapped particle becomes passing and moves along the magnetic field line until another low angle collision reduces its parallel velocity sufficiently to trap it in a different well. The magnitude and sign of $\bv_{Mi} \cdot \nabla r$ in the new well will in general be different from the sign and magnitude of the radial drift in the original well because the particle has moved a significant distance along the magnetic field line while being a passing particle. For this reason, the radial displacement of a typical trapped particle is similar to a random walk with a characteristic step size $\bv_{Mi} \cdot \nabla r/\nu_\mathrm{eff} \sim (\rho_{i*}/\nu_{i*}) \epsilon a \ll \epsilon a$ and a characteristic time $1/\nu_\mathrm{eff} \sim \epsilon/\nu_{ii}$. The corresponding diffusion coefficient is $D \sim \sqrt{\epsilon} (\bv_{Mi} \cdot \nabla r/\nu_\mathrm{eff})^2 \nu_\mathrm{eff}$, where $\sqrt{\epsilon}$ is the estimate for the trapped particle fraction. With $D \sim \epsilon^{5/2} \rho_{i*}^2 a v_{ti}/\nu_{i*}$ and using $\Gamma_i \sim D|\nabla n_i| \sim D n_i/a$ and $Q_i \sim n_i D |\nabla T_i| \sim D n_i T_i/a$, we recover estimates~\refe{eq:fluxes1nu}.

\section{The $\nu$ regime in large aspect ratio stellarators with mirror ratios close to unity} \label{sec:nuregime}

In this section, we study the limit $\nu_{i*} \ll \rho_{i*}$. We expand $g_{i, 1, W}$ in $\nu_{i*}/\rho_{i*} \ll 1$, 
\begin{equation}
g_{i, 1, W} = g_{i,1,W}^{\{0\}} + g_{i,1,W}^{\{1\}} + \dots,
\end{equation}
with $g_{i,1,W}^{\{n\}} \sim (\nu_{i*}/\rho_{i*})^n \epsilon f_{Mi}$. We proceed to expand equation~\refe{eq:eqgi1} in $\nu_{i*}/\rho_{i*} \ll 1$. In subsection~\ref{sub:nu0}, we solve the lowest order version of equation~\refe{eq:eqgi1} and we show that the lowest order distribution function does not give rise to radial fluxes of particles or energy. In subsection~\ref{sub:nu1}, we go to next order in $\nu_{i*}/\rho_{i*} \ll 1$. Finally, in subsection~\ref{sub:GammaQnu}, we calculate the particle and energy fluxes.

\subsection{Lowest order distribution function} \label{sub:nu0}

Equation \refe{eq:eqgi1} becomes, to lowest order in $\nu_{i*}/\rho_{i*} \ll 1$,
\begin{equation} \label{eq:eqginu0}
\frac{c\phi_0^\prime}{\Psi_t^\prime} \partial_{\alpha} g_{i,1,W}^{\{0\}} = - \frac{c\phi_0^\prime}{\Psi_t^\prime} \partial_\alpha r_{i, 1, W}\, \Upsilon_i f_{Mi}.
\end{equation}
Using the fact that $\Upsilon_i f_{Mi}$ does not vary with $\alpha$, we obtain the expression 
\begin{equation} \label{eq:ginu0}
g_{i,1,W}^{\{0\}} (r, \alpha, v, J, t) = - r_{i, 1, W} \Upsilon_i f_{Mi} + K_{i, W} (r, v, J, t),
\end{equation}
where the function $K_{i, W}$ does not depend on $\alpha$.

Due to the existence of junctures between different types of wells such as the one sketched in figure~\ref{fig:exampleU}, the function $K_{i, W}$ can be independent of $J$ in large regions of velocity space. Before we explain why, we need to consider what happens with well junctures in the limit  $\nu_{i*}/\rho_{i*} \ll 1$. In this limit, it is in general impossible to impose continuity of $g_{i,1,W}$ across the juncture, or condition~\refe{eq:eqgi1transition} that relates the derivatives $\partial_J g_{i, 1, W} \simeq \partial_J g_{i,1,W}^{\{0\} }$ on different sides of a juncture to each other. This is hardly surprising since continuity of $g_{i,1,W}$ and condition~\refe{eq:eqgi1transition} are a result of collisions, and we have neglected collisions to lowest order. In reality, continuity and condition~\refe{eq:eqgi1transition} are not satisfied in appearance only, because boundary layers where collisions become important can form around well junctures.    

The rest of this subsection is split into three parts. In the first one, we show that collisional boundary layers only form around certain junctures, and we find a condition that $K_{i, W}$ must satisfy when these boundary layers form. In the second part, we study an example that illustrates the shape of $K_{i, W}$. We finish with a general discussion.

\subsubsection{Junctures of different types of wells for very small collision frequencies}

Not all well junctures are problematic and need a boundary layer. Whether a juncture has a boundary layer or not depends on the direction of the velocity in $\alpha$, $c\phi_0^\prime/\Psi_t^\prime$, and the derivatives of the juncture coordinates $J_{c, W} (r, \alpha, v)$ with respect to $\alpha$. To explain this further, we consider the juncture in figure~\ref{fig:exampleU}. Some particles in well $I$ leave this well if the time derivative of $J_{c, I} (r, \alpha, v)$ is negative. Indeed, if a particle with $J = J_{c, I}(r(t), \alpha(t), v(t))$ at time $t$ were to stay in well $I$, it would have to do so keeping its second adiabatic invariant constant. However, such a particle would find itself with $J > J_{c, I}(r(t+\Delta t), \alpha(t + \Delta t), v(t + \Delta t))$ at $t + \Delta t$. Since there are no possible orbits with $J > J_{c, I}$ in well $I$, the particle must have moved into well $II$ or well $III$, and in doing so, the value of $J$ of the particle has changed abruptly. This is consistent with $J$ being an adiabatic invariant: it is only constant when the motion is a slowly changing periodic orbit, a description that does not apply to a transition from one well to another. These transitions between different types of wells can be understood to be a result of the conservation of phase-space volume. The second adiabatic invariant is the phase-space volume inside a trapped orbit. When $J_{c, I}$ becomes lower than the $J$ of a particle, the phase-space volume of the particle orbit cannot be contained inside well $I$ and the particle ``spills over" into wells $II$ and $III$ \citep{dobrott71, cary86}. 

Equivalently to particles moving out of well $I$ when the time derivative of $J_{c, I}$ is negative, some particles move into well $I$ if the time derivative of $J_{c, I}$ is positive. As we will see shortly, the fact that the time derivative of $J_{c, I}$ is positive implies that particles are leaving well $II$, well $III$ or both, and since these particles must move into another well, it is intuitive that conservation of phase space volume will force these particles to move to a well where there is ``space". Well $I$, with a positive time derivative of $J_{c, I}$, is such a well because at time $t$ all particles in that well have $J \leq J_{c, I} (r(t), \alpha(t), v(t))$, and hence, if no particle were to move into well $I$, at time $t + \Delta t$ there would be no particles that have second adiabatic invariant between $J_{c, I} (r(t), \alpha(t), v(t))$ and $J_{c, I} (r(t + \Delta t), \alpha(t + \Delta t), v(t + \Delta t))$.

To summarize, some particle move out of well $I$ if the time derivative of $J_{c,I}$ is negative and some particles move into well $I$ if the time derivative of $J_{c, I}$ is positive. Similarly, some particles transition out of well $II$ if the time derivative of $J_{c, II}$ is negative, and some transition into it if the time derivative of $J_{c, II}$ is positive. The signs reverse for well $III$ because the orbits in this well have $J$ larger than $J_{c, III}$. Thus, particles move out of well $III$ if the time derivative of $J_{c, III}$ is positive, and transition into it if the time derivative of $J_{c, III}$ is negative.

Since both $r$ and $v$ have time derivatives that are small in $\epsilon$, the time derivative of $J_{c, W} (r, \alpha, v)$ is simply
\begin{equation} \label{eq:dDJdt}
\frac{\rmd J_{c, W}}{\rmd t} \simeq \bv_E \cdot \nabla \alpha\, \partial_\alpha J_{c, W} \simeq \frac{c\phi_0^\prime}{\Psi_t^\prime}\, \partial_\alpha J_{c, W}.
\end{equation} 
Importantly, condition~\refe{eq:Jjuncturecond} imposes that $\partial_\alpha J_{c, I} + \partial_\alpha J_{c, II} = \partial_\alpha J_{c, III}$. With this equality, it is easy to check that, if some particles are moving into well $I$, i.e. $(c\phi_0^\prime/\Psi_t^\prime)\, \partial_\alpha J_{c, I} > 0$, some particles must be leaving well $II$, $(c\phi_0^\prime/\Psi_t^\prime)\, \partial_\alpha J_{c, II} < 0$, well $III$, $(c\phi_0^\prime/\Psi_t^\prime)\, \partial_\alpha J_{c, III} > 0$, or both. It cannot be the case that particles are moving into all three wells, or leaving all three wells.

Armed with the results above, we can start discussing the collisional boundary layers. There are two distinct cases to consider:
\begin{itemize}
\item the signs of $c\phi_0^\prime/\Psi_t^\prime$ and $\partial_\alpha J_{c, W}$ are such that some particles leave one of the wells and move into the other two wells; in this case, there is no boundary layer around the juncture; or

\item the signs of $c\phi_0^\prime/\Psi_t^\prime$ and $\partial_\alpha J_{c, W}$ are such that some particles leave two of the wells and move into the third well; in this case, a thin boundary layer forms around the juncture.
\end{itemize}

To illustrate what happens in the case that some particle leave one of the wells and enter the other two, we consider the following example. With the sign choices $(c\phi_0^\prime/\Psi_t^\prime)\, \partial_\alpha J_{c, I} < 0$, $(c\phi_0^\prime/\Psi_t^\prime)\, \partial_\alpha J_{c, II} > 0$ and $(c\phi_0^\prime/\Psi_t^\prime)\, \partial_\alpha J_{c, III} < 0$, some particles leave well $I$ and enter well $II$ and $III$. For small collisionality, kinetic equation \refe{eq:eqginu0} imposes that the piece of the distribution function $K_{i, W}$ is continuous along particle trajectories. Since particles are leaving well $I$ and moving into well $II$ and well $III$, the distribution function in wells $II$ and $III$ must be the same as in well $I$, that is, $K_{i,II} = K_{i,I}$ and $K_{i,III} = K_{i,I}$. Thus, there is no need for a boundary layer because the distribution is continuous. Note that this process is irreversible: if we reverse time, particles leave wells $II$ and $III$ and move into well $I$, and we will see shortly that this leads, in principle, to discontinuities in the distribution function. In our equations, collisions are the only element that introduces irreversibility, so it is surprising that they do not seem to play a role in the discussion. In reality, collisions have led to this solution indirectly. If one assumes that, due to some unknown mechanism, more particles move into well $II$ than into well $III$, giving $K_{i, II} > K_{i, III}$, the distribution function is discontinuous across the juncture. Fokker-Planck collisions impose continuity in the distribution function, and as a result we need a boundary layer around such a hypothetical juncture. Appendix~\ref{app:boundarylayer} shows that it is impossible to construct such a boundary layer, a result that implies that the distribution function $K_{i, W}$ must be continuous across junctures where particles leave one well to move into the other two wells.  

We proceed to consider a juncture in which particles leave two wells and enter into the third well. For example, consider the situation that arises if the signs of $\phi_0^\prime$, $\Psi_t^\prime$ and the derivatives of $J_{c,I}$, $J_{c,II}$ and $J_{c,III}$ with respect to $\alpha$ imply that particles leave wells $I$ and $II$ to enter well $III$. In this case, in general, $K_{i,I}$ and $K_{i,II}$ are different and we need to obtain a value for $K_{i,III}$. One can think of this problem as particles flowing into the juncture from wells $I$ and $II$, mixing inside the juncture, and leaving through well $III$ with a new distribution function $K_{i, III}$. A collisional boundary layer, described in Appendix~\ref{app:boundarylayer}, forms around any juncture in which particles move out of two wells and into the third. The result of this boundary layer is simply a balance among the phase-space fluxes of particles across the juncture. The infinitesimal element of phase-space volume integrated over the gyrophase and $l$ can be constructed from equations~\refe{eq:d3x} and \refe{eq:d3vvJvarphi}, and it is $(2\pi \Psi_t^\prime v/B_0)\, \rmd r\, \rmd \alpha\, \rmd v \, \rmd J$ to lowest order in $\epsilon$. With this phase space volume element, we can calculate the number of particles crossing $J = J_{c,W}$. One first replaces the variable $J$ with $\Delta J_W := J - J_{c, W}$. With this new set of coordinates, the infinitesimal phase space volume becomes $(2\pi \Psi_t^\prime v/B_0)\, \rmd r\, \rmd \alpha\, \rmd v \, \rmd \Delta J_W$. Then, the rate at which phase space volume crosses the boundary $J = J_{c, W}$, equivalent to $\Delta J_W = 0$, is
\begin{equation} \label{eq:junctfluxnu}
\frac{2\pi \Psi_t^\prime v}{B_0}\frac{\rmd \Delta J_W}{\rmd t}\, \rmd r\, \rmd \alpha\, \rmd v = - \frac{2\pi c \phi_0^\prime v}{B_0} \partial_\alpha J_{c, W}\, \rmd r\, \rmd \alpha\, \rmd v,
\end{equation}
where we have used equation~\refe{eq:dDJdt} and the fact that the time derivative of $J$ is zero to obtain $\rmd \Delta J_W/\rmd t \simeq - (c \phi_0^\prime/\Psi_t^\prime) \, \partial_\alpha J_{c, W}$. Using equation~\refe{eq:junctfluxnu}, we find that the number of particles crossing the boundary is $- (2\pi c \phi_0^\prime v g_{i,1,W}^{\{ 0 \} }/B_0)\, \partial_\alpha J_{c, W} \, \rmd r\, \rmd \alpha\, \rmd v$. Balance between the three fluxes leaving and entering the juncture in figure~\ref{fig:exampleU} gives
\begin{align}
- \frac{2\pi c \phi_0^\prime v}{B_0} &\, g_{i,1,I}^{\{ 0 \} }\, \partial_\alpha J_{c, I} \, \rmd r\, \rmd \alpha\, \rmd v - \frac{2\pi c \phi_0^\prime v}{B_0} \, g_{i,1,II}^{\{ 0 \} }\, \partial_\alpha J_{c, II} \, \rmd r\, \rmd \alpha\, \rmd v \nonumber\\ & = - \frac{2\pi c \phi_0^\prime v}{B_0} \, g_{i,1,III}^{\{ 0 \} }\, \partial_\alpha J_{c, III} \, \rmd r\, \rmd \alpha\, \rmd v.
\end{align}
We can simplify this expression by dividing by $- (2\pi c \phi_0^\prime v/B_0)\, \rmd r\, \rmd \alpha\, \rmd v$. Furthermore, using equation~\refe{eq:ginu0}, we find the relationship
\begin{align}
\left ( K_{i, I} - r_{i, 1, I}\, \Upsilon_i f_{Mi} \right ) &\, \partial_\alpha J_{c, I} + \left ( K_{i, II} - r_{i, 1, II}\, \Upsilon_i f_{Mi} \right )\, \partial_\alpha J_{c, II} \nonumber\\ & =\left ( K_{i, III} - r_{i, 1, III}\, \Upsilon_i f_{Mi} \right )\, \partial_\alpha J_{c, III}.
\end{align}
The definition $r_{i, 1, W}$ in equation~\refe{eq:r1def} implies that $r_{i, I} = r_{i, II} = r_{i, III}$ at the juncture because all the particles on the juncture have the same value of $\lambdabar \simeq \lambda$. With this result and the fact that condition~\refe{eq:Jjuncturecond} gives $\partial_\alpha J_{c, I} + \partial_\alpha J_{c, II} = \partial_\alpha J_{c, III}$, all the terms that contain $r_{i, 1, W}$ cancel each other, leaving
\begin{equation} \label{eq:mixrule}
K_{i, I}\, \partial_\alpha J_{c, I} + K_{i,II}\, \partial_\alpha J_{c,II} = K_{i, III} \, \partial_\alpha J_{c, III}.
\end{equation}
Equation~\refe{eq:mixrule} can be used to calculate $K_{i,III}$ given $K_{i,I}$ and $K_{i,II}$. Note that, due to condition~\refe{eq:Jjuncturecond}, $K_{i,III}$ is equal to both $K_{i,I}$ and $K_{i,II}$ if $K_{i,I} = K_{i,II}$. 

The discussion in this section is related to the probabilities of transition calculated by \cite{cary86}. If instead of the particle distribution function, one considers individual particle motion across a juncture, the problem with junctures is reversed. When a particle leaves one well to move into two other wells, the particle cannot be split into two. The exact position of the trapped particle along its quasi-periodic motion (that is, its phase) when it reaches the juncture determines the well that it moves into. Orbit-averaged motion ignores this phase, but one can assume that particles are equally distributed along this phase and, using this assumption, obtain the probability of transitioning into one well or the other. Following \cite{cary86}, if a particle leaves well $I$, $(c\phi_0^\prime/\Psi_t^\prime)\, \partial_\alpha J_{c, I} < 0$, to move into either well $II$, $(c\phi_0^\prime/\Psi_t^\prime)\, \partial_\alpha J_{c, II} > 0$, or well $III$, $(c\phi_0^\prime/\Psi_t^\prime)\, \partial_\alpha J_{c, III} < 0$, the probability that it moves into well $II$ is $- \partial_\alpha J_{c, II}/\partial_\alpha J_{c, I}$, whereas the probability that it transitions to well $III$ is $\partial_\alpha J_{c, III}/\partial_\alpha J_{c, I}$. Thus, for single particle motion, the derivatives of $J_{c, W}$ with respect to $\alpha$ enter in the problem, but they apply to the junctures in which the particle leaves one well to move into one of the other two wells, instead of applying to the case in which particles leave two wells to transition into the third well. In this latter type of juncture, a single particle leaving one of the wells does not have any other choice but to move into the well that is accepting particles. To summarize, our formulation does not require transition probabilities because it is based on the distribution function instead of single particle motion and because, by keeping collisions, it develops the collisional boundary layers described in Appendix~\ref{app:boundarylayer}.

\subsubsection{Example} \label{subsub:nuexample}

\begin{figure}
\begin{center}
\includegraphics[width=12cm]{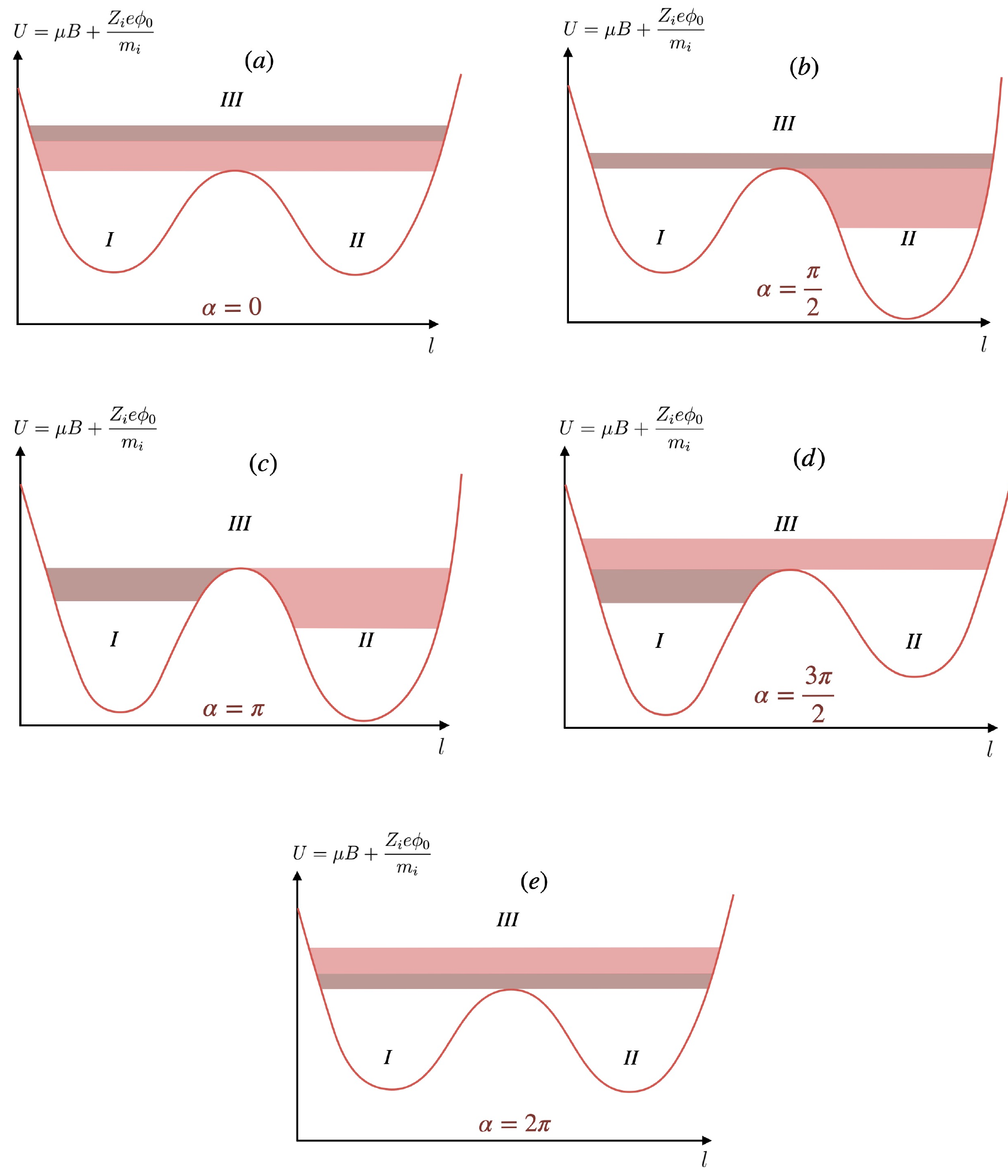}
\end{center}
\caption{\label{fig:configurationwell1} Example of effective potential $U(\alpha, l)$. Each panel corresponds to a particular value of $\alpha$.}
\end{figure}

After explaining how the distribution function behaves around junctures of different types of wells, we proceed to argue that the function $K_{i, W}$ is independent of $J$ in large regions of velocity space. To illustrate the problem, we consider the simple situation sketched in figure~\ref{fig:configurationwell1}. In this figure, we show the dependence of the effective potential $U \simeq \mu B + Z_i e \phi_0/m_i$ on $l$ for several values of $\alpha$. We consider particle motion in this effective potential for a radial electric field that forces particles to move in the direction of increasing $\alpha$, that is, $c\phi_0^\prime/\Psi_t^\prime > 0$. Due to the changes in the $U$ profile with $\alpha$, we can divide the motion into four steps: 
\begin{enumerate}
\item Step (a) $\to$ (b). The minimum of well $II$ decreases with $\alpha$, while well $I$ does not change. Thus, $\partial_\alpha J_{c, II} > 0$ and $\partial_\alpha J_{c, III} > 0$, but $\partial_\alpha J_{c, I} = 0$. Since $c\phi_0^\prime/\Psi_t^\prime > 0$, the discussion of the previous section implies that particles in well $III$ with low values of $J$ transition into well $II$. Particles cannot transition into well $I$ because $\partial_\alpha J_{c, I} = 0$.
\item Step (b) $\to$ (c). The minimum of well $I$ decreases, while well $II$ does not change. Hence, $\partial_\alpha J_{c, I} > 0$, $\partial_\alpha J_{c, II} = 0$ and $\partial_\alpha J_{c, III} > 0$, and particles in well $III$ with low values of $J$ transition into well $I$. Particles cannot transition into well $II$ because $\partial_\alpha J_{c, II} = 0$.
\item Step (c) $\to$ (d). The minimum of well $II$ increases until it reaches the value it had in (a), while well $I$ does not change. As a result, $\partial_\alpha J_{c, I} = 0$, $\partial_\alpha J_{c, II} < 0$ and $\partial_\alpha J_{c, III} < 0$, and particles that had transitioned into well $III$ during step (i) go back into well $III$, but their value of $J$ is greater than their initial one. Particles cannot transition into well $I$ because $\partial_\alpha J_{c, I} = 0$.
\item Step (d) $\to$ (e). The minimum of well $I$ increases until it reaches the value it had in (a), while well $II$ does not change. Then, $\partial_\alpha J_{c, I} < 0$, $\partial_\alpha J_{c, II} = 0$ and $\partial_\alpha J_{c, III} < 0$, and particles that had transitioned into well $I$ during step (ii) go back into well $III$, but their value of $J$ is lower than their initial one. Particles cannot transition into well $II$ because $\partial_\alpha J_{c, II} = 0$.
\end{enumerate}

\begin{figure}
\begin{center}
\includegraphics[width=11cm]{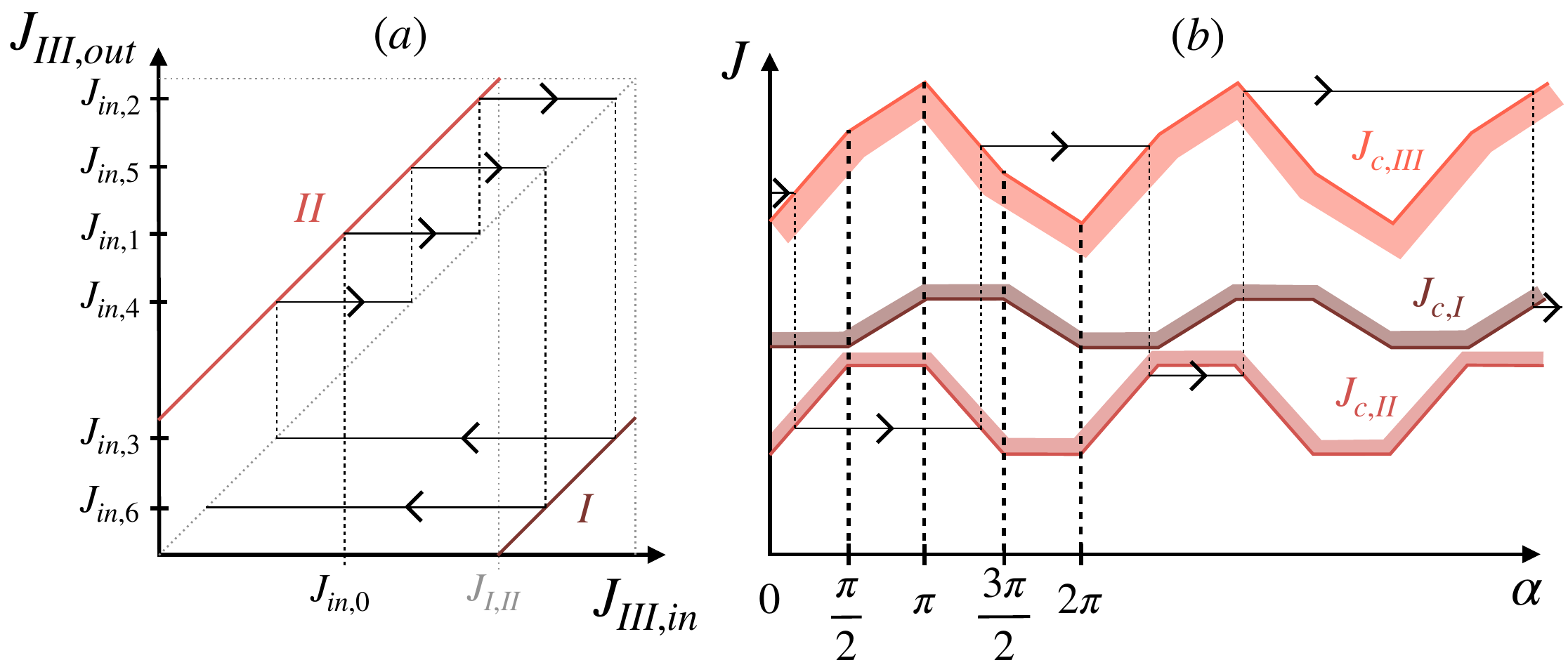}
\end{center}
\caption{\label{fig:examplewells} Evolution of the value of $J$ of a particle in the magnetic field sketched in figure~\ref{fig:configurationwell1}. We start with a particle in well $III$ with initial value of the second adiabatic invariant $J = J_{\mathrm{in},0}$. Figure (a) shows the second adiabatic invariant of particles with $J = J_{III, \mathrm{in}}$ after these particles has transition into well $I$ (dark pink straight line) or well $II$ (light pink straight line), and then back to well $III$. The trajectory of the particle with initial second adiabatic invariant $J_{\mathrm{in}, 0}$ is represented by the black staircase-like line. Figure (b) shows the initial part of the same trajectory in the plane $(\alpha, J)$. The pink lines are $J_{c, I} (\alpha)$, $J_{c, II} (\alpha)$ and $J_{c, III} (\alpha)$.}
\end{figure}

In the configuration in figure~\ref{fig:configurationwell1}, there is only one juncture between wells $I$, $II$ and $III$. This juncture is characterized by the functions $J_{c, I} (r, \alpha, v)$, $J_{c, II} (r, \alpha, v)$ and $J_{c, III} (r, \alpha, v)$, sketched in figure~\ref{fig:examplewells}(b). Each of these functions has a maximum and a minimum in $\alpha$ that we denote $J_{c, W, M} (r, v)$ and $J_{c, W, m} (r, v)$, respectively. Particles in well $I$ with values of $J$ smaller than $J_{c, I, m} (r, v)$ never transition to wells $II$ or $III$. Similarly, particles in well $II$ with $J$ smaller than $J_{c, II, m}(r, v)$ and particles in well $III$ with $J$ larger than $J_{c, III, M} (r, v)$ do not transition into the other two wells. In these regions of velocity space, the dependence of $K_{i, W}$ on $J$ will be determined in section~\ref{sub:nu1}. 

The situation is very different for particles in well $III$ that have $J_{c, III, m} (r, v) < J < J_{c, III, M} (r, v)$. These particles transition into wells $I$ and $II$ and eventually transition back into well $III$. There is a value of $J$, $J_{I, II} (r, v) := J_{c, I, m} (r, v) + J_{c, II, M}(r, v)$, such that particles in well $III$ with $J_{c, III, m} < J < J_{I, II}$ transition into well $II$, and particles with $J_{I, II} < J < J_{c, III, M}$ transition into well $I$. 

To understand what happens for particles with $J_{c, III, m} < J < J_{c, III, M}$, we define the function $J_{III, \mathrm{out}} (J_{III, \mathrm{in}})$. This function determines the value of the second adiabatic invariant of a particle in well $III$ after it has transitioned into well $I$ or well $II$ and has then transitioned back to well $III$. The value $J_{III, \mathrm{out}}$ of the adiabatic invariant after the particle has transition in and out of well $I$ or well $II$ depends only on the value that the adiabatic invariant had before the particle transitioned, $J_{III, \mathrm{in}}$. For the magnetic field configuration in figure~\ref{fig:configurationwell1}, the function $J_{III, \mathrm{out}} (J_{III, \mathrm{in}})$ is sketched in figure~\ref{fig:examplewells}(a) as a light pink straight line for particles that transition into well $II$ ($J_{c, III, m} < J < J_{I, II}$), and as a dark pink straight line for particles that transition into well $I$ ($J_{I, II} < J < J_{c, III, M}$). To sketch the function $J_{III, \mathrm{out}} (J_{III, \mathrm{in}})$, we have used the fact that
\begin{equation}
J_{III, \mathrm{out}} (J_{III, \mathrm{in}}) = \left \{
\begin{array}{l l}
J_{III, \mathrm{in}} + J_{c, I, M} - J_{c, I, m} & \text{for }J_{c, III, m} < J < J_{I, II}, \\
J_{III, \mathrm{in}} - J_{c, II, M} + J_{c, II, m} & \text{for }J_{I, II} < J < J_{c, III, M}.
\end{array}
\right.
\end{equation}

Using the function $J_{III, \mathrm{out}} (J_{III, \mathrm{in}})$, we can schematically describe particle motion for particles with $J_{c, III, m} < J < J_{c, III, M}$. This motion is shown in figure~\ref{fig:examplewells}(a) as a black staircase-like line with arrows. We start with a particle with initial second adiabatic invariant $J_{\mathrm{in, 0}} < J_{I, II}$. After its transitions in and out of well $II$, this particle has acquired the second adiabatic invariant $J_{III, \mathrm{out}} (J_{\mathrm{in}, 0})$, which is necessarily in the interval $[J_{c, III, m}, J_{c, III, M}]$. As a result, the particle will transition again into well $I$ or well $II$. This fact is shown in figure~\ref{fig:examplewells}(a) by noting that $J_{III, \mathrm{out}} (J_{\mathrm{in}, 0})$ becomes $J_{\mathrm{in}, 1}$, and this particle in turn will transition in and out of well $II$ to acquire the second adiabatic invariant $J_{III, \mathrm{out}} (J_{\mathrm{in}, 1}) = J_{III, \mathrm{out}} (J_{III, \mathrm{out}} (J_{\mathrm{in}, 0}))$. In figure~\ref{fig:examplewells}(a), the fact that $J_{\mathrm{in}, 1} = J_{III, \mathrm{out}} (J_{\mathrm{in}, 0})$ is represented as a horizontal solid line that joins $J_{III, \mathrm{out}} (J_{\mathrm{in}, 0})$ with the diagonal line $J_{III, \mathrm{out}} = J_{III, \mathrm{in}}$, shown as a dotted line, and the fact that $J_{\mathrm{in}, 1}$ becomes $J_{III, \mathrm{out}} (J_{\mathrm{in}, 1})$ is represented as a vertical dashed line from the diagonal to $J_{III, \mathrm{out}} (J_{\mathrm{in}, 1})$. Thus, in the $J_{III, \mathrm{out}}$ vs $J_{III, \mathrm{in}}$ graph, particle trajectories are the staircase-like lines sketched in figure~\ref{fig:examplewells}(a). The initial part of the same trajectory is sketched in figure~\ref{fig:examplewells}(b) in the $(\alpha, J)$ plane.

\begin{figure}
\begin{center}
\includegraphics[width=6cm]{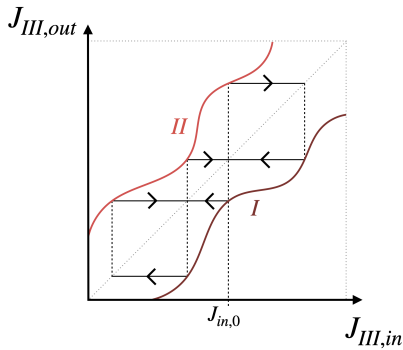}
\end{center}
\caption{\label{fig:specialwells} Function $J_{III, \mathrm{out}}(J_{III, \mathrm{in}})$ for which particle trajectories describe a loop and do not sample a finite interval of $J$.}
\end{figure}

\begin{figure}
\begin{center}
\includegraphics[width=9cm]{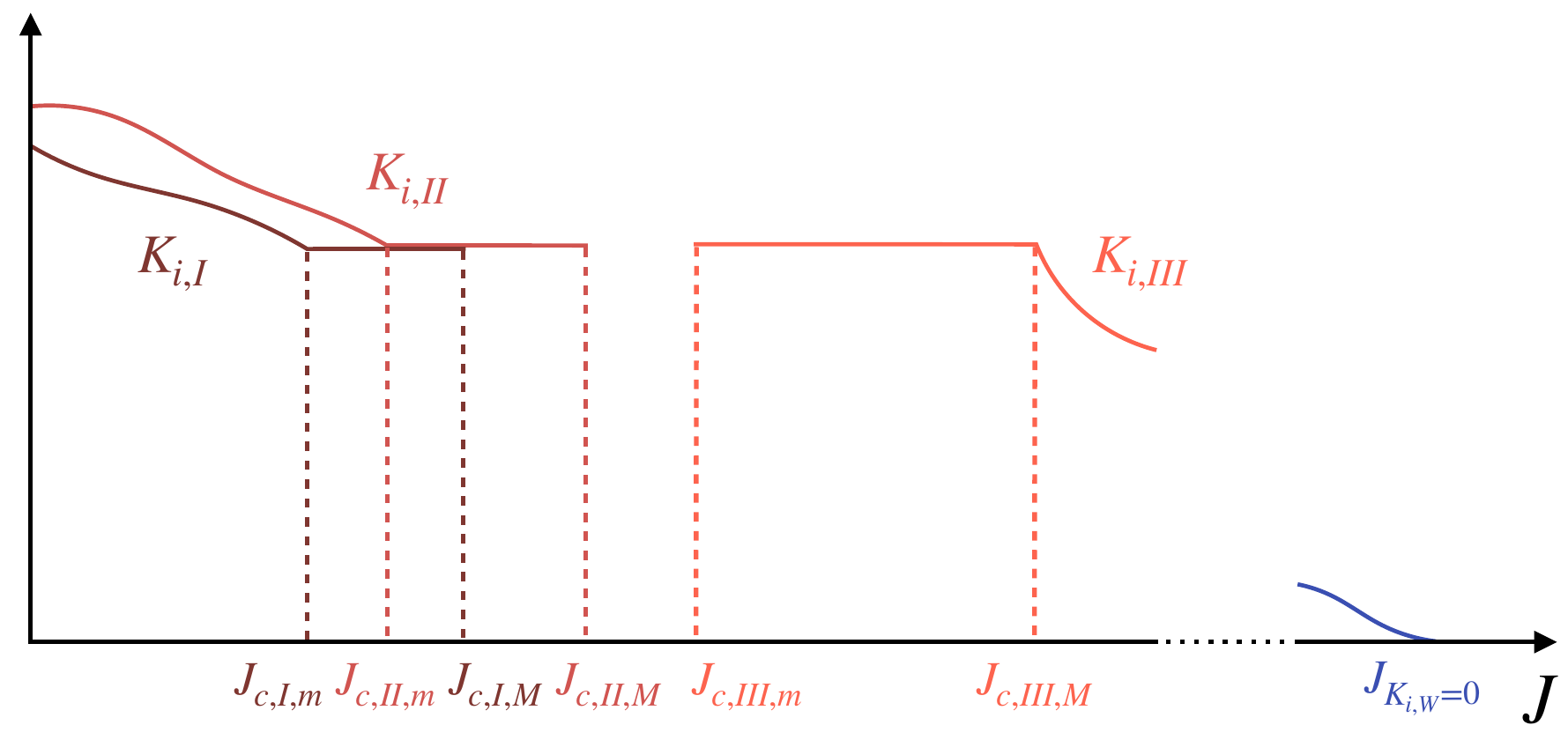}
\end{center}
\caption{\label{fig:sketchKi} Sketch of the function $K_{i, W}$ for the case in figure~\ref{fig:configurationwell1}.}
\end{figure}

Figure~\ref{fig:examplewells}(a) shows that any particle that starts with $J_{c, III, m} < J < J_{c, III, M}$ ends up sampling all possible values of the second adiabatic invariant between $J_{c, III, m}$ and $J_{c, III, M}$ unless the function $J_{III, \mathrm{out}} (J_{III, \mathrm{in}})$ is very specific -- a possible special case of this kind is represented in figure~\ref{fig:specialwells}, where particle trajectories close on themselves in a loop in the $(J, \alpha)$ plane. We expect magnetic field magnitude profiles to satisfy the necessary constraints that lead to configurations similar to the one in figure~\ref{fig:specialwells} only in a countable number of flux surfaces. Thus, in general, particles in well $III$ with $J$ values between $J_{c, III, m}$ and $J_{c, III, M}$ sample all possible values of $J$ between $J_{c, III, m}$ and $J_{c, III, M}$. Since $K_{i, W}$ is constant along particle trajectories (it is independent of $\alpha$ and it is continuous at the junctures of several wells because both $g_{i, 1, W}$ and $r_{i, 1, W}$ are continuous there), we find that $K_{i, III}$ does not depend on $J$ for $J_{c, III, m} < J < J_{c, III, M}$. For the same reasons, $K_{i, I}$ in the interval $[J_{c, I, m}, J_{c, I, M}]$ and $K_{i, II}$ in the interval $[J_{c, II, m}, J_{c, II, M}]$ are both independent of $J$ and equal to the value of $K_{i, III}$ in the interval $[J_{c, III, m}, J_{c, III, M}]$. In figure~\ref{fig:sketchKi}, we sketch $K_{i, W}$ for the configuration in figure~\ref{fig:configurationwell1}.

\subsubsection{General considerations}

We have used the example in figure~\ref{fig:configurationwell1} to illustrate the dependence of $K_{i, W}$ on $J$ because of its simplicity. In this particular configuration, particles do not have the option to choose between two wells, and particles from two different wells are not forced to move into the same well at the same time (this last case requires using equation~\refe{eq:mixrule}). Despite the simplicity of the example, we believe that only very specific configurations allow for a $K_{i, W}$ that is not constant in $J$ in regions with junctures of different types of wells. We note that equation~\refe{eq:mixrule} accepts a constant $K_{i, W}$ as a solution, and our believe is that in most cases this is the only solution. Interestingly, this $K_{i, W}$ solution does not have discontinuities in $K_{i, W}$ across junctures, and hence does not have the collisional boundary layers described in Appendix~\ref{app:boundarylayer}. These layers disappear only to lowest order, as we will see that they are necessary for the next order correction in the $\nu_{i*}/\rho_{i*} \ll 1$ expansion.

\begin{figure}
\begin{center}
\includegraphics[width=8cm]{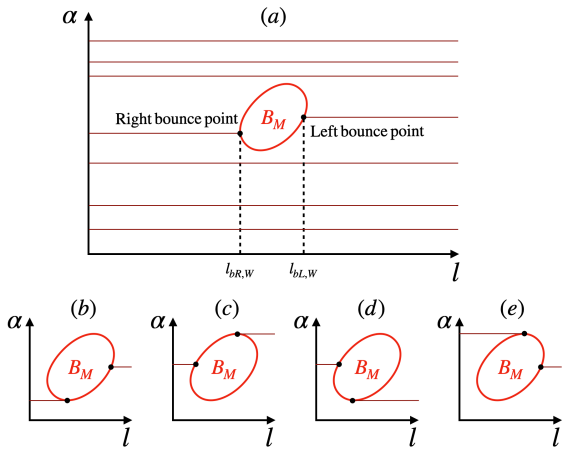}
\end{center}
\caption{\label{fig:Jlarge} Sketch of trajectories with large $J$ in the $(l, \alpha)$ plane. For large $J$, $\lambda$ must be close to $1/B_M$, and hence it must have bounce points at a value of $B$ close to $B_M$. In the figure, we sketch the contour $B = 1/\lambda \approx B_M$ as a thick red line (we have assumed that there is only one maximum of $B$). The total trajectory of the particle is sketched in panel (a) as a thin red line. The best way to identify a trajectory with a given $\lambda$ is to determine the location of the bounce points (note that the left bounce point $l_{bL,W}$ is on the right of the figure and the right bounce point $l_{bR,W}$ is on the left). Particles with the kind of trajectory shown in panel (a) are exposed to four junctures with other wells, as demonstrated by panels (b)-(e). In panel (b), if the particle moves towards negative $\alpha$, it will transition to another type of well, similarly to panel (d). In panels (c) and (e), the particle transitions when it moves towards positive $\alpha$. }
\end{figure}

For large values of $J$, there are always junctures of different types of wells (see figure~\ref{fig:Jlarge} and the Appendix of \cite{boozer90}), so we expect $K_{i, W}$ to be independent of $J$ above certain value of $J$. Moreover, since $r_{i, 1, W} \rightarrow 0$ for $J \rightarrow \infty$ by definition~\refe{eq:r1def}, boundary condition~\refe{eq:gi1bc} imposes that $K_{i, W} = 0$ in this region. From here on, the value of $J$ at which $K_{i, W}$ vanishes is denoted by $J_{K_{i, W} = 0}$ -- see figure~\ref{fig:sketchKi} for an example of $J_{K_{i, W} = 0}$.

We finish by proving that the particle and energy fluxes due to $g_{i,1,W}^{\{0\}}$ vanish. From equation~\refe{eq:Gammaidef}, we obtain that the flux of particles is
\begin{align}
\Gamma_i \simeq \frac{2\pi c \phi_0^\prime }{B_0 V^\prime} \int^{\infty}_0 \rmd v \sum_W \int_{\alpha_{L, W}}^{\alpha_{R, W}}\rmd \alpha \int^{J_{M, W}}_{J_{m, W}} \mathrm{d} J\,
v \Bigg [ \partial_\alpha ( r_{i, 1, W} K_{i, W} ) \nonumber\\ - \partial_\alpha \left ( \frac{1}{2} r_{i, 1, W}^2 \Upsilon_i f_{Mi} \right ) \Bigg ].
\end{align}
The particle flux vanishes due to the integral over $\alpha$ and the fact that $r_{i, 1, W}$ and $K_{i, W}$ are continuous across junctures between several wells. A similar proof shows that the ion heat flux vanishes to lowest order in $\nu_{i*}/\rho_{i*} \ll 1$.

\subsection{Next order distribution function} \label{sub:nu1}
To next order in $\nu_{i*}/\rho_{i*} \ll 1$, we would expect equation~\refe{eq:eqgi1} to give
\begin{equation} \label{eq:eqginu1}
\frac {c \phi_0^\prime}{\Psi_t^\prime} \partial_{\alpha} g_{i, 1, W}^{\{1\}}  = \frac{v^2 \nu_{ii,\perp} } {4 }  \partial_J \left ( \tau_W J\,  \partial_J g_{i,1,W}^{\{0\}}  \right ).
\end{equation}
We explain at the end of this section that, in the intervals of $J$ where there are junctures between different wells and hence $\partial_J K_{i, W} = 0$, equation~\refe{eq:eqginu1} is not valid. Before discussing this case, we consider the intervals of $J$ where particles never transition into other wells. Then, equation~\refe{eq:eqginu1} gives
\begin{equation} \label{eq:eqginu1normal}
\partial_{\alpha} g_{i, 1, W}^{\{1\}}  = \frac{ \Psi_t^\prime v^2 \nu_{ii,\perp} } {4 c \phi_0^\prime} \partial_J \left [ \tau_W J\left ( \partial_J K_{i, W} - \partial_J r_{i, 1, W}\,  \Upsilon_i f_{Mi}  \right ) \right ].
\end{equation}
For these values of $J$, $ g_{i, 1, W}^{\{1\}}$ is defined for all $\alpha$. For this reason, if we average equation~\refe{eq:eqginu1normal} over $\alpha$, the first term vanishes, leaving the equation for $K_{i, W}$
\begin{equation} \label{eq:eqKi}
\partial_J \left [ J\left ( \langle \tau_W \rangle_\alpha \partial_J K_{i, W} - \langle \tau_W\, \partial_J r_{i, 1, W} \rangle_\alpha \Upsilon_i f_{Mi}  \right ) \right ] = 0,
\end{equation}
where
\begin{equation}
\langle \cdot \rangle_\alpha := \frac{1}{2\pi} \int_0^{2\pi} (\ldots)\, \rmd \alpha.
\end{equation}
Solving equation~\refe{eq:eqKi} for $K_{i,W}$ allows one to integrate equation~\refe{eq:eqginu1normal} with periodic boundary conditions in $\alpha$.

Equation~\refe{eq:eqKi} determines the dependence of $K_{i, W}$ on $J$ within the intervals of $J$ where there are no junctures of different types of wells. If the interval of $J$ considered includes $J = 0$, where $\tau_W = 0$, the integral of equation~\refe{eq:eqKi} is
\begin{equation} \label{eq:dJKsolJ0}
\partial_J K_{i, W} = \frac{\langle \tau_W\, \partial_J r_{i, 1, W} \rangle_\alpha}{\langle \tau_W \rangle_\alpha} \Upsilon_i f_{Mi}.
\end{equation}
In the example in figures~\ref{fig:configurationwell1} and \ref{fig:examplewells}, this solution is valid in well $I$ for $J < J_{c, I, m}$, and in well $II$ for $J < J_{c, II, m}$. Note that solution~\refe{eq:dJKsolJ0} leads to a discontinuity in $\partial_J K_{i, W}$ at $J_{c, I, m}$ and $J_{c, II, m}$ because intervals of $J$ with $\partial_J K_{i, W} = 0$ start at these values of $J$. These discontinuities in the derivatives of $K_{i, W}$ mean that there are boundary layers where the lowest order equation~\refe{eq:eqginu0} is not valid and one needs to keep the collision operator to lowest order (see Appendix~\ref{subapp:blKconst} for a discussion of these boundary layers). More importantly, the finite values of $\partial_J K_{i, W}$ at $J = J_{c, I, m}$ and $J_{c, II, m}$ also mean that there is collisional flux into the region $J > J_{c,I,m}$ of well $I$ and the region $J > J_{c,II,m}$ of well $II$. These phase-space fluxes are rapidly transported into well $III$ by the $\bE \times \bB$ drift in the direction $\alpha$ and through jumps in $J$ due transitions to other wells. This phase-space flux must then leave at $J_{c, III, M}$, where the region with $\partial_J K_{i, W} = 0$ ends. Conservation of collisional flux determines $\partial_J K_{i, III}$ at $J_{c, III, M}$,
\begin{align} \label{eq:collfluxnu1}
J_{c, I, m} & \lim_{J \rightarrow J_{c, I, m}^-} \left (\langle \tau_I \rangle_\alpha \partial_J K_{i, I} - \langle \tau_I \, \partial_J r_{i, 1, I} \rangle_\alpha \Upsilon_i f_{Mi} \right ) \nonumber\\ & + J_{c, II, m} \lim_{J \rightarrow J_{c, II, m}^-} \left (\langle \tau_{II} \rangle_\alpha \partial_J K_{i, II} - \langle \tau_{II} \, \partial_J r_{i, 1, II} \rangle_\alpha \Upsilon_i f_{Mi} \right ) \nonumber\\ = & J_{c, III, M} \lim_{J \rightarrow J_{c, III, M}^+} \left (\langle \tau_{III} \rangle_\alpha \partial_J K_{i, III} - \langle \tau_{III}\, \partial_J r_{i, 1, III} \rangle_\alpha \Upsilon_i f_{Mi} \right ).
\end{align}
(see Appendix~\ref{subapp:blKconst} for a derivation of this condition using the equation for the higher order correction $g_{i, 1, W}^{\{1\}}$). Since solution~\refe{eq:dJKsolJ0} is valid in wells $I$ and $II$, condition~\refe{eq:collfluxnu1} simply becomes
\begin{equation}
\lim_{J \rightarrow J_{c, III, M}^+} \langle \tau_{III} \rangle_\alpha \partial_J K_{i, III} = \lim_{J \rightarrow J_{c, III, M}^+} \langle \tau_{III}\, \partial_J r_{i, 1, III} \rangle_\alpha \Upsilon_i f_{Mi}.
\end{equation}
With this condition, we can integrate equation~\refe{eq:eqKi} to obtain
\begin{equation} \label{eq:dJKsolJany}
\partial_J K_{i, III} = \frac{\langle \tau_{III}\, \partial_J r_{i, 1, III} \rangle_\alpha}{\langle \tau_{III} \rangle_\alpha} \Upsilon_i f_{Mi},
\end{equation}
that is, solution~\refe{eq:dJKsolJ0} is valid in well $III$ even though this well does not have particles with $J = 0$. Following this procedure for higher and higher values $J$, one can see that solution~\refe{eq:dJKsolJ0} is in fact valid in every well. With $\partial_J K_{i, W}$ calculated, we can integrate it to obtain $K_{i, W}$ imposing $K_{i, W} = 0$ at $J = J_{K_{i, W} = 0}$. 

The fact that $K_{i, W}$ vanishes for $J > J_{K_{i, W} = 0}$ implies that the solution for $g_{i, 1, W}^{\{ 0 \} }$ for large values of $J$ is simply
\begin{align} \label{eq:ginu0tp}
g_{i,1,W}^{ \{ 0 \} } = - r_{i, 1, W} \, \Upsilon_i f_{Mi} = - \frac{m_i v^2 B_0}{2Z_i e \phi_0^\prime} \left ( \lambda_W - \frac{1}{B_M} \right ) \Upsilon_i f_{Mi} \nonumber\\ \simeq - \frac{m_i v^2 B_0}{2Z_i e \phi_0^\prime} \left ( \lambda_W - \frac{1}{B_0} + \frac{B_{1,M}}{B_0^2} \right ) \Upsilon_i f_{Mi},
\end{align}
where we have substituted the definition of $r_{i, 1, W}$ (see equation~\refe{eq:r1def}) and we have used the approximation $\lambdabar_W \simeq \lambda_W$. Surprisingly, this solution does not satisfy property~\refe{eq:dJgi1Jinfty} (note that $\partial_J \lambda_W \simeq - 2/v^2 \tau_W B_0$) even though property~\refe{eq:dJgi1Jinfty} is a consequence of equation~\refe{eq:eqgi1} for $g_{i, 1, W}$. This apparent contradiction is resolved by a boundary layer where the lowest order equation~\refe{eq:eqginu0} is not valid because one needs to keep the collision operator to lowest order (see Appendix~\ref{subapp:bltp} for a brief discussion of this boundary layer). 

We finish our discussion of the next order correction $g_{i,1,W}^{\{1\}}$ by considering its value in the regions where $\partial_J K_{i, W} = 0$. Naively, for this region of velocity space, equation~\refe{eq:eqginu1} gives 
\begin{equation} \label{eq:eqginu1ergodic}
\frac{c \phi_0^\prime}{\Psi_t^\prime}\, \partial_{\alpha} g_{i, 1, W}^{\{1\}}  = - \frac{v^2 \nu_{ii,\perp} \Upsilon_i f_{Mi}} {4}  \partial_J \left ( \tau_W J\,  \partial_J r_{i, 1, W} \right ).
\end{equation}
This equation has to be solved in regions where there are junctures of different types of well, and as a result, there can be discontinuities in $g_{i,1,W}^{\{1\}}$ across the junctures. Following the same procedure that we developed in Appendix~\ref{app:boundarylayer}, we find that there are boundary layers in the junctures where particles leave two of the wells to go into the third, and that the result of these boundary layers is that $g_{i,1,W}^{\{1\}}$ must satisfy an equation similar to equation~\refe{eq:mixrule},
\begin{equation} \label{eq:mixrulenu1}
g^{\{1\}}_{i,1, I}\, \partial_\alpha J_{c, I} + g^{\{1\}}_{i,1,II}\, \partial_\alpha J_{c,II} = g^{\{1\}}_{i,1, III}\, \partial_\alpha J_{c, III}.
\end{equation}
Despite their apparent validity, equations~\refe{eq:eqginu1ergodic} and \refe{eq:mixrulenu1} cannot be solved. Using the definition of $r_{i,1, W}$ in equation~\refe{eq:r1def} and $\partial_J \lambda_W \simeq - 2/v^2 \tau_W B_0$, we find that equation~\refe{eq:eqginu1ergodic} is
\begin{equation} \label{eq:eqginu1ergodicfinal}
\frac{c \phi_0^\prime}{\Psi_t^\prime}\, \partial_{\alpha} g_{i, 1, W}^{\{1\}}  = \frac{m_i v^2 \nu_{ii,\perp} } {4Z_i e \phi_0^\prime} \Upsilon_i f_{Mi}.
\end{equation}
Then, $g_{i, 1, W}^{\{ 1 \}}$ increases with $\alpha$, posing a problem of continuity: due to the ergodic nature of the trajectories in the regions of velocity space where $\partial_J K_{i, W} = 0$, if we calculate $g_{i,1,W}^{\{ 1 \}}$ using equation~\refe{eq:eqginu1ergodicfinal}, the values of $g_{i,1,W}^{\{ 1 \}}$ at two contiguous values of $J$ would in general be very different  because the lengths of the paths in $\alpha$ needed to reach these similar values of $J$ starting from the same phase space point are very different. Such large differences for contiguous values of $J$ mean that we cannot neglect the collision operator in the regions where $\partial_J K_{i, W} = 0$, and instead of equation~\refe{eq:eqginu1ergodicfinal}, we need to integrate
\begin{equation} \label{eq:eqginu1ergodicreal}
\frac{c \phi_0^\prime}{\Psi_t^\prime}\, \partial_{\alpha} g_{i, 1, W}^{\{1\}} - \frac{v^2 \nu_{ii,\perp} } {4}  \partial_J \left ( \tau_W J\,  \partial_J g_{i, 1, W}^{\{ 1 \}} \right )= \frac{m_i v^2 \nu_{ii,\perp} } {4Z_i e \phi_0^\prime} \Upsilon_i f_{Mi}.
\end{equation}
For this equation to be valid, the characteristic size of $\partial_J$ must be at least $\partial_J \sim \sqrt{\rho_{i*}/\nu_{i*}}/\sqrt{\epsilon} v_{ti} R \gg 1/\sqrt{\epsilon} v_{ti} R$, and hence we expect solution $g_{i, 1, W}^{\{ 1 \}}$ to have oscillatory character in $J$. Due to the inclusion of collisions, the solutions to equation~\refe{eq:eqginu1ergodicreal} contain the boundary layers that appear around junctures of different types of wells, described in Appendix~\ref{app:boundarylayer}. At the junctures of different wells, we need to apply continuity of $g_{i, 1, W}^{\{ 1 \}}$ and condition~\refe{eq:eqgi1transition}, which at this order is
\begin{equation}
J_{c, I} \lim_{J \rightarrow J_{c, I}}  \tau_I\, \partial_J g_{i,1,I}^{\{ 1 \}} + J_{c, II} \lim_{J \rightarrow J_{c, II}}  \tau_{II}\, \partial_J g_{i,1,II}^{\{ 1 \}} = J_{c, III} \lim_{J \rightarrow J_{c, III}}  \tau_{III}\, \partial_J g_{i,1,III}^{\{ 1 \}}.
\end{equation}
We discuss the boundary conditions for $g_{i,1, W}^{\{ 1 \}}$ in regions with $\partial_J K_{i, W} = 0$ in Appendix~\ref{subapp:blKconst}.

\subsection{Radial fluxes} \label{sub:GammaQnu}

In section~\ref{sub:nu0} we showed that substituting $g_{i,1,W}^{\{0\}}$ into equation~\refe{eq:Gammaidef} for the particle flux gives zero. Thus, the particle flux is determined by $g_{i,1,W}^{\{1\}}$ and the boundary layers that we have described in Appendices~\ref{app:boundarylayer} and \ref{app:bldJK}. To account for these higher order effects, we manipulate equation~\refe{eq:Gammaidef}: we integrate by parts in $\alpha$ and we use equation~\refe{eq:eqgi1} to rewrite $\partial_\alpha g_{i, 1, W}$ in terms of the collision operator and $r_{i, 1, W}$. We find
\begin{align}
\Gamma_i = - \frac{2\pi c \phi_0^\prime }{B_0 V^\prime} \int^{\infty}_0 \rmd v \sum_W \int_{\alpha_{L, W}}^{\alpha_{R, W}}\rmd \alpha \int^{J_{M, W}}_{J_{m, W}} \mathrm{d} J\,
v \Bigg [ \frac{v^2 \nu_{ii,\perp} \Psi_t^\prime}{4c\phi_0^\prime} r_{i, 1, W} \, \partial_{J} \left ( \tau_W J \, \partial_{J} g_{i, 1, W} \right ) \nonumber\\ - \partial_\alpha \left ( \frac{1}{2} r_{i, 1, W}^2 \Upsilon_i f_{Mi} \right ) \Bigg ].
\end{align}
This expression is not useful because, inside the boundary layers described in Appendices~\ref{app:boundarylayer} and \ref{app:bldJK} and in the regions of phase space where particles undergo transitions between different types of wells, the collision operator applied on $g_{i, 1, W}$ gives large contributions to the integrand that vanish upon integration. To avoid this delicate cancellation, we integrate by parts in $J$ to find
\begin{equation}
\Gamma_i = \frac{\pi \Psi_t^\prime}{2 B_0 V^\prime} \int^{\infty}_0 \rmd v \sum_W \int_{\alpha_{L, W}}^{\alpha_{R, W}}\rmd \alpha \int^{J_{M, W}}_{J_{m, W}} \mathrm{d} J\,
v^3 \nu_{ii,\perp} \tau_W J \, \partial_J r_{i, 1, W} \, \partial_{J} g_{i, 1, W}.
\end{equation}
In this expression, we can neglect the higher order corrections to $g_{i,1,W}^{\{ 1 \}}$, and hence
\begin{align}
\Gamma_i = - \frac{\pi \Psi_t^\prime}{2 B_0 V^\prime} \int^{\infty}_0 \rmd v\, v^3  \nu_{ii,\perp}  \Upsilon_i f_{Mi} \sum_W \int_{\alpha_{L, W}}^{\alpha_{R, W}}\rmd \alpha \Bigg [ \int_{\partial_J K_{i, W} = 0} \mathrm{d} J\,  \tau_W J \, (\partial_J r_{i, 1, W})^2 \nonumber\\ + \int_{\partial_J K_{i, W} \neq 0} \mathrm{d} J\, \tau_W J \, \left ( (\partial_J r_{i, 1, W})^2 - \frac{\partial_{J} r_{i, 1, W}}{\langle \tau_W \rangle_\alpha} \left \langle \tau_W \partial_{J} r_{i, 1, W} \right \rangle_\alpha \right ) \Bigg ],
\end{align}
where we have used equation~\refe{eq:ginu0}, and in the integral in $J$, we have distinguished the region where $\partial_J K_{i, W}$ vanishes from the region where it is equal to the value given in equation~\refe{eq:dJKsolJ0}. Using $\partial_J \lambda_W \simeq - 2/v^2 B_0 \tau_W$ to write $\partial_J r_{i, 1, W} \simeq - m_i/Z_i e \phi_0^\prime \tau_W$, the particle flux simplifies to
\begin{align} \label{eq:Gammanu}
\Gamma_i = - \frac{\pi m_i^2 \Psi_t^\prime}{2 Z_i^2 e^2 \phi_0^{\prime 2} B_0 V^\prime} \int^{\infty}_0 \rmd v \, v^3  \nu_{ii,\perp}  \Upsilon_i f_{Mi}  \sum_W \int_{\alpha_{L, W}}^{\alpha_{R, W}}\rmd \alpha \Bigg [ \int_{\partial_J K_{i, W} = 0} \mathrm{d} J\, \frac{J}{\tau_W} \nonumber\\ + \int_{\partial_J K_{i, W} \neq 0} \mathrm{d} J\, J \left (\frac{1}{\tau_W} - \frac{1}{\langle \tau_W \rangle_\alpha} \right ) \Bigg ].
\end{align}
Note that the contribution from the particles in phase space regions where $\partial_J K_{i, W} \neq 0$ is reduced by a factor of $(1 - \tau_W/\langle \tau_W \rangle_\alpha)$, that is, this expression shows that particles that suffer transitions between different types of wells cause higher fluxes.

A similar procedure to the one that we have shown above gives the energy flux
\begin{align} \label{eq:Qnu}
Q_i = - \frac{\pi m_i^3 \Psi_t^\prime}{4Z_i^2 e^2 \phi_0^{\prime 2} B_0 V^\prime} \int^{\infty}_0 \rmd v\, v^5  \nu_{ii,\perp} \Upsilon_i f_{Mi} \sum_W \int_{\alpha_{L, W}}^{\alpha_{R, W}}\rmd \alpha \Bigg [ \int_{\partial_J K_{i, W} = 0} \mathrm{d} J\, \frac{J}{\tau_W} \nonumber\\ + \int_{\partial_J K_{i, W} \neq 0} \mathrm{d} J\, J \left (\frac{1}{\tau_W} - \frac{1}{\langle \tau_W \rangle_\alpha} \right ) \Bigg ].
\end{align}

The fluxes in equations~\refe{eq:Gammanu} and \refe{eq:Qnu} are inversely proportional to $\phi_0^{\prime 2}$ and hence diverge when $\phi_0^\prime$ vanishes. The reason for this divergence is that the $\bE \times \bB$ drift was assumed to be much larger than the radial component of the drifts in equation~\refe{eq:vdreps}. For $e a \phi_0^\prime/T_i \sim \epsilon$, this assumption is not satisfied, and the first term in equation~\refe{eq:ginu0} becomes of the same order as the Maxwellian $f_{Mi}$, giving $g_{i, 1, W}^{\{ 0 \}} \sim f_{Mi}$, that is, the distribution function is not close to a Maxwellian. This is a manifestation of the fact that orbit widths become comparable to the minor radius of the stellarator. To summarize, formulas~\refe{eq:Gammanu} and \refe{eq:Qnu} are not valid for $e a \phi_0^\prime/T_e \lesssim \epsilon$. At these values of the electric field, the radially global equations~\refe{quasineutrality}, \refe{eq:maxwboltz}, \refe{rho2orbit}, \refe{passingfi0} and \refe{rho2transition} must be used. Importantly, even for $e a \phi_0^\prime/T_e \sim 1$, energetic particles with energies larger than $e R \phi_0^\prime \sim m_i v_{ti}^2/\epsilon$ have $\nabla B$ and curvature drifts that are comparable to or larger than the $\bE \times \bB$ drift. This means that, unless the stellarator is close to omnigeneity \citep{calvo17, catto19}, energetic particles produced by fusion reactions, RF heating or neutral beams have to be modeled using the radially global equations~\refe{quasineutrality}, \refe{eq:maxwboltz}, \refe{rho2orbit}, \refe{passingfi0} and \refe{rho2transition}. These radially global orbit averaged equations will be valid as long as the gyroradius and the banana orbit width of energetic particles are sufficiently small. In particular, the finite banana orbit width mechanism for energetic particle transport proposed by \cite{goldston81} is negligible in the limit in which the orbit averaged equations~\refe{quasineutrality}, \refe{eq:maxwboltz}, \refe{rho2orbit}, \refe{passingfi0} and \refe{rho2transition} are valid. 

We finish by estimating the size of the particle and the energy flux. Using $\Psi_t^\prime \sim a B_0$, $\phi_0^\prime \sim T_i/ea$, $V^\prime \sim Ra$, $J \sim \sqrt{\epsilon} v_{ti} R$, $\tau_W \sim R/\sqrt{\epsilon} v_{ti}$ and $\Upsilon_i \sim 1/a$, we find
\begin{equation} \label{eq:fluxesnu}
\Gamma_i \sim \epsilon^{5/2} \nu_{i*} n_i v_{ti} , \quad Q_i \sim \epsilon^{5/2} \nu_{i*} n_i T_i v_{ti}.
\end{equation}
These estimates can be obtained using a random walk estimate. As we explained in the introduction, in equation~\refe{eq:wso}, the width of a typical trapped particle orbit is $w \sim \epsilon a$. Since the effective collision frequency for trapped particles is $\nu_\mathrm{eff} \sim \nu_{ii}/\epsilon$, trapped particles undergo a random walk with steps of length $w$ every time $\epsilon/\nu_{ii}$. The resulting diffusion coefficient is $D \sim \sqrt{\epsilon} w^2 \nu_\mathrm{eff}$, where $\sqrt{\epsilon}$ is an estimate for the trapped particle fraction. The estimates for the fluxes are then obtained from $D \sim \epsilon^{3/2} a^2 \nu_{ii}$, $\Gamma_i \sim D|\nabla n_i| \sim D n_i/a$ and $Q_i \sim n_i D |\nabla T_i| \sim D n_i T_i/a$.

\section{Large aspect ratio stellarators close to omnigeneity} \label{sec:omnigeneity}

In this section we study large aspect ratio stellarators close to omnigeneity. A magnetic field is omnigeneous when it satisfies $\partial_\alpha J_W = 0$ for all trapped particles, that is,
\begin{equation} \label{eq:omnigeneity}
\partial_\alpha J_W = - v \int_{l_{bL,W}}^{l_{bR,W}} \frac{\lambda\, \partial_\alpha B}{\sqrt{1 - \lambda B}}\, \rmd l = 0
\end{equation}
for all $\lambda$. As explained by \cite{cary97a}, \cite{cary97b} and \cite{parra15a}, this condition imposes stringent constraints on how the magnitude $B$ depends on $\alpha$ and $l$. Of these constraints, two are particularly important for our discussion. 
\begin{itemize}
\item The maxima of $B$ on a flux surface are not on isolated points, but on lines that close on themselves. Moreover, those lines where the maxima lie are separated from each other by such a distance that any particle that travels between these maxima has only one possible value of $J$, which we denote by $J_{\mathrm{om}, M}$ because it is also the maximum value that $J$ can take. 

\item In junctures of several types of wells like the one depicted in figure~\ref{fig:exampleU} (see \cite{parra15a} for a discussion of the existence of omnigeneous magnetic fields with junctures of different types of wells), the values of the second adiabatic invariant at the juncture are independent of $\alpha$, that is, $\partial_\alpha J_{c, W} = 0$. This means that there are no transitions from one type of well to another.
\end{itemize}

Collisional transport in omnigeneous stellarators is very small, and for this reason designing stellarators close to omnigeneity is of interest. We consider large aspect ratio stellarators that are close to omnigeneity. We find two distinct types. In subsection~\ref{sub:largemirror}, we show that large aspect ratio stellarators with large mirror ratios are close to omnigeneity, and hence one can use the equations derived by \cite{calvo17} to calculate neoclassical fluxes in these stellarators. In subsection~\ref{sub:optimized}, we study optimized large aspect ratio stellarators with mirror ratios close to unity.

\subsection{Large aspect ratio stellarators with large mirror ratios} \label{sub:largemirror}

In previous sections, we have discussed in detail large aspect ratio stellarators with mirror ratios close to unity, an approximation that describes well many modern stellarators. However, it is possible that increasing the mirror ratio will be of interest in the future. For this reason, we consider large aspect ratio stellarators with $B_0(l)$ a general function of $l$. Note that such stellarators satisfy equation~\refe{eq:omnigeneity} to lowest order because $B_0(l)$ is independent of $\alpha$. This means that $B_1(r, \alpha, l)$ can be considered to be the deviation from omnigeneity. We can then use the formulation by \cite{calvo17} by replacing the expansion parameter $\delta$, which measures the size of the deviation from omnigeneity, with $\epsilon$. 

Using the same techniques as \cite{calvo17}, it is possible to prove that the lowest order solution for the ion distribution function is a stationary Maxwellian $f_{Mi}$ with density and temperature that are flux functions, and that the potential is a flux function to lowest order, $\phi(r, \alpha, l, t) \simeq \phi_0(r, t) \sim T_i/e$. The correction to the Maxwellian is only significantly different from zero for trapped particles, for which it satisfies $g_{i, 1, W} \sim \epsilon f_{Mi}$. As for large aspect ratio stellarators with mirror ratios close to unity, the typical radial width of a trapped particle orbit is $\epsilon a$ because the radial magnetic drift is small by $\epsilon$ compared to the component of the $\bE \times \bB$ drift parallel to the flux surface.  Apart from these similarities, the formulation by \cite{calvo17} differs from the equations presented in previous sections of this paper in significant ways. For $B_{0, M}/B_{0, m} - 1 \sim 1$, the typical parallel velocity of trapped particles is not small compared to the thermal speed, and the fraction of trapped particles is of order unity. The energy gained or lost by work done by the radial electric field during a trapped particle's radial displacement is insufficient to affect trapped particles with $v_\| \sim v_{ti}$, unlike in large aspect ratio stellarators with mirror ratios close to unity, where trapped particles have $v_\| \sim \sqrt{\epsilon} v_{ti}$. For this reason, the velocity coordinates $\mathcal{E}$ and $\mu$ are appropriate -- in \cite{calvo17}, these coordinates are replaced by the equivalent coordinates $v$ and $\lambda := v_\perp^2/v^2 B$ for convenience. Another important difference with large aspect ratio stellarators with mirror ratios of order unity is that the correction to the lowest order potential is $\phi_1 (r, \alpha, l, t) \sim \epsilon T_i/e$ instead of $\phi_{3/2} (r, \alpha, l, t) \sim \epsilon^{3/2} T_i/e$, and as a result the radial component of the $\bE \times \bB$ drift contributes significantly to the radial displacement of trapped particles.

We proceed to discuss the different transport regimes predicted by the formulation of \cite{calvo17}. For collision frequencies $\nu_{ii}$ larger than the characteristic frequency associated to the $\bE \times \bB$ drift, $|\bv_E|/a \sim \rho_{i*} v_{ti}/a$, that is, for $\rho_{i*}/\epsilon \ll \nu_{i*} \ll 1$, the equations derived by \cite{calvo17} predict a $1/\nu$ regime. In this regime, the particle and energy fluxes are of order
\begin{equation} \label{eq:fluxmirror1nu}
\Gamma_i \sim \frac{\rho_{i*}}{\nu_{i*}} \epsilon \rho_{i*} n_i v_{ti}, \quad Q_i \sim \frac{\rho_{i*}}{\nu_{i*}} \epsilon \rho_{i*} n_i T_i v_{ti}. 
\end{equation}
Here, the characteristic length $L_0$ and the parameter $\delta$ in \cite{calvo17} have been replaced with $a$ and $\epsilon$, and our definition of $\nu_{i*}$ differs from the one used by \cite{calvo17}, $\nu_{i*}^\mathrm{Ca}$, by a factor of $\epsilon$, $\nu_{i*} = \nu_{i*}^\mathrm{Ca}/\epsilon$. Note that, for large aspect stellarators with large mirror ratios, the transition to the lower collisionality transport regime happens for $\nu_{i*} \sim \rho_{i*}/\epsilon$ -- in comparison, for large aspect ratio stellarators with mirror ratios close to unity, the transition happens for $\nu_{i*} \sim \rho_{i*}$. The difference is due to the typical parallel velocity of the trapped particles: for large aspect ratio stellarators with mirror ratios close to unity, trapped particles have $v_\| \sim \sqrt{\epsilon} v_{ti}$ and hence it is easy for collisions to detrap these particles, giving a large effective collision frequency $\nu_{ii}/\epsilon$ for trapped particles.

The estimates for the fluxes in equation~\refe{eq:fluxmirror1nu} can also be obtained from a random walk argument similar to the one that gives the estimates in equation~\refe{eq:fluxes1nu}. In large aspect ratio stellarators with large mirror ratios, the effective collision frequency for trapped particles is $\nu_{ii}$ and hence the random walk is composed of steps of length $\bv_{Mi} \cdot \nabla r/\nu_{ii} \sim (\rho_{i*}/\nu_{i*}) a \ll \epsilon a$ that happen with frequency $\nu_{ii}$. The corresponding diffusion coefficient is $D \sim (\bv_{Mi} \cdot \nabla r/\nu_{ii})^2 \nu_{ii} \sim \epsilon \rho_{i*}^2 a v_{ti}/\nu_{i*}$, where we have noted that the fraction of trapped particles is of order unity. This diffusion coefficient gives the estimates in equation~\refe{eq:fluxmirror1nu}.

For $\nu_{i*} \ll \rho_{i*}/\epsilon$, the derivation of \cite{calvo17} does not assume that the component of the $\nabla B$ and curvature drifts parallel to the flux surface is smaller than the same component of the $\bE \times \bB$ drift. For this reason, it allows for cancellation of the average drift parallel to the flux surface for certain particles. In large aspect ratio stellarators with large mirror ratios and an electric field $|\phi_0^\prime|$ that is much larger than $T_i/eR$, the $\bE \times \bB$ drift is larger than the magnetic drifts by $e|\phi_0^\prime| R/T_i \gg 1$, and hence only a exponentially small number of particles with energies above $e R \phi_0^\prime \gg m_i v_{ti}^2$ can have vanishing drifts parallel to the flux surface. These particles can be neglected, and as a result the transport is dominated by boundary layers in velocity space of width $\Delta v_\| \ll v_{ti}$ that appear at the trapped-passing boundary or around junctures of different types of wells. In these layers, the effective collision frequency is 
\begin{equation} \label{eq:nuefflayer}
\nu_\mathrm{eff} \sim \left ( \frac{v_{ti}}{\Delta v_\|} \right )^2 \frac{\nu_{ii}}{\ln (v_{ti}/\Delta v_\|)}. 
\end{equation}
The effective collision frequency becomes larger than $\nu_{ii}$ because very small angle collisions are sufficient to change the parallel velocity by a small amount $\Delta v_\| \ll v_{ti}$. The logarithmic reduction of the effective collision frequency is due to the fact that particles in these layers have bounce points that are very close to local maxima of $B$, where they spend logarithmically long times. Collisions near these maxima of $B$ are ineffective because $v_\|$ is very close to zero, $v_\| \ll \sqrt{\epsilon} v_{ti}$, and a collision that changes the parallel velocity of a particle by $\Delta v_\|$ changes its parallel kinetic energy by $m_i v_\| \Delta v_\| \ll \sqrt{\epsilon} m_i v_{ti} \Delta v_\|$ only. With such a small change in kinetic energy, the particle cannot leave the layer. The collisions that expel the particles from the boundary layer happen away from the bounce points, where $v_\| \sim \sqrt{\epsilon} v_{ti}$ and particles spend only a fraction of its bounce period of order $1/\ln(v_{ti}/\Delta v_\|)$. The effective collision frequency $\nu_\mathrm{eff}$ has to be comparable to the inverse of the time that it takes a particle to drift poloidally around the stellarator, $|\bv_E|/a \sim \rho_{i*} v_{ti}/a$, giving 
\begin{equation} \label{eq:Dvparsqrt}
\Delta v_\| \sim v_{ti} \sqrt{\frac{\epsilon \nu_{i*}}{\rho_{i*}} \frac{1}{\ln (\rho_{i*}/\epsilon \nu_{i*})}}. 
\end{equation}
The transport due to these boundary layers is of order
\begin{equation} \label{eq:fluxmirrorsqrtnu}
\Gamma_i \sim \sqrt{\frac{\epsilon \nu_{i*}}{\rho_{i*}} \ln \left ( \frac{\rho_{i*}}{\epsilon \nu_{i*}} \right )} \epsilon^2 \rho_{i*} n_i v_{ti}, \quad Q_i \sim \sqrt{ \frac{\epsilon \nu_{i*}}{\rho_{i*}}  \ln \left ( \frac{\rho_{i*}}{\epsilon \nu_{i*}} \right )} \epsilon^2 \rho_{i*} n_i T_i v_{ti}. 
\end{equation}
The formulation by \cite{calvo17} does not work for $\nu_{ii}a/|\bv_E| \lesssim \epsilon^2 |\ln \epsilon|$ -- this limit is derived in section 6 of \cite{calvo17} without the logarithmic correction (recall that $\delta$ must be replaced with $\epsilon$). This means that, in normalized quantities, the $\sqrt{\nu}$ regime does not extend below $\nu_{i*} \sim \epsilon |\ln \epsilon| \rho_{i*}$. We leave the study of what happens for collisionalities lower than this limit for another publication. 

As other estimates given in previous sections, the estimates in equation~\refe{eq:fluxmirrorsqrtnu} can be obtained from a random walk argument. Only particles in velocity space boundary layers with width $\Delta v_\|$, given in equation~\refe{eq:Dvparsqrt}, participate in the transport. These particles have an effective collision frequency $\nu_\mathrm{eff} \sim |\bv_E|/a \sim \rho_{i*} v_{ti}/a$, and hence typical radial displacements $\bv_{Mi} \cdot \nabla r/\nu_\mathrm{eff} \sim \epsilon a$. Thus, the diffusion coefficient is 
\begin{equation} \label{eq:Dsqrt}
D \sim \frac{\Delta v_\|}{v_{ti}} \ln \left ( \frac{v_{ti}}{\Delta v_\|} \right ) \left ( \frac{\bv_{Mi} \cdot \nabla r}{\nu_\mathrm{eff}} \right )^2 \nu_\mathrm{eff}, 
\end{equation}
where the prefactor $(\Delta v_\|/v_{ti}) \ln (v_{ti}/\Delta v_\|)$ is due to the small fraction of particles that are in the boundary layers and cause transport. The logarithmic correction in $(\Delta v_\|/v_{ti}) \ln (v_{ti}/\Delta v_\|)$ is due to the long bounce periods of the orbits of interest: the particles in the boundary layers spend logarithmically long times around local maxima of $B$, where they accumulate, giving a density large by $\ln (v_{ti}/\Delta v_\|)$. The diffusion coefficient in equation~\refe{eq:Dsqrt} gives $D \sim \sqrt{(\epsilon \nu_{i*}/ \rho_{i*}) \ln (\rho_{i*}/\epsilon \nu_{i*})} \epsilon^{2} \rho_{i*} a v_{ti}$, and with it we can obtain the estimates in equation~\refe{eq:fluxmirrorsqrtnu}.

\begin{figure}
\begin{center}
\includegraphics[width=10cm]{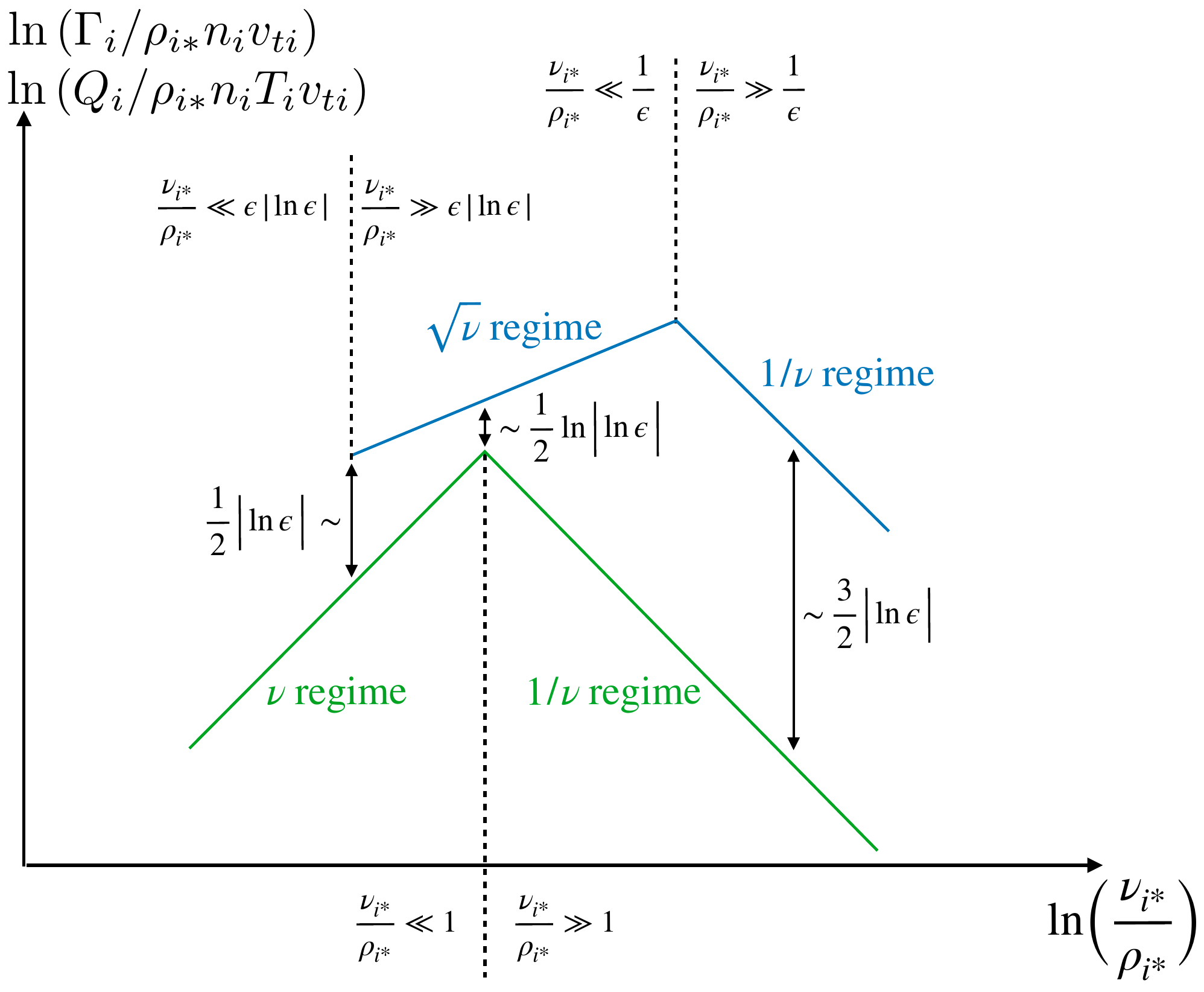}
\end{center}
\caption{\label{fig:fluxeslargemirror} Particle and energy fluxes as a function of $\nu_{i*}/\rho_{i*}$ for large aspect ratio stellarators with large mirror ratios (blue line) and mirror ratios close to unity (green line). Note that, for $\nu_{i*} \sim \rho_{i*}$, the difference between the fluxes of large aspect ratio stellarators with large mirror ratios and the fluxes of large aspect ratio stellarators with mirror ratios close to unity is only a factor of $\sqrt{|\ln \epsilon|}$.}
\end{figure}

A comparison between expressions \refe{eq:fluxmirror1nu} and \refe{eq:fluxmirrorsqrtnu} and expressions \refe{eq:fluxes1nu} and \refe{eq:fluxesnu}, sketched in figure~\ref{fig:fluxeslargemirror}, reveals that neoclassical fluxes in large aspect ratio stellarators with large mirror ratios are larger than the fluxes in equivalent large aspect ratio stellarator with mirror ratios close to unity. This is perhaps part of the reason why no stellarators with large mirror ratios have been built.

\subsection{Optimized large aspect ratio stellarators with mirror ratios close to unity} \label{sub:optimized}

It is possible to optimize large aspect ratio stellarators with mirror ratios close to unity to achieve lower neoclassical transport. We treat such cases by using the expansion proposed by \cite{calvo18a}, that is, we consider a magnetic field that can be written as
\begin{equation}
B(r, \alpha, l) = B_0 + B_1^{[0]} (r, \alpha, l)+ B_1^{[1]} (r, \alpha, l) + \dots
\end{equation}
Here $B_0 + B_1^{[0]}$, with $B_1^{[0]} \sim \epsilon B_0$, is omnigeneous, that is,
\begin{equation}
\int_{l_{bL,0,W}^{[0]}}^{l_{bR,0,W}^{[0]}} \frac{\partial_\alpha B_1^{[0]}}{\sqrt{1 - \lambda B_0 - B_1^{[0]}/B_0}}\, \rmd l = 0,
\end{equation}
where we have used the large aspect ratio approximation to equation~\refe{eq:omnigeneity}. Here $l_{bL,0,W}^{[0]}(r, \alpha, \lambda)$ and $l_{bR,0,W}^{[0]}(r, \alpha, \lambda)$ are the bounce points that correspond to the omnigeneous magnetic field, determined by the equations $B_1^{[0]} (r, \alpha, l_{bL,0,W}^{[0]})/B_0 = 1 - \lambda B_0 = B_1^{[0]} (r, \alpha, l_{bR,0,W}^{[0]})/B_0$. The correction $B_1^{[1]} \sim \delta \epsilon B_0$ is the perturbation away from omnigeneity. We assume that $\delta \ll 1$. Note that we are using the superscripts with square brackets $^{[n]}$ for the expansion in $\delta$ to distinguish it from the expansion in section~\ref{sec:nuregime}. 

We proceed to expand equation~\refe{eq:eqgi1} in $\delta \ll 1$. We first obtain $r_{i, 1, W}$ by writing 
\begin{equation}
J_W (r, \alpha, v, \lambda)  \simeq J_W^{[0]} (r, v, \lambda) + J_W^{[1]} (r, \alpha, v, \lambda) + \ldots,
\end{equation}
with 
\begin{equation}
J_W^{[0]} (r, v, \lambda) := 2 v \int_{l_{bL,0,W}^{[0]}}^{l_{bR,0,W}^{[0]}} \sqrt{1 - \lambda B_0 - B_1^{[0]}/B_0}\, \rmd l \sim \sqrt{\epsilon} v_{ti} R
\end{equation}
and
\begin{equation}
J_W^{[1]} (r, \alpha, v, \lambda) := - \frac{v}{B_0} \int_{l_{bL,0,W}^{[0]}}^{l_{bR,0,W}^{[0]}} \frac{B_1^{[1]}}{\sqrt{1 - \lambda B_0 - B_1^{[0]}/B_0}}\, \rmd l \sim \delta \sqrt{\epsilon} v_{ti} R.
\end{equation}
When we invert the relation $J_W (r, \alpha, v, \lambda)$ to obtain $\lambda_W(r, \alpha, v, J)$, we find that $\lambda_W$ does not depend on $\alpha$ to lowest order in the expansion in $\delta \ll 1$. If we continue to next order in $\delta$, we obtain
\begin{equation} \label{eq:lambdaJom}
\lambda_W(r, \alpha, v, J) = \lambda_W^{[0]} (r, v, J) + \lambda_W^{[1]} (r, \alpha, v, J) + \ldots
\end{equation}
where
\begin{equation}
\lambda_W^{[1]} (r, \alpha, v, J) := - \frac{J_W^{[1]}}{\partial_\lambda J_W^{[0]}} = \frac{2J_W^{[1]}}{v^2 B_0 \tau_W^{[0]}} \sim \frac{\delta \epsilon}{B_0}.
\end{equation}
Here
\begin{equation}
\tau_W^{[0]} (r, v, \lambda) := \frac{2}{v} \int_{l_{bL,0,W}^{[0]}}^{l_{bR,0,W}^{[0]}} \frac{1}{\sqrt{1 - \lambda B_0 - B_1^{[0]}/B_0}}\, \rmd l
\end{equation}
is the lowest order bounce time.

With the result in equation~\refe{eq:lambdaJom}, equation~\refe{eq:eqgi1} becomes
\begin{equation} \label{eq:eqgiom}
\frac{c\phi_0^\prime}{\Psi_t^\prime} \partial_\alpha g_{i, 1, W}^{[1]} - \frac{v^2 \nu_{ii,\perp}}{4} \partial_J \left ( \tau_W^{[0]} J \partial_J g_{i, 1, W}^{[1]} \right ) = - \frac{c\phi_0^\prime}{\Psi_t^\prime} \partial_\alpha r_{i, 1, W}^{[1]}\, \Upsilon_i f_{Mi},
\end{equation}
where $g_{i, 1, W}^{[1]} \sim \delta \epsilon f_{Mi}$ and
\begin{equation}
r_{i, 1, W}^{[1]} (r, \alpha, v, J) := \frac{m_i J_W^{[1]}}{Z_ie \phi_0 ^\prime \tau_W^{[0]}} \sim \delta \epsilon a.
\end{equation}
At the junctures of several wells, we need to impose condition~\refe{eq:eqgi1transition}, which in the expansion in $\delta$ becomes
\begin{equation}
J_{c, I}^{[0]} \lim_{J \rightarrow J_{c, I}^{[0]}}  \tau_I^{[0]} \, \partial_J g_{i,1,I}^{[1]} + J_{c, II}^{[0]} \lim_{J \rightarrow J_{c, II}^{[0]}}  \tau_{II}^{[0]}\, \partial_J g_{i,1,II}^{[1]} = J_{c, III}^{[0]} \lim_{J \rightarrow J_{c, III}^{[0]}}  \tau_{III}^{[0]}\, \partial_J g_{i,1,III}^{[1]}.
\end{equation}
Importantly, since the omnigeneous stellarator has a maximum value of $J$, $J_{\mathrm{om}, M}$, the boundary condition for $g_{i, 1, W}^{[1]}$ is not imposed at $J \rightarrow \infty$, but at $J = J_{\mathrm{om}, M}^{[0]}$,
\begin{equation}
g_{i, 1, W}^{[1]} (r, \alpha, v, J_{\mathrm{om}, M}^{[0]}, t) = 0.
\end{equation}

The particle and energy fluxes simplify to
\begin{equation} \label{eq:Gammaom}
\Gamma_i = \frac{2\pi c \phi_0^\prime}{B_0 V^\prime} \int_0^\infty \rmd v \sum_W \int_0^{2\pi} \rmd \alpha \int_{J_{m,W}}^{J_{M,W}} \rmd J \, v g_{i, 1, W}^{[1]}\, \partial_\alpha r_{i, 1, W}^{[1]} \sim \delta^2 \epsilon^{5/2} \rho_{i*} n_i v_{ti}
\end{equation}
and
\begin{equation} \label{eq:Qom}
Q_i = \frac{\pi m_i c \phi_0^\prime}{B_0 V^\prime} \int_0^\infty \rmd v \sum_W \int_0^{2\pi} \rmd \alpha \int_{J_{m, W}}^{J_{M, W}} \rmd J \, v ^3 g_{i, 1, W}^{[1]}\, \partial_\alpha r_{i, 1, W}^{[1]} \sim \delta^2 \epsilon^{5/2} \rho_{i*} n_i T_i v_{ti}.
\end{equation}
Note that we have used the fact that there are no transitions between wells in an omnigeneous magnetic field to extend the integrals in $\alpha$ to the whole interval $[0, 2\pi]$. 

Comparing the estimates for the fluxes in equations~\refe{eq:Gammaom} and \refe{eq:Qom} with the ones for a generic large aspect ratio stellarator with mirror ratio close to unity in equations~\refe{eq:Gammaidef} and \refe{eq:Qidef}, we see that a near-omnigeneous stellarator does indeed confine better than a generic one by a factor of $\delta^2 \ll 1$. This factor, however, is not uniform for all values of $\nu_{i*}$. While in the $1/\nu$ regime, $\nu_{i*}/\rho_{i*} \gg 1$, we can follow the arguments in section~\ref{sec:1nuregime} to obtain
\begin{equation}  \label{eq:fluxom1nu}
\Gamma_i \sim \delta^2 \frac{\rho_{i*}}{\nu_{i*}} \epsilon^{5/2} \rho_{i*} n_i v_{ti}, \quad Q_i \sim \delta^2 \frac{\rho_{i*}}{\nu_{i*}} \epsilon^{5/2} \rho_{i*} n_i T_i v_{ti},
\end{equation}
for $\nu_{i*} \ll \rho_{i*}$ the scaling with $\delta$ is not a simple $\delta^2$, as we explain below. Before considering the regime $\nu_{i*} \ll \rho_{i*}$, we discuss how to obtain the estimates in equations~\refe{eq:Gammaom}, \refe{eq:Qom} and \refe{eq:fluxom1nu} using a random walk argument. For $\nu_{i*} \gtrsim \rho_{i*}$, in between collisions, trapped particles move a radial distance $\langle \bv_{Mi} \cdot \nabla r \rangle_\tau/\nu_\mathrm{eff}$, where $\nu_\mathrm{eff} \sim \nu_{ii}/\epsilon$ is the effective collision frequency for trapped particles. Since $\langle \bv_{Mi} \cdot \nabla r \rangle_\tau \sim \delta \epsilon \rho_{i*} v_{ti}$, the typical radial displacement is $\langle \bv_{Mi} \cdot \nabla r \rangle_\tau/\nu_\mathrm{eff} \sim (\rho_{i*}/\nu_{i*}) \delta \epsilon a \lesssim \delta \epsilon a$. With this estimate, we obtain a random walk diffusion coefficient $D \sim \sqrt{\epsilon} ( \langle \bv_{Mi} \cdot \nabla r\rangle_\tau/\nu_\mathrm{eff})^2 \nu_\mathrm{eff}$, where the factor of $\sqrt{\epsilon}$ is due to the trapped particle fraction. This diffusion coefficient $D \sim \delta^2 \epsilon^{5/2} \rho_{i*}^2 a v_{ti}/\nu_{i*}$ gives the estimates~\refe{eq:fluxom1nu} for $\nu_{i*} \gg \rho_{i*}$, and the estimates in equations~\refe{eq:Gammaom} and \refe{eq:Qom} for $\nu_{i*} \sim \rho_{i*}$.

If we try to solve equation~\refe{eq:eqgiom} for $\nu_{i*} \ll \rho_{i*}$ following the procedure in section~\ref{sec:nuregime}, we run into problems at the largest value of $J$, $J_{\mathrm{om}, M}^{[0]}$, and at junctures of different types of wells. Indeed, if we neglect the collision operator in equation~\refe{eq:eqgiom}, the solution is
\begin{equation} \label{eq:solgiomnu}
g_{i, 1, W}^{[1]} (r, \alpha, v, J, t) = K_{i, W}^{[1]} (r, v, J, t) - r_{i, 1, W}^{[1]} \, \Upsilon_i f_{Mi},
\end{equation}
where the function $K_{i, W}^{[1]} (r, v, J, t)$ does not depend on $\alpha$. Solution~\refe{eq:solgiomnu} cannot vanish at $J = J_{\mathrm{om}, M}^{[0]}$ in general because $r_{i, 1, W}^{[1]}$ does not vanish there. The contribution proportional to $r_{i, 1, W}^{[1]}$ cannot be cancelled by a judicious choice of $K_{i, W}^{[1]}$ because $r_{i, 1, W}^{[1]}$ depends on $\alpha$ while $K_{i, W}^{[1]}$ does not. Similarly, it is not possible to impose continuity of $g_{i,1,W}^{[1]}$ across junctures. The problem at $J = J_{\mathrm{om}, M}^{[0]}$ is a manifestation of the fact that there are orbits with second adiabatic invariants larger than $J_{\mathrm{om}, M}^{[0]}$. The perturbation $B_1^{[1]}$ will in general introduce isolated local maxima in $B$ above those of the omnigeneous magnetic field, and these maxima will not be equal to each other. The fact that there are individual maxima means that we are back to the case that we have studied in sections \ref{sec:fiepsilon} and \ref{sec:nuregime}. Thus, one needs to includes $J > J_{\mathrm{om}, M}^{[0]}$. The problem at junctures of different types of wells is due to the fact that approximation~\refe{eq:eqgiom} does not allow transitions between different types of wells.

We proceed to study the distribution functions and fluxes for $\nu_{i*} \ll \rho_{i*}$. There are two limits of interest that we consider: $\delta^2 |\ln \delta| \ll \nu_{i*}/\rho_{i*} \ll 1$ and $\nu_{i*}/\rho_{i*} \ll \delta^2 |\ln \delta|$.

\subsubsection{The $\sqrt{\nu}$ regime in optimized large aspect ratio stellarators with mirror ratios close to unity}

We start by considering the regime $\delta^2 |\ln \delta | \ll \nu_{i*}/\rho_{i*} \ll 1$, where we will find the $\sqrt{\nu}$ regime \citep{galeev69, ho87}. In the region $J_{\mathrm{om}, M}^{[0]} - J \lesssim \delta |\ln \delta | \sqrt{\epsilon} v_{ti} R$, particles have bounce points at the new maxima introduced by the perturbation $B_1^{[1]}$ above the maxima of the omnigeneous magnetic field. To understand estimate $J_{\mathrm{om}, M}^{[0]} - J \lesssim \delta |\ln \delta| \sqrt{\epsilon} v_{ti} R$, note that the particles that bounce in the new maxima of $B$ are in an interval of $\lambda$ of order $\Delta \lambda \sim \delta \epsilon/B_0$. This interval corresponds to an interval in $J$ of order $\Delta \lambda \, \partial_\lambda J_W$, where $\partial_\lambda J_W \sim \epsilon^{-1/2} v_{ti} R B_0 |\ln \delta|$. The logarithm is due to the fact that the bounce points are always a small distance $\sqrt{\delta} R$ away from a local maximum of $B$. Particles spend logarithmically long times near these local maxima, giving $\partial_\lambda J_W \simeq - v^2 B_0 \tau_W/2 \sim \epsilon^{-1/2} v_{ti} B_0 R |\ln \delta|$. 

\begin{figure}
\begin{center}
\includegraphics[width=9cm]{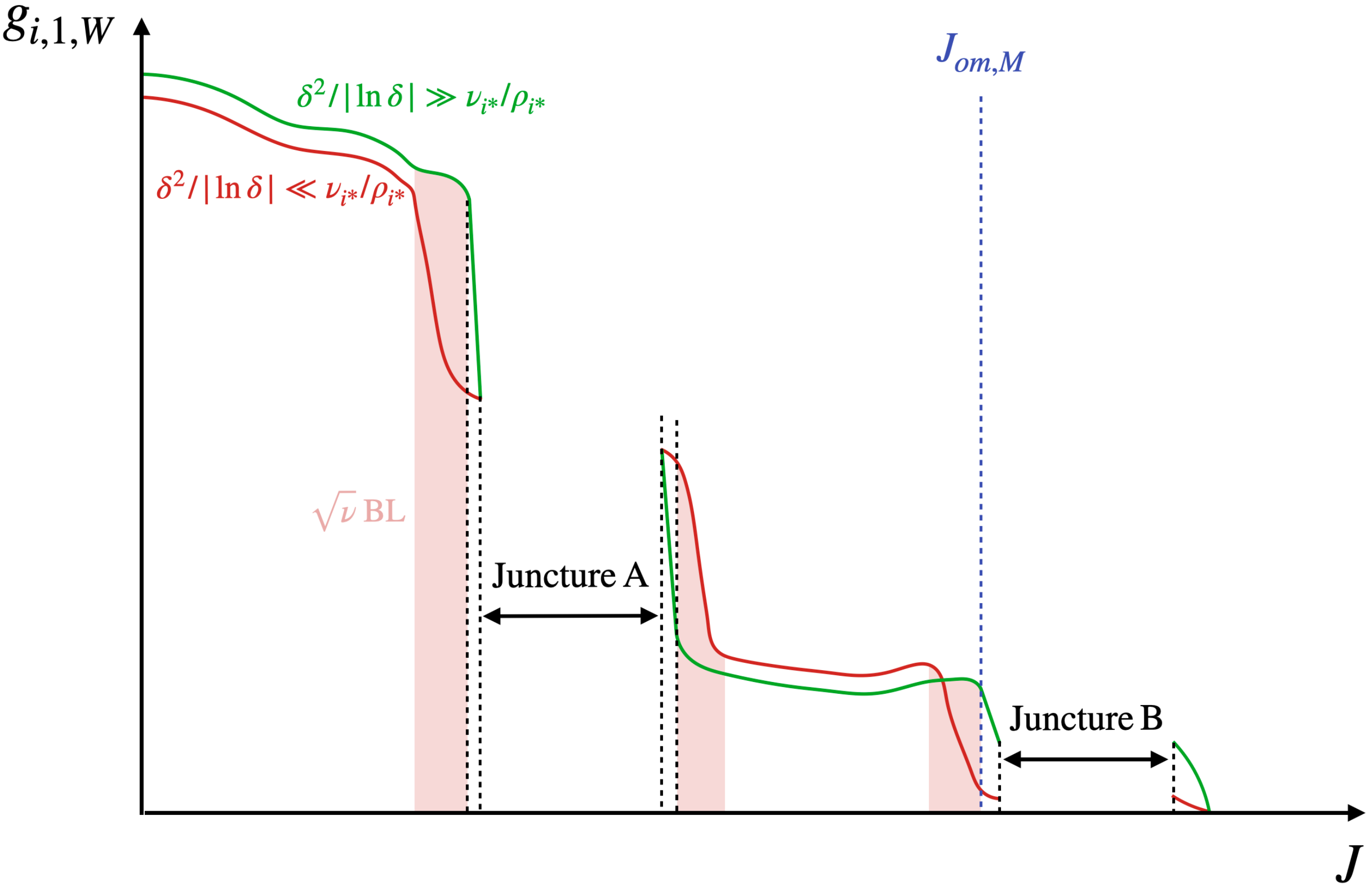}
\end{center}
\caption{\label{fig:giom} Sketch of the ion distribution function $g_{i, 1, W}$ in near-omnigeneous large aspect ratio stellarators with mirror ratios close to unity for $\delta^2 |\ln \delta | \ll \nu_{i*}/\rho_{i*} \ll 1$ (red line) and $\nu_{i*}/\rho_{i*} \ll \delta^2 |\ln \delta |$ (green line). The maximum value of the second adiabatic invariant in the omnigeneous magnetic field $J_{\mathrm{om}, M}^{[0]}$ is represented as a blue dashed line. We consider a case where there is a juncture in the region $J_{\mathrm{om}, M}^{[0]} - J \gg \delta |\ln \delta | \sqrt{\epsilon} v_{ti} R$ (juncture A). Of all the junctures that appear for $J_{\mathrm{om}, M}^{[0]} - J \lesssim \delta |\ln \delta | \sqrt{\epsilon} v_{ti} R$, we only sketch one (juncture B). The $\sqrt{\nu}$ boundary layers that appear in the case $\delta^2 |\ln \delta | \ll \nu_{i*}/\rho_{i*} \ll 1$ (red line) are highlighted as pink areas. Note that, outside of the $\sqrt{\nu}$ boundary layers and the junctures, the derivative of the distribution function $g_{i,1,W}$ with respect to $J$ remains the same as $\nu_{i*}/\rho_{i*}$ changes.}
\end{figure}

To determine whether the particles in the region $J_{\mathrm{om}, M}^{[0]} - J \lesssim \delta |\ln \delta| \sqrt{\epsilon} v_{ti} R$ can make up for the difference between $g_{i, 1, W}^{[1]}$ and the distribution function in the passing particle region, we estimate the size of the change in the distribution function $\Delta g_{i, 1, \mathrm{om}}$  for $J_{\mathrm{om}, M}^{[0]} - J \lesssim \delta |\ln \delta| \sqrt{\epsilon} v_{ti} R$. For $J_{\mathrm{om}, M}^{[0]} - J \lesssim \delta |\ln \delta| \sqrt{\epsilon} v_{ti} R$, we find
\begin{equation}
\partial_J g_{i, 1, W} \sim \frac{\Delta g_{i,1, \mathrm{om}}}{\delta |\ln \delta| \sqrt{\epsilon} v_{ti} R}.
\end{equation}
Using this estimate in equation~\refe{eq:eqgi1}, we find that the dominant balance is between the collision operator and the term proportional to $\partial_\alpha r_{i, 1, W}$, giving
\begin{equation} \label{eq:eqDgom}
\frac{\nu_{ii}}{\delta^2 |\ln \delta| \epsilon} \Delta g_{i,1, \mathrm{om}} \sim \frac{c\phi_0^\prime}{\Psi_t^\prime} \partial_\alpha r_{i, 1, W} \, \Upsilon_i f_{Mi}.
\end{equation}
Here, one of the logarithms $|\ln \delta|$ due to a derivative with respect to $J$ has cancelled with another logarithm $|\ln \delta|$ in $\tau_W$, which is of order $R |\ln \delta|/\sqrt{\epsilon} v_{ti}$ because the bounce points of the orbits are always a small distance $\sqrt{\delta} R$ away from a local maximum of $B$. In equation~\refe{eq:eqDgom}, $r_{i, 1, W}$ is proportional to $\lambda_W  - 1/B_M \sim \delta \epsilon /B_0$ and hence of order $\delta \epsilon a$. As a result, we estimate $\Delta g_{i,1, \mathrm{om}}$ to be
\begin{equation} \label{eq:solgom}
\Delta g_{i,1, \mathrm{om}} \sim \delta^3 |\ln \delta| \frac{\rho_{i*}}{\nu_{i*}} \epsilon f_{Mi} \ll \delta \epsilon f_{Mi}.
\end{equation}
This means that the solution for $g_{i, 1, W}$ in the region $J_{\mathrm{om}, M}^{[0]} - J \lesssim \delta |\ln \delta| \sqrt{\epsilon} v_{ti} R$ cannot match with $g_{i, 1, W}^{[1]}$ and we still have a discontinuity --  see figure~\ref{fig:giom} for a sketch of the situation. A similar reasoning can be made for $g_{i,1,W}$ near a juncture.

Collisional boundary layers appear at the discontinuity between $g_{i, 1, W}^{[1]}$ and the distribution function at $J_{\mathrm{om}, M}^{[0]} - J \lesssim \delta |\ln \delta | \sqrt{\epsilon} v_{ti} R$, and at the discontinuities on junctures. In these boundary layers, $g_{i, 1, W}$ changes significantly over a width $\Delta J_{\sqrt{\nu}} \ll \sqrt{\epsilon} v_{ti} R$, giving $\partial_J g_{i, 1, W} \sim g_{i, 1, W}/\Delta J_{\sqrt{\nu}} \gg g_{i, 1, W}/\sqrt{\epsilon} v_{ti} R$. This large derivatives in $J$ make the collision operator large and comparable to the $\bE \times \bB$ drift \citep{ho87, calvo17, calvo18a},
\begin{equation} \label{eq:eqwidthsqrtnu}
\frac{\nu_{ii}}{\epsilon} \left ( \frac{\sqrt{\epsilon} v_{ti} R}{\Delta J_{\sqrt{\nu}}} \right )^2 \ln \left ( \frac{\sqrt{\epsilon} v_{ti} R}{\Delta J_{\sqrt{\nu}}} \right ) \sim \frac{c\phi_0^\prime}{\Psi_t^\prime}.
\end{equation}
The logarithm in this estimate comes from the fact that the bounce points of the particles in this layer are close to local maxima of the omnigeneous magnetic field. Solving equation~\refe{eq:eqwidthsqrtnu} for $\Delta J_{\sqrt{\nu}}$ gives
\begin{equation} \label{eq:DJsqrt}
\frac{\Delta J_{\sqrt{\nu}}}{\sqrt{\epsilon} v_{ti} R} \sim \sqrt{\frac{\nu_{i*}}{\rho_{i*}} \ln \left ( \frac{\rho_{i*}}{\nu_{i*}} \right )} \ll 1.
\end{equation}
The distribution function in these boundary layers is not of the form in equation~\refe{eq:solgiomnu}, and hence does not vanish under the integrals for the particle and energy fluxes, given in equations~\refe{eq:Gammaom} and \refe{eq:Qom}. Thus, the fluxes are mostly due to particles within the boundary layer, giving
\begin{equation} \label{eq:fluxomsqrtnu}
\Gamma_i \sim \delta^2 \sqrt{\frac{\nu_{i*}}{\rho_{i*}} \ln \left ( \frac{\rho_{i*}}{\nu_{i*}} \right )} \epsilon^{5/2} \rho_{i*} n_i v_{ti}, \quad Q_i \sim \delta^2 \sqrt{\frac{\nu_{i*}}{\rho_{i*}} \ln \left ( \frac{\rho_{i*}}{\nu_{i*}} \right )} \epsilon^{5/2} \rho_{i*} n_i T_i v_{ti}.
\end{equation}
These are the particle and energy fluxes of the $\sqrt{\nu}$ regime \citep{ho87, calvo17, calvo18a}. This regime only appears in large aspect ratio stellarators with mirror ratios close to unity when they are close to omnigeneity.

The estimates in equation~\refe{eq:fluxomsqrtnu} can be obtained from a random walk argument. The particles that contribute most to radial transport are the particles in velocity space boundary layers of width $\Delta v_\| \ll \sqrt{\epsilon} v_{ti}$. We start by estimating the width of the boundary layers. The effective collision frequency in these layers is 
\begin{equation} \label{eq:nuefflayereps}
\nu_\mathrm{eff} \sim \left ( \frac{v_{ti}}{\Delta v_\|} \right )^2 \frac{\nu_{ii}}{\ln (\sqrt{\epsilon} v_{ti}/\Delta v_\|)}. 
\end{equation}
We have obtained this effective collision frequency using the same arguments that we used to find equation~\refe{eq:nuefflayer}. Note that the only difference between equations~\refe{eq:nuefflayer} and \refe{eq:nuefflayereps} is in the logarithm, where we find an extra factor of $\sqrt{\epsilon}$ in equation~\refe{eq:nuefflayereps}. In the layer, the effective collision frequency is of the same order as the inverse of the time that it takes for a particle to drift poloidally around the stellarator, $\nu_\mathrm{eff} \sim |\bv_E|/a \sim \rho_{i*} v_{ti}/a$. This balance gives a boundary layer width
\begin{equation} \label{eq:Dvparsqrteps}
\Delta v_\| \sim \sqrt{\epsilon} v_{ti} \sqrt{\frac{\nu_{i*}}{\rho_{i*}} \frac{1}{\ln (\rho_{i*}/\nu_{i*})}}. 
\end{equation} 
Comparing this result with equation~\refe{eq:DJsqrt}, we see that the width of the layer in $v_\|$ is smaller than in $J$, $\Delta J_{\sqrt{\nu}}/\sqrt{\epsilon} v_{ti} R \sim \ln (\rho_{i*}/\nu_{i*}) (\Delta v_\|/\sqrt{\epsilon} v_{ti}) \gg \Delta v_\|/\sqrt{\epsilon} v_{ti}$. This is consistent with the transformation between the velocity space variables $\vbar$ and $J$ and the cartesian velocity $\bv$. Indeed, we can take $\Delta v_\|/\sqrt{\epsilon} v_{ti} \sim B_0 \Delta \lambda/\epsilon$, with $\Delta \lambda$ the size of the layer in $\lambda := v_\perp^2/v^2 B$, and $\Delta J \sim \Delta \lambda \, \partial_\lambda J_W \sim \sqrt{\epsilon} v_{ti} R (B_0 \Delta \lambda/\epsilon) \ln ( \epsilon/ B_0 \Delta \lambda )$, where the logarithmic divergence is a result of particles spending a long time around the local maxima of $B$. 

To obtain estimates for the fluxes, in addition to the layer width $\Delta v_\|$ and the effective collision frequency $\nu_\mathrm{eff}$, we need the typical radial displacement between collisions of particles in the boundary layers, given by $\langle \bv_{Mi}\cdot \nabla r\rangle_\tau/\nu_\mathrm{eff}$. Since $\langle \bv_{Mi} \cdot \nabla r \rangle_\tau \sim \delta \epsilon \rho_{i*} v_{ti}$ and $\nu_\mathrm{eff} \sim \rho_{i*} v_{ti}/a$, we find $\langle \bv_{Mi}\cdot \nabla r\rangle_\tau/\nu_\mathrm{eff} \sim \delta \epsilon a$. With all these results, we can write the random walk diffusion coefficient as
\begin{equation} \label{eq:Dsqrteps}
D \sim \frac{\Delta v_\|}{v_{ti}} \ln \left ( \frac{\sqrt{\epsilon} v_{ti}}{\Delta v_\|} \right ) \left ( \frac{\langle \bv_{Mi} \cdot \nabla r \rangle_\mathrm{\tau}}{\nu_\mathrm{eff}} \right )^2 \nu_\mathrm{eff}, 
\end{equation}
where we have used that the fraction of particles in the boundary layers is of the order of $(\Delta v_\|/v_{ti}) \ln (\sqrt{\epsilon} v_{ti}/\Delta v_\|)$. The logarithmic correction in the particle fraction is due to the accumulation of slow particles around the local maxima of $B$. The diffusion coefficient in equation~\refe{eq:Dsqrteps} gives $D \sim \delta^2 \sqrt{\nu_{i*} \rho_{i*} \ln (\rho_{i*}/\nu_{i*})} \epsilon^{5/2} a v_{ti}$, and with it we can obtain the estimates in equation~\refe{eq:fluxomsqrtnu}.

\subsubsection{The $\nu$ regime in optimized large aspect ratio stellarators with mirror ratios close to unity}

The $\sqrt{\nu}$ regime stops being valid for $\nu_{i*}/\rho_{i*} \sim \delta^2 |\ln \delta|$ because the distribution function at $J_{\mathrm{om}, M}^{[0]} - J \lesssim \delta |\ln \delta| \sqrt{\epsilon} v_{ti} R$, given in equation~\refe{eq:solgom}, becomes comparable to $g_{i, 1, W}^{[1]}$ and there is no need for a boundary layer any longer -- see figure~\ref{fig:giom} for a sketch. The same happens for the discontinuities on junctures of different types of wells.

For $\nu_{i*}/\rho_{i*} \lesssim \delta^2 |\ln \delta|$, we need to use the procedure laid out in section~\ref{sec:nuregime}. The resulting fluxes are given in equations~\refe{eq:Gammanu} and \refe{eq:Qnu}. Particles with $J_{\mathrm{om}, M}^{[0]} - J \gg \delta |\ln \delta| \sqrt{\epsilon} v_{ti} R$ that are outside of the regions affected by junctures move without having to transition between different types of wells. Thus, particles with $J_{\mathrm{om}, M}^{[0]} - J \gg \delta |\ln \delta| \sqrt{\epsilon} v_{ti} R$ that are outside of regions affected by junctures are in the region of velocity space with $\partial_J K_{i, W} \neq 0$. The integral in this region has an integrand proportional to $1/\tau_W - 1/\langle \tau_W \rangle_\alpha$, and this difference is small because to lowest order, $\tau_W$ does not depend on $\alpha$, $1/\tau_W - 1/\langle \tau_W \rangle_\alpha \sim \delta R/\sqrt{\epsilon} v_{ti}$. Moreover, $1/\tau_W - 1/\langle \tau_W \rangle_\alpha \sim \delta \sqrt{\epsilon} v_{ti}/R$ is integrated over $\alpha$, and the integral vanishes to lowest order, finally giving
\begin{equation} \label{eq:cancelom}
\int_0^{2\pi} \left ( \frac{1}{\tau_W} - \frac{1}{\langle \tau_W \rangle_\alpha} \right )\, \rmd \alpha \simeq \int_0^{2\pi} \frac{(\tau_W - \langle \tau_W \rangle_\alpha)^2}{\langle \tau_W \rangle_\alpha^3} \, \rmd \alpha \sim \frac{\delta^2 \sqrt{\epsilon} v_{ti}}{R}.
\end{equation}
Then, the contribution to $\Gamma_i$ and $Q_i$ from particles with $J_{\mathrm{om}, M}^{[0]} - J \gg \delta |\ln \delta| \sqrt{\epsilon} v_{ti} R$ and outside of the regions affected by junctures are $\delta^2 \epsilon^{5/2} \nu_{i*} n_i v_{ti}$ and $\delta^2 \epsilon^{5/2} \nu_{i*} n_i T_i v_{ti}$, respectively. Note that these are consistent with estimates that one can find using the approximate integrals~\refe{eq:Gammaom} and \refe{eq:Qom} for the fluxes.

Conversely, in the region $J_{\mathrm{om}, M}^{[0]} - J \lesssim \delta |\ln \delta| \sqrt{\epsilon} v_{ti} R$ or in the region where particles are affected by junctures, particles have multiple transitions, and hence $\partial_J K_{i, W} = 0$. There is then no cancellation similar to the one in equation~\refe{eq:cancelom}. Using that the typical size of the intervals of interest in $J$ is $\delta |\ln \delta| \sqrt{\epsilon} v_{ti} R$ and that $\tau_W \sim R|\ln \delta|/\sqrt{\epsilon} v_{ti}$, the integrals in equations~\refe{eq:Gammanu} and \refe{eq:Qnu} give
\begin{equation} \label{eq:fluxomnu}
\Gamma_i \sim \delta \epsilon^{5/2} \nu_{i*} n_i v_{ti}, \quad Q_i \sim \delta \epsilon^{5/2} \nu_{i*} n_i T_i v_{ti}.
\end{equation}
These contributions are larger than the ones from the particles that satisfy $J_{\mathrm{om}, M}^{[0]} - J \gg \delta |\ln \delta| \sqrt{\epsilon} v_{ti} R$ and are outside of the regions affected by junctures, and hence dominate the fluxes.

We proceed to obtain the estimates in equation~\refe{eq:fluxomnu} using a random walk argument. For $\nu_{i*}/\rho_{i*} \ll \delta^2 |\ln \delta|$, particles remain close to the flux surface in which they started because their radial drift averages out after several poloidal turns around the stellarator. The time for several poloidal turns is of the order of $a/|\mathbf{v}_E|$ and, in that time, particles drift radially a distance $w \sim \langle \mathbf{v}_{Mi} \cdot \nabla r \rangle_\tau a/|\mathbf{v}_E| \sim \delta \epsilon a$. Transport is dominated by particles in thin velocity layers of width $\Delta v_\| /\sqrt{\epsilon} v_{ti} \sim \delta$ located at the trapped-passing boundary or around junctures of several wells. For particles in these layers, we can apply the formula for the effective collision frequency in a thin layer in equation~\refe{eq:nuefflayereps}, finding $\nu_\mathrm{eff} \sim \nu_{ii}/\delta^2|\ln \delta|$. With these results, we find the diffusion coefficient $D \sim \delta |\ln \delta| \sqrt{\epsilon} w^2 \nu_\mathrm{eff}$, where the fraction of particles in the boundary layer is $\delta |\ln \delta| \sqrt{\epsilon}$. The logarithmic correction to the fraction of particles in the layer is due to the accumulation of particles around the local maxima of $B$. Using all our estimates, we finally obtain $D \sim \delta \epsilon^{5/2} a v_{ti}$, and this diffusion coefficient leads to the estimates in equation~\refe{eq:fluxomnu}.

\begin{figure}
\begin{center}
\includegraphics[width=10cm]{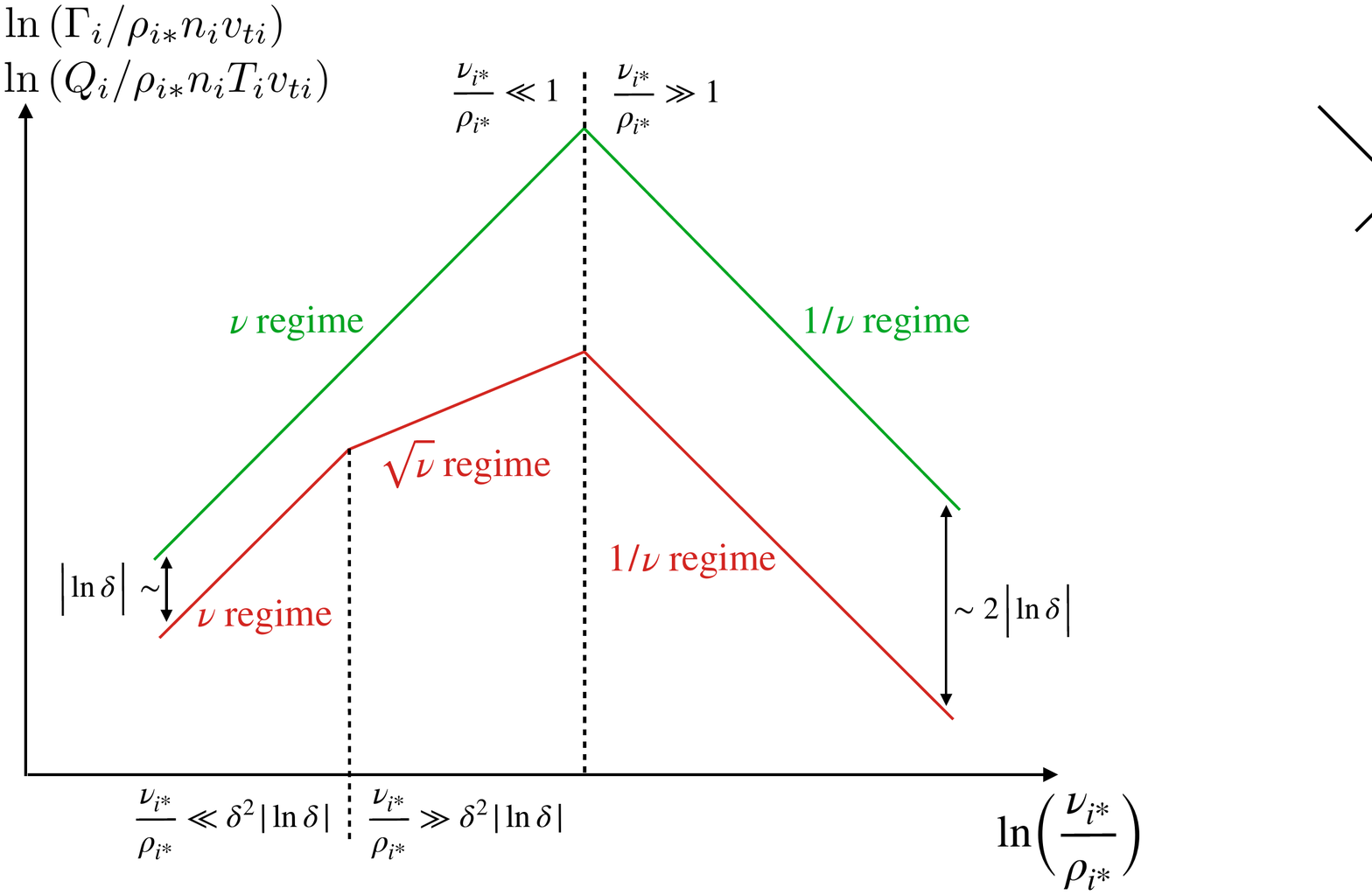}
\end{center}
\caption{\label{fig:fluxes} Particle and energy fluxes as a function of $\nu_{i*}/\rho_{i*}$ for generic large aspect ratio stellarators with mirror ratios close to unity (green line) and near-omnigeneous large aspect ratio stellarators with mirror ratios close to unity (red line).}
\end{figure}

In figure~\ref{fig:fluxes}, we sketch the estimates in equations~\refe{eq:fluxom1nu}, \refe{eq:fluxomsqrtnu} and \refe{eq:fluxomnu} for the fluxes in near-omnigeneous large aspect ratio stellarators with mirror ratios close to unity. For comparison, we sketch the corresponding estimates~\refe{eq:fluxes1nu} and \refe{eq:fluxesnu} for generic large aspect ratio stellarators with mirror ratios close to unity. Note that the optimization towards omnigeneity is much more effective in the $1/\nu$ regime. The reduced effectiveness of the optimization in the $\sqrt{\nu}$ and $\nu$ regimes is due to the barely trapped particles and the particles near junctures of different types of wells.

\section{Conclusion} \label{sec:conclusion}

There are four results in this article that are worth emphasizing. The first one is the set of orbit-averaged equations for neoclassical transport at low collisionality in large aspect ratio stellarators with mirror ratios close to unity derived in section~\ref{sec:fiepsilon}. It consists of a kinetic equation for trapped particles, given in equation~\refe{eq:eqgi1}, and the corresponding boundary conditions to be imposed around junctures of different types of wells (see equation~\refe{eq:eqgi1transition}) and at the trapped-passing boundary (see equation~\refe{eq:gi1bc}). The trapped particle distribution function obtained from these equations can then be integrated to give the particle and energy fluxes (see equations~\refe{eq:Gammaidef} and \refe{eq:Qidef}). To our knowledge, this is the most detailed derivation of a model for large aspect ratio stellarators with mirror ratios close to unity in the limit $\nu_{i*} \sim \rho_{i*}$, and in conjunction with the model described in \cite{calvo17} and \cite{calvo18a} for near-omnigeneous stellarators, is the only self-consistent local model for stellarator neoclassical transport at low collisionality that we are aware of. The set of equations has been implemented in KNOSOS \citep{velasco21}, and it produces fluxes close to those calculated by DKES but with much less computational effort. The fact that the model matches DKES neoclassical fluxes is unsurprising, as we have shown in Appendix~\ref{app:DKES} that the model and DKES kinetic equations are the same to lowest order in the inverse aspect ratio. Interestingly, the derivation in Appendix~\ref{app:DKES} also shows that there is an $O(\sqrt{\epsilon})$ difference between our model and DKES equations, and DKES is not correct to that order. Thus, when the results from the equations in this paper differ from those of DKES, one cannot assume that DKES is correct by default. More theoretical and numerical work on these higher order corrections to the aspect ratio expansion is needed as $\sqrt{\epsilon}$ is not very small.

The second interesting result in this article is the limit $\nu_{i*} \ll \rho_{i*}$ for generic large aspect ratio stellarators with mirror ratios close to unity, described in section~\ref{sec:nuregime}. We have been able to show that there is no $\sqrt{\nu}$ regime when this type of stellarator is not close to omnigeneity. Instead, generic large aspect ratio stellarators with mirror ratios close to unity enter the $\nu$ regime directly for $\nu_{i*} \lesssim \rho_{i*}$. We have also examined the $\nu$ regime in great detail, explaining how the distribution function behaves in this limit. The transitions from one type of well to another are a crucial aspect. While these transitions make the particle motion stochastic and diffuse particles in velocity space, there is no stochastic diffusion in real space and the neoclassical diffusion coefficient remains proportional to the collision frequency. We believe that transitions between different types of wells do not lead to stochastic real space diffusion independent of the collision frequency because ions in the regimes that we consider are confined by the electric field, and while transitions can cause jumps in the second adiabatic invariant $J$, they do not change the total energy of the particle. When the total energy is conserved, particles cannot move long distances ($\gg \epsilon a$) radially because they cannot overcome the barrier set by the electric potential, that is, particles remain within a distance of order $\epsilon a$ of the flux surface that they started at without collisions. Collisions are needed to break conservation of total energy for each particle. While collisions conserve energy overall, in each collision, individual trapped particles can gain or lose energy. In our calculation, they mostly exchange energy with a large population of passing particles. We should add that, even though transitions between different types of wells do not make the neoclassical diffusion independent of the collision frequency, they do enhance neoclassical diffusion, as we explain below equation~\refe{eq:Gammanu}.

Part of the outcome of the $\nu$ regime calculation is the explicit formula for the fluxes in the $\nu$ regime, given in equations~\refe{eq:Gammanu} and \refe{eq:Qnu}. With these formulas, the fluxes can be calculated from the magnetic field of the stellarator without solving a kinetic equation. These explicit formulas might be useful for optimization routines if they can be evaluated fast. We believe that this is possible using techniques similar to those developed to perform bounce averages for KNOSOS \citep{velasco20}.

The third result that we want to emphasize is that large aspect ratio stellarators with large mirror ratios are close to omnigeneous and can be treated using the formulation developed by \cite{calvo17}. These stellarators experience a $\sqrt{\nu}$ regime for $\epsilon |\ln \epsilon| \ll \nu_{i*}/\rho_{i*} \ll 1/\epsilon$. The neoclassical fluxes in large aspect ratio stellarators with large mirror ratios are significantly larger than the ones that one would obtain in an equivalent large aspect ratio stellarator with mirror ratio close to unity.

Finally, the fourth result of note is the limit of near-omnigeneity for large aspect ratio stellarators with mirror ratios close to unity. When we consider this type of stellarator, we find the $\sqrt{\nu}$ regime for a range of collisionalities that depends on the deviation from omnigeneity $\delta$: $\delta^2 |\ln \delta| \ll \nu_{i*}/\rho_{i*} \ll 1$. This interval of validity disappears when $\delta \sim 1$, explaining why we could not find a $\sqrt{\nu}$ regime in generic large aspect ratio stellarators with mirror ratios close to unity. The derivation for near-omnigeneous large aspect ratio stellarators with mirror ratios close to unity gives another interesting result, namely, optimization is less effective for $\nu_{i*} \ll \rho_{i*}$, and it is worse in the $\nu$ regime, where the fluxes are reduces only by a factor of $\delta$ instead of by a factor of $\delta^2$. For this reason, it is probably worth considering the use of the $\nu$ regime fluxes in equations~\refe{eq:Gammanu} and \refe{eq:Qnu} as optimization targets. If one can reduce effectively the fluxes in the regime that is less responsive to optimization, the fluxes in the other regimes will likely reduce even more. In particular, we expect that the optimization of the fluxes in equations~\refe{eq:Gammanu} and \refe{eq:Qnu} will target problematic particles at the trapped-passing boundary and in regions where there are transitions between different types of wells.

\section*{Acknowledgments}

The authors would like to thank Xinyu Liu, Peter J. Catto and the two anonymous referees for their suggestions.

\section*{Funding}

This work was supported by the U.S. Department of Energy under contract number DE-AC02-09CH11466. The United States Government retains a non-exclusive, paid-up, irrevocable, world-wide license to publish or reproduce the published form of this manuscript, or allow others to do so, for United States Government purposes. This work has been carried out within the framework of the EUROfusion Consortium, funded by the European Union via the Research and Training Programme (Grant Agreement No 101052200 -- EUROfusion). Views and opinions expressed are however those of the authors only and do not necessarily reflect those of the European Union or the European Commission. Neither the European Union nor the European Commission can be held responsible for them. This research was supported in part by Grant PGC2018-095307-B-I00, Ministerio de Ciencia, Innovaci\'on y Universidades, Spain.

\section*{Declaration of Interests}

The authors report no conflict of interest.

\section*{Author ORCID}

F. I. Parra, https://orcid.org/0000-0001-9621-7404; I. Calvo, https://orcid.org/0000-0003-3118-3463. 

\appendix

\section{Discontinuities in the derivatives of the distribution function at junctures of different types of wells} \label{app:junctures}
In this Appendix, we show how to obtain equation~\refe{rho2transition} by imposing, in the example juncture of figure~\ref{fig:exampleU}, that the collisional flux of particles out of wells $I$ and $II$ enters well $III$. 

For trapped particles, the collision operator in equation~\refe{eq:CiiEmu} can be written as $C_{ii} [ g_{i, W}, \overline{f}_i^{(0)}] = - |v_\| | ( \partial_\mathcal{E} F_\mathcal{E} [ g_{i, W}, \overline{f}_i^{(0)}] + \partial_\mu F_\mu [ g_{i, W}, \overline{f}_i^{(0)}] )$, where
\begin{equation}
F_\mathcal{E} [ g_{i, W}, \overline{f}_i^{(0)}] := - \frac{\gamma_{ii}}{|v_\| |} \left ( H_{\mathcal{E} \mathcal{E}} [ \overline{f}_i^{(0)} ] \,  \partial_\mathcal{E} g_{i, W} + H_{\mathcal{E} \mu} [ \overline{f}_i^{(0)} ] \, \partial_\mu g_{i, W} - L_\mathcal{E} [ \overline{f}_i^{(0)} ] \, g_{i, W} \right ) 
\end{equation}
is the phase space particle flux in the $\mathcal{E}$ direction and
\begin{align}
F_\mu [ g_{i, W}, \overline{f}_i^{(0)}] :=  - \frac{\gamma_{ii}}{|v_\| |} \left ( H_{\mu \mathcal{E}} [ \overline{f}_i^{(0)} ] \,  \partial_\mathcal{E} g_{i, W} + H_{\mu \mu} [ \overline{f}_i^{(0)} ] \, \partial_\mu g_{i, W} - L_\mu [ \overline{f}_i^{(0)} ] \, g_{i, W} \right )
\end{align}
is the phase space particle flux in the $\mu$ direction. We need the particle flux perpendicular to the boundary $\mathcal{E} = \mathcal{E}_c (r, \alpha, \mu, t)$. To obtain this flux, we use coordinates in which the boundary $\mathcal{E} = \mathcal{E}_c (r, \alpha, \mu, t)$ becomes trivial: instead of using $r$, $\alpha$, $\mathcal{E}$, $\mu$ and $\varphi$, we replace the variable $\mathcal{E}$ by $\Delta \mathcal{E} := \mathcal{E} - \mathcal{E}_c (r, \alpha, \mu, t)$ such that $\Delta \mathcal{E} = 0$ gives the desired boundary. With these new variables, the collision operator in equation~\refe{eq:CiiEmu} becomes 
\begin{align} \label{eq:Ciiacrossjunct}
C_{ii} [ g_{i, W}, \overline{f}_i^{(0)}] = & - |v_\| | \bigg \{ \partial_{\Delta \mathcal{E}} \left ( F_\mathcal{E} [ g_{i, W}, \overline{f}_i^{(0)}] + \partial_\mu \Delta \mathcal{E} \, F_\mu [ g_{i, W}, \overline{f}_i^{(0)}] \right ) \nonumber\\ & + \partial_\mu F_\mu [ g_{i, W}, \overline{f}_i^{(0)}] \bigg \},
\end{align}
where we have used the fact that $\partial_\mu \Delta \mathcal{E} = - \partial_\mu \mathcal{E}_c$ satisfies $\partial_{\Delta \mathcal{E}} ( \partial_\mu \Delta \mathcal{E} ) = 0$. Equation~\refe{eq:Ciiacrossjunct} indicates that the flux of particles across the boundary $\Delta \mathcal{E} = 0$ is given by $F_\mathcal{E} - \partial_\mu \mathcal{E}_c \, F_\mu$. Integrating this flux over the gyrophase $\varphi$ and along trapped orbits, we find that the phase space flux across the boundary $\mathcal{E} = \mathcal{E}_c(r, \alpha, \mu, t)$ can finally be written as
\begin{align}
- 2\pi \gamma_{ii} \sum_\sigma & \int_{l_{bL,W}}^{l_{bR,W}} \Bigg [ H_{\mathcal{E} \mathcal{E}} [ \overline{f}_i^{(0)}]\,  \partial_\mathcal{E} g_{i,W} + H_{\mathcal{E} \mu}[ \overline{f}_i^{(0)}] \, \partial_\mu g_{i,W} - L_\mathcal{E}[ \overline{f}_i^{(0)}] \, g_{i,W}  \nonumber\\ & - \partial_\mu \mathcal{E}_c \left ( H_{\mu \mathcal{E}}[ \overline{f}_i^{(0)}] \,  \partial_\mathcal{E} g_{i,W} + H_{\mu \mu}[ \overline{f}_i^{(0)}] \, \partial_\mu g_{i,W} - L_\mu [ \overline{f}_i^{(0)}] \, g_{i,W} \right ) \Bigg ]  \frac{\rmd l}{|v_\| |}. 
\end{align}

Using the definition of the transit average in equation~\refe{eq:transitave}, the balance between the three collisional fluxes in and out of wells $I$, $II$ and $III$ in figure~\ref{fig:exampleU} is
\begin{align} \label{eq:rho2transitionv1}
 - & 2\pi \gamma_{ii} \tau_I \Bigg [ \langle H_{\mathcal{E} \mathcal{E}} [ \overline{f}_i^{(0)}] \rangle_{\tau, I} \,  \partial_\mathcal{E} g_{i, I} + \langle H_{\mathcal{E} \mu}[ \overline{f}_i^{(0)}] \rangle_{\tau, I} \, \partial_\mu g_{i,I} - \langle L_\mathcal{E}[ \overline{f}_i^{(0)}] \rangle_{\tau, I} \, g_{i,I}  \nonumber\\ & - \partial_\mu \mathcal{E}_c \left ( \langle H_{\mu \mathcal{E}}[ \overline{f}_i^{(0)}] \rangle_{\tau, I} \,  \partial_\mathcal{E} g_{i,I} + \langle H_{\mu \mu}[ \overline{f}_i^{(0)}] \rangle_{\tau, I} \, \partial_\mu g_{i,I} - \langle L_\mu [ \overline{f}_i^{(0)}] \rangle_{\tau, I} \, g_{i,I} \right ) \Bigg ] \nonumber\\ & - 2\pi \gamma_{ii} \tau_{II} \Bigg [ \langle H_{\mathcal{E} \mathcal{E}} [ \overline{f}_i^{(0)}] \rangle_{\tau, II} \,  \partial_\mathcal{E} g_{i, II} + \langle H_{\mathcal{E} \mu}[ \overline{f}_i^{(0)}] \rangle_{\tau, II} \, \partial_\mu g_{i,II} - \langle L_\mathcal{E}[ \overline{f}_i^{(0)}] \rangle_{\tau, II} \, g_{i,II}  \nonumber\\ & - \partial_\mu \mathcal{E}_c \left ( \langle H_{\mu \mathcal{E}}[ \overline{f}_i^{(0)}] \rangle_{\tau, II} \,  \partial_\mathcal{E} g_{i,II} + \langle H_{\mu \mu}[ \overline{f}_i^{(0)}] \rangle_{\tau, II} \, \partial_\mu g_{i,II} - \langle L_\mu [ \overline{f}_i^{(0)}] \rangle_{\tau, II} \, g_{i,II} \right ) \Bigg ]  \nonumber \\ = & - 2\pi \gamma_{ii} \tau_{III} \Bigg [ \langle H_{\mathcal{E} \mathcal{E}} [ \overline{f}_i^{(0)}] \rangle_{\tau, III} \,  \partial_\mathcal{E} g_{i, III} + \langle H_{\mathcal{E} \mu}[ \overline{f}_i^{(0)}] \rangle_{\tau, III} \, \partial_\mu g_{i,III} - \langle L_\mathcal{E}[ \overline{f}_i^{(0)}] \rangle_{\tau, III} \, g_{i,III}  \nonumber\\ & - \partial_\mu \mathcal{E}_c \left ( \langle H_{\mu \mathcal{E}}[ \overline{f}_i^{(0)}] \rangle_{\tau, III} \,  \partial_\mathcal{E} g_{i,III} + \langle H_{\mu \mu}[ \overline{f}_i^{(0)}] \rangle_{\tau, III} \, \partial_\mu g_{i,III} - \langle L_\mu [ \overline{f}_i^{(0)}] \rangle_{\tau, III} \, g_{i,III} \right ) \Bigg ].
\end{align}
This long expression can be simplified due to the continuity of $g_{i,W}$ and $\partial_\mu \mathcal{E}_c \, \partial_\mathcal{E} g_{i,W} + \partial_\mu g_{i,W}$ across $\mathcal{E} = \mathcal{E}_c (r, \alpha, \mu, t)$. The continuity of $g_{i, W}$ implies that $\tau_I \langle H_{pq} [ \overline{f}_i^{(0)} ] \rangle_{\tau, I} + \tau_{II} \langle H_{pq} [ \overline{f}_i^{(0)} ] \rangle_{\tau, II} = \tau_{III} \langle H_{pq} [ \overline{f}_i^{(0)} ] \rangle_{\tau, III}$ and that $\tau_I \langle L_p [ \overline{f}_i^{(0)} ] \rangle_{\tau, I} + \tau_{II} \langle L_p [ \overline{f}_i^{(0)} ] \rangle_{\tau, II} = \tau_{III} \langle L_p [ \overline{f}_i^{(0)} ] \rangle_{\tau, III}$ for $p = \mathcal{E}, \mu$ and $q = \mathcal{E}, \mu$. Using these results and dividing by $-2\pi\gamma_{ii}$, expression~\refe{eq:rho2transitionv1} simplifies to 
\begin{align} \label{eq:rho2transitionv2}
\tau_I & \Bigg [ \langle H_{\mathcal{E} \mathcal{E}} [ \overline{f}_i^{(0)}] \rangle_{\tau, I} \,  ( \partial_\mathcal{E} g_{i, I} - \partial_\mathcal{E} g_{i, III}) + \langle H_{\mathcal{E} \mu}[ \overline{f}_i^{(0)}] \rangle_{\tau, I} \,( \partial_\mu g_{i,I} - \partial_\mu g_{i, III}) \nonumber\\ & - \partial_\mu \mathcal{E}_c \left ( \langle H_{\mu \mathcal{E}}[ \overline{f}_i^{(0)}] \rangle_{\tau, I} \,  (\partial_\mathcal{E} g_{i,I} - \partial_\mathcal{E} g_{i, III}) + \langle H_{\mu \mu}[ \overline{f}_i^{(0)}] \rangle_{\tau, I} \, ( \partial_\mu g_{i,I} - \partial_\mu g_{i, III} ) \right ) \Bigg ] \nonumber\\ & + \tau_{II} \Bigg [ \langle H_{\mathcal{E} \mathcal{E}} [ \overline{f}_i^{(0)}] \rangle_{\tau, II} \,  (\partial_\mathcal{E} g_{i, II} - \partial_\mathcal{E} g_{i, III}) + \langle H_{\mathcal{E} \mu}[ \overline{f}_i^{(0)}] \rangle_{\tau, II} \, (\partial_\mu g_{i,II} - \partial_\mu g_{i, III}) \nonumber\\ & - \partial_\mu \mathcal{E}_c \left ( \langle H_{\mu \mathcal{E}}[ \overline{f}_i^{(0)}] \rangle_{\tau, II} \,  (\partial_\mathcal{E} g_{i,II} - \partial_\mathcal{E} g_{i, III}) + \langle H_{\mu \mu}[ \overline{f}_i^{(0)}] \rangle_{\tau, II} \, (\partial_\mu g_{i,II} - \partial_\mu g_{i, III} \right ) \Bigg ] \nonumber \\ = & \, 0.
\end{align}
Using the continuity of the combination of $\partial_\mu \mathcal{E}_c\, \partial_\mathcal{E} g_{i, W} + \partial_\mu g_{i, W}$, we can write $\partial_\mu g_{i,I} - \partial_\mu g_{i, III} = - \partial_\mu \mathcal{E}_c\, ( \partial_\mathcal{E} g_{i,I} - \partial_\mathcal{E} g_{i, III} )$ and $\partial_\mu g_{i,II} - \partial_\mu g_{i, III} = - \partial_\mu \mathcal{E}_c\, ( \partial_\mathcal{E} g_{i,II} - \partial_\mathcal{E} g_{i, III} )$. With these results, and employing the fact that $H_{\mathcal{E} \mu} = H_{\mu \mathcal{E}}$, expression \refe{eq:rho2transitionv2} becomes
\begin{align} \label{eq:rho2transitionv3}
& \tau_I \left [ \langle H_{\mathcal{E} \mathcal{E}} [ \overline{f}_i^{(0)}] \rangle_{\tau, I} - 2 \langle H_{\mathcal{E} \mu}[ \overline{f}_i^{(0)}] \rangle_{\tau, I} \, \partial_\mu \mathcal{E}_c + \langle H_{\mu \mu}[ \overline{f}_i^{(0)}] \rangle_{\tau, I} \, (\partial_\mu \mathcal{E}_c)^2 \right ] \,  ( \partial_\mathcal{E} g_{i, I} - \partial_\mathcal{E} g_{i, III}) \nonumber\\ & \, + \tau_{II} \left [ \langle H_{\mathcal{E} \mathcal{E}} [ \overline{f}_i^{(0)}] \rangle_{\tau, I} - 2 \langle H_{\mathcal{E} \mu}[ \overline{f}_i^{(0)}] \rangle_{\tau, I} \, \partial_\mu \mathcal{E}_c + \langle H_{\mu \mu}[ \overline{f}_i^{(0)}] \rangle_{\tau, I} \, (\partial_\mu \mathcal{E}_c)^2 \right ] \,  ( \partial_\mathcal{E} g_{i, I} - \partial_\mathcal{E} g_{i, III}) \nonumber \\ & = 0.
\end{align}
Using the identities $\tau_I \langle H_{pq} [ \overline{f}_i^{(0)} ] \rangle_{\tau, I} + \tau_{II} \langle H_{pq} [ \overline{f}_i^{(0)} ] \rangle_{\tau, II} = \tau_{III} \langle H_{pq} [ \overline{f}_i^{(0)} ] \rangle_{\tau, III}$ again, we obtain the compact expression in equation~\refe{rho2transition}.

\section{Conditions on the flux surface shape imposed by the MHD equilibrium equations} \label{app:MHDeq}

In this Appendix, we give the constraints that $\bx_1(r, \alpha, l)$ must satisfy. In addition to equation~\refe{eq:BJacob0}, $\bx_1$ must satisfy the first order version of equation~\refe{eq:arc},  
\begin{equation} \label{eq:arc0}
\bun_0 \cdot \partial_l \bx_1 = 0,
\end{equation}
and a solvability constraint of the MHD force balance equations that we proceed to obtain. The second order terms in the $\epsilon$ expansion of equations~\refe{eq:MHDeqpsi} and \refe{eq:MHDeqalpha} are given by
\begin{align} \label{eq:MHDeqpsi1}
\partial_r \left ( \frac{B_0 B_2}{4\pi} + \frac{B_1^2}{8\pi} - \frac{B_0^2}{4\pi} \bkappa_0 \cdot \bx_2 - \frac{B_0}{4\pi} \frac{\rmd B_0}{\rmd l} \bun_0 \cdot \bx_2 \right ) - \frac{\partial_l (B_0 B_1)}{4\pi} \bun_0 \cdot \partial_r \bx_1 \nonumber\\  - \frac{B_0}{4\pi} \frac{\rmd B_0}{\rmd l} \partial_l \bx_1 \cdot \partial_r \bx_1 = \frac{B_0 B_1}{2\pi} \bkappa_0 \cdot \partial_r \bx_1 + \frac{B_0^2}{4\pi} \partial^2_{ll} \bx_1 \cdot \partial_r \bx_1
\end{align}
and
\begin{align} \label{eq:MHDeqalpha1}
\partial_\alpha \left ( \frac{B_0 B_2}{4\pi} + \frac{B_1^2}{8\pi} - \frac{B_0^2}{4\pi} \bkappa_0 \cdot \bx_2 - \frac{B_0}{4\pi} \frac{\rmd B_0}{\rmd l} \bun_0 \cdot \bx_2 \right ) - \frac{\partial_l (B_0 B_1)}{4\pi} \bun_0 \cdot \partial_\alpha \bx_1 \nonumber\\ - \frac{B_0}{4\pi} \frac{\rmd B_0}{\rmd l} \partial_l \bx_1 \cdot \partial_\alpha \bx_1 = \frac{B_0 B_1}{2\pi} \bkappa_0 \cdot \partial_\alpha \bx_1 + \frac{B_0^2}{4\pi} \partial^2_{ll} \bx_1 \cdot \partial_\alpha \bx_1.
\end{align}
To eliminate $B_2$ and $\bx_2$ from these equations, we differentiate equation~\refe{eq:MHDeqpsi1} with respect to $\alpha$ and equation~\refe{eq:MHDeqalpha1} with respect to $r$, and we subtract these two derivatives from each other to obtain
\begin{align} \label{eq:MHDconstraint}
\frac{B_0}{4\pi} \partial_l \Bigg [ & B_0 \Big ( \partial^2_{l\alpha} \bx_1 \cdot \partial_r \bx_1 - \partial^2_{l r} \bx_1 \cdot \partial_\alpha \bx_1 \nonumber\\  & + \left ( \bkappa_0 \cdot \partial_\alpha \bx_1 \right ) \left ( \bun_0 \cdot \partial_r \bx_1 \right ) - \left ( \bkappa_0 \cdot \partial_r \bx_1 \right ) \left ( \bun_0 \cdot \partial_\alpha \bx_1 \right ) \Big ) \Bigg ] = - 2 P^\prime \bkappa_0 \cdot \partial_\alpha \bx_1,
\end{align}
where we have used equation~\refe{eq:B1} to write $B_1$ in terms of $\bx_1$ and equation~\refe{eq:arc0} to obtain $\bun_0 \cdot \partial^2_{lr} \bx_1 = 0$ and $\bun_0 \cdot \partial^2_{l\alpha} \bx_1 = 0$. Note that, if $\bx_1$ is a solution to equations~\refe{eq:BJacob0}, \refe{eq:arc0} and \refe{eq:MHDconstraint}, the function $\bx_1 (r, \alpha, l) +  \lambda(r, \alpha) \bun_0 (l)$ is also a solution. Here, $\lambda(r, \alpha)$ can be any function of $\alpha$ and $r$. This set of equivalent solutions arises from the fact that one is free to choose where $l$ vanishes. 

\section{Lowest order ion distribution function in large aspect ratio stellarators with mirror ratios close to unity} \label{app:Maxwellian}

In this Appendix, we first solve for the lowest order trapped particle distribution function $g_{i, 0, W}$ in subsection~\ref{subapp:trapped0} and we then obtain the lowest order passing particle distribution function $h_{i,0}$ in subsection~\ref{subapp:passing0}. We finish with a discussion of the distribution function in the barely passing region in subsection~\ref{subapp:barelypassing0}.

\subsection{Lowest order trapped particle distribution function} \label{subapp:trapped0}
We proceed to rewrite equation~\refe{rho2orbit} using the coordinates $\vbar$ and $J$. We neglect the time derivatives and the source $S_i$ employing the estimates in equation~\refe{eq:Sieps}. Thus, equation~\refe{rho2orbit} becomes 
\begin{align} \label{eq:eqgiv1}
\langle (\bv_{E}+\bv_{Mi}) & \cdot \nabla \alpha \rangle_{\tau, W}\, \left ( \partial_\alpha g_{i,W} + \partial_\alpha J_W \, \partial_J g_{i,W} \right ) \nonumber\\ &+ \langle (\bv_{E}+\bv_{Mi}) \cdot \nabla r \rangle_{\tau, W}\, \left ( \partial_r g_{i,W} + \partial_r \vbar \, \partial_{\vbar} g_{i,W} + \partial_r J_W \, \partial_J g_{i,W} \right ) \nonumber \\ \simeq & \langle C_{ii} [ g_{i,W}, \overline{f}_i^{(0)} ] \rangle_{\tau, W} ,
\end{align}
where the derivatives of $g_{i,W}$ with respect to $r$ and $\alpha$ are performed holding $\vbar$ and $J$ fixed, whereas the derivatives of $\vbar$ and $J_W$ with respect to the same variables are performed holding $\mathcal{E}$ and $\mu$ constant. Employing equations~\refe{eq:dJdt0} and \refe{eq:dvbardt}, to lowest order in $\epsilon$, equation~\refe{eq:eqgiv1} can be rewritten as
\begin{equation} \label{eq:eqgi0v1}
\frac{c \phi_0^\prime}{\Psi_t^\prime} \partial_\alpha g_{i, 0, W} - \langle C_{ii} [ g_{i, 0, W}, \overline{f}_i^{(0)} ] \rangle_{\tau, W} = 0.
\end{equation}
The terms proportional to $\partial_{\vbar} g_{i, 0, W}$ and $\partial_r g_{i, 0, W}$ have been neglected because we assume that the derivatives of $g_{i, 0, W}$ with respect to $\vbar$ and $r$ satisfy $\partial_{\vbar} \ln g_{i, 0, W} \sim v_{ti}^{-1}$ and $\partial_r \ln g_{i, 0, W} \sim a^{-1}$. Recall that the estimate $\partial_r \ln g_{i, 0, W} \sim a^{-1} \ll (\epsilon a)^{-1}$ is valid because the derivative is performed holding $\vbar$ and $J$ constant.

The fact that the derivative of $g_{i, 0, W}$ with respect to $J$ is larger than its derivative with respect to $\vbar$ (compare equations~\refe{eq:dvbargiestimate} and \refe{eq:dJgiestimate}) simplifies significantly the collision operator in equation~\refe{eq:eqgi0v1} because the term with two derivatives with respect to $J$ becomes the largest,
\begin{align} \label{eq:CiiJv1}
C_{ii} [ g_{i, 0, W}, \overline{f}_i^{(0)} ] = \gamma_{ii}\, \nabla_v J_W \cdot \partial_J \left (  \nabla_v \nabla_v H [h_{i,0}] \cdot \nabla_v J_W \, \partial_J g_{i, 0, W} \right ) [1 + O(\epsilon^{1/2})] \nonumber \\ \sim \frac{\nu_{ii}}{\epsilon} g_{i, 0, W}.
\end{align}
Here the Rosenbluth potential $H[h_{i,0}]$ is evaluated using only $h_{i,0}$ because the contribution to the integral from $g_{i, 0, W}$ is smaller by $\sqrt{\epsilon}$ due to the smallness of the region of velocity space where $g_{i, 0, W}$ is defined. Since $\nabla_v J_W \simeq \tau_W v_\| \bun$ for trapped particles (see equation~\refe{eq:gradvJbar}), we need the component $\bun \cdot \nabla_v \nabla_v H [h_{i,0}] \cdot \bun$ of $\nabla_v \nabla_v H [h_{i,0}]$ for the lowest order version of the collision operator. We find
\begin{equation} \label{eq:bggHbv1}
\bun \cdot \nabla_v \nabla_v H [h_{i,0}] \cdot \bun = \int \frac{|\bv_\perp - \bv_\perp^\prime|^2 h_{i,0} (\bv^\prime)}{|\bv - \bv^\prime|^3} \, \rmd^3 v^\prime.
\end{equation}
Note that, for trapped particles, the parallel component of the velocity is small in $\epsilon$, giving $\bv \simeq \vbar\, \eun_\perp(\bx, \varphi)$, where $\eun_\perp$ is defined in equation~\refe{eq:eperp}. Using this result and writing $\bv^\prime \simeq \vbar^\prime \xi^\prime \bun + \vbar^\prime \sqrt{1 - \xi^{\prime2} } ( \cos \varphi^\prime\, \eun_\perp + \sin \varphi^\prime\, \bun \times \eun_\perp )$, equation~\refe{eq:bggHbv1} can be written as 
\begin{equation} \label{eq:bggHb}
\bun \cdot \nabla_v \nabla_v H [h_{i,0}] \cdot \bun \simeq H_{bb} [h_{i,0}](\vbar),
\end{equation}
where we have defined the new functional
\begin{equation}
H_{bb} [h_{i,0}](\vbar) : = \int \frac{\vbar^2 - 2 \vbar\, \vbar^\prime \sqrt{1 - \xi^{\prime 2}} \cos \varphi^\prime + \vbar^{\prime 2} (1 - \xi^{\prime 2} )}{( \vbar^2 - 2 \vbar\, \vbar^\prime \sqrt{1 - \xi^{\prime 2}} \cos \varphi^\prime + \vbar^{\prime 2} )^{3/2}} \, h_{i,0} (\vbar^\prime, \xi^\prime) \vbar^{\prime 2} \, \rmd \vbar^\prime\, \rmd \xi^\prime \, \rmd \varphi^\prime.
\end{equation}
This result demonstrates that $\bun \cdot \nabla_v \nabla_v H [h_{i,0}] \cdot \bun$ only depends on $\vbar$ to lowest order in $\epsilon$. With all these results, equation~\refe{eq:CiiJv1} becomes 
\begin{equation} \label{eq:CiiJ}
C_{ii} [ g_{i, 0, W}, \overline{f}_i^{(0)} ] = \gamma_{ii} H_{bb} [h_{i,0}] (\vbar)\, \tau_W v_\| \partial_J \left ( \tau_W v_\| \, \partial_J g_{i, 0, W} \right ) [1 + O(\epsilon^{1/2})] \sim \frac{\nu_{ii}}{\epsilon} g_{i, 0, W}.
\end{equation}
Transit averaging equation~\refe{eq:CiiJ}, we obtain
\begin{equation} \label{eq:CiiJave}
\langle C_{ii} [ g_{i, 0, W}, \overline{f}_i^{(0)} ] \rangle_{\tau, W} = \gamma_{ii} H_{bb} [h_{i,0}] (\vbar)\, \partial_J \left ( \tau_W J\, \partial_J g_{i, 0, W} \right ) [1 + O(\epsilon^{1/2})] \sim \frac{\nu_{ii}}{\epsilon} g_{i, 0, W}.
\end{equation}

At the junctures of several wells, we need to impose continuity of $g_{i, 0, W}$ and use condition~\refe{rho2transition} to relate the discontinuous derivatives on different sides of the well. We proceed to rewrite equation~\refe{rho2transition} in the new coordinates $\vbar$ and $J$. The derivative of $g_{i, 0, W}$ with respect to $\mathcal{E}$ is $\partial_\mathcal{E} g_{i, 0, W} = \partial_\mathcal{E} \vbar\, \partial_{\vbar} g_{i, 0, W} + \partial_\mathcal{E} J_W\, \partial_J g_{i, 0, W}$. Noting that $\partial_\mathcal{E} \vbar = 1/\vbar \sim 1/v_{ti}$, $\partial_\mathcal{E} J_W = \tau_W \simeq R/\sqrt{\epsilon} v_{ti}$, $\partial_{\vbar} g_{i, 0, W} \sim g_{i, 0, W}/v_{ti}$ and $\partial_J g_{i, 0, W} \sim g_{i, 0, W}/\sqrt{\epsilon} v_{ti} R$, the derivative simplifies to $\partial_\mathcal{E} g_{i, 0, W} \simeq \tau_W\, \partial_J g_{i, 0, W}$. Equation~\refe{rho2transition} can be simplified further by noting that $\mathcal{E}_c (r, \alpha, \mu, t)$ is given by equation~\refe{eq:EcLAR} and hence $\partial_\mu \mathcal{E}_c \simeq B_{lM}$. With these results, equation~\refe{rho2transition} becomes
\begin{align} \label{eq:eqgi0transitionv1}
& \lim_{J \rightarrow J_{c, I}} \tau_I\, \partial_J g_{i,0, I} \sum_\sigma \int_{l_{bL, I}}^{l_{bR, I}} ( H_{\mathcal{E} \mathcal{E}}[ h_{i, 0} ] - 2  B_{lM} H_{\mathcal{E} \mu}[ h_{i, 0} ] + B_{lM}^2 H_{\mu \mu}[ h_{i, 0} ] ) \, \frac{\rmd l}{|v_\| |} \nonumber\\ & + \lim_{J \rightarrow J_{c, II}} \tau_{II}\, \partial_J g_{i,0, II} \sum_\sigma \int_{l_{bL, II}}^{l_{bR, II}} ( H_{\mathcal{E} \mathcal{E}}[ h_{i, 0} ] - 2  B_{lM} H_{\mathcal{E} \mu}[ h_{i, 0} ] + B_{lM}^2 H_{\mu \mu}[ h_{i, 0} ] ) \, \frac{\rmd l}{|v_\| |}\nonumber\\ & = \lim_{J \rightarrow J_{c, III}} \tau_{III}\, \partial_J g_{i,0, III} \sum_\sigma \int_{l_{bL, III}}^{l_{bR, III}} ( H_{\mathcal{E} \mathcal{E}}[ h_{i, 0} ] - 2  B_{lM} H_{\mathcal{E} \mu}[ h_{i, 0} ] + B_{lM}^2 H_{\mu \mu}[ h_{i, 0} ] ) \, \frac{\rmd l}{|v_\| |}.
\end{align}
Note that $H_{\mathcal{E} \mathcal{E}} = \nabla_v \mathcal{E} \cdot \nabla_v \nabla_v H \cdot \nabla_v \mathcal{E} = \bv \cdot \nabla_v \nabla_v H \cdot \bv$, $H_{\mathcal{E} \mu} = \nabla_v \mathcal{E} \cdot \nabla_v \nabla_v H \cdot \nabla_v \mu = \bv \cdot \nabla_v \nabla_v H \cdot \bv_\perp/B$ and $H_{\mu \mu} = \nabla_v \mu \cdot \nabla_v \nabla_v H \cdot \nabla_v \mu = \bv_\perp \cdot \nabla_v \nabla_v H \cdot \bv_\perp/B^2$. Thus, 
\begin{align}
H_{\mathcal{E} \mathcal{E}} & [ h_{i, 0} ] - 2  B_{lM} H_{\mathcal{E} \mu}[ h_{i, 0} ] + B_{lM}^2 H_{\mu \mu}[ h_{i, 0} ] \nonumber\\ & = \left ( v_\| \bun + \bv_\perp \left ( 1 - \frac{B_{lM}}{B} \right ) \right ) \cdot \nabla_v \nabla_v H [ h_{i, 0} ] \cdot \left ( v_\| \bun + \bv_\perp \left ( 1 - \frac{B_{lM}}{B} \right ) \right ).
\end{align}
Since $1 - B_{lM}/B \sim \epsilon$ but $v_\| \sim \sqrt{\epsilon} v_{ti}$, $H_{\mathcal{E} \mathcal{E}} [ h_{i, 0} ] - 2  B_{lM} H_{\mathcal{E} \mu}[ h_{i, 0} ] + B_{lM}^2 H_{\mu \mu}[ h_{i, 0} ] = v_\|^2 H_{bb} [h_{i, 0}] ( 1 + O(\epsilon^{1/2}) )$. Then, by dividing equation~\refe{eq:eqgi0transitionv1} by $H_{bb} [h_{i, 0}] (\vbar)$, we find 
\begin{equation} \label{eq:eqgi0transition}
J_{c, I} \lim_{J \rightarrow J_{c, I}}  \tau_I\, \partial_J g_{i,0,I} + J_{c, II} \lim_{J \rightarrow J_{c, II}}  \tau_{II}\, \partial_J g_{i,0,II} = J_{c, III} \lim_{J \rightarrow J_{c, III}}  \tau_{III}\, \partial_J g_{i,0,III}.
\end{equation}
Equation~\refe{eq:eqgi0transition} could have also been obtained by following procedure similar to the one in Appendix~\ref{app:junctures}. Note that equation~\refe{eq:CiiJave} implies that the phase-space particle flux across the boundary $J = J_{c, W} (r, \alpha, \vbar, t)$ is $- 2\pi \gamma_{ii} H_{bb} [h_{i,0}] (\vbar)\, J_{c, W} \tau_W \partial_J g_{i, 0, W}$.

Using equations~\refe{eq:eqgi0v1} and \refe{eq:CiiJave}, we find that that the equation for $g_{i, 0, W}$ is
\begin{equation} \label{eq:eqgi0}
\frac{c \phi_0^\prime}{\Psi_t^\prime} \partial_\alpha g_{i, 0, W} - \gamma_{ii} H_{bb} [h_{i,0}](\vbar)\, \partial_{J} \left ( \tau_W J \, \partial_{J} g_{i, 0, W} \right ) = 0.
\end{equation}
We need to impose the continuity condition~\refe{eq:condJinftyv1} that implies that $g_{i, 0, W}$ for $J \rightarrow \infty$ must be equal to $h_{i,0}$ at the trapped-passing boundary, where $\xi \sim \sqrt{\epsilon} \ll 1$. Then,
\begin{equation} \label{eq:condJinftyv2}
\lim_{J \rightarrow \infty} g_{i, 0, W_{\mathrm{bt}}} (r, \alpha, \vbar, J, t) = h_{i,0} (r, \vbar, 0, t).
\end{equation}
Note that we have approximated the trapped-passing boundary by $\xi = 0$. We discuss this approximation in detail in Appendix~\ref{subapp:barelypassing0}.

To find the solution to equation~\refe{eq:eqgi0}, we first need to show that
\begin{equation} \label{eq:dJgi0Jinfty}
\lim_{J \rightarrow \infty} \tau_{W_\mathrm{bt}} J \, \partial_{J} g_{i, 0, W_\mathrm{bt}} (r, \alpha, \vbar, J, t) = 0,
\end{equation}
that is, there is no collisional flux from the trapped region into the passing region. Property~\refe{eq:dJgi0Jinfty} can be shown to be true following the same procedure that we employed to find property~\refe{eq:dJgi1Jinfty}. With property~\refe{eq:dJgi0Jinfty}, we can find the solution $g_{i, 0, W}$ given in equation~\refe{eq:gi0sol}. We multiply equation~\refe{eq:eqgi0} by $g_{i, 0, W}$ to write
\begin{align}
\frac{c \phi_0^\prime}{\Psi_t^\prime} \partial_\alpha \left ( \frac{g_{i, 0, W}^2}{2} \right ) + \gamma_{ii} H_{bb} [h_{i,0}](\vbar) \tau_W J \, (\partial_{J} g_{i, 0, W} )^2 \nonumber \\ - \gamma_{ii} H_{bb} [h_{i,0}](\vbar)\, \partial_{J} \left ( g_{i, 0, W} \tau_W J \, \partial_{J} g_{i, 0, W} \right ) = 0.
\end{align}
We integrate this equation over $J$ and $\alpha$ and we sum over $W$. We use equation~\refe{eq:eqgi0transition} at the junctures of different wells, and we employ property~\refe{eq:dJgi0Jinfty} to show that a boundary term at $J \rightarrow \infty$ vanishes. After all these calculations, we find
\begin{equation}
\sum_W  \int_{\alpha_{L, W}}^{\alpha_{R, W}} \rmd \alpha \int_{J_{m, W}}^{J_{M, W}} \rmd J\, \tau_W J \left ( \partial_{J} g_{i, 0, W} \right )^2 = 0.
\end{equation}
This equation implies that $g_{i, 0, W}$ is independent of $J$. Using this result in equation~\refe{eq:eqgi0}, we also find that $g_{i, 0, W}$ is independent of $\alpha$. We obtain solution~\refe{eq:gi0sol} by applying the boundary condition~\refe{eq:condJinftyv2}.

\subsection{Lowest order passing particle distribution function} \label{subapp:passing0}
To lowest order in $\epsilon$, according to equation~\refe{eq:Sieps}, we can neglect the time derivatives and the source in equation~\refe{passingfi0}, finding
\begin{equation} \label{eq:eqhiv1}
\left \langle \frac{B}{|v_\| |} C_{ii}[h_i , h_i] \right \rangle_\mathrm{fs} \simeq 0.
\end{equation}
For most passing particles, $v_\| \simeq \sigma \sqrt{2( \mathcal{E} - \mu B_0 - Z_ie\phi_0 (r, t)/m_i)}$ is very close to being independent of $\alpha$ and $l$. Thus, for most values of $\xi$, equation~\refe{eq:eqhiv1} becomes
\begin{equation} \label{eq:eqhi0}
C_{ii}[h_{i,0} , h_{i,0}] = 0.
\end{equation}

We need to impose boundary condition~\refe{eq:diffcondJinftyv1} that requires that the derivatives of $g_{i,W}$ and $h_i$ with respect to velocity are continuous across the trapped-passing boundary. The continuity of the derivatives imposes that the collisional flux across the trapped-passing boundary be continuous. This collisional flux is dominated by pitch-angle scattering events, and the pitch-angle scattering flux is controlled by the derivative of $g_{i,W}$ with respect to $J$ in the trapped region of velocity space. Since we have shown that $\partial_J g_{i, 0, W} = 0$, only the derivatives of $g_{i, 1, W}$ play a role in the collisional flux across the trapped-passing boundary. Due to the smallness of $|v_\| | \sim \sqrt{\epsilon} v_{ti}$ in the trapped region, $\nabla_v g_{i, 1, W} \sim \epsilon^{-1/2} g_{i, 1, W}/v_{ti} \sim \epsilon^{1/2} h_{i,0}/v_{ti}$, where we have used $g_{i, 1, W} \sim \epsilon h_{i,0}$. Hence, the collisional flux across the trapped-passing boundary is small in $\epsilon$ because $\nabla_v g_{i, 1, W}$ is small in $\epsilon$. The lack of collisional flux across the trapped-passing boundary gives the boundary condition 
\begin{equation} \label{eq:bchi0}
\partial_\xi h_{i,0} (r, \vbar, 0, t) = 0
\end{equation}
for $h_{i,0}$. A more rigorous derivation of this boundary condition can be found in Appendix~\ref{subapp:barelypassing0}.

To obtain boundary condition~\refe{eq:bchi0}, the fact that there is no piece of $g_{i,W}$ of order $\epsilon^{1/2} g_{i, 0, W}$ is important. Using equations~\refe{eq:vdreps}, \refe{eq:vdalphaeps} and \refe{eq:dvbardt}, we find that the next order corrections to equation~\refe{eq:eqgi0} are
\begin{align} \label{eq:eqgi1v1}
\frac{c \phi_0^\prime}{\Psi_t^\prime} \partial_\alpha g_{i, 1, W} - \gamma_{ii} H_{bb} [h_{i,0}] (\vbar)\, \partial_{J} \left ( \tau_W J \, \partial_{J} g_{i, 1, W} \right ) = \langle C_{ii} [ g_{i, 0, W}, h_{i,0} ] \rangle_{\tau, W}  \nonumber\\ + \frac{m_i c \phi_0^\prime \vbar^2}{2 Z_i e B_0 \Psi_t^\prime} \left \langle \partial_\alpha B_1 \right \rangle_{\tau, W} \left ( \partial_r g_{i, 0, W} - \frac{Z_i e \phi_0^\prime}{m_i \vbar} \partial_{\vbar} g_{i, 0, W} \right ) .
\end{align}
To obtain this result, we have used the fact that $\langle C_{ii} [ g_{i, 0, W}, h_{i,0} ] \rangle_{\tau, W}$ is of order $\nu_{ii} g_{i, 0, W}$ instead of $\nu_{ii} g_{i, 0, W}/\epsilon$ because $g_{i, 0, W}$ only depends on $\vbar$. Thus, there is no correction to $g_{i, 0, W}$ of order $\epsilon^{1/2} g_{i, 0, W}$, and we can use boundary condition~\refe{eq:bchi0}.

The only possible solution to equation~\refe{eq:eqhi0} with condition~\refe{eq:bchi0} is a stationary Maxwellian with density and temperature that are flux functions, as shown in equation~\refe{eq:hifM}.

\subsection{Lowest order barely passing particle distribution function} \label{subapp:barelypassing0}

The coordinate $\xi$ is not a constant of the motion for barely passing particles with $|\xi| \sim \sqrt{\epsilon} \ll 1$. In this small region of velocity space, it is more convenient to use as a coordinate the value of $\xi$ at the location of the maximum of $U$, 
\begin{align}
\xibar (r, \mathcal{E}, \mu, \sigma, t) &:= \sigma \sqrt{\frac{\mathcal{E} - U_M (r, \mu, t)}{\mathcal{E} - Z_i e \phi_{U_M} (r, \mu, t)/m_i}} \nonumber\\ & = \frac{\sigma \sqrt{2( \mathcal{E} - \mu B_M (r) - Z_i e \phi_0 (r, t)/m_i)}}{\vbar} \, [ 1 + O (\epsilon^{1/2}) ],
\end{align} 
where $\phi_{U_M}$ is the value of the potential at the location where $U$ is maximum, and $B_M(r)$ is the maximum of $B$ on flux surface $r$. 

Using $\vbar$ and $\xibar$ as coordinates, the parallel velocity can be written as 
\begin{equation} \label{eq:vparbp}
v_\| = \frac{\xibar}{|\xibar|} \vbar \sqrt{\xibar^2 + \frac{B_{1, M} (r) - B_1(r, \alpha, l)}{B_0}} \, [ 1 + O(\epsilon^{1/2}) ], 
\end{equation}
where $B_{1,M} (r)$ is the maximum value of $B_1$ on flux surface $r$. According to this approximation, $v_\|$ and hence $\xi$ are constant for $|\xibar| \gg \sqrt{\epsilon}$. Moreover, $\xibar \simeq \xi$ for $|\xibar| \gg \sqrt{\epsilon}$. Unfortunately, we need to consider $|\xibar| \sim \sqrt{\epsilon}$ because it is in this region that we find the trapped-passing boundary: at the trapped-passing boundary, $\xibar = 0$. 

In the barely trapped region $|\xibar| \sim \sqrt{\epsilon}$, we can write the distribution function as 
\begin{equation} \label{eq:hibpdef}
h_i (r, \vbar, \xibar, t) + \Delta h_{i, \mathrm{bp}} (r, \vbar, \xibar, t) \simeq h_i (r, \vbar, \xi, t) + (\xibar - \xi)\, \partial_\xi h_i (r, \vbar, \xi, t) + \Delta h_{i, \mathrm{bp}} (r, \vbar, \xibar, t), 
\end{equation}
where $h_i (r, \vbar, \xibar, t)$ is the function $h_i (r, \vbar, \xi, t)$ but with $\xi$ replaced by $\xibar$, and $h_i(r, \vbar, \xi, t) \simeq h_{i, 0} (r, \vbar, \xi, t)$ is the passing particle distribution function discussed in Appendix~\ref{subapp:passing0}. The passing particle distribution function $h_i(r, \vbar, \xi, t) \simeq h_{i, 0} (r, \vbar, \xi, t)$ is determined by equation~\refe{eq:eqhi0} and satisfies 
\begin{equation} \label{eq:estimdxihxi0}
\partial_\xi h_i (r, \vbar, \xi, t) \lesssim \sqrt{\epsilon} h_{i, 0} 
\end{equation}
for $|\xi| \sim \sqrt{\epsilon}$, as we will see at the end of this Appendix. The piece $\Delta h_{i, \mathrm{bp}} (r, \vbar, \lambdabar, t)$ is small in $\epsilon$ and vanishes for $|\xibar| \gg \sqrt{\epsilon}$, leaving the solution $h_i (r, \vbar, \xi, t)$ for freely passing particles with $|\xi| \sim 1$. We proceed to show that $\Delta h_{i, \mathrm{bp}}$ simply ensures that there is continuity in the collisional fluxes across the barely passing region.

In Appendix~\ref{subapp:trapped0}, we show that $\partial_J g_{i, 0, W} = 0$. With this result, we can give a bound for the size of $\partial_{\xibar} \Delta h_{i,\mathrm{bp}}$. Using the continuity condition~\refe{eq:condJinftyv1} to write
\begin{equation} \label{eq:condJinftyv3}
\lim_{J \rightarrow \infty} g_{i,W_\mathrm{bt}} (r, \alpha, \vbar, J, t) = \lim_{\xibar \rightarrow 0} \left [ h_i (r, \vbar, \xibar, t) + \Delta h_{i, \mathrm{bp}} (r, \vbar, \xibar, t) \right ],
\end{equation}
we can deduce that 
\begin{equation}
\lim_{J \rightarrow \infty} \partial_{\vbar} g_{i,W_\mathrm{bt}} (r, \alpha, \vbar, J, t) = \lim_{\xibar \rightarrow 0} \left [ \partial_{\vbar} h_i (r, \vbar, \xibar, t) + \partial_{\vbar} \Delta h_{i, \mathrm{bp}} (r, \vbar, \xibar, t) \right ].
\end{equation}
Applying this result to equation~\refe{eq:diffcondJinftyv1}, where $\partial_\mathcal{E} = \partial_\mathcal{E} \vbar \, \partial_{\vbar} + \partial_\mathcal{E} J_W \, \partial_J$ for trapped particles and $\partial_\mathcal{E} = \partial_\mathcal{E} \vbar \, \partial_{\vbar} + \partial_\mathcal{E} \xibar \, \partial_{\xibar}$ for passing particles, we find
\begin{equation}
\lim_{J \rightarrow \infty} \partial_\mathcal{E} J_{W_\mathrm{bt}} \, \partial_J g_{i,W_\mathrm{bt}} (r, \alpha, \vbar, J, t)  = \lim_{\xibar \rightarrow 0}  \partial_\mathcal{E} \xibar \left [ \partial_{\xibar} h_i (r, \vbar, \xibar, t) + \partial_{\xibar} \Delta h_{i, \mathrm{bp}} (r, \vbar, \xibar, t) \right ].
\end{equation}
Using $\partial_\mathcal{E} J_W = \tau_W$, $\partial_\mathcal{E} \xibar = (1 - \xibar^2)/\vbar^2 \xibar$ and $\partial_J g_{i, 0, W} = 0$, this condition becomes
\begin{equation} \label{eq:diffcondJinftyv2}
\lim_{\xibar \rightarrow 0}  \frac{\partial_{\xibar} h_i (r, \vbar, \xibar, t) + \partial_{\xibar} \Delta h_{i, \mathrm{bp}} (r, \vbar, \xibar, t)}{\xibar} = \vbar^2\, \lim_{J \rightarrow \infty} \tau_{W_\mathrm{bt}}\, \partial_J g_{i, 1, W_\mathrm{bt}} (r, \alpha, \vbar, J, t).
\end{equation}
Using $|\xibar| \sim \sqrt{\epsilon}$, $\tau_W \sim \epsilon^{-1/2} R/v_{ti}$, $\partial_J \sim \epsilon^{-1/2}/v_{ti} R$ and equation~\refe{eq:estimdxihxi0}, we find the estimate
\begin{equation}
\partial_{\xibar} \Delta h_{i, \mathrm{bp}} \sim \sqrt{\epsilon} h_{i,0}.
\end{equation}
To calculate $\Delta h_{i, \mathrm{bp}}$, we need to integrate $\partial_{\xibar} \Delta h_{i, \mathrm{bp}} \sim \sqrt{\epsilon} h_{i,0}$ over $\xibar$. Since the region of interest is $|\xibar| \sim \sqrt{\epsilon}$, the integral is over an interval of $\xibar$ of order $O(\sqrt{\epsilon})$, and hence $\Delta h_{i, \mathrm{bp}} \sim \sqrt{\epsilon}\, \partial_{\xibar} \Delta h_{i, \mathrm{bp}} \sim \epsilon h_{i,0}$, justifying the neglect of $\Delta h_{i, \mathrm{bp}}$ in equation~\refe{eq:condJinftyv3} to obtain equation~\refe{eq:condJinftyv2}. In Appendix~\ref{app:barelypassing} we will see that $\Delta h_{i, \mathrm{bp}} \sim \epsilon h_{i,0}$ is in fact only a bound as $\Delta h_{i,\mathrm{bp}}$ is smaller than $\epsilon h_{i,0}$.

We proceed to solve equation~\refe{eq:eqhiv1} in the barely passing region. Using $B \simeq B_0$ and equation~\refe{eq:eqhi0}, we find that equation~\refe{eq:eqhiv1} becomes, to lowest order in $\epsilon$,
\begin{equation} \label{eq:eqDhibpv1}
B_0 \left \langle |v_\| |^{-1} C_{ii} \left [(\xibar - \xi)\, \partial_\xi h_i (r, \vbar, \xi, t) + \Delta h_{i,\mathrm{bp}}, h_{i,0} \right ] \right \rangle_\mathrm{fs} = 0,
\end{equation}
where we have neglected the contributions of the difference $\xibar - \xi$ and of $\Delta h_{i,\mathrm{bp}}$ to the Rosenbluth potentials because they are small in $\epsilon$. The derivatives of $\xibar - \xi$ and $\Delta h_{i, \mathrm{bp}}$ with respect to $\xibar$ in the barely passing region are large, $\partial_{\xibar} \ln ( \xibar - \xi) \sim \partial_{\xibar} \ln \Delta h_{i, \mathrm{bp}} \sim \epsilon^{-1/2}$. Thus, in the collision operator applied to $\xibar - \xi$ and $\Delta h_{i,\mathrm{bp}}$, the term that contains two derivatives with respect to $\xibar$ dominates,
\begin{align}
C_{ii} & \left [ (\xibar - \xi)\, \partial_\xi h_i (r, \vbar, \xi, t) + \Delta h_{i, \mathrm{bp}}, h_{i,0} \right ] \nonumber \\ & \simeq \gamma_{ii} \nabla_v \xibar \cdot \partial_{\xibar} \left [ \nabla_v \nabla_v H [h_{i,0}] \cdot \nabla_v \xibar\, \partial_{\xibar} \left ( (\xibar - \xi)\, \partial_\xi h_i (r, \vbar, \xi, t) + \Delta h_{i, \mathrm{bp}} \right ) \right ] \sim \nu_{ii} h_{i,0}.
\end{align}
Using $\nabla_v \xibar \simeq (\xi/ \xibar) \bun$, $\partial_{\xibar} \xi \simeq \xibar/\xi$ and equation~\refe{eq:bggHb}, and employing the fact that the derivative of $\partial_\xi h_i (r, \vbar, \xi, t)$ with respect to $\bv$ is small compared to the derivatives of $\xibar - \xi$ and $\Delta h_{i, \mathrm{bp}}$ with respect to $\xibar$, the collision operator simplifies to
\begin{align}
C_{ii} & \left [ (\xibar - \xi)\, \partial_\xi h_i (r, \vbar, 0, t) + \Delta h_{i, \mathrm{bp}}, h_{i,0} \right ] \nonumber \\ & \simeq \frac{\gamma_{ii} H_{bb} [h_{i,0}]}{\vbar^2 \xibar}\, v_\| \, \partial_{\xibar} \left [ \frac{v_\|}{\xibar} \, \left ( \partial_{\xibar} \Delta h_{i, \mathrm{bp}} + \left ( 1 - \frac{\vbar \xibar}{v_\|} \right ) \partial_\xi h_i (r, \vbar, \xi, t) \right ) \right ] \sim \nu_{ii} h_{i,0}.
\end{align}
Employing this result, equation~\refe{eq:eqDhibpv1} becomes
\begin{equation}
\frac{\gamma_{ii} H_{bb} [h_{i,0}] (\vbar)}{\vbar^2 |\xibar|} \partial_{\xibar} \left [ \frac{\langle |v_\|| \rangle_\mathrm{fs}}{|\xibar|}\, \partial_{\xibar} \Delta h_{i, \mathrm{bp}} + \left ( \frac{\langle |v_\|| \rangle_\mathrm{fs}}{|\xibar|} - \vbar \right ) \partial_\xi h_i (r, \vbar, \xi, t) \right ] = 0.
\end{equation}
We solve for $\partial_{\xibar} \Delta h_{i,\mathrm{bp}}$ by realizing that $\partial_{\xibar} \Delta h_{i, \mathrm{bp}}$ is small for $|\xibar| \gg \sqrt{\epsilon}$ because $\Delta h_{i, \mathrm{bp}}$ vanishes for $|\xibar| \gg \sqrt{\epsilon}$, finding
\begin{align}  \label{eq:diffbarelypassing}
\partial_{\xibar} \Delta h_{i,\mathrm{bp}} (r, \vbar, \xibar, t) = \left ( \frac{|\xibar|}{\left \langle \sqrt{\xibar^2 + (B_{1, M} - B_1)/B_0} \right \rangle_\mathrm{fs}} - 1 \right ) \partial_\xi h_i (r, \vbar, \xi, t),
\end{align}
where we have used the lowest order version of expression~\refe{eq:vparbp} for the parallel velocity. We can integrate once more to obtain $\Delta h_{i, \mathrm{bp}}$,
\begin{equation}  \label{eq:barelypassing}
\Delta h_{i,\mathrm{bp}} (r, \vbar, \xibar, t) = \partial_\xi h_i (r, \vbar, \xi, t) \frac{\xibar}{|\xibar|} \int_{|\xibar|}^\infty \left ( 1 - \frac{\xi^\prime}{\left \langle \sqrt{\xi^{\prime 2} + (B_{1, M} - B_1)/B_0} \right \rangle_\mathrm{fs}} \right )\, \rmd \xi^\prime,
\end{equation}
where we have used that $\Delta h_{i, \mathrm{bp}}$ vanishes for $|\xibar| \gg \sqrt{\epsilon}$.

Substituting equation~\refe{eq:diffbarelypassing} into equation~\refe{eq:diffcondJinftyv2}, we obtain
\begin{align} \label{eq:jumphi0}
\lim_{\xi \rightarrow 0^+} \partial_\xi h_i (r, \vbar, \xi, t) & = - \lim_{\xi \rightarrow 0^-} \partial_\xi h_i (r, \vbar, \xi, t) \nonumber\\ & = \left \langle \sqrt{\frac{B_{1,M} - B_1}{B_0}} \right \rangle_\mathrm{fs} \vbar^2\, \lim_{J \rightarrow \infty} \tau_{W_\mathrm{bt}}\, \partial_J g_{i, 1, W_\mathrm{bt}} (r, \alpha, \vbar, J, t) \sim \sqrt{\epsilon} h_{i,0}.
\end{align}
Since the derivatives of $h_{i,0}$ with respect to $\xi$ at $\xi = 0$ are small in $\epsilon$, they have to be set to zero to lowest order, giving condition~\refe{eq:bchi0}.

\section{Barely passing particles} \label{app:barelypassing}

In this Appendix, we show that the correction to the distribution function due to the barely passing particles, $\Delta h_{i, \mathrm{bp}}$, is of order $\epsilon^2 f_{Mi}$ and hence several boundary conditions that we have used in this article are valid. To discuss the barely passing particles, we use the formalism developed in Appendix~\ref{subapp:barelypassing0}. In Appendix~\ref{subapp:barelypassing0}, we state that $\Delta h_{i, \mathrm{bp}}$ is at most of order $\epsilon h_{i,0}$, but in reality it is smaller. In section~\ref{sub:gi1} and Appendix~\ref{app:Etangential}, we show that $\tau_W \, \partial_J g_{i, 1, W}$ and $\tau_W \, \partial_J g_{i, 3/2, W}$ vanish for $J \rightarrow \infty$. These results mean that equation~\refe{eq:diffcondJinftyv2} must be modified to give
\begin{equation} \label{eq:diffcondJinftyv3}
\lim_{\xibar \rightarrow 0}  \frac{\partial_{\xibar} h_{i, 3/2} (r, \vbar, \xibar, t) + \partial_{\xibar} \Delta h_{i, \mathrm{bp}} (r, \vbar, \xibar, t)}{\xibar} = \vbar^2 \, \lim_{J \rightarrow \infty} \tau_{W_\mathrm{bt}}\, \partial_J g_{i, 2, W_\mathrm{bt}} (r, \alpha, \vbar, J, t).
\end{equation}
This equation gives the estimate
\begin{equation}
\partial_{\xibar} \Delta h_{i, \mathrm{bp}} \sim \epsilon^{3/2} f_{Mi}.
\end{equation}
By integrating $\partial_{\xibar} \Delta h_{i, \mathrm{bp}}$ over the barely trapped region, of width $\Delta \xibar \sim \sqrt{\epsilon}$, we obtain the size that we announced at the start of this appendix,
\begin{equation}
\Delta h_{i, \mathrm{bp}} \sim \epsilon^2 f_{Mi}.
\end{equation}
The small size of this contribution justifies neglecting $\Delta h_{i, \mathrm{bp}}$ in the continuity condition~\refe{eq:condJinftyv3} to obtain the boundary conditions~\refe{eq:gi1bc} and \refe{eq:condJinfty32}.

Solutions~\refe{eq:diffbarelypassing} and \refe{eq:barelypassing} are valid in the barely passing region even at the high order that we are considering. Following the same procedure that led to equation~\refe{eq:jumphi0} but using equation~\refe{eq:diffcondJinftyv3} instead of equation~\refe{eq:diffcondJinftyv2}, we obtain the boundary condition
\begin{align}
\lim_{\xi \rightarrow 0^+} \partial_\xi h_{i,3/2} (r, \vbar, \xi, t) & = - \lim_{\xi \rightarrow 0^-} \partial_\xi h_{i,3/2} (r, \vbar, \xi, t) \nonumber\\ & = \left \langle \sqrt{\frac{B_{1,M} - B_1}{B_0}} \right \rangle_\mathrm{fs} \vbar^2\, \lim_{J \rightarrow \infty} \tau_{W_\mathrm{bt}}\, \partial_J g_{i, 2, W_\mathrm{bt}} (r, \alpha, \vbar, J, t).
\end{align}
This is the boundary condition that connects $g_{i, 2, W}$ and $h_{i,3/2}$.

\section{Example of a problematic well juncture} \label{app:problemjuncture}

\begin{figure}
\begin{center}
\includegraphics[width=13cm]{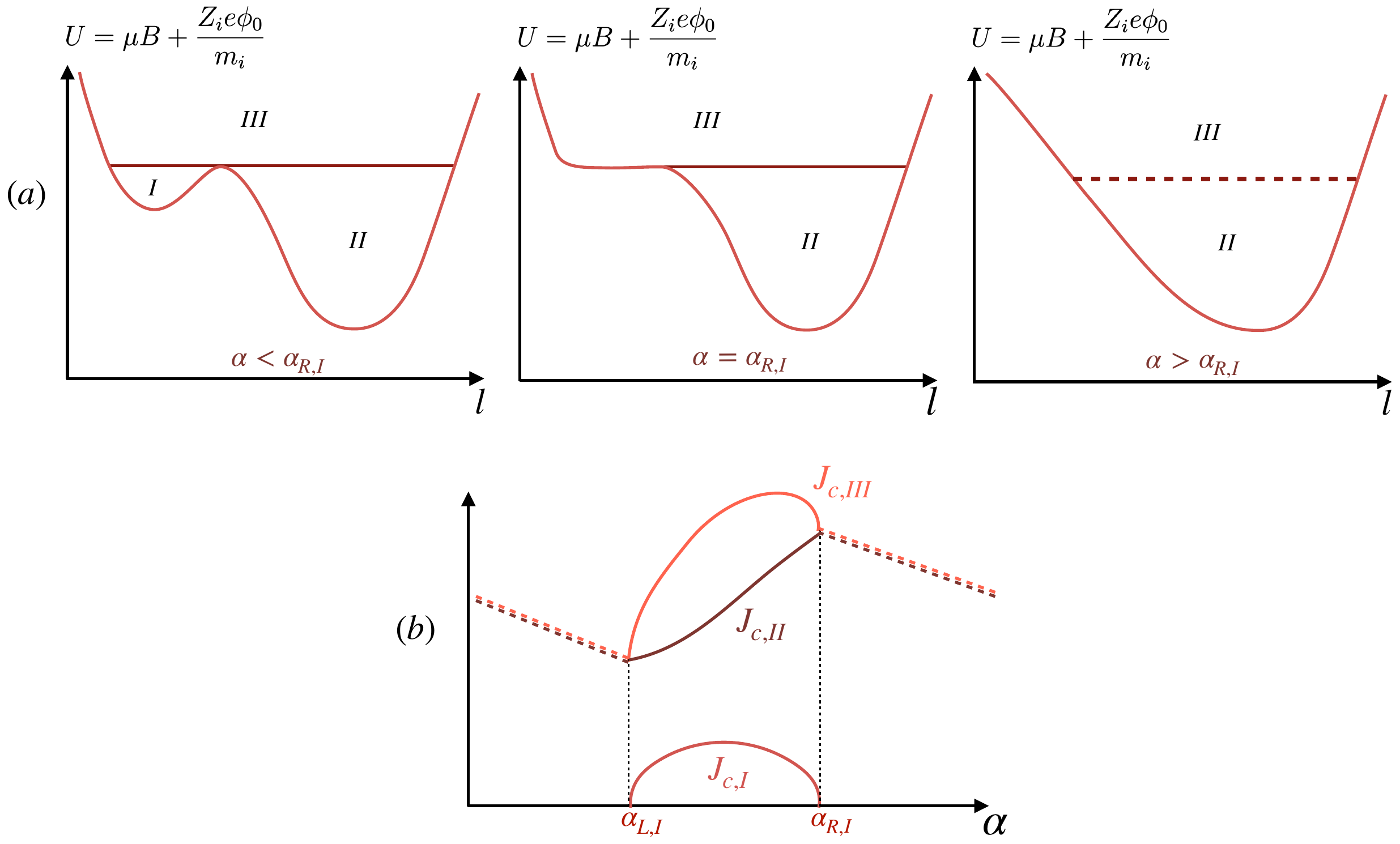}
\end{center}

\caption{\label{fig:alphaLRpatho} (a) Example of limit $\alpha_{R, I}$ in which well $I$ closes and well $II$ and $III$ become indistinguishable. There are several ways to treat such a case. We choose to define a ``false" juncture $J_{c, II} = J_{c, III}$ (represented by a dashed line) that splits particles into well $II$ and well $III$ arbitrarily for $\alpha > \alpha_{R, I}$. (b) Sketch of the functions $J_{c, I}$, $J_{c, II}$ and $J_{c, III}$ vs $\alpha$. The sketch is consistent with the $\alpha_{R, I}$ sketched in figure (a). The ``false" juncture $J_{c, II} = J_{c, III}$ is represented by dashed lines. The shape of $J_{c,II} = J_{c, III}$ in the region where well $I$ does not exist is arbitrary. We choose it such that $J_{c, II} (r, \alpha, \vbar, t)$ and $J_{c, III} (r, \alpha, \vbar, t)$ are continuous in $\alpha$.}
\end{figure}

In section~\ref{subsub:passing32}, we state that wells end at locations $\alpha_{L, W}$ and $\alpha_{R, W}$ where $J_{m, W} = J_{M, W}$. There are cases in which this property is not obvious. We show such an example in figure~\ref{fig:alphaLRpatho}(a), in which well $I$ disappears at $\alpha = \alpha_{R, I}$ because $J_{m, I} = J_{M, I} = 0$ at that value of $\alpha$. The problem with this configuration arises for $\alpha > \alpha_{R, I}$, where well $II$ and $III$ merge into a new larger well. There are different ways to address this problem, and we comment on two: one could terminate wells $II$ and $III$ at $\alpha = \alpha_{R, I}$ and start a new well $IV$ at that value of $\alpha$, or one could terminate well $III$ and declare that the merged well beyond $\alpha_{R, I}$ is well $II$. In either of these options, a well disappears without $J_{m, W}$ being equal to $J_{M, W}$. Moreover, if we were to choose the second option and declare that the merged well is well $II$, the function $J_{c, II}$ that determines the location of the juncture in velocity space becomes discontinuous in $\alpha$. Both of these features are undesirable, and instead we choose to add a ``false" two-well juncture for $\alpha > \alpha_{R, I}$, represented by a red dashed line in figure~\ref{fig:alphaLRpatho}(a). We can locate the ``false" juncture $J_{c, II} = J_{c, III}$ that artificially separates particles into wells $II$ and $III$ wherever we wish. The only condition that we impose for convenience is that the extended functions $J_{c, II}$ and $J_{c, III}$ be continuous, as shown in the example of figure~\ref{fig:alphaLRpatho}(b). Note that such a ``false" juncture does not affect the final result because junctures impose continuity of the distribution function and continuity of the collisional velocity space flux, and hence continuity of the distribution function derivatives. With these ``false" junctures, we ensure that wells end only at values of $\alpha$ where $J_{m, W} = J_{M, W}$. 

\section{Correction $g_{i, 3/2, W}$ to the trapped particle distribution function} \label{app:Etangential}
In this Appendix, we show that $g_{i, 3/2, W}$ does not depend on $\alpha$ or $J$ and hence $\tau_W\, J\, \partial_J g_{i, 3/2, W} = 0$ for all $J$ and not only at the trapped-passing boundary. 

We start by discussing the contributions to $g_{i, 3/2, W}$ due to the collision operator. One might think that the differential piece of the linearized collision operator applied on $g_{i, 1, W}$ has pieces of order $\epsilon^{1/2} \nu_{ii} f_{Mi}$ that would have to be included in an equation for $g_{i, 3/2, W}$, but after careful manipulations using $\nabla_v \vbar = \bv/\vbar$ and equation~\refe{eq:gradvJbar} for $\nabla_v J_W$, we find that this is not the case. Indeed, since $\nabla_v = \nabla_v \vbar\, \partial_{\vbar}  + \nabla_v J_W\, \partial_J$ and
\begin{align}
\nabla_v \cdot ( \ldots ) & = \frac{\tau_W \langle B \rangle_{\tau, W} |v_\||}{\vbar} \Bigg [ \partial_{\vbar} \left ( \frac{\vbar}{\tau_W \langle B \rangle_{\tau, W} |v_\||} (\ldots) \cdot \nabla_v \vbar \right ) \nonumber\\ & + \partial_J \left ( \frac{\vbar}{\tau_W \langle B \rangle_{\tau, W} |v_\||} (\ldots) \cdot \nabla_v J_W \right ) \Bigg ], 
\end{align}
we obtain
\begin{align} \label{eq:DCDgi1}
\langle C_{ii, D}^{\ell} & [ g_{i, 1, W} ] \rangle_{\tau, W} - \frac{\vbar^2 \nu_{ii,\perp}(r, \vbar, t)}{4}\, \partial_{J} \left ( \tau_W J \, \partial_{J} g_{i, 1, W} \right ) \nonumber \\ = &  - \frac{\vbar^2 \tau_W J \nu_{ii,\perp}(r, \vbar, t)}{4} \, \partial_{J} g_{i, 1, W}\, \partial_J \ln \langle B \rangle_{\tau, W} \nonumber \\ & + \langle B \rangle_{\tau, W} \, \partial_J \Bigg [ \frac{\tau_W^2}{\langle B \rangle_{\tau, W}} \left \langle v_\|^2 \left ( \frac{v^2\nu_{ii, \perp}(r, v, t)}{4} - \frac{\vbar^2\nu_{ii, \perp}(r, \vbar, t)}{4}\right ) \right \rangle_{\tau, W} \, \partial_J g_{i, 1, W} \nonumber \\ & + \frac{2 \mu \tau_W^2}{\langle B \rangle_{\tau, W}} \left \langle \frac{v^2 B \nu_{ii, \perp}}{4} \left (1 - \frac{\langle B \rangle_{\tau, W}}{B} \right )^2 \right \rangle_{\tau, W} \, \partial_J g_{i, 1, W} \nonumber\\ & + \frac{\tau_W^2}{\langle B \rangle_{\tau, W}} \left \langle \left ( \frac{\nu_{i,\|}}{2} - \frac{\nu_{i, \perp}}{4} \right ) \left ( v_\|^2 + 2\mu \left (B - \langle B \rangle_{\tau, W} \right ) \right )^2 \right \rangle_{\tau, W} \, \partial_J g_{i, 1, W} \nonumber \\ & + \frac{\tau_W}{\vbar \langle B \rangle_{\tau, W}} \left \langle \frac{v^2 \nu_{i,\|}}{2} \left ( v_\|^2 + 2\mu \left (B - \langle B \rangle_{\tau, W} \right ) \right ) \right \rangle_{\tau, W} \, \partial_{\vbar} g_{i, 1, W} \Bigg ] \nonumber\\ & + \frac{\langle B \rangle_{\tau, W}}{\vbar} \partial_{\vbar} \left ( \frac{\langle v^4 \nu_{i, \|} \rangle_{\tau, W}}{2\vbar \langle B \rangle_{\tau, W}} \, \partial_{\vbar} g_{i, 1, W} \right ) \sim \nu_{ii} g_{i, 1, W} \sim \epsilon^2 \rho_{i*} \frac{v_{ti} f_{Mi}}{a}.
\end{align}
The contributions of the integral part of the linearized collision operator to the trapped particle equation can be neglected to the order that determines $g_{i, 3/2, W}$ because their size is given by $C_{ii, I}^{\ell} [h_{i,3/2}] \sim \nu_{ii} h_{i,3/2} \sim \epsilon^{5/2} \rho_{i*} v_{ti} f_{Mi}/a$ and $C_{ii,I}^{\ell} [ g_{i, 1, W} ] \sim  \epsilon^{1/2} \nu_{ii} g_{i, 1, W} \sim \epsilon^{5/2} \rho_{i*} v_{ti} f_{Mi}/a$. Here the integral operator applied on $g_{i, 1, W}$ has an extra factor of $\sqrt{\epsilon}$ because the region of velocity space where $g_{i, 1, W}$ is defined is small in $\sqrt{\epsilon}$. 

We proceed to consider the contributions to the equation for $g_{i, 3/2, W}$ due to the $\bE \times \bB$ drift $(c/B)\, \bun \times \nabla \phi_{3/2}$ (see equations~\refe{eq:vdreps} and \refe{eq:dvbardt}). This $\bE \times \bB$ contribution has already been accounted for in equation~\refe{eq:eqgi1} in the term proportional to $(c \phi_0^\prime/\Psi_t^\prime)\, \partial_\alpha r_{i, 1, W}$. Keeping higher order corrections in equation~\refe{eq:Jbarlambda}, we find
\begin{equation} \label{eq:Jbarlambdaho}
J_W (r, \alpha, \vbar, \lambdabar, t) \simeq 2 \vbar \int_{l_{bL,W}}^{l_{bR,W}} \sqrt{ 1 - \lambdabar B_0 - \frac{B_1(r, \alpha, l)}{B_0} - \frac{2Z_i e \phi_{3/2}(r, \alpha, l, t)}{m_i \vbar^2} } \, \rmd l.
\end{equation}
We calculate $\partial_\alpha \lambdabar_W$ by differentiating equation~\refe{eq:Jbarlambdaho} with respect to $\alpha$ holding $r$, $\vbar$ and $J$ constant. We find
\begin{equation}
\left \langle \frac{\vbar^2}{2B_0} \partial_\alpha B_1 + \frac{Z_i e}{m_i} \partial_\alpha \phi_{3/2} \right \rangle_{\tau, W} \simeq - \frac{\vbar^2 B_0}{2} \partial_\alpha \lambdabar_W.
\end{equation}
This result shows that the $\bE \times \bB$ drift due to $\phi_{3/2}$ is included naturally in the term $(c \phi_0^\prime/\Psi_t^\prime)\, \partial_\alpha r_{i, 1, W}$ because, using equation~\refe{eq:vdreps}, we can write the radial drifts as
\begin{align}
\left \langle (\bv_E + \bv_{Mi}) \cdot \nabla r \right \rangle_{\tau, W} = - \frac{m_i c}{Z_i e \Psi_t^\prime} \left \langle \frac{\vbar^2}{2B_0} \partial_\alpha B_1 + \frac{Z_i e}{m_i} \partial_\alpha \phi_{3/2} \right \rangle_{\tau, W} \ [1 + O ( \epsilon ) ] \nonumber\\ = \frac{c\phi_0^\prime}{\Psi_t^\prime}\, \partial_\alpha r_{i, 1, W} \ [1 + O ( \epsilon ) ].
\end{align}

Due to all the considerations listed above, the equation for $g_{i, 3/2, W}$ is
\begin{equation}  \label{eq:eqgi32}
\frac{c \phi_0^\prime}{\Psi_t^\prime} \partial_\alpha g_{i, 3/2, W} - \frac{\vbar^2 \nu_{ii,\perp}}{4}\, \partial_{J} \left ( \tau_W J \, \partial_{J} g_{i, 3/2, W} \right ) = 0.
\end{equation}
At this order, the equation for the junctures of different wells is still the simple result
\begin{equation} \label{eq:eqgi32transition}
J_{c, I} \lim_{J \rightarrow J_{c, I}}  \tau_I\, \partial_J g_{i,3/2,I} + J_{c, II} \lim_{J \rightarrow J_{c, II}}  \tau_{II}\, \partial_J g_{i,3/2,II} = J_{c, III} \lim_{J \rightarrow J_{c, III}}  \tau_{III}\, \partial_J g_{i,3/2,III}.
\end{equation}
Equations~\refe{eq:eqgi32} and \refe{eq:eqgi32transition} have to be solved in conjunction with the boundary condition
\begin{equation} \label{eq:condJinfty32}
\lim_{J \rightarrow \infty} g_{i, 3/2, W_\mathrm{bt} } (r, \alpha, \vbar, J, t) = h_{i,3/2} (r, \vbar, 0, t)
\end{equation}
that we obtain from condition~\refe{eq:condJinftyv1} (see Appendix~\ref{app:barelypassing} for more detail). Using the same methodology that led to solution~\refe{eq:gi0sol}, we can show that the solution to equations~\refe{eq:eqgi32}, \refe{eq:eqgi32transition} and \refe{eq:condJinfty32} is given by equation~\refe{eq:gi32sol} and hence $g_{i, 3/2, W}$ does not depend on $\alpha$ or $J$.

\section{Derivation of equations~\refe{eq:eqgi1}, \refe{eq:eqgi1transition} and \refe{eq:gi1bc} from the DKES kinetic equation} \label{app:DKES}
In this appendix, we derive equations~\refe{eq:eqgi1}, \refe{eq:eqgi1transition} and \refe{eq:gi1bc} from the DKES kinetic equation. DKES assumes that the ion distribution function is the Maxwellian distribution function $f_{Mi}$ corrected by the small piece $\hat{f}_i$,
\begin{equation}
f_i (r, \alpha, l, v, \gamma, t) = f_{Mi} (r, v, t) + \hat{f}_i (r, \alpha, l, v, \gamma, t).
\end{equation}
The velocity space coordinates are the magnitude of the velocity $v$ and the angle $\gamma$ between the velocity $\bv$ and the magnetic field, $\gamma := \arccos( \bun \cdot \bv/v)$. The DKES kinetic equation in these variables is
\begin{align} \label{eq:DKESeq}
\left [ v \cos \gamma + \frac{c \phi_0^\prime}{\langle B^2 \rangle_\mathrm{fs}} (\bB \times \nabla r) \cdot \nabla l \right ] \partial_l \hat{f}_i + \frac{c \phi_0^\prime}{\Psi_t^\prime} \frac{B^2}{\langle B^2 \rangle_\mathrm{fs}} \, \partial_\alpha \hat{f}_i + \frac{1}{2B} v \sin \gamma \, \partial_l B \, \partial_\gamma \hat{f}_i \nonumber\\ - \mathcal{L}_{ii} [ \hat{f}_i ] = - \bv_{Mi} \cdot \nabla r\, \Upsilon_i f_{Mi},
\end{align}
where
\begin{equation}
\mathcal{L}_{ii} [ \hat{f}_i ] := \frac{\nu_{ii,\perp} (v)}{4 \sin \gamma} \partial_\gamma \left ( \sin \gamma\, \partial_\gamma \hat{f}_i \right )
\end{equation}
is a model pitch angle scattering operator. This equation has to be solved with magnetic fields that satisfy the MHD constraint $(\nabla \times \bB) \cdot \nabla r = 0$. In the $\{ r, \alpha, l \}$ coordinates, this constraint is
\begin{equation} \label{eq:DKESMHD}
\partial_\alpha \left ( \frac{B}{\Psi_t^\prime} \right ) + \partial_l \left [ (\bun \times \nabla r) \cdot \nabla l \right ] = 0.
\end{equation}

To obtain equations~\refe{eq:eqgi1}, \refe{eq:eqgi1transition} and \refe{eq:gi1bc} from equation~\refe{eq:DKESeq}, we need to expand in $\rho_{i*} \sim \nu_{i*} \ll 1$ and in $\epsilon \ll 1$. We first expand in $\rho_{i*} \sim \nu_{i*} \ll 1$, 
\begin{equation}
\hat{f}_i = \hat{f}_i^{(0)} + \hat{f}_i^{(1)} + \ldots,
\end{equation}
where $\hat{f}_i^{(n)} \sim \rho_{i*}^n \hat{f}_i$. For the expansion $\rho_{i*} \sim \nu_{i*} \ll 1$, we need to use the velocity space coordinates $v$, $\lambda = \sin^2\gamma/B$ and $\sigma$. In these coordinates, equation~\refe{eq:DKESeq} becomes
\begin{align} \label{eq:DKESeqv2}
\bigg [ v_\| &+ \frac{c \phi_0^\prime}{\langle B^2 \rangle_\mathrm{fs}} (\bB \times \nabla r) \cdot \nabla l \bigg ] \partial_l \hat{f}_i + \frac{c \phi_0^\prime}{\Psi_t^\prime} \frac{B^2}{\langle B^2 \rangle_\mathrm{fs}} \partial_\alpha \hat{f}_i \nonumber\\ &- \frac{\lambda c \phi_0^\prime}{\langle B^2 \rangle_\mathrm{fs}} \left [  (\bun \times \nabla r) \cdot \nabla l \, \partial_l B + \frac{B}{\Psi_t^\prime} \partial_\alpha B \right ] \partial_\lambda \hat{f}_i - \mathcal{L}_{ii} [ \hat{f}_i ] = - \bv_{Mi} \cdot \nabla r\, \Upsilon_i f_{Mi},
\end{align}
where $v_\| = v \cos \gamma = \sigma v \sqrt{1 - \lambda B}$ and
\begin{equation}
\mathcal{L}_{ii} [ \hat{f}_i ] = \frac{\nu_{ii,\perp} (v)}{v^2 B} v_\| \partial_\lambda \left ( v_\| \lambda \, \partial_\lambda \hat{f}_i \right ).
\end{equation}
To lowest order in $\rho_{i*} \ll 1$, equation~\refe{eq:DKESeqv2} simplifies to $v_\|\, \partial_l \hat{f}_i^{(0)} = 0$. Following the same procedure that we used in section~\ref{sec:dkequation}, we split $\hat{f}_i^{(0)}$ into a trapped particle distribution function $\hat{g}_{i,W}(r, \alpha, v, \lambda, t)$ and a passing particle distribution function $\hat{h}_i (r, v, \lambda, \sigma, t)$. To next order in $\rho_{i*} \ll 1$, equation~\refe{eq:DKESeqv2} becomes 
\begin{align}  \label{eq:DKESeq1v1}
v_\| \partial_l \hat{f}_i^{(1)} + \frac{c \phi_0^\prime}{\Psi_t^\prime} \frac{B^2}{\langle B^2 \rangle_\mathrm{fs}} \partial_\alpha \hat{f}_i^{(0)} - \frac{\lambda c \phi_0^\prime}{\langle B^2 \rangle_\mathrm{fs}} \left [  (\bun \times \nabla r) \cdot \nabla l \, \partial_l B + \frac{B}{\Psi_t^\prime} \partial_\alpha B \right ] \partial_\lambda \hat{f}_i^{(0)} \nonumber\\ - \mathcal{L}_{ii} [ \hat{f}_i^{(0)} ] = - \bv_{Mi} \cdot \nabla r\, \Upsilon_i f_{Mi}.
\end{align}
To eliminate the first term in this equation, we transit average in the trapped particle region, and we multiply by $B/|v_\| |$ and flux surface average in the passing particle region. To perform these operations, we use
\begin{equation}
- \frac{\lambda}{|v_\||} \left [  (\bun \times \nabla r) \cdot \nabla l \, \partial_l B + \frac{B}{\Psi_t^\prime} \partial_\alpha B \right ] = \frac{2}{v^2} \left [ (\bun \times \nabla r) \cdot \nabla l \, \partial_l |v_\| | + \frac{B}{\Psi_t^\prime} \partial_\alpha |v_\| | \right ]
\end{equation}
and equation~\refe{eq:DKESMHD} to write
\begin{equation}
- \left \langle \frac{\lambda B c \phi_0^\prime}{|v_\| |\langle B^2 \rangle_\mathrm{fs}} \left [  (\bun \times \nabla r) \cdot \nabla l \, \partial_l B + \frac{B}{\Psi_t^\prime} \partial_\alpha B \right ] \right \rangle_\mathrm{fs} = 0
\end{equation}
and
\begin{equation}
- \left \langle \frac{\lambda c \phi_0^\prime}{\langle B^2 \rangle_\mathrm{fs}} \left [  (\bun \times \nabla r) \cdot \nabla l \, \partial_l B + \frac{B}{\Psi_t^\prime} \partial_\alpha B \right ] \right \rangle_{\tau, W} = \frac{2 c \phi_0^\prime}{v^2 \tau_W \Psi_t^\prime \langle B^2 \rangle_\mathrm{fs} } \partial_\alpha \hat{J}_W,
\end{equation}
where $\hat{J}_W$ is defined in equation~\refe{eq:JDKES}. With these results, the transit average of equation~\refe{eq:DKESeq1v1} becomes
\begin{align}  \label{eq:DKES1trappedv1}
\frac{c \phi_0^\prime}{\Psi_t^\prime} \frac{\langle B^2 \rangle_{\tau, W}}{\langle B^2 \rangle_\mathrm{fs}} \, \partial_\alpha \hat{g}_{i,W} + \frac{2 c \phi_0^\prime}{v^2 \tau_W \Psi_t^\prime \langle B^2 \rangle_\mathrm{fs} } \, \partial_\alpha \hat{J}_W\, \partial_\lambda \hat{g}_{i,W} \nonumber\\ - \left \langle \mathcal{L}_{ii} [ \hat{g}_{i,W} ] \right \rangle_{\tau, W} = - \left \langle \bv_{Mi} \cdot \nabla r \right \rangle_{\tau, W} \, \Upsilon_i f_{Mi}
\end{align}
for trapped particles. The flux surface average of equation~\refe{eq:DKESeq1v1} multiplied by $B/|v_\| |$ gives
\begin{equation}  \label{eq:DKES1passing}
\left \langle \frac{B}{|v_\| |} \mathcal{L}_{ii} [ \hat{h}_i ] \right \rangle_\mathrm{fs} = 0
\end{equation}
for passing particles. 

Equation~\refe{eq:DKES1trappedv1} can be simplified even further for trapped particles by noting that
\begin{equation}
\partial_\lambda \hat{J}_W = - \frac{v^2 \tau_W}{2} \langle B^2 \rangle_{\tau, W}.
\end{equation}
Using this expression, we can rewrite equation~\refe{eq:DKES1trappedv1} as
\begin{align}  \label{eq:DKES1trappedv2}
\frac{2 c \phi_0^\prime}{v^2 \tau_W \Psi_t^\prime \langle B^2 \rangle_\mathrm{fs}} \left ( - \partial_\lambda \hat{J}_W\, \partial_\alpha \hat{g}_{i,W} + \partial_\alpha \hat{J}_W\, \partial_\lambda \hat{g}_{i,W} \right ) \nonumber\\ - \left \langle \mathcal{L}_{ii} [ \hat{g}_{i,W} ] \right \rangle_{\tau, W} = - \left \langle \bv_{Mi} \cdot \nabla r \right \rangle_{\tau, W} \, \Upsilon_i f_{Mi}.
\end{align}
This form of the equation shows that $\hat{J}_W$ is a constant of the motion in DKES. Indeed, if instead of using $v$ and $\lambda$ as velocity space coordinates, we use $v$ and $\hat{J}$, equation~\refe{eq:DKES1trappedv2} can be written as
\begin{equation}  \label{eq:DKES1trapped}
\frac{c \phi_0^\prime}{\Psi_t^\prime} \frac{\langle B^2 \rangle_{\tau, W}}{\langle B^2 \rangle_\mathrm{fs}} \, \partial_\alpha \hat{g}_{i,W} - \left \langle \mathcal{L}_{ii} [ \hat{g}_{i,W} ] \right \rangle_{\tau, W} = - \left \langle \bv_{Mi} \cdot \nabla r \right \rangle_{\tau, W} \, \Upsilon_i f_{Mi}, 
\end{equation}
where 
\begin{equation}
\left \langle \mathcal{L}_{ii} [ \hat{g}_{i,W} ] \right \rangle_{\tau, W} = \frac{v^2 \nu_{ii,\perp} (v) \langle B^2 \rangle_{\tau, W}}{4} \partial_{\hat{J}} \left (  \lambda \tau_W^2 \langle B^2 \rangle_{\tau, W} \left \langle \frac{v_\|^2}{B} \right \rangle_{\tau, W} \, \partial_{\hat{J}} \hat{g}_{i,W} \right ).
\end{equation}

In the DKES kinetic equation, there will also be junctures of different wells such as the one shown in figure~\ref{fig:exampleU}. These junctures are determined by values of $\hat{J}$ that depend on $r$, $\alpha$, $v$ and the well index $W$, $\hat{J} = \hat{J}_{c, W} (r, \alpha, v)$. These values of $\hat{J}$ satisfy
\begin{equation}
\hat{J}_{c, I} (r, \alpha, v) + \hat{J}_{c, II} (r, \alpha, v) = \hat{J}_{c, III} (r, \alpha, v),
\end{equation}
and imposing that the collisional flux across these junctures is continuous gives the condition
\begin{align} \label{eq:DKEStransition}
\lim_{\hat{J} \rightarrow \hat{J}_{c,I}}  & \tau_I^2 \langle B^2 \rangle_{\tau, I} \left \langle \frac{v_\|^2}{B} \right \rangle_{\tau, I} \, \partial_{\hat{J}} \hat{g}_{i,I} + \lim_{\hat{J} \rightarrow \hat{J}_{c,II}} \tau_{II}^2 \langle B^2 \rangle_{\tau, II}  \left \langle \frac{v_\|^2}{B} \right \rangle_{\tau, II} \, \partial_{\hat{J}} \hat{g}_{i,II} \nonumber\\ = & \lim_{\hat{J} \rightarrow \hat{J}_{c,III}} \tau_{III}^2 \langle B^2 \rangle_{\tau, III}  \left \langle \frac{v_\|^2}{B} \right \rangle_{\tau, III} \, \partial_{\hat{J}} \hat{g}_{i,III}.
\end{align}

Equations~\refe{eq:DKES1passing}, \refe{eq:DKES1trapped} and \refe{eq:DKEStransition} can be further simplified using the expansion in $\epsilon \ll 1$. Using a procedure similar to the one we follow in section~\ref{sub:gi1}, we find
\begin{equation}
\hat{g}_{i,W} = \hat{g}_{i,1, W} + \hat{g}_{i,3/2, W} + \hat{g}_{i,2, W} + \ldots
\end{equation}
and 
\begin{equation}
\hat{h}_i = \hat{h}_{i,3/2} + \ldots,
\end{equation}
where $\hat{g}_{i, n, W} \sim \epsilon^n f_{Mi}$ and $\hat{h}_{i, n} \sim \epsilon^n f_{Mi}$. Using the fact that $B \simeq B_0$ to lowest order, we obtain that $\hat{J}$ and $J$ are approximately the same coordinate, $\hat{J} \simeq B_0 J$. As a result, equations~\refe{eq:DKES1trapped} and \refe{eq:DKEStransition} become equations~\refe{eq:eqgi1} and \refe{eq:eqgi1transition} to lowest order. As explained in section~\ref{sub:gi1}, the continuity of derivatives across the trapped-passing boundary implies that the size of $\hat{h}_i$ is $\epsilon^{3/2} f_{Mi}$, and as a result we obtain the boundary condition~\refe{eq:gi1bc} for $\hat{g}_{i,1,W}$.

\section{Boundary layer at the juncture of several magnetic wells} \label{app:boundarylayer}

In junctures where particles leave two of the wells and enter a third one, a collisional boundary layer appears in the region 
\begin{equation} \label{eq:widthbljunct}
\frac{|J - J_{c, W}|}{\sqrt{\epsilon} v_{ti} R} \sim  \frac{\nu_{i*}}{\rho_{i*}} \ln \left ( \frac{\rho_{i*}}{\nu_{i*}} \right ) \ll 1
\end{equation}
(we justify this estimate below). The distribution function $g_{i, 1, W}$ in this boundary layer can be written as
\begin{equation} \label{eq:gi1junctlayer}
g_{i, 1, W} = K_{i,\mathrm{junct}, W} (r, \alpha, v, \Delta J_\mathrm{junct}, t) - r_{i, 1, W}\, \Upsilon_i f_{Mi},
\end{equation}
where $\Delta J_\mathrm{junct} := J - J_{c, W} (r, \alpha, v)$. The function $K_{i,\mathrm{junct}, W} (r, \alpha, v, \Delta J_\mathrm{junct}, t)$ changes rapidly in $\Delta J_\mathrm{junct}$, and slowly in $\alpha$. 

We can find the equation for $K_{i, \mathrm{junct}, W}$ by substituting equation~\refe{eq:gi1junctlayer} into equation~\refe{eq:eqgi1} and neglecting small terms
\begin{equation} \label{eq:eqjunctlayer}
- \frac{c \phi_0^\prime}{\Psi_t^\prime} \partial_\alpha J_{c, W}\, \partial_{\Delta J_\mathrm{junct}} K_{i,\mathrm{junct}, W} - \frac{v^2 \nu_{ii,\perp} J_{c, W}}{4}\, \partial_{\Delta J_\mathrm{junct}} \left (\tau_{c,W}  \, \partial_{\Delta J_\mathrm{junct}} K_{i,\mathrm{junct}, W} \right ) = 0,
\end{equation}
where the logarithmically diverging function 
\begin{equation}
\tau_{c, W} (r, \alpha, v, \Delta J_\mathrm{junct}) := - \hat{\tau}_{c, W, \log} (r, \alpha, v) \ln |\Delta J_\mathrm{junct}| + \hat{\tau}_{c, W, 0} (r, \alpha, v) 
\end{equation}
is the asymptotic approximation for $\tau_W$ near $J = J_{c, W}$. We have neglected the derivative of $K_{i, \mathrm{junct}, W}$ with respect to $\alpha$ because it is small compared to the first term in equation~\refe{eq:eqjunctlayer}. We can obtain estimate~\refe{eq:widthbljunct} by balancing the two terms in equation~\refe{eq:eqjunctlayer}, 
\begin{equation}
\frac{c \phi_0^\prime}{\Psi_t^\prime} \frac{\partial_\alpha J_{c, W}}{\Delta J_\mathrm{junct}} \sim  \frac{v^2 \nu_{ii,\perp} J_{c, W} \hat{\tau}_{c, W, \log}}{(\Delta J_\mathrm{junct})^2} \ln |\Delta J_\mathrm{junct}|.
\end{equation}
Indeed, this equation gives
\begin{equation} \label{eq:estimDJjunct}
|\Delta J_\mathrm{junct}| \sim \left | \frac{\Psi_t^\prime v^2 \nu_{ii,\perp} \hat{\tau}_{c, W, \log}}{c \phi_0^\prime \, \partial_\alpha \ln J_{c, W}} \right | \ln  \left | \frac{c \phi_0^\prime \,\partial_\alpha \ln J_{c, W}}{\Psi_t^\prime v^2 \nu_{ii,\perp} \hat{\tau}_{c, W, \log}} \right | \sim \frac{\nu_{i*}}{\rho_{i*}} \ln \left ( \frac{\rho_{i*}}{\nu_{i*}} \right ).
\end{equation}
For the order of magnitude estimate, we have used $\phi_0^\prime \sim T_i/ea$, $\Psi_t^\prime \sim a B_0$, $J_{c, W} \sim \sqrt{\epsilon} v_{ti} R$ and $\hat{\tau}_{c, W, \log} \sim R/\sqrt{\epsilon} v_{ti}$. With equation~\refe{eq:estimDJjunct}, we can simplify $\tau_{c, W}$ to
\begin{equation}
\tau_{c, W} \simeq \hat{\tau}_{c, W, \log}  \ln  \left | \frac{c \phi_0^\prime \,\partial_\alpha \ln J_{c, W}}{\Psi_t^\prime v^2 \nu_{ii,\perp} \hat{\tau}_{c, W, \log}} \right |. 
\end{equation}
From here on, we use this approximation that makes $\tau_{c, W}$ independent of $\Delta J_\mathrm{junct}$.

Equation~\refe{eq:eqjunctlayer} has to be solved in conjunction with continuity of $K_{i,\mathrm{junct},W}$ across the juncture and along with condition~\refe{eq:eqgi1transition}, which in the layer becomes
\begin{align} \label{eq:junctlayercond}
J_{c, I} \tau_{c, I} \lim_{\Delta J_\mathrm{junct} \rightarrow 0} \partial_{\Delta J_\mathrm{junct}} K_{i, \mathrm{junct}, I} + J_{c, II} \tau_{c, II} \lim_{\Delta J_\mathrm{junct} \rightarrow 0} \partial_{\Delta J_\mathrm{junct}} K_{i,\mathrm{junct}, II} \nonumber\\ = J_{c, III}  \tau_{c, III} \lim_{\Delta J_\mathrm{junct} \rightarrow 0} \partial_{\Delta J_\mathrm{junct}} K_{i,\mathrm{junct}, III}.
\end{align}

For large $\Delta J_\mathrm{junct}$, the function $K_{i,\mathrm{junct}, W} (r, \alpha, v, \Delta J_\mathrm{junct}, t)$ must tend to the solutions outside of the layer,
\begin{equation} \label{eq:bcjunct}
\lim_{|\Delta J_\mathrm{junct}| \rightarrow \infty} K_{i, \mathrm{junct}, W} (r, \alpha, v, \Delta J_\mathrm{junct}, t) = K_{i, W} (r, \alpha, v, J_{c, W}, t).
\end{equation}
With boundary conditions~\refe{eq:bcjunct}, the solutions to equation~\refe{eq:eqjunctlayer} are
\begin{align} \label{eq:soljunctlayer}
K_{i,\mathrm{junct}, W} & (r, \alpha, v, \Delta J_\mathrm{junct}, t) = K_{i, W} (r, \alpha, v, J_{c, W}, t) \nonumber\\& + A_W (r, \alpha, v, t) \exp \left ( - \frac{4 c \phi_0^\prime\, \partial_\alpha \ln J_{c, W}}{ \Psi_t^\prime v^2 \nu_{ii, \perp} \tau_{c, W}} \Delta J_\mathrm{junct} \right ),
\end{align}
where the function $A_W(r, \alpha, v, t)$ is a constant of integration.

We proceed to discuss how to use solution~\refe{eq:soljunctlayer} to solve the problem. We consider a juncture like the the one in figure~\ref{fig:exampleU} such that particles leave well $I$ and $II$ to enter well $III$, that is, $- (c\phi_0^\prime/\Psi_t^\prime)\, \partial_\alpha J_{c, W} > 0$ for $W = I, II, III$. This means that the exponential in solution~\refe{eq:soljunctlayer} diverges for well $III$, where $\Delta J_\mathrm{junct} > 0$, and thus the solution in well $III$ can tend to $K_{i, III} (r, \alpha, v, J_{c, III}, t)$ only if $A_{III} (r, \alpha, v, t) = 0$. As a result, the derivative of $K_{i, \mathrm{junct}, III}$ with respect to $\Delta J_\mathrm{junct}$ vanishes, and condition~\refe{eq:junctlayercond} simply gives
\begin{equation}
J_{c, I}  \tau_{c, I} \lim_{\Delta J_\mathrm{junct} \rightarrow 0} \partial_{\Delta J_\mathrm{junct}} K_{i, \mathrm{junct}, I} + J_{c, II} \tau_{c, II} \lim_{\Delta J_\mathrm{junct} \rightarrow 0} \partial_{\Delta J_\mathrm{junct}} K_{i,\mathrm{junct}, II} = 0.
\end{equation}
Imposing this condition and the fact that $K_{i, \mathrm{junct}, I}$ and $K_{i, \mathrm{junct}, II}$ have to be equal to each other at $\Delta J_\mathrm{junct} = 0$ to be continuous with $K_{i, \mathrm{junct}, III}$, we find the functions $A_I$ and $A_{II}$. The final solutions are 
\begin{align} 
K_{i,\mathrm{junct}, I}(r, \alpha, v, \Delta J_\mathrm{junct}, t) & = K_{i, I} (r, \alpha, v, J_{c, I}, t) \nonumber\\ + & \frac{\partial_\alpha J_{c, II} [K_{i, II}(r, \alpha, v, J_{c, II}, t) - K_{i,I}(r, \alpha, v, J_{c, I}, t)]}{\partial_\alpha J_{c, I} + \partial_\alpha J_{c, II}} \nonumber\\ & \times \exp \left ( - \frac{4 c \phi_0^\prime \, \partial_\alpha \ln J_{c, I}}{ \Psi_t^\prime v^2 \nu_{ii, \perp} \tau_{c, I}} \Delta J_\mathrm{junct} \right ), 
\end{align}
and
\begin{align} 
K_{i,\mathrm{junct}, II}(r, \alpha, v, \Delta J_\mathrm{junct}, t) & = K_{i, II} (r, \alpha, v, J_{c, II}, t) \nonumber\\ + & \frac{\partial_\alpha J_{c, I} [K_{i, I}(r, \alpha, v, J_{c, I}, t) - K_{i,II}(r, \alpha, v, J_{c, II}, t)]}{\partial_\alpha J_{c, I} + \partial_\alpha J_{c, II}} \nonumber\\ & \times \exp \left ( - \frac{4 c \phi_0^\prime\, \partial_\alpha \ln J_{c, II}}{ \Psi_t^\prime v^2 \nu_{ii, \perp} \tau_{c, II}} \Delta J_\mathrm{junct} \right ).
\end{align}
With these solutions, we can find $K_{i, III} (r, \alpha, v, J_{c, III}, t) = K_{i, \mathrm{junct}, III} (r, \alpha, v, 0, t)$ because $K_{i, \mathrm{junct}, III} (r, \alpha, v, 0, t)$ has to be equal to the values $K_{i, \mathrm{junct}, I}$ and $K_{i, \mathrm{junct}, II}$ at $\Delta J_\mathrm{junct} = 0$. The solution for $K_{i, III} (r, \alpha, v, J_{c, III}, t)$ is the same one that we found by applying conservation of particles, given in equation~\refe{eq:mixrule}.

It is worth noting that if we apply this method to a hypothetical boundary layer in a juncture where particles leave one well to enter two other wells, we find that the exponential in solution~\refe{eq:soljunctlayer} diverges in two of the wells (the two that the particles enter), and as a result, the functions $A_W$ have to vanish in those wells. One ends up finding that the solution is constant, recovering the result that $K_{i, W}$ is constant across this kind of juncture.

\section{Boundary layers around discontinuities in $\partial_J K_{i, W}$} \label{app:bldJK}

There are two types of boundary layers that form around discontinuities in $\partial_J K_{i, W}$: the layers on the boundaries of the regions where $K_{i, W}$ is independent of $J$, and the layer on the trapped-passing boundary. We describe both types of layers in this appendix.

\subsection{Boundary layers on the boundaries of the regions where $\partial_J K_{i, W} = 0$} \label{subapp:blKconst}

To discuss the boundary layers that appear on the boundaries of the regions where $K_{i, W}$ does not depend on $J$, we use the example of the layer that arises around $J = J_{c, III, M}$ in the example discussed in section~\ref{subsub:nuexample}. We use the variable $\Delta J_\mathrm{bl} := J - J_{c, III, M}$. The typical size of $\Delta J_\mathrm{bl}$ is 
\begin{equation} \label{eq:widthblKconst}
\frac{\Delta J_\mathrm{bl}}{\sqrt{\epsilon} v_{ti} R} \sim \sqrt{\frac{\nu_{i*}}{\rho_{i*}}} \ll 1,
\end{equation}
as we will show below. 

Particles behave differently for $\Delta J_\mathrm{bl}$ positive or negative. For $\Delta J_\mathrm{bl} > 0$, we can write the distribution function in the boundary layer region as
\begin{equation} \label{eq:giblKconst}
g_{i, 1, III} = K_{i, III} - r_{i, 1, III} \, \Upsilon_i f_{Mi} + g_{i, 1, III}^{\{\mathrm{bl}\}}(r, \alpha, v, \Delta J_\mathrm{bl}, t) + O \left ( \frac{\nu_{i*}}{\rho_{i*}} \epsilon f_{Mi} \right ),
\end{equation}
where
\begin{equation} \label{eq:heightblKconst}
\frac{g_{i, 1, III}^{\{\mathrm{bl}\}}}{\epsilon f_{Mi}} \sim \sqrt{\frac{\nu_{i*}}{\rho_{i*}}} \ll 1.
\end{equation}
Substituting equation~\refe{eq:giblKconst} into equation~\refe{eq:eqgi1}, we obtain the equation for the boundary layer for $\Delta J_\mathrm{bl} > 0$,
\begin{equation} \label{eq:eqblKconst}
\frac{c \phi_0^\prime}{\Psi_t^\prime} \partial_\alpha  g_{i, 1, III}^{\{\mathrm{bl}\}} - \frac{v^2 \nu_{ii,\perp} J_{c, III, M} \tau_{III}}{4}\, \partial^2_{\Delta J_\mathrm{bl} \Delta J_\mathrm{bl}}  g_{i, 1, III}^{\{\mathrm{bl}\}} = 0,
\end{equation}
where $\tau_{III}$ is evaluated at $J = J_{c, III, M}$. For $J > J_{c, III, M}$, particles move uninterruptedly along $\alpha$. Thus, we impose periodic boundary conditions in $\alpha$ for $g_{i, 1, III}^{\{\mathrm{bl}\}}$. We also impose 
\begin{equation} \label{eq:bcblinfty}
\lim_{\Delta J_\mathrm{bl} \rightarrow \infty} g_{i, 1, III}^{\{\mathrm{bl}\}} (r, \alpha, v, \Delta J_\mathrm{bl}, t) = 0
\end{equation}
and continuity and continuity of derivatives with $g_{i, 1, III}$ for $J < J_{c, III, M}$.

The distribution function in the region $J < J_{c, III, M}$ is simply $g_{i, 1, III} = K_{i, III} - r_{i, 1, III} \, \Upsilon_i f_{Mi} + g_{i, 1, III}^{\{ 1 \}}$, where $g_{i,1,III}^{\{1\}}$ is determined by equation~\refe{eq:eqginu1ergodicreal}. In other words, unlike for $J > J_{c, III, M}$, we do not distinguish between $g_{i, 1, III}^{\{ 1 \}}$ and the boundary layer piece of the distribution function because the equation for $g_{i, 1, III}^{\{ 1 \}}$ contains collisions. As a result, the continuity of $g_{i, 1, III}$ across $J = J_{c, III, M}$ gives
\begin{equation} \label{eq:bcbl}
g_{i, 1, III}^{\{\mathrm{bl}\}} (r, \alpha, v, 0, t) = g_{i, 1, III}^{\{1\}} (r, \alpha, v, J_{c, III, M}, t),
\end{equation} 
and the continuity of the derivative with respect to $J$ gives
\begin{equation} \label{eq:dJbcbl}
\partial_J K_{i, III} (r, v, J_{c, III, M}^+, t) + \partial_{\Delta J_\mathrm{bl}} g_{i, 1, III}^{\{\mathrm{bl}\}} (r, \alpha, v, 0, t) = \partial_J g_{i, 1, III}^{\{1\}} (r, \alpha, v, J_{c, III, M}, t),
\end{equation}
where $\partial_J K_{i, III} (r, v, J_{c, III, M}^+, t)$ is the derivative of $K_{i, III}$ with respect to $J$ evaluated at a value of $J$ slightly above $J = J_{c, III, M}$. Recall that $\partial_J K_{i, III}$ vanishes for $J < J_{c, III, M}$.

By balancing the different terms in equation~\refe{eq:eqblKconst}, we obtain estimate~\refe{eq:widthblKconst} for the boundary layer width. The estimate for the size of $g_{i, 1, III}^{\{\mathrm{bl}\}}$ in equation~\refe{eq:heightblKconst} is obtained from the fact that $g_{i, 1, III}^{\{\mathrm{bl}\}}$ is the integral of $\partial_{\Delta J_\mathrm{bl}} g_{i, 1, III}^{\{\mathrm{bl}\}}$ over an interval of $\Delta J_\mathrm{bl}$ of the size given in equation~\refe{eq:widthblKconst}. The size of $\partial_{\Delta J_\mathrm{bl}} g_{i, 1, III}^{\{\mathrm{bl}\}}$, $\partial_{\Delta J_\mathrm{bl}} g_{i, 1, III}^{\{\mathrm{bl}\}} \sim \sqrt{\epsilon} f_{Mi}/v_{ti} R$, is determined by boundary condition~\refe{eq:dJbcbl}.

Equations~\refe{eq:bcbl} and \refe{eq:dJbcbl} are also the boundary conditions for $g_{i, 1, III}^{\{ 1 \}}$ in the region where $\partial_J K_{i, III}$ vanishes. Note that this means that $g_{i, 1, III}^{\{1\}}$ is larger than expected by a factor of $\sqrt{\rho_{i*}/\nu_{i*}}$ near $J = J_{c, III, M}$, but it becomes of order $(\nu_{i*}/\rho_{i*}) \epsilon f_{Mi}$ for $(J_{c, III, M} - J)/\sqrt{\epsilon} v_{ti} R \gg \sqrt{\nu_{i*}/\rho_{i*}}$, except for some regions of small phase space volume that we proceed to discuss. 

In the example that we are considering (see figures~\ref{fig:configurationwell1} and \ref{fig:examplewells}), $g_{i, 1, III}^{\{1\}}$ for $\alpha$ slightly smaller than $\pi$ ($\alpha = \pi^-$) is different from $g_{i, 1, III}^{\{1\}}$ for $\alpha$ slightly larger than $\pi$ ($\alpha = \pi^+$). The particles at $\alpha = \pi^+$ are particles that used to have a second adiabatic invariant close to $J = J_{c, II, M}$ and hence are determined by the value of $g_{i, 1, II}^{\{1\}}$ around $J = J_{c, II, M}$. The particles at $\alpha = \pi^-$ transition into well $I$ where they have a second adiabatic invariant very similar to $J_{c, I, M}$. This means that $g_{i, 1, I}^{\{1\}}$ around $J = J_{c, I, M}$ is determined by $g_{i, 1, III}^{\{1\}}$ at $J = J_{c, III, M}$, and hence is large by a factor of $\sqrt{\rho_{i*}/\nu_{i*}}$ in a region of width $\sqrt{\epsilon\nu_{i*}/\rho_{i*}} v_{ti} R$ around $J_{c, I, M}$. The particles in this region around $J_{c, I, M}$, in turn, transition back into well $III$ at some value of $\alpha$ between $\pi$ and $2\pi$. This means that there is another region around another value of $J$ that connects to $J_{c, I, M}$ where $g_{i, 1, III}^{\{1\}}$ is larger than expected. These regions of large $g_{i, 1, W}^{\{1\}}$ go on until $g_{i, 1, W}^{\{1\}}$ becomes sufficiently small due to diffusion in $J$.

Boundary condition~\refe{eq:dJbcbl} gives the velocity space flux conservation condition~\refe{eq:collfluxnu1}. Indeed, integrating equation~\refe{eq:eqginu1ergodicreal} in $\alpha$ and in $J$ for the juncture sketched in figures~\ref{fig:configurationwell1} and \ref{fig:examplewells}, we find
\begin{align} \label{eq:integralKconst}
- & \frac{v^2 \nu_{ii, \perp}}{4} \Bigg [J_{c, III, M} \lim_{J \rightarrow J_{c, III, M}^+} \left \langle \tau_{III}\, \partial_J g_{i, 1, III}^{\{1\}}\right \rangle_\alpha - J_{c, I, m} \lim_{J \rightarrow J_{c, I, m}^-} \left \langle \tau_{I}\, \partial_J g_{i, 1, I}^{\{1\}}\right \rangle_\alpha \nonumber\\ & - J_{c, II, m} \lim_{J \rightarrow J_{c, II, m}^-} \left \langle \tau_{II}\, \partial_J g_{i, 1, II}^{\{1\}}\right \rangle_\alpha \Bigg ] = \frac{m_i v^2 \nu_{ii,\perp}}{4Z_i e \phi_0^\prime} (J_{c, III, M} - J_{c, I, m} - J_{c, II, m}) \Upsilon_i f_{Mi}.
\end{align}
Using the boundary condition~\refe{eq:dJbcbl}, the definition of $r_{i, 1, W}$ in equation~\refe{eq:r1def} and $\partial_J \lambda_W \simeq - 2/v^2 B_0 \tau_W$, equation~\refe{eq:integralKconst} becomes
\begin{align}
J_{c, III, M} & \lim_{J \rightarrow J_{c, III, M}^+} \left \langle \tau_{III}\, \left ( \partial_J K_{i, III} - \partial_J r_{i, 1, III}\, \Upsilon_i f_{Mi} \right ) \right \rangle_\alpha \nonumber\\ & - J_{c, I, m} \lim_{J \rightarrow J_{c, I, m}^-} \left \langle \tau_{I}\, \left ( \partial_J K_{i, I} - \partial_J r_{i, 1, III}\, \Upsilon_i f_{Mi} \right ) \right \rangle_\alpha \nonumber\\ & - J_{c, II, m} \lim_{J \rightarrow J_{c, II, m}^-} \left \langle \tau_{II}\, \left ( \partial_J K_{i, II} - \partial_J r_{i, 1, II}\, \Upsilon_i f_{Mi} \right ) \right \rangle_\alpha \nonumber\\ = & -  J_{c, III, M} \lim_{\Delta J_\mathrm{bl} \rightarrow 0} \left \langle \tau_{III}\, \partial_{\Delta J_\mathrm{bl}} g_{i, 1, III}^{\{\mathrm{bl}\}}\right \rangle_\alpha + J_{c, I, m} \lim_{\Delta J_\mathrm{bl} \rightarrow 0} \left \langle \tau_{I}\, \partial_{\Delta J_\mathrm{bl}} g_{i, 1, I}^{\{\mathrm{bl}\}}\right \rangle_\alpha \nonumber\\ & + J_{c, II, m} \lim_{\Delta J_\mathrm{bl} \rightarrow 0} \left \langle \tau_{II}\, \partial_{\Delta J_\mathrm{bl}} g_{i, 1, II}^{\{\mathrm{bl}\}}\right \rangle_\alpha.
\end{align}
We obtain the velocity space flux conservation condition~\refe{eq:collfluxnu1} from this expression by employing the fact that equations~\refe{eq:eqblKconst} and \refe{eq:bcblinfty} and the equivalent equations for the boundary layers at $J = J_{c, I, m}$ and $J_{c, II, m}$ give
\begin{equation}
\lim_{\Delta J_\mathrm{bl} \rightarrow 0} \left \langle \tau_W\, \partial_{\Delta J_\mathrm{bl}} g_{i, 1, W}^{\{\mathrm{bl}\}}\right \rangle_\alpha = 0.
\end{equation}

Incidentally, $\tau_{III}$ in equation~\refe{eq:eqblKconst} diverges logarithmically at $\alpha = \pi$, where $J_{c, III} = J_{c, III, M}$. This divergence is integrable and does not affect the previous discussion.

\subsection{Boundary layer at the trapped-passing boundary} \label{subapp:bltp}

The collisional layer that appears at the trapped-passing boundary is a result of the lowest order solution~\refe{eq:ginu0tp} not satisfying property~\refe{eq:dJgi1Jinfty} at the trapped-passing boundary. This layer is best described in the coordinate $\Delta \lambda := \lambda - 1/B_M \simeq \lambda - 1/B_0 + B_{1,M}/B_0^2$, which is of order
\begin{equation} \label{eq:widthbltp}
\frac{|\Delta \lambda| B_0}{\epsilon} \sim \sqrt{ \frac{\nu_{i*}}{\rho_{i*}} \frac{1}{\ln (\rho_{i*}/\nu_{i*})}} \ll 1
\end{equation}
(we will argue that this estimate is correct below). We consider $g_{i, 1, W}^{\{1\}}$, determined by equation~\refe{eq:eqginu1ergodicreal}, as a function of $\Delta \lambda$. We find below that in the region of velocity space that satisfies equation~\refe{eq:widthbltp}, $g_{i, 1, W}^{\{1\}}$ is not of order $(\nu_{i*}/\rho_{i*}) \epsilon f_{Mi}$, but much larger,
\begin{equation} \label{eq:heightbltp}
\frac{g_{i, 1, W}^{\{1\}}}{\epsilon f_{Mi}} \sim \sqrt{ \frac{\nu_{i*}}{\rho_{i*}} \frac{1}{\ln (\rho_{i*}/\nu_{i*})}} \ll 1.
\end{equation}

Using estimates~\refe{eq:widthbltp} and \refe{eq:heightbltp} in equation~\refe{eq:eqginu1ergodicreal}, we find
\begin{equation} \label{eq:eqbltpv1}
\frac{c\phi_0^\prime}{\Psi_t^\prime} \left ( \partial_\alpha g_{i, 1, W}^{\{1\}} + \partial_\alpha \lambda_W\, \partial_{\Delta \lambda} g_{i, 1, W}^{\{1\}} \right ) - \frac{v^2 \nu_{ii, \perp}}{4} \tau_W J_W\, ( \partial_J \lambda_W )^2 \, \partial^2_{\Delta \lambda \Delta \lambda} g_{i, 1, W}^{\{1\}} = 0.
\end{equation}
Here we can use $\partial_J \lambda_W \simeq - 2/v^2 B_0 \tau_W$ and $\partial_\alpha \lambda_W \simeq - B_0^{-2} \left \langle \partial_\alpha B_1 \right \rangle_{\tau, W}$ to simplify equation~\refe{eq:eqbltpv1} to
\begin{equation} \label{eq:eqbltpv2}
\frac{c\phi_0^\prime}{\Psi_t^\prime} \left ( \partial_\alpha g_{i, 1, W}^{\{1\}} - \frac{\left \langle \partial_\alpha B_1 \right \rangle_{\tau, W}}{B_0^2}\, \partial_{\Delta \lambda} g_{i, 1, W}^{\{1\}} \right ) - \frac{\nu_{ii, \perp} J_W}{v^2 B_0^2 \tau_W} \partial_{\Delta \lambda \Delta \lambda}^2 g_{i, 1, W}^{\{1\}} = 0.
\end{equation}
We need to evaluate the different coefficients in this equation for small $\Delta \lambda$. We will use the property that, for $\Delta \lambda \rightarrow 0$, the particle trajectory covers the whole flux surface in such a way that the average of any function over the length of the line is related to the flux surface average of the function by
\begin{equation}
\lim_{\Delta \lambda \rightarrow 0} \frac{\int_{l_{bL,W}}^{l_{bR,W}} (\ldots)\, \rmd l}{l_{bR,W} - l_{bL,W}} = \frac{\langle B (\ldots) \rangle_\mathrm{fs}}{\langle B \rangle_\mathrm{fs}}.
\end{equation}
Then, we can write
\begin{equation}
\frac{J_W}{\tau_W} \simeq \frac{\langle B |v_{\|, \mathrm{tp}}| \rangle_\mathrm{fs}}{\langle B |v_{\|, \mathrm{tp}}|^{-1} \rangle_\mathrm{fs}} \simeq \frac{\langle |v_{\|, \mathrm{tp}}| \rangle_\mathrm{fs}}{\langle |v_{\|, \mathrm{tp}}|^{-1} \rangle_\mathrm{fs}},
\end{equation}
where $v_{\|, \mathrm{tp}}$ is the parallel velocity of the particles at the trapped-passing boundary. Note that the average $\langle |v_{\|, \mathrm{tp}}|^{-1} \rangle_\mathrm{fs}$ does not diverge logarithmically despite $|v_{\|, \mathrm{tp}}|$ depending linearly on $l$ near the maximum of $B_1$ because the integral over $\alpha$ eliminates de divergence. With this result, equation~\refe{eq:eqbltpv2} simplifies to
\begin{equation} \label{eq:eqbltpfinal}
\frac{c\phi_0^\prime}{\Psi_t^\prime} \left ( \partial_\alpha g_{i, 1, W}^{\{1\}} - \frac{\left \langle \partial_\alpha B_1 \right \rangle_{\tau, W}}{B_0^2}\ \partial_{\Delta \lambda} g_{i, 1, W}^{\{1\}} \right ) - \frac{\nu_{ii, \perp} \langle |v_{\|, \mathrm{tp}}| \rangle_\mathrm{fs}}{v^2 B_0^2 \langle |v_{\|, \mathrm{tp}}|^{-1} \rangle_\mathrm{fs}} \partial_{\Delta \lambda \Delta \lambda}^2 g_{i, 1, W}^{\{1\}} = 0.
\end{equation}

\begin{figure}
\begin{center}
\includegraphics[width=8cm]{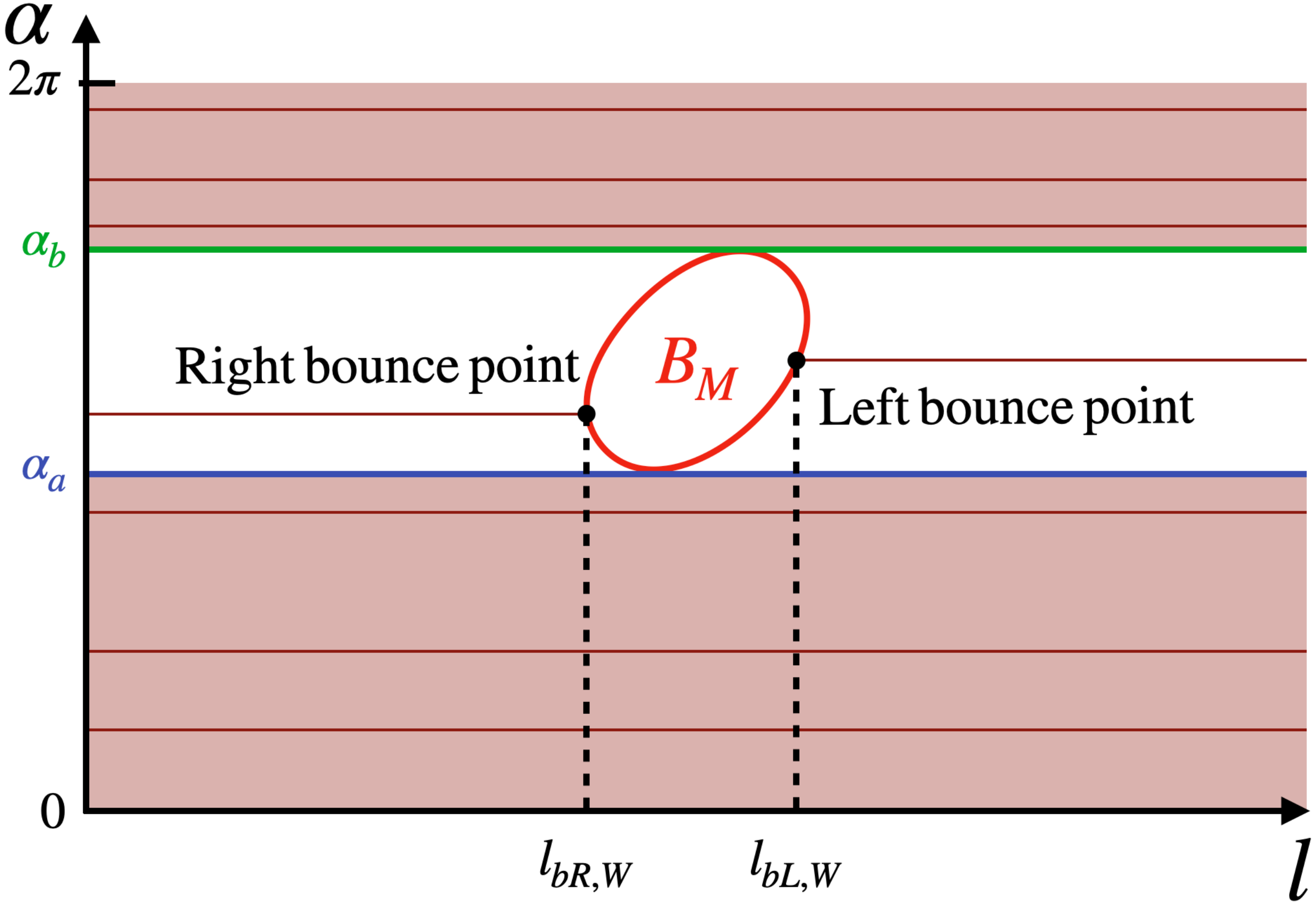}
\end{center}
\caption{\label{fig:expansiontp} Path of integration of the integral in equation~\refe{eq:dalphaB1avev1} in the $(l, \alpha)$ plane (thin dark pink line), and area of integration for the integral in equation~\refe{eq:dalphaB1avev2} (shaded region). For large $J$, $\lambda$ must be close to $1/B_M$, and hence it must have bounce points at a value of $B$ close to $B_M$. In the figure, we sketch the contour $B = 1/\lambda \approx B_M$ as a red line (we have assumed that there is only one maximum of $B$). The lines $\alpha_a$ and $\alpha_b$ are marked in blue and green, respectively.}
\end{figure}

To obtain an estimate for $\left \langle \partial_\alpha B_1 \right \rangle_{\tau, W}$, we need to consider the integral
\begin{equation} \label{eq:dalphaB1avev1}
\left \langle \partial_\alpha B_1 \right \rangle_{\tau, W} \simeq \frac{2}{v \tau_W} \int_{l_{bL,0,W}}^{l_{bR,0,W}} \frac{\partial_\alpha B_1(l)}{\sqrt{(B_{1,M} -  B_1(l))/B_0- \Delta \lambda B_0 }}\, \rmd l.
\end{equation}
We first note that, for $\Delta \lambda \rightarrow 0$, this integral vanishes,
\begin{align} \label{eq:dalphaB1avelimit0}
\lim_{\Delta \lambda \rightarrow 0} & \left \langle \partial_\alpha B_1 \right \rangle_{\tau, W} \simeq \frac{2}{v \int_0^{2\pi} \rmd \alpha \int _0^{L_0}\rmd l\, |v_{\|, \mathrm{tp}}|^{-1} } \int_0^{2\pi} \rmd \alpha \int_0^{L_0} \rmd l\, \frac{\partial_\alpha B_1(l)}{\sqrt{(B_{1,M} -  B_1(l))/B_0}} \nonumber\\ & = - \frac{4B_0}{v \int_0^{2\pi} \rmd \alpha \int_0^{L_0} \rmd l |v_{\|, \mathrm{tp}}|^{-1} } \int_0^{2\pi} \rmd \alpha \, \partial_\alpha \left ( \int_0^{L_0} \sqrt{(B_{1,M} -  B_1(l))/B_0} \, \rmd l \right ) = 0.
\end{align}
Here we have used the fact that the length $L$, defined below equation~\refe{eq:fsavedef}, is independent of $\alpha$ to lowest order in the aspect ratio expansion, $L (r, \alpha) \simeq L_0 (r)$. We proceed to determine how integral \refe{eq:dalphaB1avev1} goes to zero for small $\Delta \lambda$. A sketch of the integral path is shown in figure~\ref{fig:expansiontp}: the path of integration is composed of an almost surface-covering piece and a piece of short length in the region $\alpha_a < \alpha < \alpha_b$, where $\alpha_a$ and $\alpha_b$ are the values of $\alpha$ for which the maximum of $B$ on the line $\alpha$ is equal to $1/\lambda$. For most values of $\alpha$ and $\Delta \lambda$, the bulk of the integral is the almost surface-covering piece, and its size can be estimated by replacing the line integral by an integral over the surface that the line covers, shaded in figure~\ref{fig:expansiontp}, 
\begin{align} \label{eq:dalphaB1avev2}
\left \langle \partial_\alpha B_1 \right \rangle_{\tau, W} \sim \frac{1}{v\int_0^{2\pi} \rmd \alpha \int_0^{L_0} \rmd l \, |v_{\|, \mathrm{tp}}|^{-1}} \Bigg [ \int_0^{\alpha_a} \rmd \alpha \int_0^{L_0} \rmd l\, \frac{\partial_\alpha B_1(l)}{\sqrt{ (B_{1,M} -  B_1(l))/B_0- \Delta \lambda B_0 }} \nonumber\\ + \int_{\alpha_b}^{2\pi} \rmd \alpha \int_0^{L_0} \rmd l\, \frac{\partial_\alpha B_1(l)}{\sqrt{ (B_{1,M} -  B_1(l))/B_0- \Delta \lambda B_0 }} \Bigg ].
\end{align}
The value of this surface integral is dominated by the region near the maximum of $B_1$ because the rest of surface integral vanishes as shown by equation~\refe{eq:dalphaB1avelimit0}. To avoid cluttering notation, we choose $\alpha$ and $l$ such that $\alpha = 0$ and $l = 0$ at the maximum of $B_1$, that is, $B_{1, M} = B_1(r, 0, 0)$. Then, around the maximum, 
\begin{equation}
B_1(r, \alpha, l) \simeq B_{1, M} + \frac{1}{2} l^2\, \partial_{ll}^2 B_1 (r, 0, 0) + \alpha l\, \partial^2_{\alpha l} B_1 (r, 0, 0) + \frac{1}{2} \alpha^2\, \partial_{\alpha \alpha}^2 B_1 (r, 0, 0), 
\end{equation}
with $\partial_{ll}^2 B_1 (r, 0, 0) < 0$, $\partial_{\alpha \alpha}^2 B_1 (r, 0, 0) < 0$ and 
\begin{equation}
\partial_{ll}^2 B_1 (r, 0, 0) \, \partial_{\alpha \alpha}^2 B_1 (r, 0, 0) - [ \partial_{\alpha l}^2 B_1 (r, 0, 0) ]^2 > 0. 
\end{equation}
This Taylor expansion around the maximum $l = 0$ and $\alpha = 0$ implies that $\alpha_a \sim \alpha_b \sim \sqrt{\epsilon^{-1}\Delta \lambda B_0}$, and that the integrals over $l$ in equation~\refe{eq:dalphaB1avev2} are of order
\begin{equation}
\int_0^L \frac{\partial_\alpha B_1(l)}{\sqrt{ (B_{1,M} -  B_1(l))/B_0- \Delta \lambda B_0 }}\, \rmd l \sim \sqrt{\epsilon} B_0 R \alpha \left | \ln \left ( \alpha^2 + \epsilon^{-1} \Delta \lambda B_0 \right ) \right |
\end{equation}
for small $\alpha$. Thus, we find 
\begin{align} \label{eq:dalphaB1avefinal}
\left \langle \partial_\alpha B_1 \right \rangle_{\tau, W} \sim B_0^2 \Delta \lambda \ln \left ( \frac{\epsilon}{B_0\Delta \lambda} \right ).
\end{align}
This is only an estimate. The real value and sign of $\langle \partial_\alpha B_1 \rangle_{\tau, W}$ depends strongly on the particular orbit under consideration, determined by its $\alpha$. Before we proceed, we point out that it is important not to confuse the $\alpha$ shown in figure~\ref{fig:expansiontp}, where a single orbit samples several values of $\alpha$, with the value of $\alpha$ that we assign to the orbit (for example, we can assign to an orbit the value of $\alpha$ where it has its left bounce point). Going back to the dependence of $\langle \partial_\alpha B_1 \rangle_{\tau, W}$ on the particular orbit, consider what happens as the bounce points move from $\alpha_a$ to $\alpha_b$. When a bounce point approaches $\alpha_a$, the part of the path near the bounce, which we ignored in our estimate~\refe{eq:dalphaB1avefinal}, becomes dominant due to a logarithmic divergence at the bounce point. As a result, $\langle \partial_\alpha B_1 \rangle_{\tau, W}$ is the value of $\partial_\alpha B_1$ at the bounce point, which is of order $\alpha_a\, \partial_{\alpha \alpha}^2 B_1(r, 0, 0) \sim \sqrt{\epsilon \Delta \lambda} B_0^{3/2}$ -- importantly, note that the logarithmic divergence can only dominate when the bounce point is very close to $\alpha_b$, that is, close by $\exp( - R/(l_{bR,0,W} - l_{bL,0,W})) \ll 1$, and hence our estimate~\refe{eq:dalphaB1avefinal} is valid for most orbits. The situation for a bounce point in $\alpha_b$ is similar, but in this case the sign of $\partial_\alpha B_1$ at the bounce point is the opposite to the one that we consider above, showing that $\langle \partial_\alpha B_1 \rangle_{\tau, W}$ changes sign.

Using estimate~\refe{eq:dalphaB1avefinal} in equation~\refe{eq:eqbltpfinal}, we find the estimate~\refe{eq:widthbltp} for the width of the layer. The estimate for the size of $g_{i, 1, W}^{\{1\}}$ in equation~\refe{eq:heightbltp} is obtained from the estimate for the width of the layer and the fact that the derivative of $g_{i, 1, W}^{\{1\}}$ with respect to $\Delta \lambda$ must be such that property~\refe{eq:dJgi1Jinfty} is satisfied. Note, however, that property~\refe{eq:dJgi1Jinfty} is not imposed on the equation as a boundary condition. The boundary condition at $\Delta \lambda = 0$ is
\begin{equation}
g_{i, 1, W}^{\{1\}} (r, \alpha, v, 0, t) = 0.
\end{equation}
The phase space flows along $\alpha$ and across $J$ at large values of $\Delta \lambda$ are the ones that ensure that property~\refe{eq:dJgi1Jinfty} is satisfied. These phase space flows, determined by equation~\refe{eq:eqginu1ergodicreal} and boundary conditions~\refe{eq:bcbl} and \refe{eq:dJbcbl}, are the boundary condition that we need to impose for large $\Delta \lambda$ to solve equation~\refe{eq:eqbltpfinal}.


\bibliographystyle{jpp}

\bibliography{stellarator}

\end{document}